\documentclass[12pt]{iopart}
\usepackage{iopams}
\usepackage{epsfig}
\usepackage{epstopdf}
\usepackage{epsf}
\usepackage{subfigure}
\usepackage{graphicx}
\usepackage{bm}
\usepackage{float}
\usepackage{sidecap}
\usepackage{wrapfig}
\usepackage{hyperref}
\usepackage{xcolor}
\usepackage{ulem}

\def \bfx {{\bf x}}
\def \bfu {{\bf u}}
\def \bff {{\bm f}}
\def \bfb {{\bf b}}
\def \bfa {{\bf a}}
\def \bfk {{\bf k}}
\def \bfl {{\bf l}}
\def \bfz {{\bf z}}
\def \bfj {{\bf j}}

\def \bomega {{\bm \omega}}

\def \f {{\bm f}}

\def \lap {\nabla^2}

\begin{document}
\title[Systematics of the ${\rm Pr_M}$ dependence of
homogeneous, isotropic MHD turbulence]{Systematics of the 
magnetic-Prandtl-number dependence of
homogeneous, isotropic magnetohydrodynamic turbulence}
\author{Ganapati Sahoo~$^1$, Prasad Perlekar~$^2$, and Rahul Pandit~$^{1,3}$}
\address{$^1$ Centre for Condensed Matter
Theory, Department of Physics,
Indian Institute of Science,
Bangalore 560012, India. \\
$^2$ Department of Mathematics and Computer Science,
Technische Universiteit Eindhoven, Postbus 513, 5600 MB, Eindhoven,
Netherlands. \\
$^3$ Also at: Jawaharlal Nehru Centre for
Advanced Scientific Research, Jakkur,
Bangalore, India}
\ead{\mailto{ganapati@physics.iisc.ernet.in},
\mailto{p.perlekar@tue.nl},
\mailto{rahul@physics.iisc.ernet.in}}

\begin{abstract} 

We present the results of our detailed pseudospectral direct numerical
simulation (DNS) studies, with up to $1024^3$ collocation points, of
incompressible, magnetohydrodynamic (MHD) turbulence in three dimensions,
without a mean magnetic field. Our study concentrates on the dependence of
various statistical properties of both decaying and statistically steady MHD
turbulence on the magnetic Prandtl number ${\rm Pr_M}$ over a large range,
namely, $0.01 \leq {\rm Pr_M} \leq 10$. We obtain data for a wide variety of
statistical measures such as probability distribution functions (PDFs) of
moduli of the vorticity and current density, the energy dissipation rates,
and velocity and magnetic-field increments, energy and other spectra,
velocity and magnetic-field structure functions, which we use to characterise
intermittency, isosurfaces of quantities such as the
moduli of the vorticity and current, and joint PDFs such as those of fluid
and magnetic dissipation rates. Our systematic study uncovers interesting
results that have not been noted hitherto. In particular, we find a crossover
from larger intermittency in the magnetic field than in the
velocity field, at large ${\rm Pr_M}$, to smaller intermittency
in the magnetic field than in the velocity field, at low ${\rm Pr_M}$.
Furthermore, a comparison of our results for decaying MHD turbulence and its
forced, statistically steady analogue suggests that we have strong
universality in the sense that, for a fixed value of ${\rm Pr_M}$,
multiscaling exponent ratios agree, at least within our errorbars, for both
decaying and statistically steady homogeneous, isotropic MHD turbulence.  

\end{abstract}

\maketitle

\section{Introduction}

The hydrodynamics of conducting fluids is of great importance in many
terrestrial and astrophysical phenomena. Examples include the generation of
magnetic fields via dynamo action in the interstellar medium, stars, and
planets~\cite{book-arnab,book-vkrishan,book-rudiger,book-goedbloed,book-biskamp,schekochihin02,
verma04,elmegreen04,njpspecial,dormy08,sahoo10}, and liquid-metal
systems~\cite{lehnert55,roberts00,peffley00,riga00,karl01,fauve01,bourgoin02}
that are studied in laboratories. The flows in such settings, which can be
described at the simplest level by the equations of magnetohydrodynamics
(MHD), are often turbulent~\cite{book-biskamp}. The larger the kinetic and
magnetic Reynolds numbers, ${\rm Re}=UL/\nu$ and ${\rm Re_M}=UL/\eta$,
respectively, the more turbulent is the motion of the conducting fluid; here
$L$ and $U$ are typical length and velocity scales in the flow, $\nu$ is the
kinematic viscosity, and $\eta$ is the magnetic diffusivity. The statistical
characterization of turbulent MHD flows, which continues to pose challenges
for experiments~\cite{pinton07}, direct numerical simulations~\cite{ponty05},
and theory~\cite{boldyrev06}, is even harder than its analogue in fluid
turbulence because (a) we must control both ${\rm Re}$ and ${\rm Re_M}$ and
(b) we must obtain the statistical properties of both velocity and magnetic
fields.

The kinematic viscosity $\nu$ and the magnetic diffusivity $\eta$ can differ
by several orders of magnitude, so the magnetic Prandtl number ${\rm
Pr_M}\equiv {\rm Re_M}/{\rm Re} = \nu/\eta$ can vary over a large range. For
example, ${\rm Pr_M}\simeq 10^{-5}$ in the liquid-sodium
system~\cite{riga00,karl01}, ${\rm Pr_M}\simeq 10^{-2}$ at the base of the
Sun's convection zone~\cite{schekochihin04prl}, and ${\rm Pr_M}\simeq
10^{14}$ in the interstellar medium~\cite{elmegreen04,ponty05}. Furthermore,
two dissipative scales play an important role in MHD; they are the Kolmogorov
scale $\ell_d$ ($\sim \nu^{3/4}$ at the level of Kolmogorov 1941 (K41)
phenomenology~\cite{k41,book-frisch}) and the magnetic-resistive scale
$\ell_d^M$ ($\sim \eta^{3/4}$ in K41). A thorough study of the statistical
properties of MHD turbulence must resolve both these dissipative scales.
Given current computational resources, this is a daunting task at large ${\rm
Re}$ especially if ${\rm Pr_M}$ is significantly different from unity. Thus,
most direct numerical simulations (DNS) of MHD
turbulence~\cite{biskamp00,kalelkar04,mininni06,mason08,baerenzung08,mininni09}
have been restricted to ${\rm Pr_M}\simeq 1$.  Some DNS studies have now
started moving away from the ${\rm Pr_M}\simeq 1$ regime especially in the
context of the dynamo problem~\cite{ponty07,brandenburg09}.

Here we initiate a detailed DNS study of the statistical properties of
incompressible, homogeneous, and isotropic MHD turbulence for a large range
of the magnetic Prandtl number, namely, $0.01 \leq {\rm Pr_M} \leq 10$. There is
no mean magnetic field in our DNS~\cite{goldreich95}; and we restrict
ourselves to Eulerian measurements~\cite{homann07}. Before we give the
details of our DNS study, we highlight a few of our qualitative, principal
results. Elements of some of our results, for the case ${\rm Pr_M} = 1$ and
for quantities such as energy spectra, exist in the MHD-turbulence literature
as can be seen from the representative
Refs.~\cite{book-biskamp,schekochihin02,biskamp00,brandenburg95,cao99,chou01,mininni07}.
However, to the best of our knowledge, no study has attempted as detailed and
systematic an investigation of the statistical properties of MHD turbulence
as we present here, especially with a view to elucidating their dependence on
${\rm Pr_M}$. Our study uncovers interesting trends that have not been noted
hitherto.  These emerge from our detailed characterisation of 
intermittency, via a variety of measures which include various probability
distribution functions (PDFs) such as those of the modulus of the vorticity
and the energy dissipation rates,  velocity and magnetic-field structure
functions that can be used to characterise intermittency,
isosurfaces of quantities such as the moduli of the vorticity and current,
and joint PDFs such as those of fluid and magnetic dissipation rates. Earlier
DNS studies~\cite{mininni09} have suggested that intermittency,
as characterised say by the multiscaling exponents for velocity- and
magentic-field structure functions, is more intense for the magnetic field
than for the velocity field when ${\rm Pr_M}=1$. Our study confirms this and
suggests, in addition, that this result is reversed as we lower ${\rm Pr_M}$.
This crossover from larger intermittency in the magnetic field
than in the velocity field, at large ${\rm Pr_M}$, to smaller 
intermittency in the magnetic field than in the velocity field, at low ${\rm
Pr_M}$, shows up not only in the values of multiscaling exponent ratios,
which we obtain from a detailed local-slope analysis of
extended-self-similarity (ESS) plots~\cite{benzi93,chakraborty10} of one
structure function against another, but also in the behaviours of tails of
PDFs of dissipation rates, the moduli of vorticity and current density, and
velocity and magnetic-field increments. Furthermore, a comparison of our
results for decaying MHD turbulence and its forced, statistically steady
analogue suggest that, at least given our conservative errors, the
homogeneous, isotropic MHD turbulence that we study here displays strong
universality~\cite{lvov03,ray08} in the sense that multiscaling exponent
ratios agree for both decaying and statistically steady cases.

The remaining part of this paper is organised as follows.  In
Sec.~\ref{sec:equations} we describe the MHD equations, the details of the
numerical schemes we use (Subsection~\ref{sec:dns}), and the statistical
measures we use to characterise MHD turbulence
(Subsection~\ref{sec:statmeasures}). In Sec.~\ref{sec:results} we present our
results; these are described in the seven Subsections~\ref{sec:NS},
\ref{sec:tseries}, \ref{sec:spectra}, \ref{sec:pdfs}, \ref{sec:stfn},
\ref{sec:isosurf}, and \ref{sec:jpdf} that are devoted, respectively, to (a)
a summary of well-known results from fluid turbulence that are relevant to
our study, (b) the temporal evolution of quantities such as the energy and
energy-dissipation rates, (c) energy, dissipation-rate, Els\"asser-variable,
and effective-pressure spectra, (d) various probability distribution
functions (PDFs) that can be used, {\it inter alia}, to characterise the
alignments of vectors such as the vorticity with, say, the eigenvectors of
the rate-of-strain tensor, (e) velocity and magnetic-field structure
functions that can be used to characterise intermittency,
(f) isosurfaces of quantities such as the moduli of the vorticity and
current, and (g) and joint PDFs such as those of fluid and magnetic
dissipation rates.  Section~\ref{sec:conclusions} contains a discussion of
our results.

\section{MHD Equations \label{sec:equations}}

The hydrodynamics of a conducting fluid is governed by the MHD
equations~\cite{book-arnab,book-vkrishan,book-rudiger,book-goedbloed,book-biskamp,verma04},
in which the Navier-Stokes equation for a fluid is coupled to the induction
equation for the magnetic field:
\begin{eqnarray}
\frac{\partial\bfu}{\partial t}+(\bfu\cdot\nabla)\bfu=\nu\lap\bfu
-\nabla\bar{p}+(\bfb\cdot\nabla)\bfb+\f_u, \\
\label{magone}
\frac{\partial\bfb}{\partial t}+(\bfu\cdot\nabla)\bfb
=(\bfb\cdot\nabla)\bfu+\eta\lap\bfb+\f_b ;
\label{eq:magtwo}
\end{eqnarray}
here $\bfu$, $\bfb$, $\bomega=\nabla\times\bfu$ and $\bfj=\nabla\times\bfb$
are, respectively, the velocity field, the magnetic field, the vorticity, and
the current density at the point $\bf{x}$ and time $t$; $\nu$ and $\eta$ are
the kinematic viscosity and the magnetic diffusivity, respectively, and the
effective pressure is $\bar{p}=p+(b^2/8\pi)$, where $p$ is the pressure;
$\f_u$ and $\f_b$ are the external forces; while studying decaying MHD
turbulence we set $\f_u=\f_b=0$. The MHD equations can also be written in
terms of the Els\"asser variables $\bfz^\pm =
\bfu\pm\bfb$~\cite{verma04,biskamp00}.  We restrict ourselves to
low-Mach-number flows so we use the incompressibility condition
$\nabla\cdot\bfu({\bf x},t)=0$; and we must, of course, impose
$\nabla\cdot\bfb({\bf x},t)=0$. By using the incompressibility condition, we
can eliminate the effective pressure and obtain the velocity and magnetic
fields via a pseudospectral method that we describe in the next Subsection.
The effective pressure then follows from the solution of the Poisson equation
\begin{equation}
\nabla^2{\bar p} =
\nabla\cdot[(\bfb\cdot\nabla)\bfb-(\bfu\cdot\nabla)\bfu].
\end{equation}

\subsection{Direct Numerical Simulation \label{sec:dns}}
\begin{table}
\caption{List of parameters for our 16 DNS runs R1-R5, R3B-R5B, R1C-R4C, and
R1D-R4D: $N^3$ is the number of collocation points in our simulation, $\nu$
is the kinematic viscosity, ${\rm Pr_M}$ is magnetic Prandtl number, $\delta
t$ is time step; $u_{\rm rms}$, $\lambda$, and ${\rm Re}_{\lambda}$ are the
root-mean-sqare velocity, the Taylor microscale, and the Taylor-microscale
Reynolds number, respectively.  These are obtained at $t_c$ for our
decaying-MHD-turbulence runs R1-R5, R3B-R5B, and R1C-R4C; and for
statistically steady MHD turbulence (runs R1D-R4D) these are averaged over
the statistically steady state; here $t_c$ (iteration steps multiplied by
$\delta t$) is the time at which the cascades for both the fluid and magnetic
fields are completed (see text); 
$\eta_d^u$ and $\eta_d^b$ are, respectively, the Kolmogorov dissipation
length scales for the fluid and magnetic fields. $k_{\rm max}$ is
the magnitude of
the largest-magnitude wave vectors resolved in our DNS studies which use the
$2/3$ dealiasing rule; $k_{\rm max} \simeq 170.67$ and $341.33$ for $N=512$ and
$1024$, respectively.}
\label{table:param}
\begin{center}
\begin{tabular}{|l|c|c|c|c|c|c|c|c|c|c|c|c|c|c|}
\hline \hline
Runs & $N$ & $\nu$ & ${\rm Pr_M}$ & $\delta t$ & $u_{\rm rms}$ & $\lambda$ & ${\rm Re}_{\lambda}$ & $t_c$ & $k_{\rm max}\eta_d^u$
& $k_{\rm max}\eta_d^b$\\ \hline \hline
R1 & $512$ & $10^{-4}$ & $0.1$ & $10^{-3}$ & $0.34$ & $0.18$ & $610$ & $9.7$ & $0.629$ & $2.280$\\ \hline
R2 & $512$ & $5\times10^{-4}$ & $0.5$ & $10^{-3}$ & $0.34$ & $0.27$ & $187$ & $9.1$ & $1.752$ & $2.389$\\ \hline
R3 & $512$ & $10^{-3}$ & $1$ & $10^{-3}$ & $0.34$ & $0.35$ & $121$ & $8.1$ & $2.772$ & $2.444$\\ \hline
R4 & $512$ & $5\times10^{-3}$ & $5$ & $10^{-3}$ & $0.33$ & $0.60$ & $39$ & $7.0$ & $8.224$ & $2.692$\\ \hline
R5 & $512$ & $10^{-2}$ & $10$ & $10^{-3}$ & $0.31$ & $0.73$ & $23$ & $6.5$ & $13.267$ & $2.836$\\ \hline
R3B & $512$ & $10^{-3}$ & $1$ & $10^{-4}$ & $1.07$ & $0.20$ & $210$ & $3.1$ & $1.175$ & $1.052$\\ \hline
R4B & $512$ & $5\times10^{-3}$ & $5$ & $10^{-4}$ & $2.32$ & $0.24$ & $110$ & $1.4$ & $1.961$ & $0.644$\\ \hline
R5B & $512$ & $10^{-2}$ & $10$ & $10^{-4}$ & $3.21$ & $0.26$ & $85$ & $1.0$ & $2.490$ & $0.520$\\ \hline
R1C & $1024$ & $10^{-4}$ & $0.01$ & $10^{-4}$ & $0.35$ & $0.23$ & $810$ & $8.0$ & $1.431$ & $22.12$\\ \hline
R2C & $1024$ & $10^{-4}$ & $0.1$ & $10^{-4}$ & $1.11$ & $0.08$ & $890$ & $2.9$ & $0.472$ & $1.690$\\ \hline
R3C & $1024$ & $10^{-3}$ & $1$ & $10^{-4}$ & $1.14$ & $0.15$ & $172$ & $2.5$ & $1.996$ & $1.779$\\ \hline
R4C & $1024$ & $10^{-2}$ & $10$ & $10^{-4}$ & $2.37$ & $0.24$ & $57$ & $1.1$ & $5.550$ & $1.164$\\ \hline
R1D & $512$ & $10^{-4}$ & $0.01$ & $10^{-4}$ & $1.31$ & $0.18$ & $2367$ & $--$ & $0.320$ & $5.364$\\ \hline
R2D & $512$ & $10^{-4}$ & $0.1$ & $10^{-4}$ & $0.99$ & $0.14$ & $1457$ & $--$ & $0.334$ & $1.145$\\ \hline
R3D & $512$ & $10^{-3}$ & $1$ & $10^{-4}$ & $1.06$ & $0.17$ & $239$ & $--$ & $1.264$ & $1.033$\\ \hline
R4D & $512$ & $10^{-2}$ & $10$ & $10^{-4}$ & $1.04$ & $0.23$ & $61$ & $--$ & $6.505$ & $1.129$\\ \hline  \hline
\end{tabular}
\end{center}
\end{table}

Our goal is to study the statistical properties of homogeneous and isotropic
MHD turbulence so we use periodic boundary conditions and a standard
pseudospectral method~\cite{book-canuto} with $N^3$ collocation points in a
cubical simulation domain with sides of length $L=2\pi$; thus, we evaluate
spatial derivatives in Fourier space and local products of fields in real
space. We use the $2/3$ dealiasing method~\cite{book-canuto} to remove
aliasing errors; after this dealiasing $k_{\rm max}$ is the
magnitude of the
largest-magnitude wave vectors resolved in our DNS studies. We have carried
out extensive simulations with $N=512$ and $N=1024$; the parameters that we
use for different runs are given in Table~\ref{table:param} for decaying and
statistically steady turbulence.

We use a second-order, slaved, Adams-Bashforth scheme, with a time step
$\delta t$, for the time evolution of the velocity and magnetic fields; this
time step is chosen such that the Courant-Friedrichs-Lewy (CFL) condition is
satisfied~\cite{cfl}.

In our decaying-MHD-turbulence studies we have taken the initial (superscript
$0$) energy spectra $E_u^0(k)$ and $E_b^0(k)$, for velocity and magnetic
fields, respectively, to be the same; specifically, we have chosen
\begin{equation}
E_u^0(k) = E_b^0(k) = E^0k^4\exp(-2k^2) ,
\label{einitial}
\end{equation}
where $E^0$, the initial amplitude, is chosen
in such a way that we resolve both fluid and magnetic dissipation scales
$\eta^u_d$ and $\eta^b_d$, respectively: in all, except a few, of our runs
$k_{\rm max}\eta^u_d \gtrsim 1$ and $k_{\rm max}\eta^b_d \gtrsim 1$. The
initial phases of the Fourier components of the velocity and magnetic fields
are taken to be different and chosen such that they are distributed randomly
and uniformly between $0$ and $2\pi$. In such studies, it is convenient to
pick a reference time at which various statistical properties can be
compared. One such reference time is the peak that occurs in a plot of the
energy dissipation versus time; this reference time has been used in studies
of decaying fluid turbulence~\cite{kalelkar05,kalelkar06}, decaying fluid
turbulence with polymer additives~\cite{kalelkar05p,perlekar06}, and decaying
MHD turbulence~\cite{biskamp00,kalelkar04,lee10}.  Such peaks are associated
with the completion of the energy cascade from large length scales, at which
energy is injected into the system, to small length scales at which viscous
losses are significant. In the MHD case, these peaks occur at slightly
different times, $t_u$ and $t_b$, respectively, in plots of the kinetic
($\epsilon_u$) and magnetic ($\epsilon_b$) energy-dissipation rates. In our
decaying-MHD-turbulence studies we store velocity and magnetic fields at the
time $t_c$; if $t_u > t_b$, $t_c=t_u$; and $t_c=t_b$ otherwise; from these
fields we calculate the statistical properties that we present in the next
Section.

In the simulations in which we force the MHD equations to obtain
a nonequilibrium statistically steady state (NESS), we use a
generalization of the constant-energy-injection method described
in Ref.~\cite{lamorgese05}. We do not force the magnetic field
directly so we choose $\f_b=0$. The force $\f_u(\bfx,t)$ is
specified most simply in terms of ${\tilde \bff}_u(\bfk,t)$, its
spatial Fourier components, as follows: \begin{equation} {\tilde
\bff}_u(\bfk,t) = \frac{{\mathcal
P}~\Theta(k_f-k)}{2E_u(k_f,t)}{\tilde \bfu}(\bfk,t),
\end{equation} where $\Theta(k_f-k)$ is $1$ if $k\le k_f$ and $0$
otherwise, ${\mathcal P}$ is the power input, and
$E_u(k_f,t)=\sum_{k\le k_f}E_u(\bfk,t)$; in our DNS we use
$k_f=2$. This yields a
statistically steady state in which the mean value of the total
energy dissipation rate per unit volume balances the power input,
i.e., \begin{equation} \langle\epsilon\rangle = {\mathcal P};
\end{equation} once this state has been established, we save 50
representative velocity- and magnetic-field configurations
over $\simeq 36.08t_I$, $29.29t_I$, $32.61t_I$, and $30.95t_I$, for R1D,
R2D, R3D, and R4D, respectively, where $t_I=\ell_I/u_{\rm rms}$
is the integral-scale eddy-turnover time. We use these configurations to
obtain the statistical properties that we describe below. 

For decaying MHD turbulence we have carried out eight simulations with
$512^3$ collocation points and four simulations with $1024^3$ collocation
points. The parameters used in these simulations, which we have organised
into three sets, are given in Table~\ref{table:param}. 

In the first set of runs, R1-R5, we set the magnetic diffusivity
$\eta=10^{-3}$ and use five values of $\nu$, namely, $10^{-4}, \,
5.0\times10^{-4}, \, 10^{-3}, \, 5.0\times10^{-3}, $ and $10^{-2}$, which
yield ${\rm Pr_M} = 0.1, \, 0.5, \, 1, \, 5$, and $10$. These runs have been
designed to study the effects, on decaying MHD turbulence, of an increase in
${\rm Pr_M}$, {\it with the initial energy held fixed}: in particular, we use
$E^0_u=E^0_b\simeq0.32$ in Eq.~\ref{einitial} for runs R1-R5.  Given that this
initial energy and $\eta$ are both fixed, an increase in ${\rm Pr_M}$ leads
to a decrease in ${\rm Re}$, and thus an increase in $k_{\rm max}\eta^u_d$ and
$k_{\rm max}\eta^b_d$ as we discuss in detail later.

In our second set of decaying-MHD-turbulence runs, R3B, R4B and R5B, we
increase $E^0$ in Eq.~\ref{einitial} as we increase $\nu$, and thereby ${\rm
Pr_M}$, so that $k_{\rm max}\eta^u_d \simeq 1$ and $k_{\rm max}\eta^b_d
\simeq 1$.  Thus, in these runs, the inertial ranges in energy spectra extend
over comparable ranges of the wave-vector magnitude $k$.

Our third set of decaying-MHD-turbulence runs, R1C, R2C, R3C, and R4C, use
$1024^3$ collocation points and ${\rm Pr_M} = 0.01, \, 0.1, \, 1$, and $10$,
respectively. By comparing the results of these runs with those of R1-R5,
R3B, R4B, and R5B, we can check whether our qualitative results depend
significantly on the number of collocation points that we use.

We have carried out another set of four runs, R1D, R2D, R3D, and R4D, in
which we force the MHD equations, as described above, until we obtain a
nonequilibrium statistically steady state. These runs help us to compare the
statistical properties of decaying and statistically steady turbulence. In
these runs we use $512^3$ collocation points, and $\nu$ and $\eta$ such that
${\rm Pr_M}=0.01,~0.1,~1$, and $10$, respectively.

\subsection{Statistical measures \label{sec:statmeasures}}

We use several statistical measures to characterise homogeneous, isotropic
MHD turbulence. Some, but not all, of these have been used in
earlier DNS studies~\cite{biskamp00,baerenzung08,brandenburg95,chou01,
stribling90,matthaeus08,wan09} and solar-wind
turbulence~\cite{matthaeus82,podesta07,salem09}.

We calculate the kinetic, magnetic, and total energy spectra $E_u(k) =
\sum_{\bfk\ni |\bfk|=k}|{\tilde\bfu}(\bfk)|^2$, $E_b(k) = \sum_{\bfk\ni
|\bfk|=k}|{\tilde\bfb}(\bfk)|^2$, and $E(k) = E_u(k)+E_b(k)$, respectively,
the kinetic, magnetic, and total energies $E_u = \sum_k E_u(k)/2$, $E_b =
\sum_k E_b(k)/2$, and $E = E_u+E_b$, and the ratio $E_b/E_u$. Spectra for the
Els\"asser variables, energy dissipation rates, and the effective pressure
are, respectively, $E_{z^\pm}(k) = \sum_{\bfk\ni
|\bfk|=k}|{\tilde\bfz^{\pm}}(\bfk)|^2$, $\epsilon_u(k) = \nu k^2 E_u(k)$,
$\epsilon_b(k) = \nu k^2 E_b(k)$, and $P(k) = \sum_{\bfk\ni
|\bfk|=k}|\tilde{\bar p}(\bfk)|^2$. 

Our MHD simulations are characterised by the Taylor-microscale 
Reynolds number ${\rm Re}_{\lambda} = u_{\rm rms}\lambda/\nu$, the magnetic
Taylor-microscale Reynolds number ${\rm Rm}_\lambda=u_{\rm rms}\lambda/\eta$,
and the magnetic Prandtl number ${\rm Pr_M}={\rm Rm}_\lambda/{\rm
Re}_\lambda=\nu/\eta$, where the root-mean-square velocity $u_{\rm rms} =
\sqrt{2E_u/3}$ and the Taylor microscale $\lambda = \left[\sum_k
k^2E(k)/E\right]^{-1/2}$. We also calculate the integral length scale $\ell_I
= \left[\sum_k E(k)/k\right]/E$, the mean kinetic energy dissipation rate per
unit mass, $\epsilon_u = \nu \sum_{i,j} (\partial_iu_j+\partial_ju_i)^2 =
\nu\sum_k k^2 E_u(k)$, the mean magnetic energy dissipation rate per unit
mass $\epsilon_b = \eta \sum_{i,j} (\partial_ib_j+\partial_jb_i)^2 =
\eta\sum_k k^2 E_b(k)$, the mean energy dissipation rate per unit mass
$\epsilon = \epsilon_u+\epsilon_b$, and the dissipation length scales for
velocity and magnetic fields $\eta_d^u = (\nu^3/\epsilon_u)^{1/4}$ and
$\eta_d^b = (\eta^3/\epsilon_b)^{1/4}$, respectively. 

We calculate the eigenvalues $\Lambda_n^u$ and the associated eigenvectors
${\hat e}_n^u$, with $n=1, 2,$ or $3$,  of the rate-of-strain tensor $\mathbb
S$ whose components are $S_{ij}=\partial_iu_j+\partial_ju_i$.  Similarly
$\Lambda_1^b$, $\Lambda_2^b$ and $\Lambda_3^b$ denote the eigenvalues of the
tensile magnetic stress tensor $\mathbb T$, which has components $T_{ij} =
-b_ib_j$; the corresponding eigenvectors are, respectively, ${\hat e}_1^b$,
${\hat e}_2^b$, and ${\hat e}_3^b$.

For incompressible flows $\sum_n\Lambda_n^u = 0$, so at least one of the
eigenvalues $\Lambda_n^u$ must be positive and another negative; we label
them in such a way that $\Lambda_3^u$ is positive, $\Lambda_1^u$ negative,
and $\Lambda_2^u$ lies in between them; note that $\Lambda_2^u$ can be
positive or negative. We obtain probability distribution functions (PDFs) of
these eigenvalues; furthermore, we obtain PDFs of the cosines of the angles
that the associated eigenvectors make with the vectors such as $\bfu$,
$\bomega$, etc. These PDFs and those of quantities such as the local cross
helicity $H_C=\bfu\cdot\bfb$ help us to quantify the degree of alignment of
pairs of vectors such as $\bfu$ and $\bfb$~\cite{matthaeus08}.  We also
compare PDFs of magnitudes of local vorticity $\omega$, the current
density $j$, and local energy dissipation rates $\epsilon_u$ and $\epsilon_b$ to
get information about intermittency in velocity and magnetic fields.

We also obtain several interesting joint PDFs; to the best of our knowledge,
these have not been obtained earlier for MHD turbulence. We first obtain the
velocity-derivative tensor $\mathbb A$, also known as the rate-of-deformation
tensor, with components $A_{ij}=\partial_iu_j$, and thence the invariants $Q =
-\frac{1}{2}tr(\mathbb{A}^2)$ and $R = -\frac{1}{3}tr(\mathbb{A}^3)$, which
have been used frequently to characterise fluid
turbulence~\cite{book-davidson,cantwell93,biferale07}.  The zero-discriminant
line $D\equiv \frac{27}{4}R^2+Q^3=0$ and the $Q$ and $R$ axes divide the
$QR$ plane into qualitatively different regimes. In particular, regions
in a turbulent flow can be classified as follows: when $Q$ is large and
negative, local strains are high and vortex formation is not favoured;
furthermore, if $R>0$, fluid elements experience axial strain, whereas, if
$R<0$, they feel biaxial strain. In contrast, when $Q$ is large and positive,
vorticity dominates the flow; if, in addition, $R<0$, vortices are
compressed, whereas, if $R>0$, they are stretched. Thus, some properties of a
turbulent flow can be highlighted by making contour plots of the joint PDF of
$Q$ and $R$; these $QR$ plots show a characteristic, tear-drop shape. We
explore the forms of these and other joint PDFs, such as joint PDFs of
$\epsilon_u$ and $\epsilon_b$, in MHD turbulence.

To characterise intermittency in MHD turbulence we calculate the order-$p$,
equal-time, longitudinal structure functions $S_p^a(l) \equiv \langle|\delta
a_{\parallel}(\bfx,l)|^p\rangle$, where the longitudinal component of the
field $\bfa$ is given by $\delta a_{\parallel}(\bfx,l) \equiv
[\bfa(\bfx+\bfl,t)-\bfa(\bfx,t)]\cdot\frac{\bfl}{l}$, where $\bfa$ can be
$\bfu$, $\bfb$, or one of the Els\"asser variables.  From these structure
functions we also obtain the hyperflatness $F_6^a(l)=S_6^a(l)/[S_2^a(l)]^3$.
For separations $l$ in the inertial range, i.e., $\eta_d^u, \eta_d^b \ll l
\ll L$, we expect $S_p^a(l)\sim l^{\zeta_p^a}$, where $\zeta_p^a$ are the
inertial-range multiscaling exponents for the field $\bfa$; the Kolmogorov
phenomenology of 1941~\cite{k41,book-frisch,biskamp00}, henceforth referred to
as K41, yields the simple scaling result $\zeta_p^{aK41} = p/3$; but
multiscaling corrections are significant with $\zeta_p^a\neq\zeta_p^{aK41}$
[Sec.~\ref{sec:results}]. From the increments $\delta a_{\parallel}(\bfx,l)
\equiv [\bfa(\bfx+\bfl,t)-\bfa(\bfx,t)]\cdot\frac{\bfl}{l}$ we also obtain
the dependence of PDFs of $\delta a_{\parallel}$ on the scale $l$.

\section{Results \label{sec:results}}

To set the stage for the types of studies we carry out for MHD turbulence, we begin with
a very brief summary of similar and well-known results from studies of homogeneous,
isotropic Navier-Stokes turbulence, which can be found, e.g., in
Refs.~\cite{book-frisch,kalelkar05,kalelkar06,schumann78,chen93,gotoh02,kaneda03,
schumacher07,pramana09,schumacher10}.

\subsection{Overview of fluid turbulence \label{sec:NS}}

For ready reference we show here illustrative plots from a DNS study that we
have carried out for the three-dimensional Navier-Stokes equation by using a
pseudospectral method, with $512^3$ collocation points and the $2/3$ rule for
removing aliasing errors; here $\nu = 0.001$, ${\rm Re}_\lambda \simeq 340$
and $k_{\rm max}\eta_d^u \simeq 0.3$.

\begin{figure}[!htb]
\begin{center}
\includegraphics[width=0.95\textwidth]{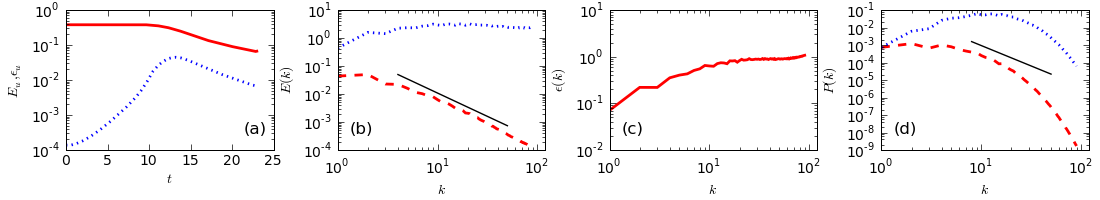}
\end{center}
\label{fig:ns-spectra}
\caption{Plots from our DNS of decaying fluid turbulence in the Navier-Stokes
equation with $512^3$ collocation points: (a) plots of the energy $E$
(red full line) and mean energy dissipation rate $\epsilon$ (blue
dotted line) versus time $t$ (given as a product of the number of
iterations and the time step $\delta t$); (b) log-log (base 10) plots of the energy spectrum
$E(k)$ (red dashed line) and the corresponding compensated spectrum
$k^{5/3}E(k)$ (blue dotted line) versus $k$. The black solid line shows
the K41 result $k^{-5/3}$
for comparison; (c) log-log (base 10) plot of spectrum of energy-dissipation
or enstrophy spectrum $\epsilon(k)$; (d)  log-log (base 10) plots of the
pressure spectrum $P(k)$ (red dashed line) and the compensated pressure spectrum
$k^{7/3}P(k)$ (blue dotted line). The black solid line shows the
K41 result $k^{-7/3}$ for comparison.} 
\end{figure}
\begin{figure}[!htb]
\begin{center}
\includegraphics[width=0.95\textwidth]{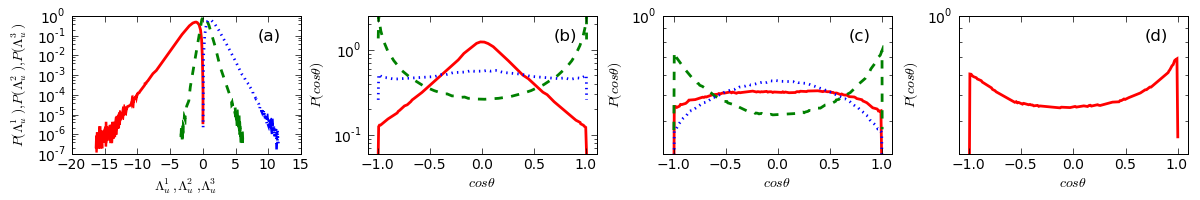}
\end{center}
\label{fig:ns-eigen}
\caption{PDFs from our DNS of decaying fluid turbulence in Navier-Stokes
equation with $512^3$ collocation points: (a) semilog (base 10) plots
of PDFs of eigenvalues of rate-of-strain tensor ${\mathbb S}$, namely,
$\Lambda_1^u$ (red full line), $\Lambda_2^u$ (green dashed line) and
$\Lambda_3^u$ (blue dotted line); (b) semilog (base 10) plots of PDFs
of cosines of angles between the vorticity $\bomega$ and eigenvectors of
$\mathbb S$, namely, ${\hat e}_1^u$ (red full line), ${\hat e}_2^u$
(green dashed line), and ${\hat e}_3^u$ (blue dotted line); 
(c) semilog (base 10) plots of PDFs of cosines of angles between
the velocity $\bfu$ and eigenvectors of $\mathbb S$, namely,
${\hat e}_1^u$ (red full line), ${\hat e}_2^u$ (green dashed line),
and ${\hat e}_3^u$ (blue dotted line); (d) semilog (base 10) plots
of PDFs of cosines of angles between the velocity $\bfu$ and
vorticity $\bomega$.}
\end{figure}
\begin{figure}[!htb]
\begin{center}
\includegraphics[width=0.9\textwidth]{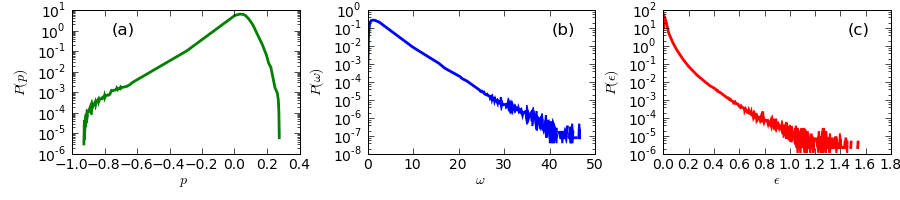}
\end{center}
\label{fig:ns-hist}
\caption[]{PDFs from our DNS of decaying fluid turbulence in Navier-Stokes
equation with $512^3$ collocation points: semilog (base 10) plots
of the PDFs of (a) the pressure $p$, (b) modulus of the vorticity $\omega$ and
(c) the local energy-dissipation rate $\epsilon$.}
\end{figure}
\begin{figure}[!htb]
\begin{center}
\includegraphics[width=0.3\textwidth]{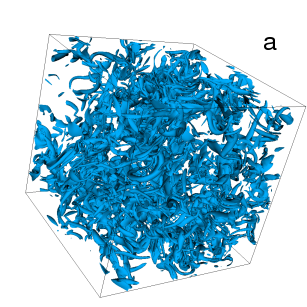}
\includegraphics[width=0.3\textwidth]{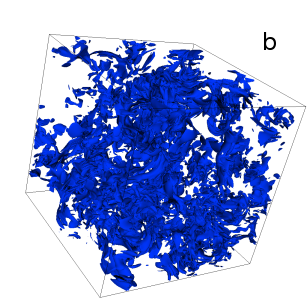}
\includegraphics[width=0.3\textwidth]{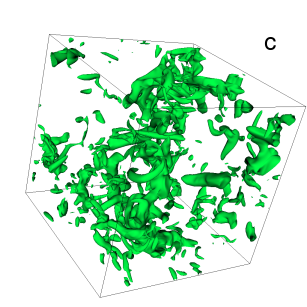}
\end{center}
\label{fig:ns-isosurfaces}
\caption{Isosurfaces of (a) the modulus of the vorticity $\omega$,
(b) the local energy-dissipation rate $\epsilon$, and (c) the local
pressure $p$, from our DNS of decaying fluid turbulence in Navier-Stokes
equation with $512^3$ collocation points. The isovalues used in these plots
are two standard deviations more than the mean values of the quantities.}
\end{figure}
\begin{figure}[!htb]
\begin{center}
\includegraphics[width=0.4\textwidth]{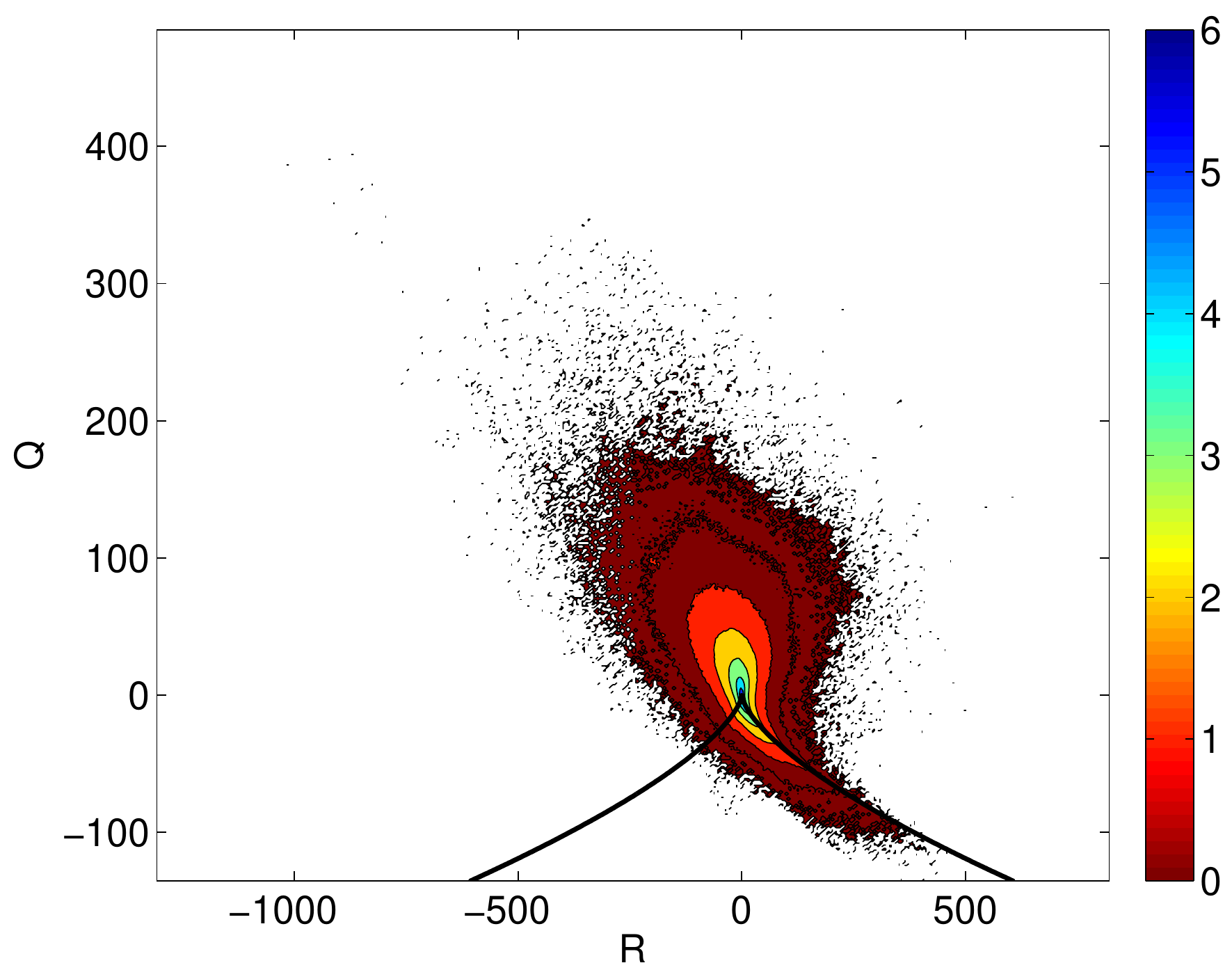}
\end{center}
\caption[]{$QR$ plot, i.e., the joint PDF of $Q$ and $R$ (see
text) shown as filled contour plot 
in our log-log (base 10) scale, obtained from a DNS of decaying fluid
turbulence in Navier-Stokes equation with $512^3$ collocation points.}
\label{fig:ns-qrplot}
\end{figure}

In decaying fluid turbulence, energy is injected at large spatial scales as
described in the previous Section for the MHD case.  This energy cascades
down till it reaches the dissipative scale at which viscous losses are
significant. We study various statistical properties; these are given in
points (i)-(vi) below: 

(i) Plots of the energy $E$ and the mean energy dissipation
rate $\epsilon$ versus time show, respectively, a gentle decay and a
maximum as shown, e.g., by the full red and dotted blue curves in
Fig.~\ref{fig:ns-spectra}(a).  This maximum in $\epsilon$ is associated with
the completion of the energy cascade at a time $t_c$; the remaining properties
(ii)-(vi) are obtained at $t_c$.
(ii) If ${\rm Re}_\lambda$ is sufficiently large and we have a well-resolved DNS
(i.e., $k_{\rm max}\eta_d^u > 1$), then, at $t_c$, the spectrum $E(k)$ shows a
well-developed inertial range, where at the K41 level $E(k) \sim k^{-5/3}$,
and a dissipation range, in which the behaviour of the energy spectrum is
consistent with $E(k) \sim k^{\alpha}\exp(-\beta k)$, where $\alpha$ and 
$\beta$ are non-universal, positive
constants~\cite{chen93,schumacher07} and 
$5 k_d < k < 10 k_d$, with
$k_d = 1/\eta_d^u$. An illustrative energy spectrum is shown by the dashed
red line in Fig.~\ref{fig:ns-spectra}(b); the blue dotted curve shows the
compensated spectrum $k^{5/3}E(k)$; the associated dissipation or enstrophy
spectrum $\epsilon(k)$ is shown in Fig.~\ref{fig:ns-spectra}(c) and the
inertial-range pressure spectrum~\cite{gotoh01}, $P(k) \sim k^{-7/3}$ at the
K41 level is shown in Fig.~\ref{fig:ns-spectra}(d). [Note that our DNS for the 
Navier-Stokes equation, which suffices for our purposes of illustration, does not have a
well-resolved dissipation range because $k_{\rm max}\eta_d^u \simeq 0.3 < 1$;
this is also reflected in the lack of a well-developed maximum in the
enstrophy spectrum of Fig.~\ref{fig:ns-spectra}(c).] 
(iii) Illustrative PDFs of the eigenvalues $\Lambda_n^u$ of the rate-of-strain
tensor  $\mathbb S$ are given for $n=1,\,2,$ and $3$, respectively, by the
full red, dashed green, and dotted blue curves in Fig.~\ref{fig:ns-eigen}(a);
PDFs of the cosines of the angles that the vorticity $\bomega$ and the
velocity $\bfu$ make with the associated eigenvectors $\hat{e}_n^{u}$ are
given, respectively, in Figs.~\ref{fig:ns-eigen}(b) and \ref{fig:ns-eigen}(c) via full red
$(n=1)$, dashed green $(n=2)$, and dotted blue $(n=3)$ curves; these show that
both $\bomega$ and $\bfu$ have a tendency to be preferentially aligned
parallel or antiparallel to $\hat{e}_2^{u}$~\cite{biferale07};
the PDF of the cosine of the
angle between $\bfu$ and $\bomega$ also indicates preferential alignment or
antialignment of these two vectors, but with a greater tendency towards
alignment as found in experiments with a small amount of helicity~\cite{kit87}
and as illustrated in our
Fig.~\ref{fig:ns-eigen}(d). Finally, we give representative PDFs of the
pressure $p$, modulus of vorticity $\omega=|\bomega|$, and the local energy dissipation
$\epsilon$ in Figs.~\ref{fig:ns-hist}(a), \ref{fig:ns-hist}(b),
and \ref{fig:ns-hist}(c), respectively; note that
the PDF of the pressure is negatively skewed. 
(iv) Inertial-range structure functions $S_p^u(l)\sim l^{\zeta_p^u}$ show
significant deviations~\cite{book-frisch} from the K41 result $\zeta_p^{uK41}=p/3$
especially for $p > 3$. From these structure functions we can obtain the
hyperflatness $F_6^u(l)$; this increases as the length scale $l$ decreases, a
clear signature of intermittency, as shown, e.g., in
Refs.~\cite{perlekar06,pramana09}. This intermittency also leads to
non-Gaussian tails, especially for small $l$, in PDFs of velocity increments
(see, e.g., Refs.~\cite{pramana09,pedrizzetti96,jung05}) such as
$\delta u_{\parallel}(l)$. 
(v) Small-scale structures in turbulent flows can be visualised via
isosurfaces~\cite{okamoto07} of, say, $\omega$, $\epsilon$, and $p$,
illustrative plots of which are given in 
Figs.~\ref{fig:ns-isosurfaces}(a)-\ref{fig:ns-isosurfaces}(c);
these show that regions of large $\omega$ are organised into
slender tubes whereas isosurfaces of $\epsilon$ look like shredded sheets;
pressure isosurfaces also show tubes~\cite{cao99,kalelkar06} but some studies have
suggested the term cloud-like for them~\cite{schumann78}.
(vi) Joint PDFs also provide useful information about turbulent flows;
in particular, contour plots of the joint PDF of $Q$ and $R$, as in the representative
Fig.~\ref{fig:ns-qrplot}, show a characteristic tear-drop structure. 

The properties of statistically steady, homogeneous, isotropic fluid
turbulence are similar to those described in points (ii)-(vi) in the preceding
paragraph for the case of decaying fluid turbulence at cascade completion at
$t_c$. In particular, the strong-universality~\cite{lvov03} hypothesis
suggests that the multiscaling exponents $\zeta_p^u$ have the same values in
decaying and statistically steady turbulence.

The remaining part of this Section is devoted to our detailed study of the
MHD-turbulence analogues of the properties (i)-(vi) summarised above; these
are discussed, respectively, in the six Subsections~\ref{sec:tseries}-\ref{sec:jpdf}.

\begin{figure}[!htb]
\begin{center}
\includegraphics[width=0.95\textwidth]{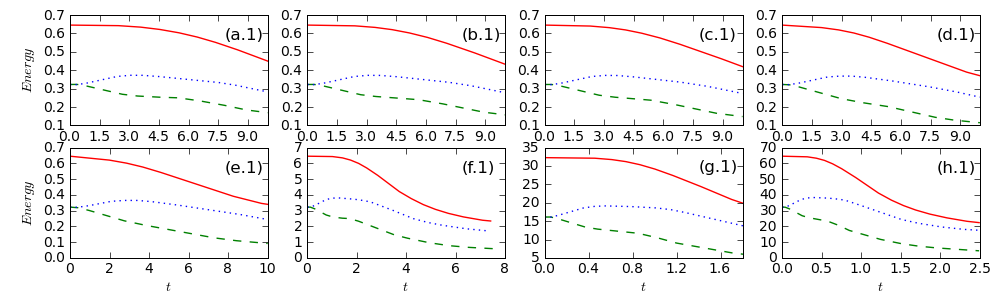}\\
\includegraphics[width=0.95\textwidth]{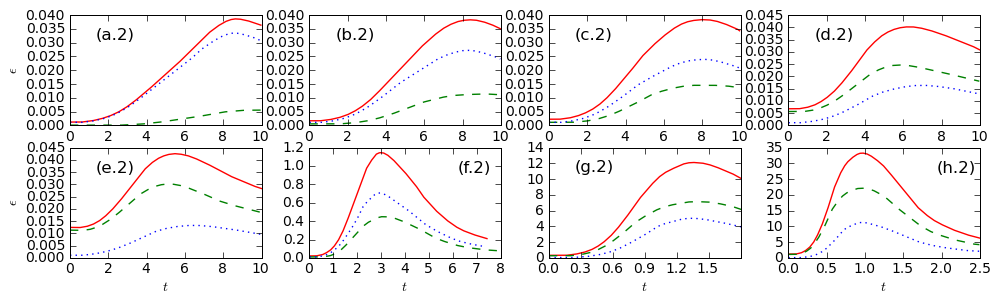}\\
\includegraphics[width=0.95\textwidth]{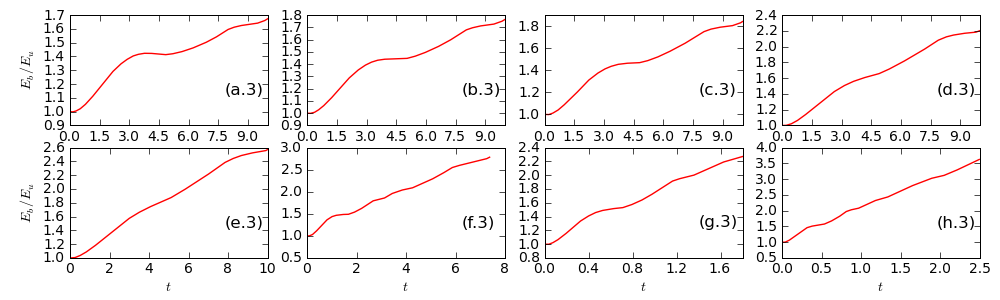}
\end{center}
\caption[]{Plots versus time $t$ (given as a product of the number of iterations
and the time step $\delta t$) of energies (a.1)-(h.1); total energy $E$ (red full line),
kinetic-energy $E_u$ (green dashed line), and magnetic-energy $E_b$ (blue dotted line),
of energy-dissipation rates (a.2)-(h.2); mean energy dissipation rate $\epsilon$
(red, full line), kinetic-energy dissipation $\epsilon_u$
(green dashed line), and magnetic-energy dissipation rate $\epsilon_b$ (blue dotted line),
and of the ratio $E_b/E_u$ (a.3)-(h.3), generically, for decaying simulations 
(a) ${\rm Pr_M}=0.1$ (R1), (b) ${\rm Pr_M}=0.5$ (R2), (c) ${\rm Pr_M}=1.0$ (R3),
(d) ${\rm Pr_M}=5.0$ (R4), (e) ${\rm Pr_M}=10.0$ (R5), (f) ${\rm Pr_M}=1.0$ (R3B),
(g) ${\rm Pr_M}=5.0$ (R4B), and (h) ${\rm Pr_M}=10.0$ (R5B).}
\label{fig:energy-decay}
\end{figure} 
\begin{figure}[htb]
\begin{center}
\includegraphics[width=0.95\textwidth]{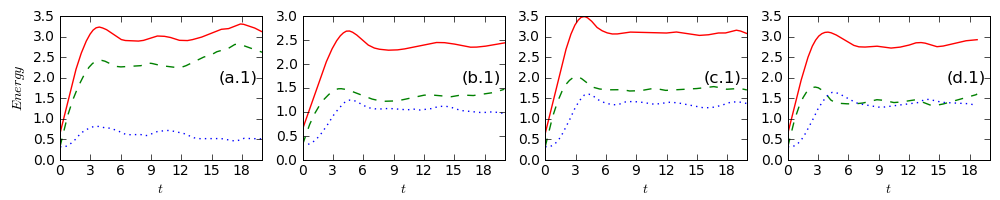}\\
\includegraphics[width=0.95\textwidth]{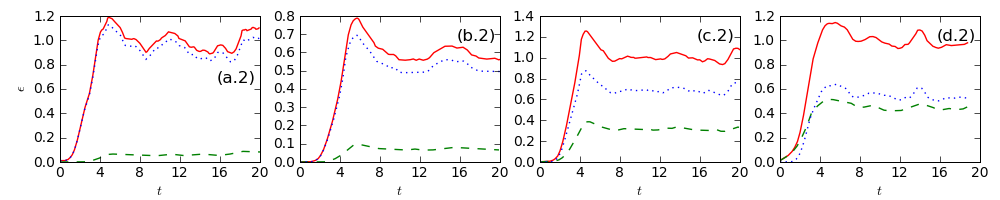}
\includegraphics[width=0.95\textwidth]{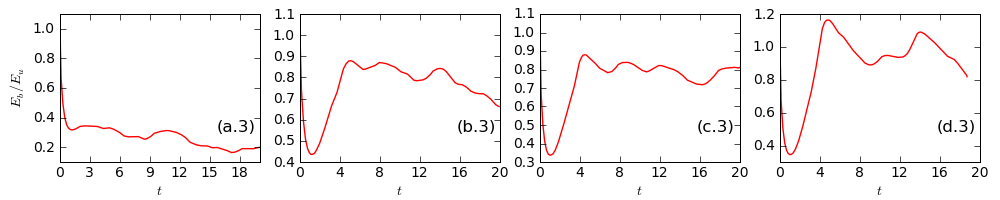}
\end{center}
\caption[]{Plots versus time $t$ (given as a product of the number of iterations
and the time step $\delta t$) of energies (a.1)-(d.1); total energy $E$ (red full line),
kinetic-energy $E_u$ (green dashed line), and magnetic-energy $E_b$ (blue dotted line),
of energy-dissipation rates (a.2)-(d.2); mean energy dissipation rate $\epsilon$
(red, full line), kinetic-energy dissipation rate $\epsilon_u$ (green dashed line),
and magnetic-energy dissipation rate $\epsilon_b$ (blue dotted line),
and of the ratio $E_b/E_u$ (a.3)-(d.3), generically, for
forced simulations (a) ${\rm Pr_M}=0.01$ (R1D), (b) ${\rm Pr_M}=0.1$ (R2D), (c) ${\rm Pr_M}=1.0$ (R3D), 
and (d) ${\rm Pr_M}=10$ (R4D).}
\label{fig:energy-forced}
\end{figure} 

\subsection{Temporal evolution \label{sec:tseries}}

We examine the time evolution of the energy, the energy-dissipation rates,
and related quantities, first for decaying and then for statistically steady
MHD turbulence.

Figure~\ref{fig:energy-decay} shows how the total energy $E$ (red
full line), the kinetic energy $E_u$ (green dashed line), and the
magnetic energy $E_b$ (blue dotted line) evolve with time $t$
(given as a product of the number of iterations and the time
step $\delta t$) for runs R1-R5
[Figs.~\ref{fig:energy-decay}(a.1)-\ref{fig:energy-decay}(e.1)]
and runs R3B-R5B
[Figs.~\ref{fig:energy-decay}(f.1)-\ref{fig:energy-decay}(h.1)]
for decaying MHD turbulence. Figure~\ref{fig:energy-decay} also
shows similar plots for the mean energy dissipation rate
$\epsilon$ (red full line), the mean kinetic-energy dissipation
rate $\epsilon_u$ (green dashed line), and the mean
magnetic-energy dissipation rate $\epsilon_b$ (blue dotted line)
versus time $t$ for runs R1-R5
[Figs.~\ref{fig:energy-decay}(a.2)-\ref{fig:energy-decay}(e.2)]
and runs R3B-R5B
[Figs.~\ref{fig:energy-decay}(f.2)-\ref{fig:energy-decay}(h.2)].
In addition Fig.~\ref{fig:energy-decay} depicts the
time-evolution of the ratio $E_b/E_u$ for runs R1-R5
[Figs.~\ref{fig:energy-decay}(a.3)-\ref{fig:energy-decay}(e.3)]
and runs R3B-R5B
[Figs.~\ref{fig:energy-decay}(f.3)-\ref{fig:energy-decay}(h.3)].
We see from these figures that, for all the values of ${\rm
Pr_M}$ we have used, the energies $E$ and $E_u$ decay gently with
$t$ but $E_b$ rises initially such that the ratio $E_b/E_u$
rises, nearly monotonically, with $t$ over the times we have
considered; this is an intriguing trend that does not seem to
have been noticed earlier. The times over which we have carried
out our DNS are comparable to the cascade-completion time $t_c$
that can be obtained from the peaks in the plots of $\epsilon$,
$\epsilon_u$, and $\epsilon_b$ versus $t$
[Figs.~\ref{fig:energy-decay}(a.2)-\ref{fig:energy-decay}(h.2)];
by comparing these plots we see that, as we move from ${\rm
Pr_M}=0.1$ to ${\rm Pr_M}=10$, with fixed $\eta$, we find that
$(\epsilon_u-\epsilon_b)$ and $(t_b-t_u)$ grow from negative
values to positive ones because $\epsilon_u$ increases with ${\rm
Pr_M}$, where $t_b$ and $t_u$ are the positions of the
cascade-completion maxima in $\epsilon_b$ and $\epsilon_u$,
respectively.  We do not pursue the time evolution of our system
well beyond $t_u$ and $t_b$ because the integral scale begins to
grow thereafter and, eventually, can become comparable to the
linear size of the simulation domain~\cite{kalelkar05}.

Figures~\ref{fig:energy-forced}(a.1)-\ref{fig:energy-forced}(d.1), show how the
total energy $E$ (red full line), the total kinetic energy $E_u$ (green dashed
line), and the total magnetic energy $E_b$ (blue dotted line) evolve with time
$t$ (given as a product of the number of iterations and the time step $\delta
t$) for, respectively, runs R1D-R4D for forced and statistically steady MHD
turbulence.  Figures~\ref{fig:energy-forced}(a.2)-\ref{fig:energy-forced}(d.2),
show similar plots for the mean energy dissipation rate $\epsilon$ (red full
line), the mean kinetic-energy dissipation rate $\epsilon_u$ (green dashed
line), and the mean magnetic-energy dissipation rate $\epsilon_b$ (blue dotted
line) versus time $t$ for, respectively, runs R1D-R4D. And
Figs.~\ref{fig:energy-forced}(a.3)-\ref{fig:energy-forced}(d.3), depict the
time-evolution of the ratio $E_b/E_u$ for these runs.  We see from these
figures that a statistically steady state is established in which the energies
$E$, $E_u$, and $E_b$, the dissipation rates $\epsilon$, $\epsilon_u$, and
$\epsilon_b$, and the ratio $E_b/E_u$ fluctuate about their mean values (after
initial transients have died out). The mean value of $E_b/E_u$ increases from
about $0.2-0.3$ to $E_b/E_u \simeq 1$ as ${\rm Pr_M}$ increases from $0.01$ to
$10$.  Furthermore, the mean values of the dissipation rates $\epsilon_u$ and
$\epsilon_b$ are such that $(\epsilon_u - \epsilon_b)$ grows from a negative
value $\simeq -1$ to a value close to zero as ${\rm Pr_M}$ increases from
$0.01$ to $10$.

\subsection{Spectra \label{sec:spectra}}

\begin{figure}[htb]
\begin{center}
\includegraphics[width=0.95\textwidth]{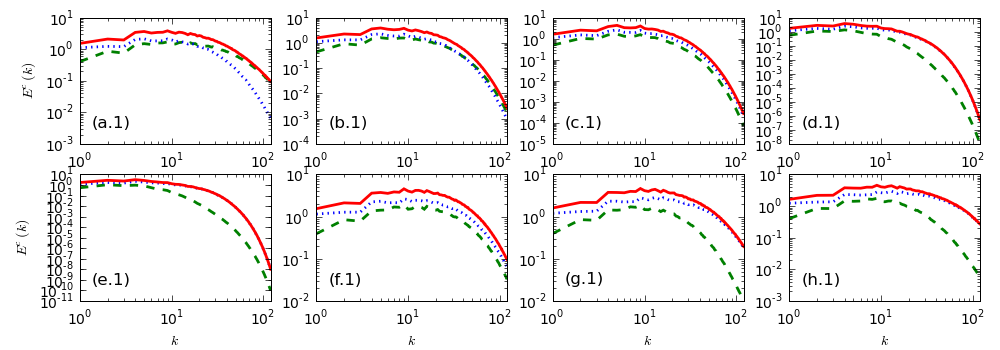}\\
\includegraphics[width=0.95\textwidth]{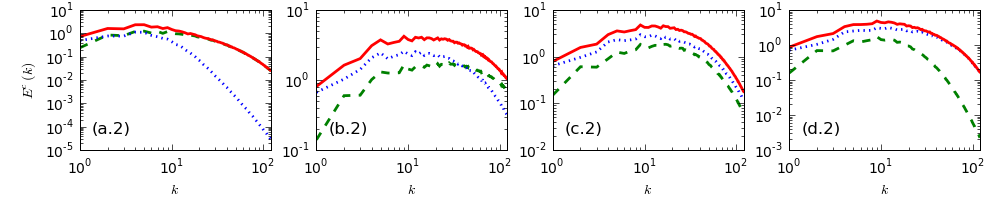}\\
\includegraphics[width=0.95\textwidth]{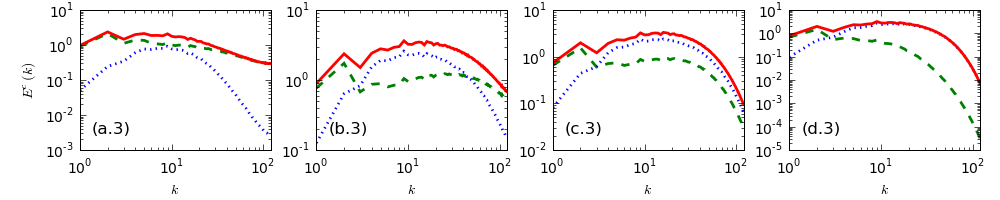}\\
\end{center}
\caption[]{Log-log (base 10) plots of the compensated energy spectra
$\epsilon^{-2/3}k^{5/3}E(k)$ (red full lines), $\epsilon_{-2/3}k^{5/3}E_u(k)$ 
(green dashed lines), and $\epsilon^{-2/3}k^{5/3}E_b(k)$ (blue dotted lines);
on the vertical axes these are denoted generically as $E^c(k)$: 
(a.1) ${\rm Pr_M}=0.1$ (R1), (b.1) ${\rm Pr_M}=0.5$ (R2), (c.1) 
${\rm Pr_M}=1.0$ (R3), (d.1) ${\rm Pr_M}=5.0$ (R4), (e.1) ${\rm Pr_M}=10.0$ 
(R5), (f.1) ${\rm Pr_M}=1.0$ (R3B), (g.1) ${\rm Pr_M}=5.0$ (R4B), (h.1) 
${\rm Pr_M}=10.0$ (R5B), (a.2) ${\rm Pr_M}=0.01$ (R1C), (b.2) 
${\rm Pr_M}=0.1$ (R2C), (c.2) ${\rm Pr_M}=1.0$ (R3C), and 
(d.2) ${\rm Pr_M}=10.0$ (R4C) for decaying MHD turbulence; and 
for statistically steady MHD turbulence (a.3) ${\rm Pr_M}=0.01$ (R1D), 
(b.3) ${\rm Pr_M}=0.1$ (R2D), (c.3) ${\rm Pr_M}=1.0$ (R3D), and 
(d.3) ${\rm Pr_M}=10.0$ (R4D).}
\label{fig:energy-spectra}
\end{figure} 
\begin{figure}[htb]
\begin{center}
\includegraphics[width=0.95\textwidth]{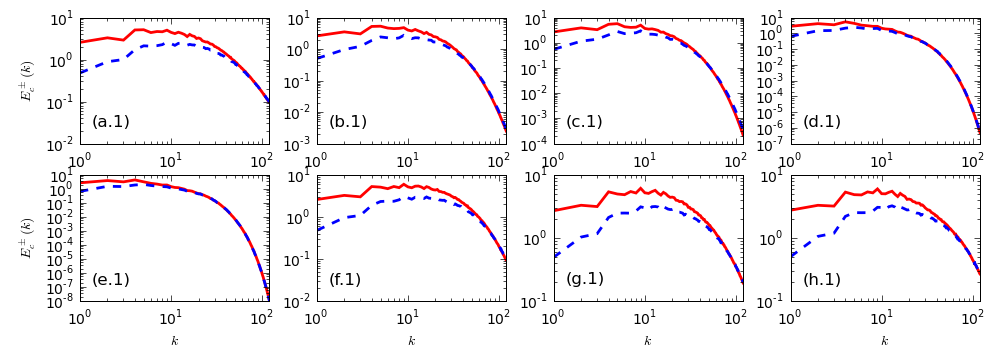}\\
\includegraphics[width=0.95\textwidth]{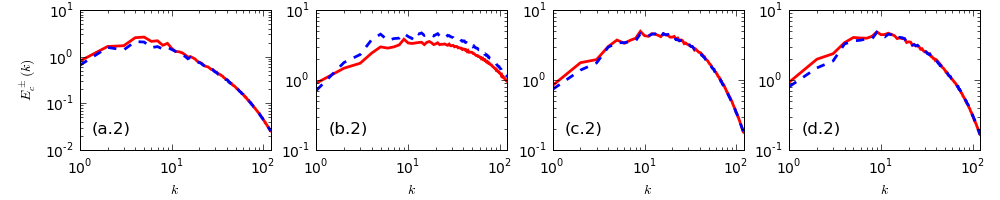}\\
\includegraphics[width=0.95\textwidth]{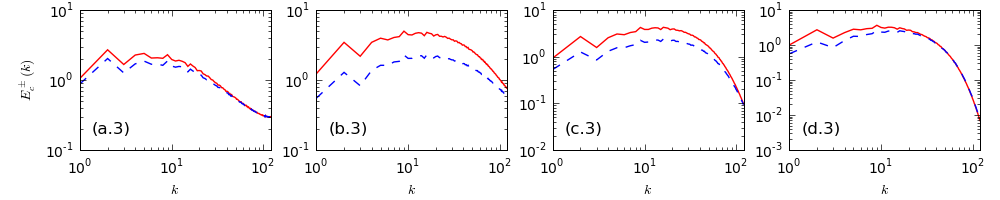}
\end{center}
\caption[]{Log-log (base 10) plots of compensated energy spectra, 
$E_c^{\pm}(k) = k^{5/3}E^\pm(k)$, with $k$ the magnitude of the wave 
vector, for the Els\"asser variables fields $z^+$ (red full line) and $z^-$ 
(blue dashed line):
(a.1) ${\rm Pr_M}=0.1$ (R1), (b.1) ${\rm Pr_M}=0.5$ (R2), (c.1) 
${\rm Pr_M}=1.0$ (R3), (d.1) ${\rm Pr_M}=5.0$ (R4), (e.1) ${\rm Pr_M}=10.0$ 
(R5), (f.1) ${\rm Pr_M}=1.0$ (R3B), (g.1) ${\rm Pr_M}=5.0$ (R4B), (h.1) 
${\rm Pr_M}=10.0$ (R5B), (a.2) ${\rm Pr_M}=0.01$ (R1C), (b.2) 
${\rm Pr_M}=0.1$ (R2C), (c.2) ${\rm Pr_M}=1.0$ (R3C), and 
(d.2) ${\rm Pr_M}=10.0$ (R4C) for decaying MHD turbulence; and 
for statistically steady MHD turbulence (a.3) ${\rm Pr_M}=0.01$ (R1D), 
(b.3) ${\rm Pr_M}=0.1$ (R2D), (c.3) ${\rm Pr_M}=1.0$ (R3D), and 
(d.3) ${\rm Pr_M}=10.0$ (R4D).}
\label{fig:zspectra}
\end{figure} 
\begin{figure}[htb]
\begin{center}
\includegraphics[width=0.95\textwidth]{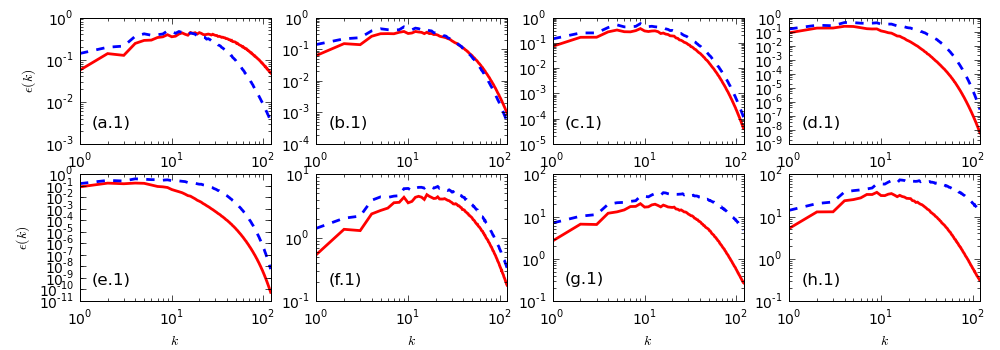}\\
\includegraphics[width=0.95\textwidth]{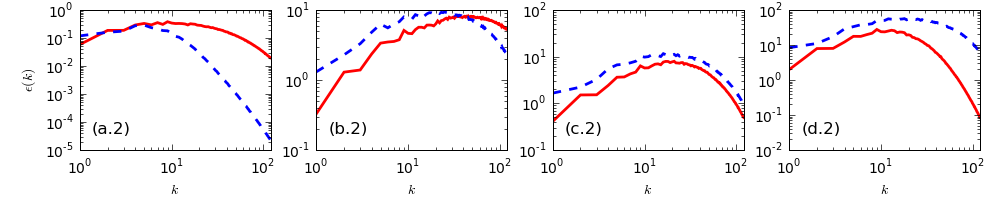}\\
\includegraphics[width=0.95\textwidth]{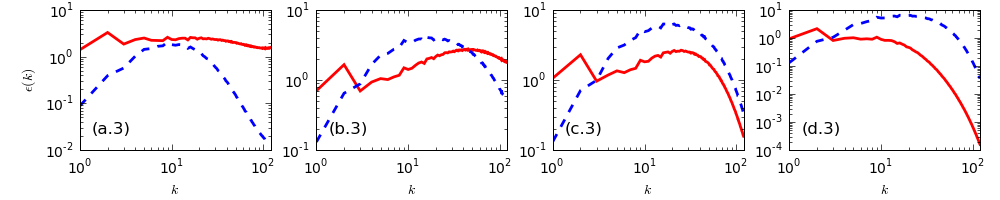}\\
\end{center}
\caption[]{Log-log (base 10) plots of energy-dissipation spectra 
for the fluid (red full lines) and magnetic (blue dashed lines) fields,
with $k$ the magnitude of the wave vector:
(a.1) ${\rm Pr_M}=0.1$ (R1), (b.1) ${\rm Pr_M}=0.5$ (R2), (c.1) 
${\rm Pr_M}=1.0$ (R3), (d.1) ${\rm Pr_M}=5.0$ (R4), (e.1) ${\rm Pr_M}=10.0$ 
(R5), (f.1) ${\rm Pr_M}=1.0$ (R3B), (g.1) ${\rm Pr_M}=5.0$ (R4B), (h.1) 
${\rm Pr_M}=10.0$ (R5B), (a.2) ${\rm Pr_M}=0.01$ (R1C), (b.2) 
${\rm Pr_M}=0.1$ (R2C), (c.2) ${\rm Pr_M}=1.0$ (R3C), and 
(d.2) ${\rm Pr_M}=10.0$ (R4C) for decaying MHD turbulence; and 
for statistically steady MHD turbulence (a.3) ${\rm Pr_M}=0.01$ (R1D), 
(b.3) ${\rm Pr_M}=0.1$ (R2D), (c.3) ${\rm Pr_M}=1.0$ (R3D), and 
(d.3) ${\rm Pr_M}=10.0$ (R4D).}
\label{fig:dissipation-spectra}
\end{figure} 

\begin{figure}[htb]
\begin{center}
\includegraphics[width=0.4\textwidth]{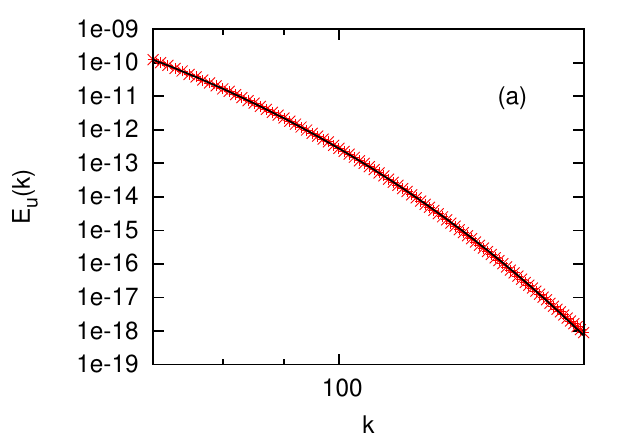}
\includegraphics[width=0.4\textwidth]{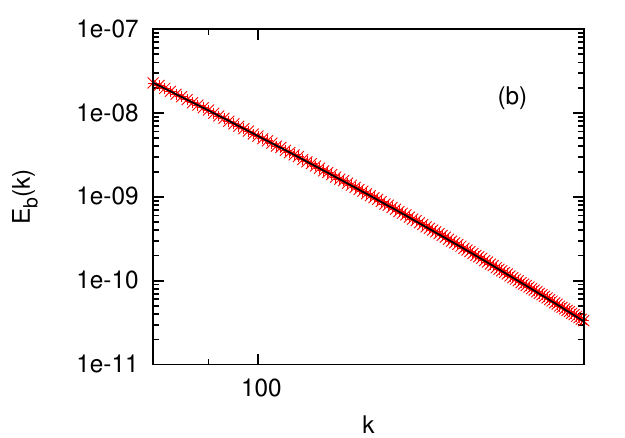}
\end{center}
\label{fig:deepdiss-spectra}
\caption{(a) The kinetic energy spectrum $E_u(k)$ (red asterisks) deep in the
dissipation range for run R5; the black line indicates the fit $E_u(k) \sim
k^{2.68}\exp(-0.235~k)$ for $5k^u_d < k < 10k^u_d$, where $k^u_d
=1/\eta_d^u$;
(b) the magnetic energy spectrum $E_b(k)$ (red asterisks) deep in the
dissipation range for run R1C; the black line indicates the fit $E_b(k) \sim
k^{-5.24}\exp(-0.014~k)$ for $5k^b_d < k < 10k^b_d$, where $k^b_d
=1/\eta_d^b$.}
\end{figure}
\begin{figure}[htb]
\begin{center}
\includegraphics[width=0.95\textwidth]{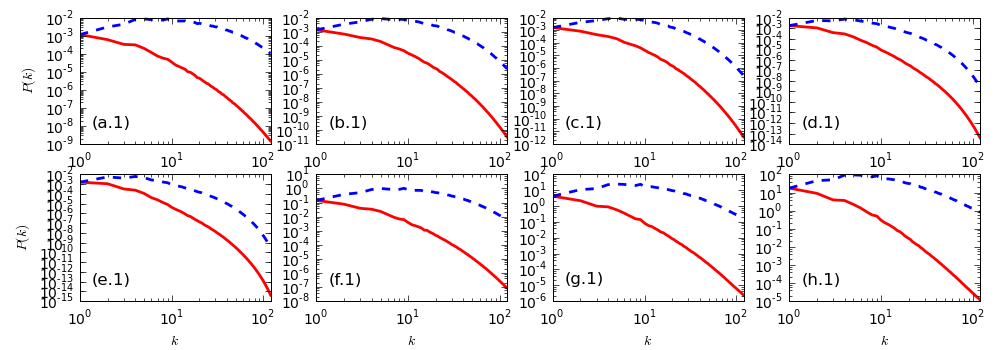}\\
\includegraphics[width=0.95\textwidth]{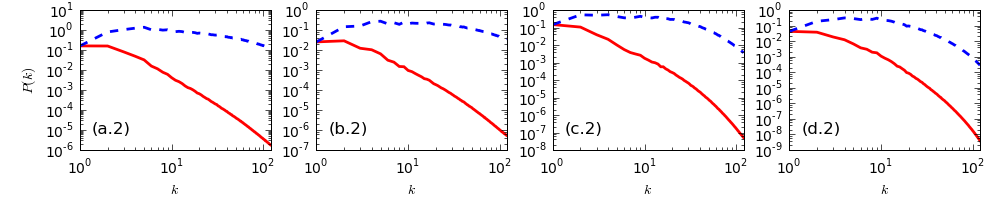}\\
\end{center}
\caption[]{Log-log (base 10) plots of effective pressure spectra
$P(k)$ (red full lines), with $k$ the magnitude of the wave vector, and the
corresponding compensated spectra $P(k) k^{7/3}$ (blue, dashed lines): (a.1)
${\rm Pr_M}=0.1$ (R1), (b.1) ${\rm Pr_M}=0.5$ (R2), (c.1) ${\rm Pr_M}=1.0$
(R3), (d.1) ${\rm Pr_M}=5.0$ (R4), (e.1) ${\rm Pr_M}=10.0$ (R5), (f.1) ${\rm
Pr_M}=1.0$ (R3B), (g.1) ${\rm Pr_M}=5.0$ (R4B), (h.1) ${\rm Pr_M}=10.0$
(R5B), 
for decaying MHD turbulence; and for statistically steady MHD turbulence
(a.2) ${\rm Pr_M}=0.01$ (R1D), (b.2) ${\rm Pr_M}=0.1$ (R2D), (c.2) ${\rm
Pr_M}=1.0$ (R3D), and (d.2) ${\rm Pr_M}=10.0$ (R4D).}
\label{fig:pressure-spectra}
\end{figure} 

We now discuss the behaviours of the energy, kinetic-energy, magnetic-energy,
Els\"asser variable, dissipation-rate, and effective-pressure spectra, first
for decaying and then for statistically steady MHD turbulence.  In the former
case, spectra are obtained at the cascade-completion time $t_c$; in the
latter, they are averaged over the statistically steady state that we obtain. 

We present compensated spectra of the total energy
$E_c(k)=\epsilon^{-2/3}k^{5/3}E(k)$ (red full line), the kinetic energy
$E_c^u(k)=\epsilon^{-2/3}k^{5/3}E_u(k)$ (green dashed line), and the total
magnetic energy $E_c^b(k)=\epsilon^{-2/3}k^{5/3}E_b(k)$ (blue dotted line) at
$t_c$ for runs R1-R5 [Figs.~\ref{fig:energy-spectra}(a.1)-\ref{fig:energy-spectra}(e.1)], R3B-R5B
[Figs.~\ref{fig:energy-spectra}(f.1)-\ref{fig:energy-spectra}(h.1)], and
R1C-R4C [Figs.~\ref{fig:energy-spectra}(a.2)-\ref{fig:energy-spectra}(d.2)]
for decaying MHD turbulence; and runs R1D-R4D
[Figs.~\ref{fig:energy-spectra}(a.3)-\ref{fig:energy-spectra}(d.3)] show these
for statistically steady MHD turbulence. From
Figs.~\ref{fig:energy-spectra}(a.1)-\ref{fig:energy-spectra}(e.1) and
Table~\ref{table:param} we see that $\eta_d^u$ increases as we increase $\nu$
to increase ${\rm Pr_M}$, because the initial energy is the same for runs
R1-R5, so the dissipation tail in $E_c^u(k)$ is drawn in towards smaller and
smaller values of $k$ as we move from ${\rm Pr_M}=0.1$ to ${\rm Pr_M}=10$;
between ${\rm Pr_M}=0.5$ and ${\rm Pr_M}=1$ the tails of $E_c^u(k)$ and
$E_c^b(k)$ and eventually $E_c^b(k)$ dominates and becomes indistinguishable
from $E_c(k)$ on the scales of Figs.~\ref{fig:energy-spectra}(d.1) and
\ref{fig:energy-spectra}(e.1). A comparison of
Figs.~\ref{fig:energy-spectra}(f.1)-\ref{fig:energy-spectra}(h.1) shows that,
if we increase ${\rm Pr_M}$ from $1$ to $10$, we can keep both $k_{\rm
max}\eta_d^u$ and $k_{\rm max}\eta_d^b$ close to $1$ so the dissipation
ranges of these spectra span comparable ranges of $k$; however, as ${\rm
Pr_M}$ increases, more and more of the energy is concentrated in the magnetic
field. These trends are not affected (a) if we increase the number of
collocation points, as can be seen from the compensated spectra in
Figs.~\ref{fig:energy-spectra}(a.2)-\ref{fig:energy-spectra}(d.2) for runs
R1C-R4C, which use $1024^3$ collocation points, or (b) if we study energy
spectra for statistically steady MHD turbulence  as can be seen from the
compensated spectra in
Figs.~\ref{fig:energy-spectra}(a.3)-\ref{fig:energy-spectra}(d.3) for runs
R1D-R4D. Figures~\ref{fig:energy-spectra}(c.1),
\ref{fig:energy-spectra}(g.1), \ref{fig:energy-spectra}(c.2), and
\ref{fig:energy-spectra}(c.3), for runs R3 (${\rm Re}_{\lambda}=121$), R3B
(${\rm Re}_{\lambda}=210$), R3C (${\rm Re}_{\lambda}=172$), and R3D (${\rm Re}_{\lambda}=239$),
respectively, all lie in one column and all have ${\rm Pr_M}=1$; so they
provide a convenient way of comparing the ${\rm Re}_{\lambda}$ dependence of these
spectra with a fixed value of ${\rm Pr_M}=1$. All the spectra in subfigures
of Fig.~\ref{fig:energy-spectra} have been compensated by the $5/3$ power of
$k$ and, to the extent that they show small, flat parts, their
inertial-range, energy-spectra scalings are consistent with $k^{-5/3}$
behaviours; other powers, such as $-3/2$, can also give small, flat parts in
compensated spectra.  A detailed error analysis is required to decide which
power is most consistent with our data; we defer such an error analysis to
the Subsection~\ref{sec:stfn} where we carry it out for structure functions.  

Compensated spectra of the Els\"asser variables, namely, $E^+_c(k) =
\epsilon^{-2/3}k^{5/3}E^+(k)$ (red full lines) and $E^-_c(k) =
\epsilon_u^{-2/3}k^{5/3}E^-(k)$ (blue dashed lines) are shown, at the
cascade-completion time $t_c$, for the decaying-MHD-turbulence runs R1-R5 in
Figs.~\ref{fig:zspectra}(a.1)-\ref{fig:zspectra}(e.1), R3B-R5B in
Figs.~\ref{fig:zspectra}(f.1)-\ref{fig:zspectra}(h.1), and R1C-R4C in
Figs.~\ref{fig:zspectra}(a.2)-\ref{fig:zspectra}(d.2); and
Figs.~\ref{fig:energy-spectra}(a.3)-\ref{fig:zspectra}(d.3) show these
spectra for statistically steady MHD turbulence in runs R1D-R4D,
respectively. Note that the dissipation ranges of $E^+_c(k)$ and $E^-_c(k)$
overlap nearly on the scales of these figures. Differences between these are
most pronounced at small $k$, where, typically, $E^-_c(k)$ lies below
$E^+_c(k)$; these differences decrease with increasing ${\rm Pr_M}$, if we
hold the initial energy fixed as in
Figs.~\ref{fig:zspectra}(a.1)-\ref{fig:zspectra}(e.1) for runs R1-R5.

Next we come to the energy-dissipation (or enstrophy) spectra $\epsilon_u(k)
= k^2E_u(k)$ (red full line) and $\epsilon_b(k) = k^2E_b(k)$ (blue dashed
line) at $t_c$. These are shown, at the cascade-completion time $t_c$, for
the decaying-MHD-turbulence runs R1-R5 in
Figs.~\ref{fig:dissipation-spectra}(a.1)-\ref{fig:dissipation-spectra}(e.1),
R3B-R5B in
Figs.~\ref{fig:dissipation-spectra}(f.1)-\ref{fig:dissipation-spectra}(h.1),
and R1C-R4C in
Figs.~\ref{fig:dissipation-spectra}(a.2)-\ref{fig:dissipation-spectra}(d.2);
and
Figs.~\ref{fig:dissipation-spectra}(a.3)-\ref{fig:dissipation-spectra}(d.3)
depict these spectra for statistically steady MHD-turbulence runs R1D-R4D. To
the extent that most of these spectra show maxima at values of $k$ at the
beginning of the dissipation range, most of our runs have well-resolved
dissipation ranges; this also follows from the values of $k_{\rm
max}\eta_d^u$ and $k_{\rm max}\eta_d^b$ in Table~\ref{table:param}. Runs R1D
and R2D have slightly under-resolved fluid-dissipation ranges with $k_{\rm
max}\eta_d^u \simeq 0.32$ and $0.33$, respectively; and, for the former, a
barely discernible, dissipation-range maximum in $\epsilon_u(k)$; however, as
shown in our Navier-Stokes DNS in Subsection~\ref{sec:NS}, reasonable results
can be obtained for various statistical properties with $k_{\rm max}\eta_d^u
\simeq 0.3$.  The elucidation of the behaviours of dissipation-range spectra
of course require large values of $k_{\rm max}\eta_d^u$ or
$k_{\rm max}\eta_d^b$; in
particular, runs R5 and R1C, with $k_{\rm max}\eta_d^u \simeq 13.3$ and
$k_{\rm max}\eta_d^b \simeq 22.1$, respectively, are well suited for
uncovering the functional forms of $E_u(k)$ and $E_b(k)$ in their dissipation
ranges. In Figs.~\ref{fig:deepdiss-spectra}(a) and
\ref{fig:deepdiss-spectra}(b) we show, respectively, the kinetic- and
magnetic-energy spectra $E_u(k)$ and $E_b(k)$ deep in their dissipation
ranges for runs R5 and R1, respectively; our data for these spectra can be
fit to the form $\sim k^{\alpha}\exp(-\beta k)$ for $k$ deep in the
dissipation range and $\alpha$ and $\beta$ nonuniversal numbers that depend
on the parameters of the simulation; similar results have been obtained for
fluid turbulence~\cite{chen93,schumacher07}. In particular, our data
[Figs.~\ref{fig:deepdiss-spectra}(a) and \ref{fig:deepdiss-spectra}(b)] for
runs R5 and R1C are consistent with $E_u(k) \sim k^{2.68}\exp(-0.235~k)$, for
$5k^u_d < k < 10k^u_d$ with $k^u_d =1/\eta_d^u$, and $E_b(k) \sim
k^{-5.24}\exp(-0.014~k)$, for $5k^b_d < k < 10k^b_d$ with $k^b_d
=1/\eta_d^b$, respectively.

We now turn to the spectra for the effective pressure $P(k)$ (red full lines)
and their compensated versions $k^{7/3}P(k)$ (blue dashed lines) that are
shown at $t_c$ for runs R1-R5
[Figs.~\ref{fig:pressure-spectra}(a.1)-\ref{fig:pressure-spectra}(e.1)] and
R3B-R5B
[Figs.~\ref{fig:pressure-spectra}(f.1)-\ref{fig:pressure-spectra}(h.1)] for
decaying MHD turbulence; and for statistically steady MHD turbulence they are
shown in
Figs.~\ref{fig:pressure-spectra}(a.2)-\ref{fig:pressure-spectra}(d.2) for
runs R1D-R4D.  Pressure spectra have been studied for fluid turbulence as,
e.g., in Refs.~\cite{kalelkar06,gotoh01}; to the best of our knowledge they
have not been obtained for MHD turbulence. The compensated spectra here show
that, for all our runs, the inertial-range behaviours of these
effective-pressure spectra are consistent with the power law $k^{-7/3}$; this
is consistent with the $k^{-5/3}$ behaviours of the energy spectra discussed
above.  Furthermore, as ${\rm Pr_M}$ increases from $0.1$ to $10$ in runs
R1-R5, $P(k)$ falls more and more rapidly as can be seen from the vertical
scales in
Figs.~\ref{fig:pressure-spectra}(a.1)-\ref{fig:pressure-spectra}(e.1).

\subsection{Probability distribution functions \label{sec:pdfs}}

We calculate several PDFs to characterise the statistical properties of
decaying and statistically steady MHD turbulence. In the former case, PDFs
are obtained at the cascade-completion time $t_c$; in the latter, they are
averaged over the statistically steady state that we obtain.  The PDFs we
consider are of two types: the first type are PDFs of the cosines of angles
between various vectors, such as $\bfu$ and $\bomega$; these help us to
quantify the degrees of alignment between such vectors; the second type are
PDFs of quantities such as $\epsilon_u$, $\epsilon_b$, and the
eigenvalues of the rate-of-strain tensor.

\begin{figure}[htb]
\begin{center}
\includegraphics[width=0.95\textwidth]{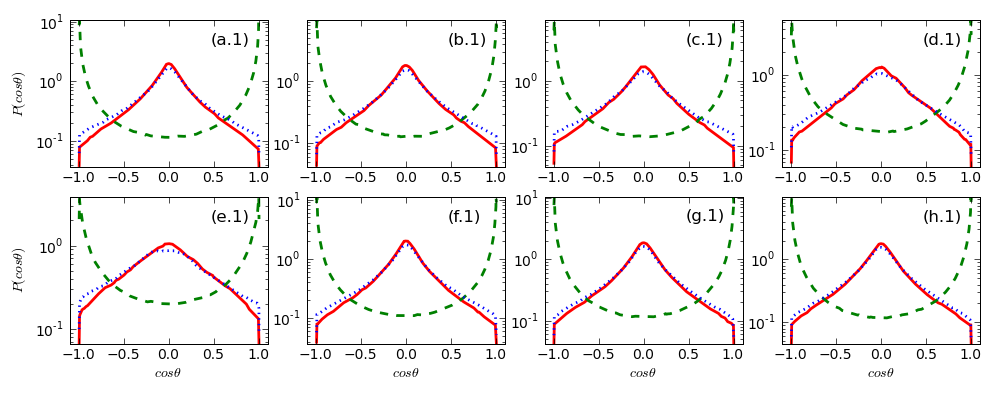}\\
\end{center}
\caption[]{Semilog (base 10) plots of the PDFs of cosines
of the angles, denoted generically by $\theta$, between
the vorticity $\bomega$ and the eigenvectors of 
the fluid rate-of-strain tensor $\mathbb S$, namely, $\hat e^1_u$ 
(red full line),$\hat e^2_u$ (green dashed line), and $\hat e^3_u$ 
(blue dotted line): (a.1) ${\rm Pr_M}=0.1$ (R1), (b.1) ${\rm Pr_M}=0.5$ (R2), 
(c.1) ${\rm Pr_M}=1.0$ (R3), (d.1) ${\rm Pr_M}=5.0$ (R4), (e.1) ${\rm Pr_M}=10.0$ 
(R5), (f.1) ${\rm Pr_M}=1.0$ (R3B), (g.1) ${\rm Pr_M}=5.0$ (R4B), and 
(h.1) ${\rm Pr_M}=10.0$ (R5B) for decaying MHD turbulence.}
\label{fig:angle-w-eu}
\end{figure}

\begin{figure}[htb]
\begin{center}
\includegraphics[width=0.95\textwidth]{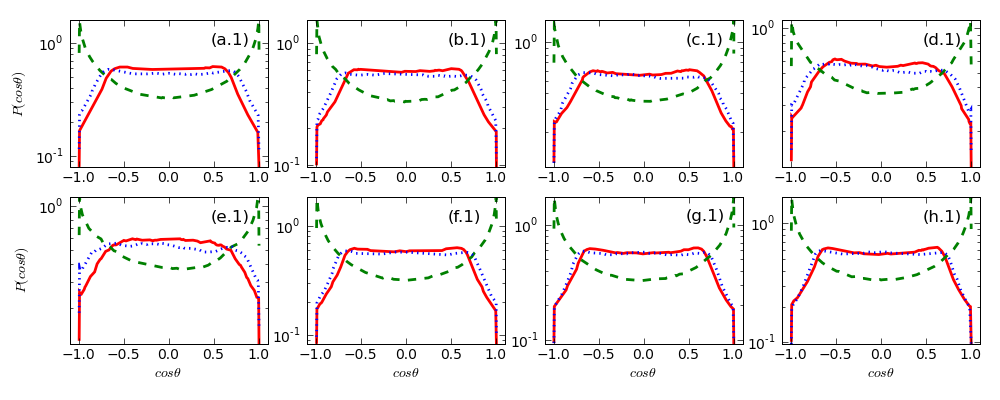}\\
\end{center}
\caption[]{Semilog (base 10) plots of the PDFs cosines of
angles, denoted generically by $\theta$,
between the current density $\bfj$ and the eigenvectors of 
fluid rate-of-strain tensor  $\mathbb S$, namely, $\hat e^1_u$ 
(red full line), $\hat e^2_u$ (green dashed line), and $\hat e^3_u$ 
(blue dotted line): (a.1) ${\rm Pr_M}=0.1$ (R1), (b.1) ${\rm Pr_M}=0.5$ (R2),
(c.1) ${\rm Pr_M}=1.0$ (R3), (d.1) ${\rm Pr_M}=5.0$ (R4),
(e.1) ${\rm Pr_M}=10.0$ (R5), (f.1) ${\rm Pr_M}=1.0$ (R3B),
(g.1) ${\rm Pr_M}=5.0$ (R4B), and (h.1) ${\rm Pr_M}=10.0$
(R5B) for decaying MHD turbulence.}
\label{fig:angle-j-eu}
\end{figure}

In Fig.~\ref{fig:angle-w-eu} we show plots of the PDFs of cosines of the
angles between the vorticity $\bomega$ and the eigenvectors of the fluid
rate-of-strain tensor $\mathbb S$, namely, $\hat e^1_u$ (red full line),
$\hat e^2_u$ (green dashed lines), and $\hat e^3_u$ (blue dotted lines) for
runs R1-R5 and R3B-R5B at the cascade-completion time $t_c$ for the case of
decaying MHD turbulence.  In Fig.~\ref{fig:angle-j-eu} we show similar plots
of the PDFs of cosines of the angles between the current density $\bfj$ and
the eigenvectors of the fluid rate-of-strain tensor $\mathbb S$.  The most
important features of these figures are sharp peaks in the green dashed
lines; these show that there is a marked tendency for the alignment or
antialignment of $\bomega$ and $\hat e^2_u$, as in fluid turbulence, and of
a similar tendency for the alignment or antialignment of $\bfj$ and $\hat
e^2_u$; these features do not depend very sensitively on ${\rm Pr_M}$.
Furthermore, the PDFs of cosines of the angles between $\bomega$ and $\hat
e^1_u$ (blue dotted lines) and $\bomega$ and $\hat e^3_u$ (red full lines)
show peaks near zero in Fig.~\ref{fig:angle-w-eu}; in contrast, analogous
PDFs for the cosines of the angles between $\bfj$ and $\hat e^1_u$ (red full
lines) and $\bomega$ and $\hat e^3_u$ (blue dotted lines) show nearly flat
plateaux in the middle with very gentle maxima near $-0.5$ and $0.5$
[Fig.~\ref{fig:angle-j-eu}]. Runs R1C-R4C and R1D-R4D yield similar PDFs,
for the cosines of these angles, so we do not give them here. 

\begin{figure}[htb]
\begin{center}
\includegraphics[width=0.95\textwidth]{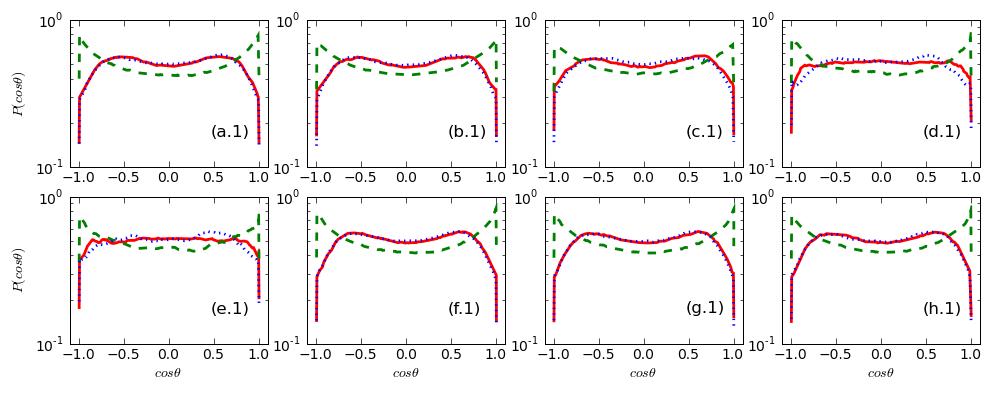}\\
\end{center}
\caption[]{Semilog (base 10) plots of the PDFs of cosines
of angles, denoted generically by $\theta$, between the velocity
$\bfu$ and the eigenvectors of the fluid rate-of-strain tensor
$\mathbb S$, namely, $\hat e^1_u$ (red full line), $\hat e^2_u$
(green dashed line), and $\hat e^3_u$ (blue dotted line):
(a.1) ${\rm Pr_M}=0.1$ (R1), (b.1) ${\rm Pr_M}=0.5$ (R2),
(c.1) ${\rm Pr_M}=1.0$ (R3), (d.1) ${\rm Pr_M}=5.0$ (R4),
(e.1) ${\rm Pr_M}=10.0$ (R5), (f.1) ${\rm Pr_M}=1.0$ (R3B),
(g.1) ${\rm Pr_M}=5.0$ (R4B), and (h.1) ${\rm Pr_M}=10.0$
(R5B) for decaying MHD turbulence.}
\label{fig:angle-u-eu}
\end{figure}

\begin{figure}[htb]
\begin{center}
\includegraphics[width=0.95\textwidth]{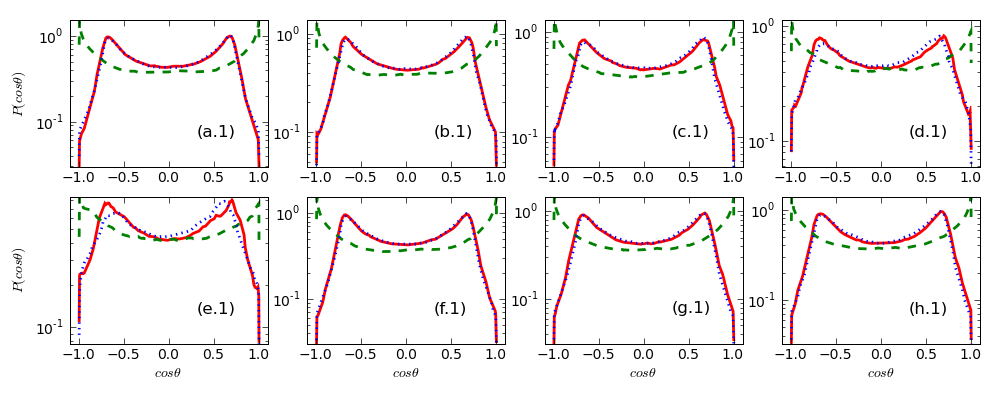}\\
\end{center}
\caption[]{Semilog (base 10) plots of the PDFs of cosines
of angles, denoted generically by $\theta$, between the magnetic
field $\bfb$ and the eigenvectors of fluid rate-of-strain tensor
$\mathbb S$, namely, $\hat e^1_u$ (red full line), $\hat e^2_u$
(green dashed line), and $\hat e^3_u$ (blue dotted line):
(a.1) ${\rm Pr_M}=0.1$ (R1), (b.1) ${\rm Pr_M}=0.5$ (R2),
(c.1) ${\rm Pr_M}=1.0$ (R3), (d.1) ${\rm Pr_M}=5.0$ (R4),
(e.1) ${\rm Pr_M}=10.0$ (R5), (f.1) ${\rm Pr_M}=1.0$ (R3B),
(g.1) ${\rm Pr_M}=5.0$ (R4B), and (h.1) ${\rm Pr_M}=10.0$ (R5B).}
\label{fig:angle-b-eu}
\end{figure}

Plots of the PDFs of cosines of the angles between the velocity
$\bfu$ and the eigenvectors of the fluid rate-of-strain tensor $\mathbb S$
are given Fig.~\ref{fig:angle-u-eu}; their analogues for $\bfb$ are given
Fig.~\ref{fig:angle-b-eu}. Again, the most prominent features of these
figures are sharp peaks in the green dashed lines; these show that there is a
marked tendency for the alignment or antialignment of $\bfu$ and $\hat
e^2_u$ and of a similar tendency for the alignment or antialignment of
$\bfj$ and $\hat e^2_u$; these features do not depend very sensitively on
${\rm Pr_M}$. The PDFs of cosines of the angles between $\bfu$ and $\hat
e^1_u$ (red full line) and $\bfu$ and $\hat e^3_u$ (blue dotted lines) show
gentle, broad peaks that imply a weak preference for angles close to
$45^\circ$ or $135^\circ$; these peaks are suppressed as we increase ${\rm
Pr_M}$ [Figs.~\ref{fig:angle-u-eu}(a.1)-\ref{fig:angle-u-eu}(e.1) for runs
R1-R5] with fixed initial energy, but they reappear if we compensate for the
increase of ${\rm Pr_M}$ by increasing the initial energy
[Figs.~\ref{fig:angle-u-eu}(f.1)-\ref{fig:angle-u-eu}(h.1)]. Similar, but
sharper, peaks appear in the PDFs of cosines of the angles between $\bfu$
and $\hat e^1_u$ (red full lines) and $\bfu$ and $\hat e^3_u$ (blue dotted
lines); these show a weak preference for angles close to $47^\circ$ or
$133^\circ$ [Fig.~\ref{fig:angle-b-eu}].  Some simulations of compressible
MHD turbulence have noted the presence of such
peaks~\cite{brandenburg95} for ${\rm Pr_M}=1$.

Only one of the eigenvalues $\Lambda_1^b$ of the tensile magnetic
stress tensor $\mathbb T$ is non-zero; and the corresponding
eigenvector ${\hat e}_b^1$ is identically aligned with $\bfb$.
Thus PDFs of cosines of angles between $\bfu$, $\bomega$, $\bfj$,
and $\bfb$ and the eigenvectors of $\mathbb T$ are simpler than
their counterparts for $\mathbb S$ and are not presented here.

\begin{figure}[htb]
\begin{center}
\includegraphics[width=0.95\textwidth]{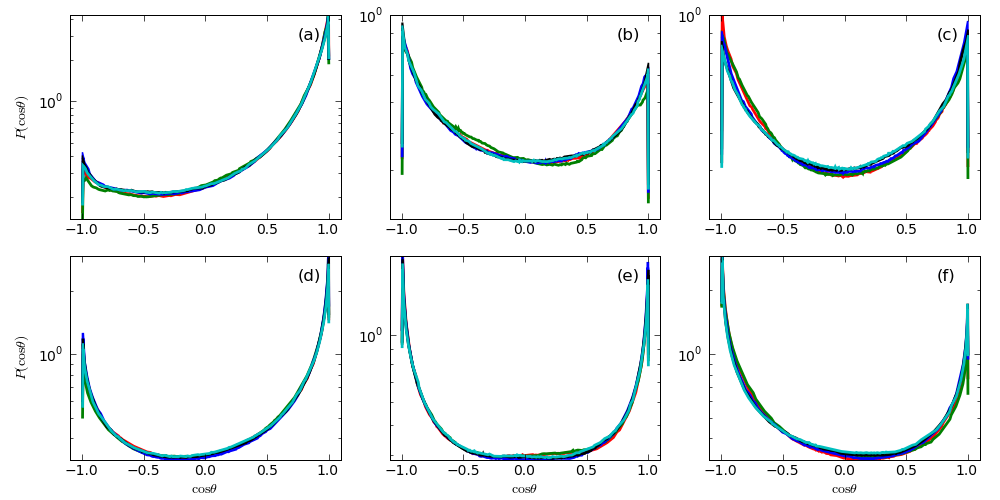}\\
\end{center}
\caption[]{Semilog (base 10) plots of PDFs of cosines of angles,
denoted generically by $\theta$, between (a) $\bfu$ and $\bfb$, (b) $\bfu$ and $\bomega$, 
(c) $\bfu$ and $\bfj$, (d) $\bomega$ and $\bfj$, (e) $\bfb$ and $\bomega$,
and (f) $\bfb$ and $\bfj$ for runs R1 (red lines), R2 (green lines), 
R3 (blue lines), R4 (black lines), and R5 (cyan lines).}
\label{fig:angle-wj-ub}
\end{figure}

Figure~\ref{fig:angle-wj-ub} shows plots of PDFs of cosines of angles,
denoted generically by $\theta$, between (a) $\bfu$ and $\bfb$, (b) $\bfu$
and $\bomega$, (c) $\bfu$ and $\bfj$, (d) $\bomega$ and $\bfj$, (e) $\bfb$
and $\bomega$, and (f) $\bfb$ and $\bfj$ for runs R1 (red lines), R2 (green
lines), R3 (blue lines), R4 (black lines), and R5 (cyan lines).  These
figures show the following: (a) $\bfu$ and $\bfb$ are more aligned than
antialigned [this is related to the small, positive, mean values of $H_C$
(see below) in our runs R1-R5]; (b) $\bfu$ and $\bomega$ and more
antialigned than aligned, as noted for decaying fluid turbulence with slight
helicity in Refs.~\cite{kalelkar06,kit87}; (c) $\bfu$ and $\bfj$ show
approximately equal tendencies for alignment and antialignment; (d) $\bomega$
and $\bfj$ display a greater tendency for alignment than antialignment; (e)
$\bfb$ and $\bomega$ have approximately equal tendencies for alignment and
antialignment; and (f) $\bfb$ and $\bfj$ are more antialigned than aligned.

\begin{figure}[htb]
\begin{center}
\includegraphics[width=0.95\textwidth]{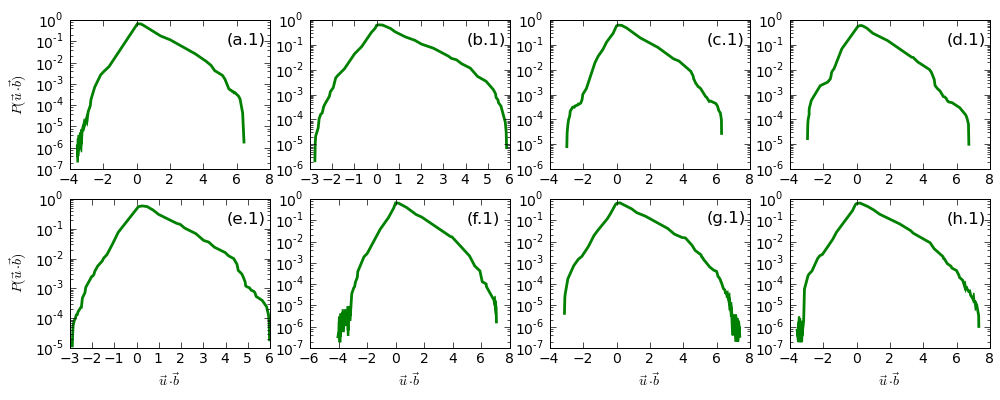}\\
\includegraphics[width=0.95\textwidth]{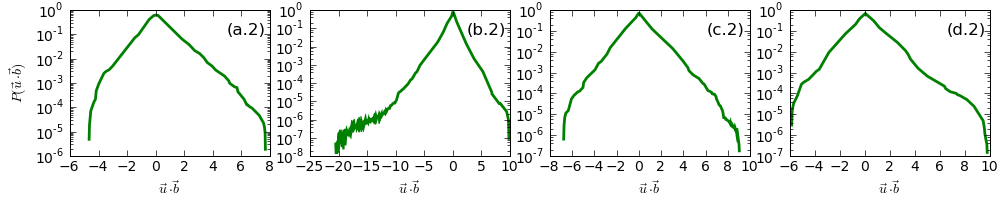}\\
\includegraphics[width=0.95\textwidth]{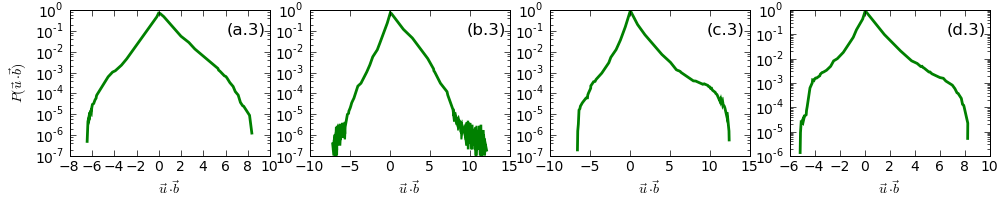}\\
\end{center}
\caption[]{Semilog (base 10) plots of the PDFs of the cross
helicity $H_C = \bfu\cdot\bfb$ for (a.1) ${\rm Pr_M}=0.1$ (R1), 
(b.1) ${\rm Pr_M}=0.5$ (R2), (c.1) ${\rm Pr_M}=1.0$ (R3),
(d.1) ${\rm Pr_M}=5.0$ (R4), (e.1) ${\rm Pr_M}=10.0$ (R5), 
(f.1) ${\rm Pr_M}=1.0$ (R3B), (g.1) ${\rm Pr_M}=5.0$ (R4B),
(h.1) ${\rm Pr_M}=10.0$ (R5B), (a.2) ${\rm Pr_M}=0.01$ (R1C), (b.2) 
${\rm Pr_M}=0.1$ (R2C), (c.2) ${\rm Pr_M}=1.0$ (R3C), and 
(d.2) ${\rm Pr_M}=10.0$ (R4C) for decaying MHD turbulence; and 
for statistically steady MHD turbulence (a.3) ${\rm Pr_M}=0.01$ (R1D), 
(b.3) ${\rm Pr_M}=0.1$ (R2D), (c.3) ${\rm Pr_M}=1.0$ (R3D), and 
(d.3) ${\rm Pr_M}=10.0$ (R4D); the arguments of the PDFs are scaled by their 
standard deviations $\sigma_{H_C}$.}
\label{fig:vdotb}
\end{figure} 
\begin{table}
\caption{The mean $\mu_{H_C}$, standard deviation $\sigma_{H_C}$,  skewness 
$\mu_{3,H_C}$, and kurtosis $\mu_{4,H_C}$ of
the PDF of the cross helicity ${H_C}$ for our runs 
R1-R5 and R3B-R5B for decaying MHD turbulence at cascade
completion; columns 6 and 7 give, respectively, the mean energy $E$
and ratio of the means of the
cross helicity and the energy, i.e., $\mu_{H_C}/E$.}
\label{table:hc}
\begin{center}
\begin{tabular}{lllllll}
\hline\noalign{\smallskip}
Run~~~ & $\mu_{H_C}$~~~ & $\sigma_{H_C}$~~~ & $\mu_{3,H_C}$ & $\mu_{4,H_C}$ & $E$ & $\mu_{H_C}/E$ \\
\noalign{\smallskip}\hline\noalign{\smallskip}
R1  &  0.118  &  0.173  &  1.103  &  4.901 & 0.461 &  0.256 \\
R2  &  0.118  &  0.169  &  1.096  &  4.685 & 0.467 &  0.252 \\
R3  &  0.120  &  0.170  &  1.096  &  4.679 & 0.490 &  0.245 \\
R4  &  0.112  &  0.153  &  1.003  &  4.579 & 0.477 &  0.235 \\
R5  &  0.105  &  0.141  &  0.934  &  4.324 & 0.460 &  0.228 \\
R3B &  1.217  &  1.804  &  1.100  &  4.912 & 4.909 &  0.248 \\
R4B &  5.915  &  8.766  &  1.097  &  4.917 & 24.50 &  0.241 \\
R5B &  11.50  &  17.05  &  1.102  &  5.000 & 48.32 &  0.238 \\
R1C &  0.014  &  0.113  &  0.615  &  5.748 & 0.358 &  0.041 \\
R2C & -0.224  &  1.994  & -0.698  &  8.441 & 5.440 & -0.041 \\
R3C &  0.130  &  2.005  &  0.313  &  5.637 & 5.969 &  0.022 \\
R4C &  0.859  &  9.156  &  0.364  &  5.747 & 29.05 &  0.029 \\
R1D &  0.169  &  0.724  &  0.737  &  6.813 & 3.090 &  0.055 \\
R2D &  0.478  &  0.886  &  1.126  &  5.954 & 2.405 &  0.199 \\
R3D &  0.454  &  1.244  &  1.904  &  13.75 & 3.039 &  0.149 \\
R4D &  0.389  &  1.110  &  1.207  &  9.225 & 2.767 &  0.140 \\
\noalign{\smallskip}\hline
\end{tabular}
\end{center}
\end{table}

Probability distribution functions of the local cross helicity $H_C =
\bfu\cdot\bfb$ are shown via green full lines in Fig.~\ref{fig:vdotb}.  The
arguments of these PDFs are scaled by their standard deviations, namely,
$\sigma_{H_C}$; data for the PDFs are obtained at $t_c$ for runs R1-R5 in
Figs.~\ref{fig:vdotb}(a.1)-\ref{fig:vdotb}(e.1), runs R3B-R5B in Figs.~\ref{fig:vdotb}(f.1)-\ref{fig:vdotb}(h.1),
and runs R1C-R4C in Figs.~\ref{fig:vdotb}(a.2)-\ref{fig:vdotb}(d.2) for decaying MHD
turbulence.  For statistically steady MHD turbulence these PDFs are shown in
Figs.~\ref{fig:vdotb}(a.3)-\ref{fig:vdotb}(d.3) for runs R1D-R4D. All these PDFs have peaks
close to $H_C=0$; this reflects the tendency for $\bfu$ and $\bfb$ to be
aligned or  antialigned that we have discussed above. However, these PDFs
are quite broad and distinctly non-Gaussian; this can be seen easily from the
values of the mean $\mu_{H_C}$, standard deviation $\sigma_{H_C}$, skewness
$\gamma_{3,H_C}$, and kurtosis $\gamma_{4,H_C}$ given in
Table~\ref{table:hc}. Thus fluctuations of $H_C$ away from the mean are very
significant. Table~\ref{table:hc} also gives the value of the mean energy $E$
and the ratio $E/\mu_{H_C}$, which does not appear to be universal; for the
runs R1-R5 and R3B-R5B it lies in the range $0.23$-$0.26$, for R1C-R2C in the
range $-0.04$-$0.04$, and for R1D-R4D in the range $0.05$-$0.2$. For all our
runs, with the exception of R2C, the mean $\mu_{H_C}$ and the skewness
$\gamma_{3,H_C}$ are positive.  Even if the PDF of $H_C$ had been a Gaussian,
its mean value would have been within one standard deviation of
0; the actual PDF is much broader than a Gaussian.
On symmetry grounds there is no reason for the
system to display a nonzero value for $\mu_{H_C}$ unless there is some bias
in the forcing or in the initial condition (the latter for the case of
decaying turbulence). In any given run, if there is some residual $H_C$, it
is reflected in a slight asymmetry in alignment (or antialignment) of $\bfu$
and $\bfb$, which we have studied above via the PDF of the cosine of the
angle between $\bfu$ and $\bfb$.  When we consider the ratio $\mu_{H_C}/E$ it
seems to be substantial in some runs but, given the arguments above, we
expect it to vanish in runs with a very large number of collocation
points; indeed, it is very small in runs R1C-R4C.

\begin{figure}[htb]
\begin{center}
\includegraphics[width=0.95\textwidth]{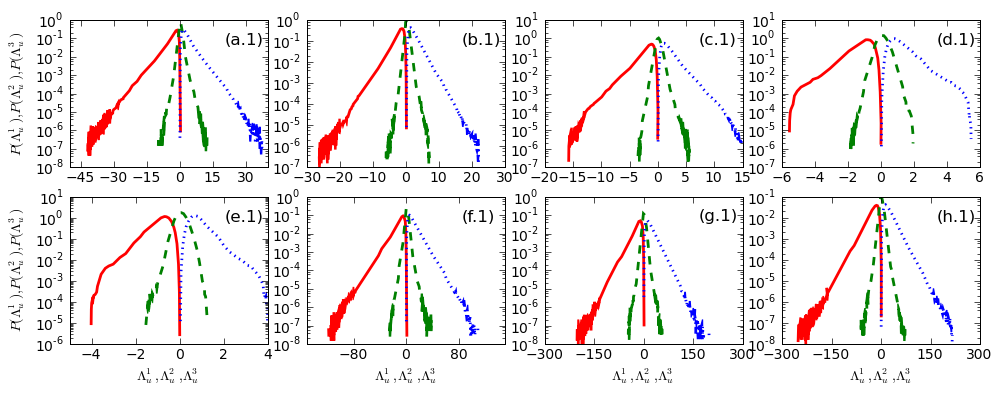}\\
\end{center}
\caption[] {Semilog (base 10) plots of PDFs of the eigenvalues
$\Lambda^1_u$ (blue dotted line), $\Lambda^2_u$ (green dashed line),
and $\Lambda^3_u$ (red full line) of the rate-of-strain tensor 
$\mathbb S$ for (a.1) ${\rm Pr_M}=0.1$ (R1), (b.1) ${\rm Pr_M}=0.5$ (R2),
(c.1) ${\rm Pr_M}=1.0$ (R3), (d.1) ${\rm Pr_M}=5.0$ (R4), 
(e.1) ${\rm Pr_M}=10.0$ (R5), (f.1) ${\rm Pr_M}=1.0$ (R3B),
(g.1) ${\rm Pr_M}=5.0$ (R4B), and (h.1) ${\rm Pr_M}=10.0$ (R5B);
the arguments of the PDFS are scaled by their standard deviations.}
\label{fig:lambda-u}
\end{figure}

Consider now the PDFs of the eigenvalues $\Lambda^1_u$ (blue dotted line),
$\Lambda^2_u$ (green dashed line), and $\Lambda^3_u$ (red full line) of the
rate-of-strain tensor $\mathbb S$ shown in
Figs.~\ref{fig:lambda-u}(a.1)-\ref{fig:lambda-u}(e.1) R1-R5 and
Figs.~\ref{fig:lambda-u}(f.1)-\ref{fig:lambda-u}(h.1) for runs R3B-R5B. Recall
that these eigenvalues provide measures of the local stretching and
compression of the fluid; also we label the eigenvalues such that
$\Lambda^1_u > \Lambda^2_u > \Lambda^3_u$. The incompressibility condition
yields $\sum_{n=1}^3 \Lambda^n_u = 0$, whence it follows that $\Lambda^1_u >
0$ and $\Lambda^3_u < 0$; the intermediate eigenvalue $\Lambda^2_u$ can be
either positive or negative.  The illustrative plots in
Figs.~\ref{fig:lambda-u}(a.1)-\ref{fig:lambda-u}(h.1) from our
decaying-MHD-turbulence runs show that the PDFs of $\Lambda^1_u$ and
$\Lambda^3_u$ have long tails on the right- and left-hand sides, respectively.
These tails shrink as we increase ${\rm Pr_M}$
[Figs.~\ref{fig:lambda-u}(a.1)-\ref{fig:lambda-u}(e.1) for runs R1-R5,
respectively], by increasing $\nu$ while holding the initial energy fixed;
thus, there is a substantial decrease in regions of large strain. However, if
we compensate for the increase in $\nu$ by increasing the energy in the
initial condition such that $k_{\rm max}\eta_d^u$ and $k_{\rm
max}\eta_d^b$ are both $\simeq 1$, we see that these tails stretch out, i.e.,
regions of large strain reappear.

\begin{figure}[htb]
\begin{center}
\includegraphics[width=0.95\textwidth]{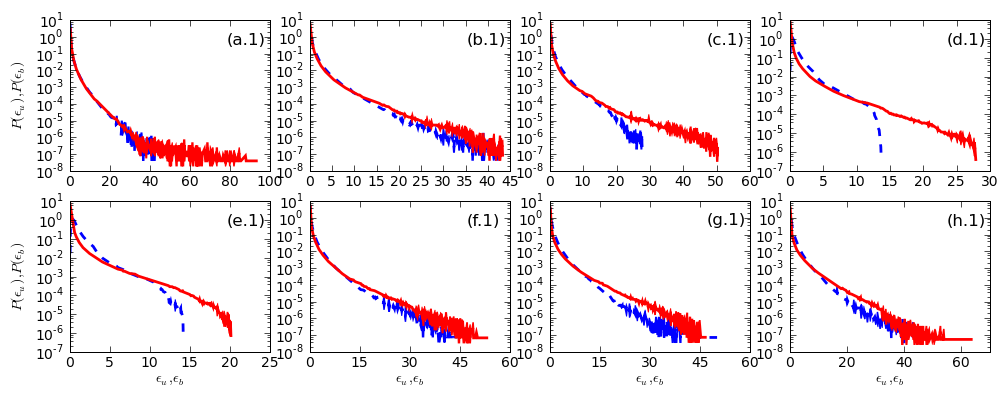}\\
\includegraphics[width=0.95\textwidth]{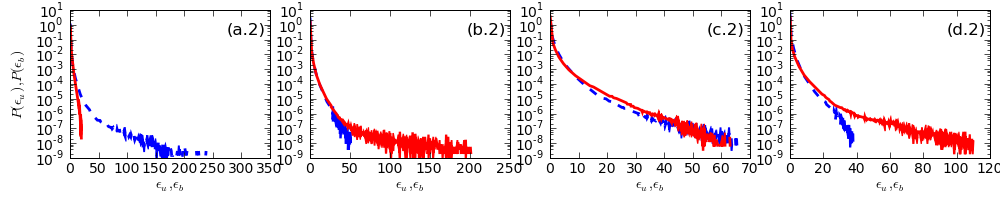}\\
\includegraphics[width=0.95\textwidth]{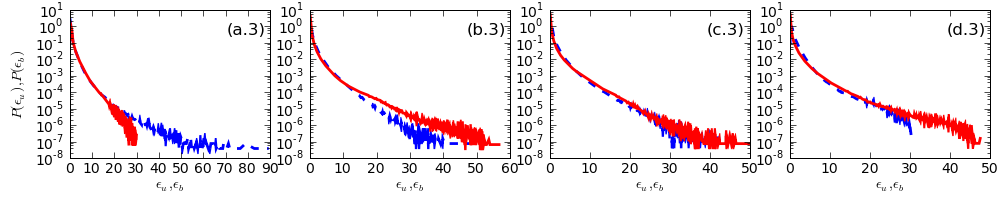}
\end{center}
\caption[] {Semilog (base 10) plots of PDFs of the local
kinetic-energy dissipation rate $\epsilon_u$ (blue dashed line) and the
magnetic-energy dissipation rate $\epsilon_b$ (red full line),
with the arguments scaled by their standard deviations, for:
(a.1) ${\rm Pr_M}=0.1$ (R1), (b.1) ${\rm Pr_M}=0.5$ (R2), (c.1) 
${\rm Pr_M}=1.0$ (R3), (d.1) ${\rm Pr_M}=5.0$ (R4), (e.1) ${\rm Pr_M}=10.0$ 
(R5), (f.1) ${\rm Pr_M}=1.0$ (R3B), 
(g.1) ${\rm Pr_M}=5.0$ (R4B), (h.1) ${\rm Pr_M}=10.0$ (R5B), 
(a.2) ${\rm Pr_M}=0.01$ (R1C), (b.2) ${\rm Pr_M}=0.1$ (R2C), 
(c.2) ${\rm Pr_M}=1.0$ (R3C), and (d.2) ${\rm Pr_M}=10.0$ (R4C) 
for decaying MHD turbulence; and for statistically steady MHD
turbulence (a.3) ${\rm Pr_M}=0.01$ (R1D), (b.3) ${\rm Pr_M}=0.1$ 
(R2D), (c.3) ${\rm Pr_M}=1.0$ (R3D), and (d.3) ${\rm Pr_M}=10.0$ (R4D).}
\label{fig:epsilon}
\end{figure} 
\begin{table}
\caption{The mean $\mu_{\epsilon_u}$, standard deviation 
$\sigma_{\epsilon_u}$, skewness $\gamma_{3,\epsilon_u}$, and kurtosis 
$\gamma_{4,\epsilon_u}$ of the PDFs of the local fluid energy dissipation 
$\epsilon_u$, and their analogues for $\epsilon_b$, for runs R1-R5, R3B-R5B,
R1C-R4C, and R1D-R4D.}
\label{table:epsu}
\begin{center}
\begin{tabular}{lllll|llll}
\hline\noalign{\smallskip}
Run~~~ & $\mu_{\epsilon_u}$~~~ & $\sigma_{\epsilon_u}$~~~ & 
$\gamma_{3, \epsilon_u}$~~~ & $\gamma_{4, \epsilon_u}$ & 
$\mu_{\epsilon_b}$~~~ & $\sigma_{\epsilon_b}$~~~ & 
$\gamma_{3, \epsilon_b}$~~~ & $\gamma_{4, \epsilon_b}$\\
\noalign{\smallskip}\hline\noalign{\smallskip}
R1  & 0.0048 & 0.0096 & 6.382 & 75.069 & 0.0302 & 0.0550 & 7.611 & 144.86 \\
R2  & 0.0109 & 0.0187 & 6.053 & 70.566 & 0.0255 & 0.0527 & 8.182 & 121.47 \\
R3  & 0.0141 & 0.0226 & 5.450 & 52.884 & 0.0233 & 0.0566 & 10.46 & 204.37 \\
R4  & 0.0231 & 0.0284 & 4.042 & 28.306 & 0.0160 & 0.0397 & 7.955 & 97.662 \\
R5  & 0.0273 & 0.0302 & 3.684 & 24.559 & 0.0130 & 0.0315 & 6.682 & 64.206 \\
R3B & 0.4165 & 0.7345 & 5.941 & 70.070 & 0.6440 & 1.5881 & 9.029 & 147.71 \\
R4B & 6.7843 & 10.898 & 5.343 & 55.676 & 4.4541 & 13.377 & 9.672 & 155.71 \\
R5B & 21.164 & 32.438 & 5.353 & 59.163 & 9.8332 & 31.177 & 9.621 & 151.64 \\
R1C & 0.0031 & 0.0076 & 18.45 & 1620.0 & 0.0566 & 0.0632 & 3.340 & 22.270 \\
R2C & 0.2354 & 0.5177 & 7.599 & 112.92 & 1.5655 & 3.5169 & 13.99 & 981.17 \\
R3C & 0.8349 & 1.6375 & 6.841 & 105.85 & 1.3186 & 3.5524 & 10.41 & 205.54 \\
R4C & 14.208 & 22.900 & 5.535 & 66.496 & 7.1624 & 24.974 & 13.33 & 406.21 \\
R1D & 0.0448 & 0.0630 & 5.799 & 89.678 & 0.8087 & 1.1004 & 4.587 & 40.328 \\
R2D & 0.0601 & 0.0933 & 5.180 & 51.311 & 0.4995 & 0.9808 & 8.488 & 142.97 \\
R3D & 0.2886 & 0.4233 & 5.230 & 55.366 & 0.6389 & 1.3120 & 6.755 & 80.154 \\
R4D & 0.4498 & 0.5536 & 5.055 & 58.832 & 0.5037 & 1.1391 & 8.077 & 129.65 \\
\noalign{\smallskip}\hline
\end{tabular}
\end{center}
\end{table}

We show PDFs of the kinetic-energy dissipation rate $\epsilon_u$ (blue dashed
lines) and the magnetic-energy dissipation rate $\epsilon_b$ (red full lines)
that are obtained at $t_c$ for runs R1-R5 in
Figs.~\ref{fig:epsilon}(a.1)-\ref{fig:epsilon}(e.1), runs R3B-R5B in
Figs.~\ref{fig:epsilon}(f.1)-\ref{fig:epsilon}(h.1), and runs R1C-R4C  in
Figs.~\ref{fig:epsilon}(a.2)-\ref{fig:epsilon}(d.2) for decaying  MHD
turbulence; and for statistically steady MHD turbulence they are shown in
Figs.~\ref{fig:epsilon}(a.3)-\ref{fig:epsilon}(d.3) for runs R1D-R4D.  All
these PDFs have long tails; the tail of the PDF for $\epsilon_b$ extends
further than the tail of that for $\epsilon_u$ for all except the smallest
values of ${\rm Pr_M}$ [Figs.~\ref{fig:epsilon}(a.1), \ref{fig:epsilon}(a.2),
\ref{fig:epsilon}(a.3) for runs R1, R1C, and R1D, respectively].  This
indicates that large values of $\epsilon_b$ are more likely to appear than
large values of $\epsilon_u$ and, given the long tails of these PDFs,
suggests that, except at the smallest values of ${\rm Pr_M}$ we have used, we
might obtain more marked intermittency for the magnetic field
than for the velocity field. Furthermore, as we expect, the tail of the PDF
of $\epsilon_u$ is drawn in towards small values of $\epsilon_u$ as we
increase ${\rm Pr_M}$ [Figs.~\ref{fig:epsilon}(a.1)-\ref{fig:epsilon}(e.1)
for runs R1-R5, respectively] while holding $\eta$ and the initial energy
fixed. However, if we compensate for the increase in $\nu$ by increasing the
initial energy so that $k_{\rm max}\eta_d^u$ and $k_{\rm max}\eta_d^b$ are
both $\simeq 1$, we see that the tails of the PDFs of $\epsilon_b$ and
$\epsilon_u$ get elongated as we increase ${\rm Pr_M}$, e.g., in
Figs.~\ref{fig:epsilon}(f.1)-\ref{fig:epsilon}(h.1) for runs R3B-R5B,
respectively.  The values of the mean $\mu_{\epsilon_u}$, standard deviation
$\sigma_{\epsilon_u}$, skewness $\gamma_{3,\epsilon_u}$, and kurtosis
$\gamma_{4,\epsilon_u}$ of the PDFs of the local fluid energy dissipation
$\epsilon_u$ are given for all our runs and their counterparts for
$\epsilon_b$ are given in Table~\ref{table:epsu}. From these values we see
that the right tails of these distributions fall much more slowly than the
tail of a Gaussian distribution.

\begin{figure}[htb]
\begin{center}
\includegraphics[width=0.95\textwidth]{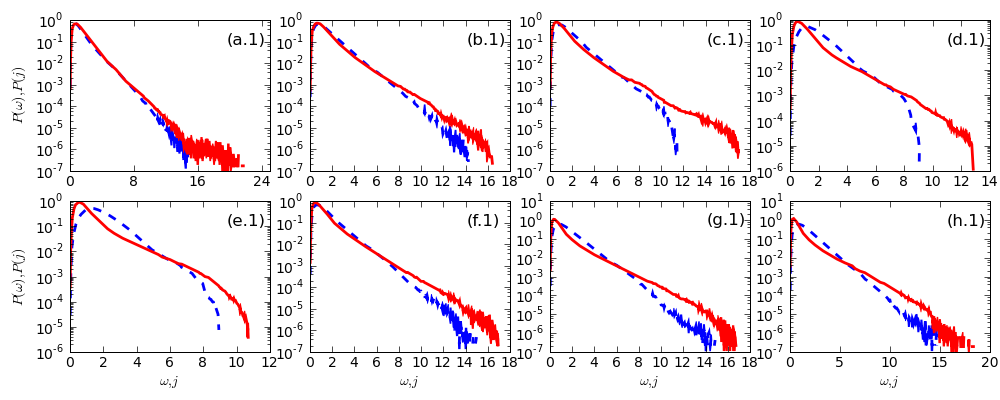}\\
\includegraphics[width=0.95\textwidth]{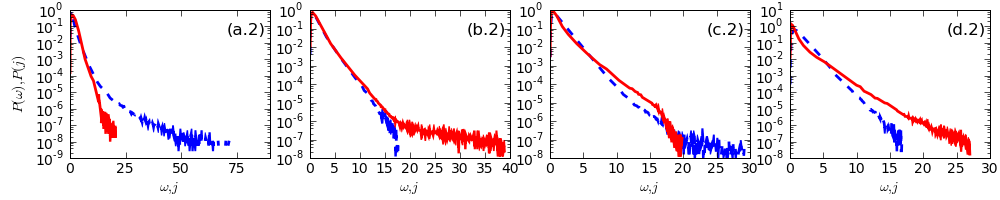}\\
\includegraphics[width=0.95\textwidth]{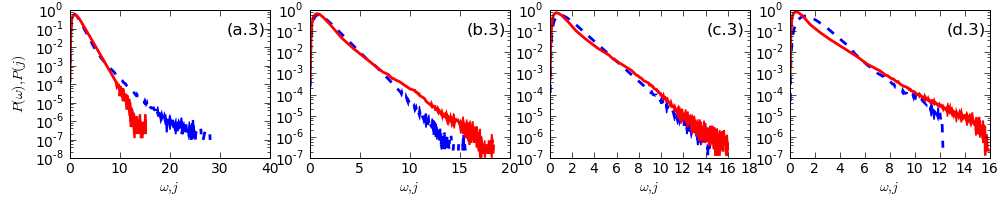}\\
\end{center}
\caption[]{Semilog (base 10) plots of PDFs of the moduli of the
local vorticity (blue dashed lines) and the current density (red full lines),
$\omega$ and $j$, respectively, with the arguments of the PDFs scaled by
their standard deviations, for (a.1) ${\rm Pr_M}=0.1$ (R1), (b.1) ${\rm
Pr_M}=0.5$ (R2), (c.1) ${\rm Pr_M}=1.0$ (R3), (d.1) ${\rm Pr_M}=5.0$ (R4),
(e.1) ${\rm Pr_M}=10.0$ (R5), (f.1) ${\rm Pr_M}=1.0$ (R3B), (g.1) ${\rm
Pr_M}=5.0$ (R4B), (h.1) ${\rm Pr_M}=10.0$ (R5B), (a.2) ${\rm Pr_M}=0.01$
(R1C), (b.2) ${\rm Pr_M}=0.1$ (R2C), (c.2) ${\rm Pr_M}=1.0$ (R3C), and (d.2)
${\rm Pr_M}=10.0$ (R4C) for decaying MHD turbulence; and for statistically
steady MHD turbulence (a.3) ${\rm Pr_M}=0.01$ (R1D), (b.3) ${\rm Pr_M}=0.1$
(R2D), (c.3) ${\rm Pr_M}=1.0$ (R3D), and (d.3) ${\rm Pr_M}=10.0$ (R4D).}
\label{fig:wj}
\end{figure} 
\begin{table}
\caption{The mean $\mu_{\omega}$, standard deviation $\sigma_{\omega}$, 
skewness $\gamma_{3,\omega}$, and kurtosis $\gamma_{4,\omega}$ of
the PDFs of the modulus of the local vorticity $\omega$, and their
analogues for $j$, for runs R1-R5, R3B-R5B, R1C-R4C, and R1D-R4D.}
\label{table:omgu}
\begin{center}
\begin{tabular}{lllll|llll}
\hline\noalign{\smallskip}
Run~~~ & $\mu_{\omega}$~~~ & $\sigma_{\omega}$~~~ & $\gamma_{3, \omega}$~~~ & $\gamma_{4, \omega}$
& $\mu_{j}$~~~ & $\sigma_{j}$~~~ & $\gamma_{3, j}$~~~ & $\gamma_{4, j}$\\
\noalign{\smallskip}\hline\noalign{\smallskip}
R1  & 3.817 & 3.138 & 2.174 & 10.25 & 3.109 & 2.340 & 2.206 & 11.42 \\
R2  & 2.680 & 1.938 & 1.961 & 9.235 & 2.794 & 2.222 & 2.525 & 13.84 \\
R3  & 2.189 & 1.512 & 1.871 & 8.598 & 2.606 & 2.215 & 2.873 & 17.16 \\
R4  & 1.312 & 0.766 & 1.534 & 6.859 & 2.121 & 1.870 & 2.903 & 15.74 \\
R5  & 1.022 & 0.567 & 1.363 & 6.098 & 1.906 & 1.695 & 2.789 & 13.95 \\
R3B & 11.50 & 8.706 & 1.949 & 8.908 & 13.27 & 12.06 & 2.791 & 15.61 \\
R4B & 21.30 & 14.98 & 1.828 & 8.246 & 32.58 & 34.11 & 3.243 & 19.05 \\
R5B & 26.93 & 18.23 & 1.770 & 8.049 & 47.08 & 51.90 & 3.360 & 19.79 \\
R1C & 25.31 & 23.10 & 2.265 & 11.17 & 21.00 & 18.47 & 2.456 & 14.56 \\
R2C & 15.70 & 13.07 & 2.062 & 9.831 & 18.35 & 17.95 & 2.899 & 17.01 \\
R3C & 21.68 & 15.49 & 1.746 & 7.910 & 38.85 & 45.50 & 3.563 & 23.00 \\
R4C & 2.945 & 2.661 & 3.027 & 26.37 & 1.446 & 0.861 & 1.451 & 6.816 \\
R1D & 12.14 & 8.700 & 2.297 & 14.16 & 5.336 & 3.456 & 1.702 & 7.683 \\
R2D & 14.26 & 9.802 & 1.745 & 7.821 & 12.58 & 9.544 & 2.336 & 12.64 \\
R3D & 10.07 & 6.544 & 1.644 & 7.463 & 13.88 & 11.24 & 2.320 & 11.36 \\
R4D & 4.119 & 2.349 & 1.388 & 6.526 & 11.92 & 10.47 & 2.429 & 12.46 \\
\noalign{\smallskip}\hline
\end{tabular}
\end{center}
\end{table}

Similar trends emerge if we examine the PDFs of the moduli of the vorticity
and the current density, $\omega$ (blue dashed lines) and $j$ (red full
lines), respectively: These are presented at $t_c$ for runs R1-R5 in
Figs.~\ref{fig:wj}(a.1)-\ref{fig:wj}(e.1), runs R3B-R5B in
Figs.~\ref{fig:wj}(f.1)-\ref{fig:wj}(h.1), and runs R1C-R4C in
Figs.~\ref{fig:wj}(a.2)-\ref{fig:wj}(d.2) for decaying MHD turbulence; and
for statistically steady MHD turbulence they are shown in
Figs.~\ref{fig:wj}(a.3)-\ref{fig:wj}(d.3) for runs R1D-R4D.  The tail of the
PDF for $j$ extends further than the tail of that for $\epsilon_u$ for all
except the smallest values of ${\rm Pr_M}$ [Figs.~\ref{fig:wj}(a.1),
\ref{fig:wj}(a.2), and \ref{fig:wj}(a.3) for runs R1, R1C, and R1D,
respectively], so large values of $j$ are more likely than large values of
$\omega$. Thus, given that these PDFs have long tails, it is reasonable to
expect that, except at the smallest values of ${\rm Pr_M}$ we have used, 
intermittency for the magnetic field might be larger than that
for the velocity field. Moreover, the tail of the PDF of $\omega$ is drawn in
towards small values of $\omega$ as we increase ${\rm Pr_M}$
[Figs.~\ref{fig:wj}(a.1)-\ref{fig:wj}(e.1) for runs R1-R5, respectively]
while holding $\eta$ and the initial energy fixed; but if, while increasing
$\nu$, we also increase the initial energy so that $k_{\rm max}\eta_d^u$ and
$k_{\rm max}\eta_d^b$ are $\simeq 1$, we see that the tails of the PDFs of
$j$ and $\omega$ get stretched out as we increase ${\rm Pr_M}$, e.g.,
in Figs.~\ref{fig:wj}(f.1)-\ref{fig:wj}(h.1) for runs R3B-R5B, respectively.
The values of the mean $\mu_{\omega}$, standard deviation $\sigma_{\omega}$,
skewness $\gamma_{3,\omega}$, and kurtosis $\gamma_{4,\omega}$ of the PDFs of
the modulus of the local vorticity $\omega$ for all our runs and their
counterparts for $j$ are given in Table~\ref{table:omgu}. From these values
we see that the right tails of these distributions fall much more slowly than
the tail of a Gaussian distribution.

\begin{figure}[htb]
\begin{center}
\includegraphics[width=0.95\textwidth]{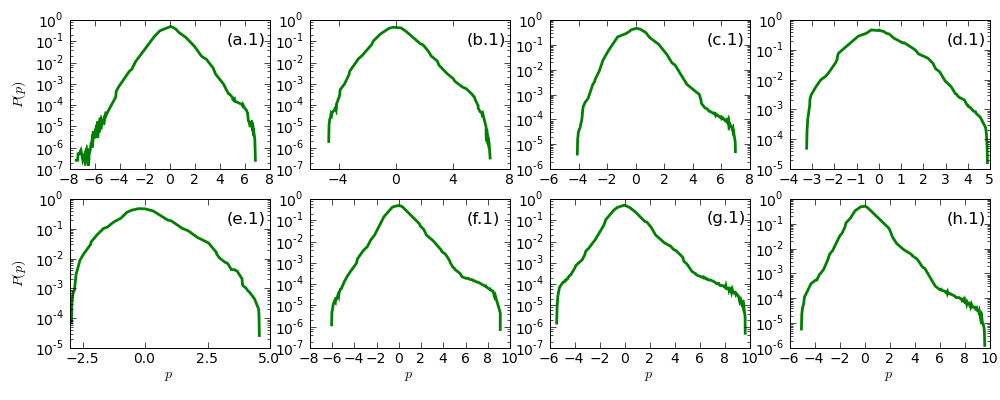}\\
\includegraphics[width=0.95\textwidth]{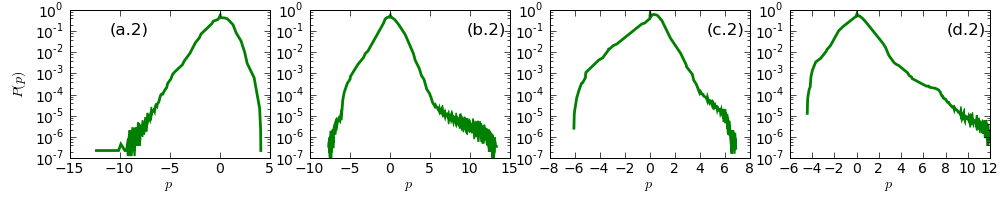}\\
\end{center}

\caption[]{Semilog (base 10) plots of PDFs of local 
effective pressure fluctuations (green full lines), with the 
arguments of the PDFs scaled by their standard deviations, for
(a.1) ${\rm Pr_M}=0.1$ (R1), (b.1) ${\rm Pr_M}=0.5$ (R2), (c.1) 
${\rm Pr_M}=1.0$ (R3), (d.1) ${\rm Pr_M}=5.0$ (R4), (e.1) ${\rm Pr_M}=10.0$
(R5), (f.1) ${\rm Pr_M}=1.0$ (R3B), 
(g.1) ${\rm Pr_M}=5.0$ (R4B), (h.1) ${\rm Pr_M}=10.0$ (R5B), 
for decaying MHD turbulence; and 
for statistically steady MHD turbulence (a.2) ${\rm Pr_M}=0.01$ (R1D), 
(b.2) ${\rm Pr_M}=0.1$ (R2D), (c.2) ${\rm Pr_M}=1.0$ (R3D), and 
(d.2) ${\rm Pr_M}=10.0$ (R4D).}
\label{fig:pressure}
\end{figure}
\begin{table}
\caption{The mean $\mu_{p}$, standard deviation 
$\sigma_{p}$, skewness $\gamma_{3,p}$, and kurtosis 
$\gamma_{4,p}$ of the PDFs of the local effective pressure 
$\bar{p}$ for runs R1-R5, R3B-R5B, and R1D-R4D.}
\label{table:press}
\begin{center}
\begin{tabular}{lllll}
\hline\noalign{\smallskip}
Run~~~ & $\mu_{p}$~~~ & $\sigma_{p}$~~~ & 
$\gamma_{3,p}$~~~ & $\gamma_{4,p}$\\
\noalign{\smallskip}\hline\noalign{\smallskip}
R1  & -3.283E-16  &  0.055  &  0.224  &  4.152 \\
R2  & -2.183E-16  &  0.057  &  0.256  &  4.052 \\
R3  &  6.606E-16  &  0.061  &  0.315  &  3.722 \\
R4  & -2.787E-16  &  0.060  &  0.397  &  3.493 \\
R5  & -6.596E-16  &  0.059  &  0.433  &  3.527 \\
R3B &  2.975E-15  &  0.609  &  0.526  &  5.283 \\
R4B &  1.323E-14  &  3.184  &  0.660  &  5.645 \\
R5B & -3.475E-14  &  6.397  &  0.719  &  5.776 \\
R1D &  9.014E-15  &  0.738  & -0.533  &  3.882 \\
R2D &  1.136E-15  &  0.313  & -0.153  &  4.697 \\
R3D &  1.244E-14  &  0.589  & -1.066  &  5.338 \\
R4D &  9.983E-16  &  0.363  &  0.221  &  5.560 \\
\noalign{\smallskip}\hline
\end{tabular}
\end{center}
\end{table}

We move now to PDFs of the local effective pressure (green full lines), which
are shown at $t_c$ for runs R1-R5 in
Figs.~\ref{fig:pressure}(a.1)-\ref{fig:pressure}(e.1) and runs R3B-R5B in
Figs.~\ref{fig:pressure}(f.1)-\ref{fig:pressure}(h.1) for decaying  MHD
turbulence; for statistically steady MHD turbulence they are shown in
Figs.~\ref{fig:pressure} a.2-d.2 for runs R1D-R4D. The values of the mean
$\mu_{p}$, standard deviation $\sigma_{p}$, skewness $\gamma_{3,p}$, and
kurtosis $\gamma_{4,p}$ of the PDFs of the local effective pressure $p$ are
given for these runs in Table~\ref{table:press}. These have mean $\mu_{p}=0$
but are distinctly non-Gaussian as can be seen from the values of
$\gamma_{3,p}$ and $\gamma_{4,p}$. Pressure PDFs are negatively skewed in
pure fluid turbulence as we have mentioned above; however, for MHD turbulence
we find that the PDFs of the effective pressure $\bar{p}$ can be positively
skewed, as in runs R1-R5, R3B-R5B, and run R4D, or negatively skewed, as in
runs R1D-R3D; negative skewness seems to arise at low values of
${\rm Pr_M}$.  

\begin{figure}[htb]
\begin{center}
\includegraphics[width=0.95\textwidth]{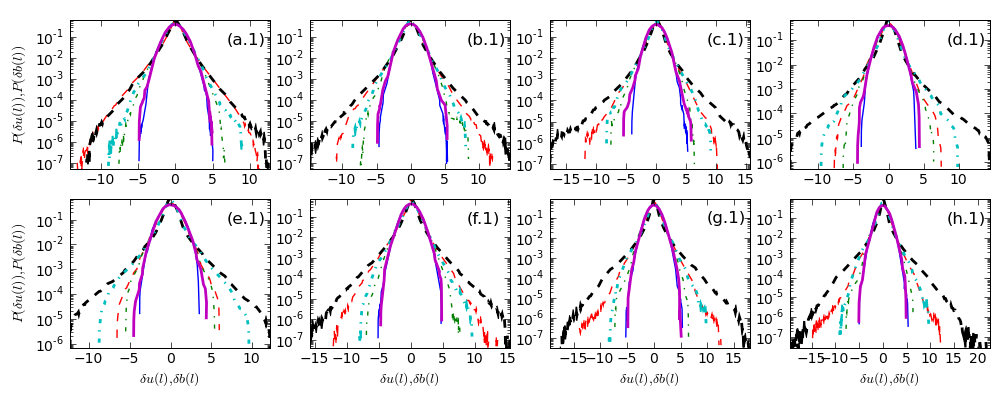}\\
\includegraphics[width=0.95\textwidth]{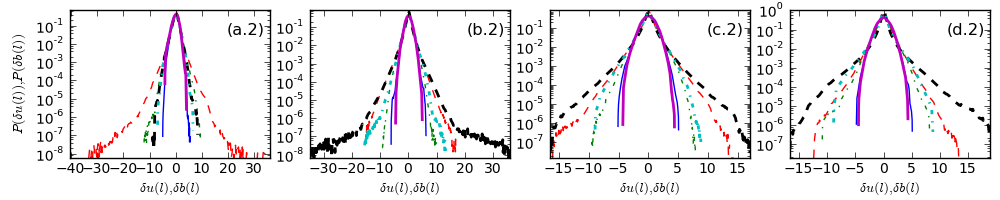}\\
\includegraphics[width=0.95\textwidth]{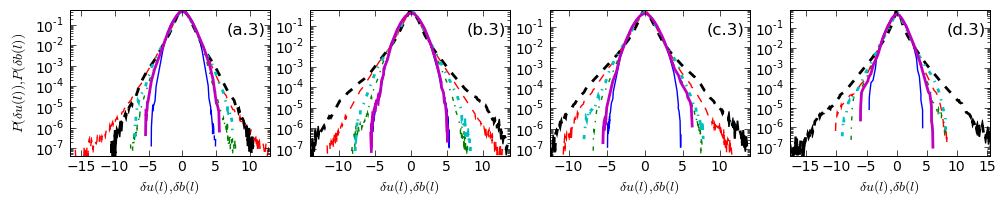}\\
\end{center}
\caption[]{Semilog (base 10) plots of PDFs of velocity 
increments $\delta u(l)$, for separations $l = 2\delta x$ (red dashed thin 
line), $10\delta x$ (green dot-dashed thin line), and $100\delta x$ (blue, 
full thin line), and of magnetic-field increments $\delta b(l)$, for 
separations $l = 2\delta x$ (black dashed line), $10\delta x$ 
(cyan dot-dashed line), and $100\delta x$ (magenta full line), with the 
arguments of the PDFs scaled by their standard deviations, for 
(a.1) ${\rm Pr_M}=0.1$ (R1),
(b.1) ${\rm Pr_M}=0.5$ (R2), (c.1) ${\rm Pr_M}=1.0$ (R3), 
(d.1) ${\rm Pr_M}=5.0$ (R4), (e.1) ${\rm Pr_M}=10.0$ (R5), 
(f.1) ${\rm Pr_M}=1.0$ (R3B), (g.1) ${\rm Pr_M}=5.0$ (R4B), 
(h.1) ${\rm Pr_M}=10.0$ (R5B), (a.2) ${\rm Pr_M}=0.01$ (R1C), 
(b.2) ${\rm Pr_M}=0.1$ (R2C), (c.2) ${\rm Pr_M}=1.0$ (R3C), and 
(d.2) ${\rm Pr_M}=10.0$ (R4C) for decaying MHD turbulence; and 
for statistically steady MHD turbulence (a.3) ${\rm Pr_M}=0.01$ (R1D), 
(b.3) ${\rm Pr_M}=0.1$ (R2D), (c.3) ${\rm Pr_M}=1.0$ (R3D), and 
(d.3) ${\rm Pr_M}=10.0$ (R4D).}
\label{fig:incrementpdfs}
\end{figure} 


The scale dependence of PDFs of velocity increments provides important clues
about the nature of intermittency in fluid turbulence. To explore
similar intermittency in MHD turbulence~\cite{weygand05}, we present data for
the scale dependence of PDFs velocity and magnetic-field increments. As
mentioned above, these increments are of the form $\delta
a_{\parallel}(\bfx,l) \equiv
\bfa(\bfx+\bfl,t)-\bfa(\bfx,t)]\cdot\frac{\bfl}{l}$, with $\bfa$ either
$\bfu$ or $\bfb$, $l = |\bfl|$ the length scale, and $\bfx$ an origin over
which we can average to determine the dependence of the PDFs of $\delta
a_{\parallel}$ on the scale $l$; for notational convenience, such velocity
and magnetic-field increments are denoted by $\delta u(l)$ and $\delta b(l)$
in our plots. These PDFs are obtained at $t_c$ for runs R1-R5 in
Figs.~\ref{fig:incrementpdfs}(a.1)-\ref{fig:incrementpdfs}(e.1), runs R3B-R5B
in Figs.~\ref{fig:incrementpdfs}(f.1)-\ref{fig:incrementpdfs}(h.1), and runs
R1C-R4C  in Figs.~\ref{fig:incrementpdfs}(a.2)-\ref{fig:incrementpdfs}(d.2)
for decaying  MHD turbulence; and for statistically steady MHD turbulence
they are shown in
Figs.~\ref{fig:incrementpdfs}(a.3)-\ref{fig:incrementpdfs}(d.3) for runs
R1D-R4D. The PDFs of velocity increments are shown for separations $l =
2\delta x$ (red dashed thin line), $l=10\delta x$ (green dot-dashed thin
line), and $l=100\delta x$ (blue full thin line), where $\delta x$ is our
real-space lattice spacing; for PDFs of magnetic-field increments we also use
the separations $l = 2\delta x$ (black dashed line), $l = 10\delta x$ (cyan
dot-dashed line), and $l = 100\delta x$ (magenta full line); the arguments of
these PDFs are scaled by their standard deviations. As in fluid turbulence, we
see that these PDFs are nearly Gaussian if the length scale $l$ is large. As
$l$ decreases, the PDFs develop, long, non-Gaussian tails, a clear signature
of intermittency. Furthermore, a comparison of the red and black
dashed lines in these plots indicates that the PDFs of the magnetic-field
increments are  broader than their velocity counterparts in most of our runs;
this suggests, as we had surmised from the PDFs of energy-dissipation rates
given above, that the magnetic field displays stronger intermittency than the
velocity field at all but the smallest values of ${\rm Pr_M}$
[Figs.~\ref{fig:incrementpdfs}(a.1), \ref{fig:incrementpdfs}(a.2), and
\ref{fig:incrementpdfs}(a.3) for runs R1, R1C, and R1D]; the general trend
that we notice from these figures is that the magnetic-field intermittency is
stronger than that of the velocity field at large magnetic Prandtl numbers
but the difference between these intermittencies decreases as ${\rm Pr_M}$ is
lowered. We will try to quantify this when we present structure functions in
Subsection~\ref{sec:stfn}.

\subsection{Structure functions \label{sec:stfn}}

We continue our elucidation of intermittency in MHD turbulence by
studying the scale dependence of order-$p$ equal-time, velocity and
magnetic-field longitudinal structure functions $S_p^u(l) \equiv
\langle|\delta u_{\parallel}(\bfx,l)|^p\rangle$ and magnetic-field
longitudinal structure functions $S_p^b(l) \equiv \langle|\delta
b_{\parallel}(\bfx,l)|^p\rangle$, respectively, where $\delta
u_{\parallel}(\bfx,l) \equiv
[\bfu(\bfx+\bfl,t)-\bfu(\bfx,t)]\cdot\frac{\bfl}{l}$ and $\delta
b_{\parallel}(\bfx,l) \equiv
[\bfb(\bfx+\bfl,t)-\bfb(\bfx,t)]\cdot\frac{\bfl}{l}$. From these structure
functions we also obtain the hyperflatnesses $F_6^u(r)=S_6^u(r)/[S_2^u(r)]^3$
and $F_6^b(r)=S_6^b(r)/[S_2^b(r)]^3$.  For the inertial range $\eta_d^u,
\eta_d^b \ll l \ll L$, we expect $S_p^u(l)\sim l^{\zeta_p^u}$ and
$S_p^b(l)\sim l^{\zeta_p^b}$, where $\zeta_p^u$ and $\zeta_p^b$ are
inertial-range multiscaling exponents for velocity and magnetic fields,
respectively; if these fields show multiscaling, we expect significant
deviations from the K41 result $\zeta_p^{uK41} = \zeta_p^{bK41} = p/3$.
[Note that we do not expect any Iroshnikov-Kraichnan~\cite{ik} scaling
because we have no mean magnetic field in our simulations.] Given large
inertial ranges, the multiscaling exponents can be extracted from slopes of
log-log plots of structure functions versus $l$. However, in practical
calculations inertial ranges are limited, so we use the
extended-self-similarity (ESS) procedure~\cite{benzi93,chakraborty10} in
which we determine the multiscaling exponent ratios $\zeta_p^u/\zeta_3^u$ and
$\zeta_p^b/\zeta_3^b$, respectively, from slopes of log-log plots of (a)
$S_p^u$ versus $S_3^u$ and (b) $S_p^b$ versus $S_3^b$; we refer to these as
ESS plots. Our data for structure functions are averaged over $51$ and $400$
origins, respectively, for simulations with $512^3$ and $1024^3$ collocation
points.

\begin{figure}[htb]
\begin{center}
\includegraphics[width=0.23\textwidth]{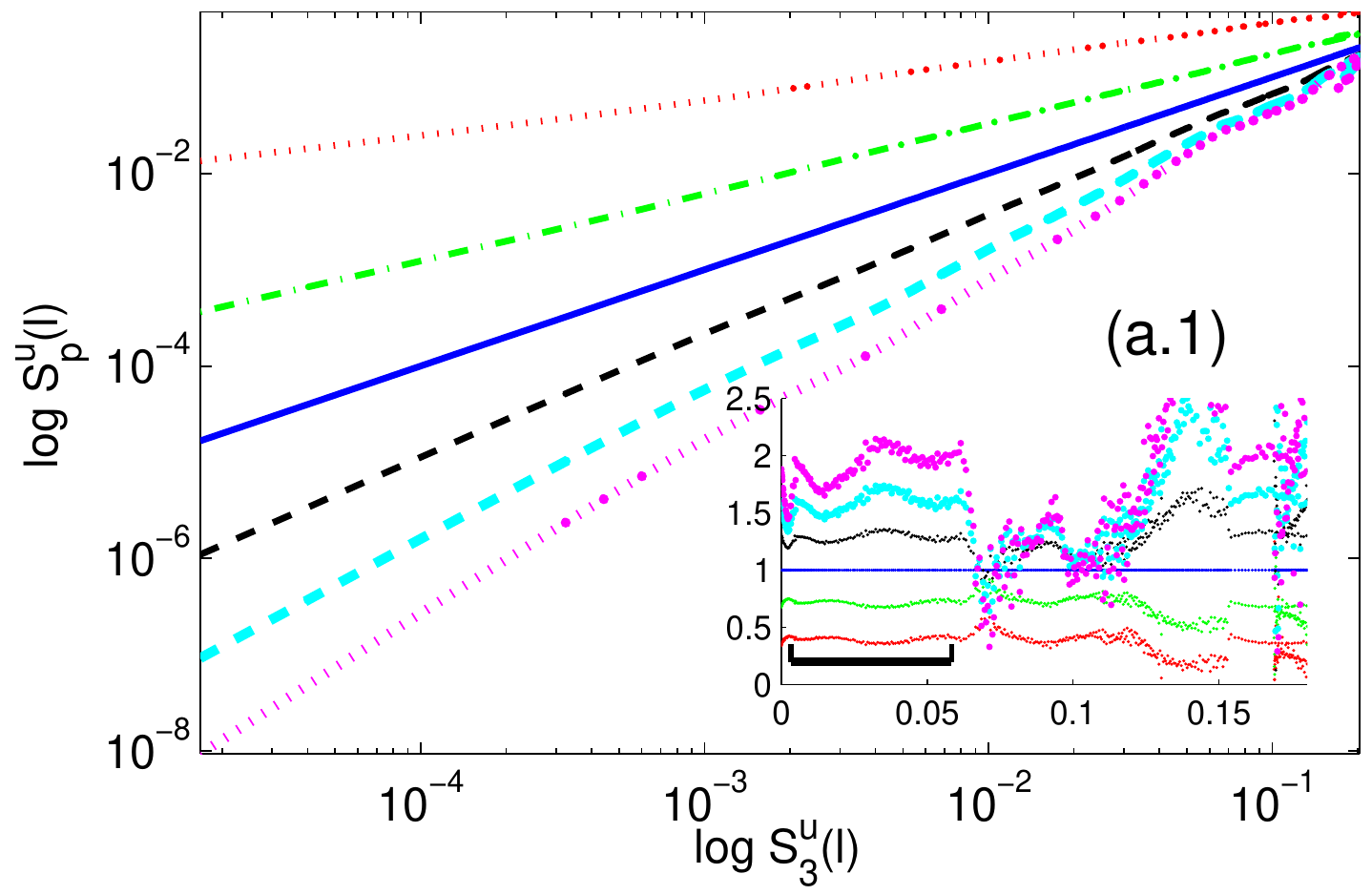}
\includegraphics[width=0.23\textwidth]{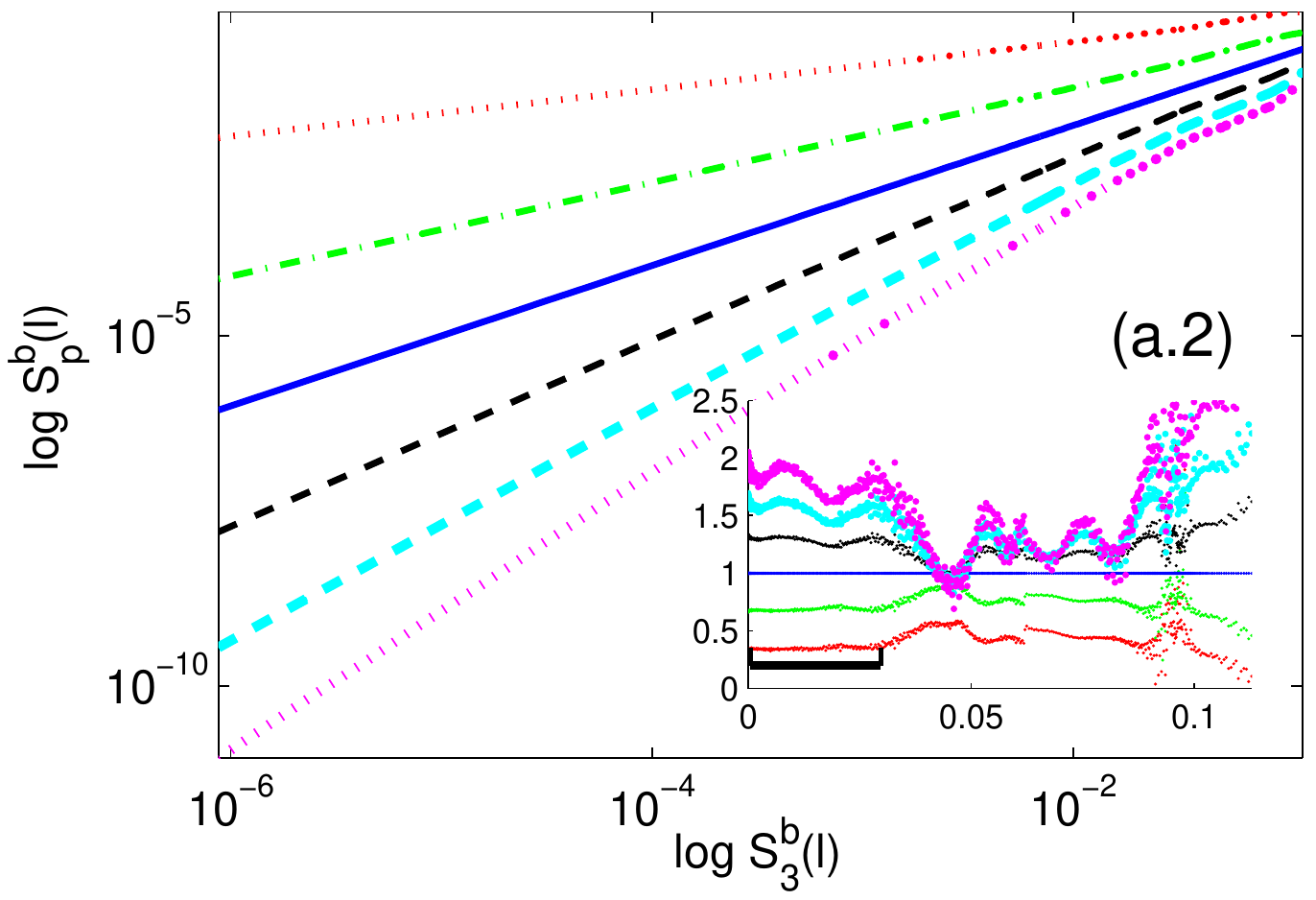}
\includegraphics[width=0.23\textwidth]{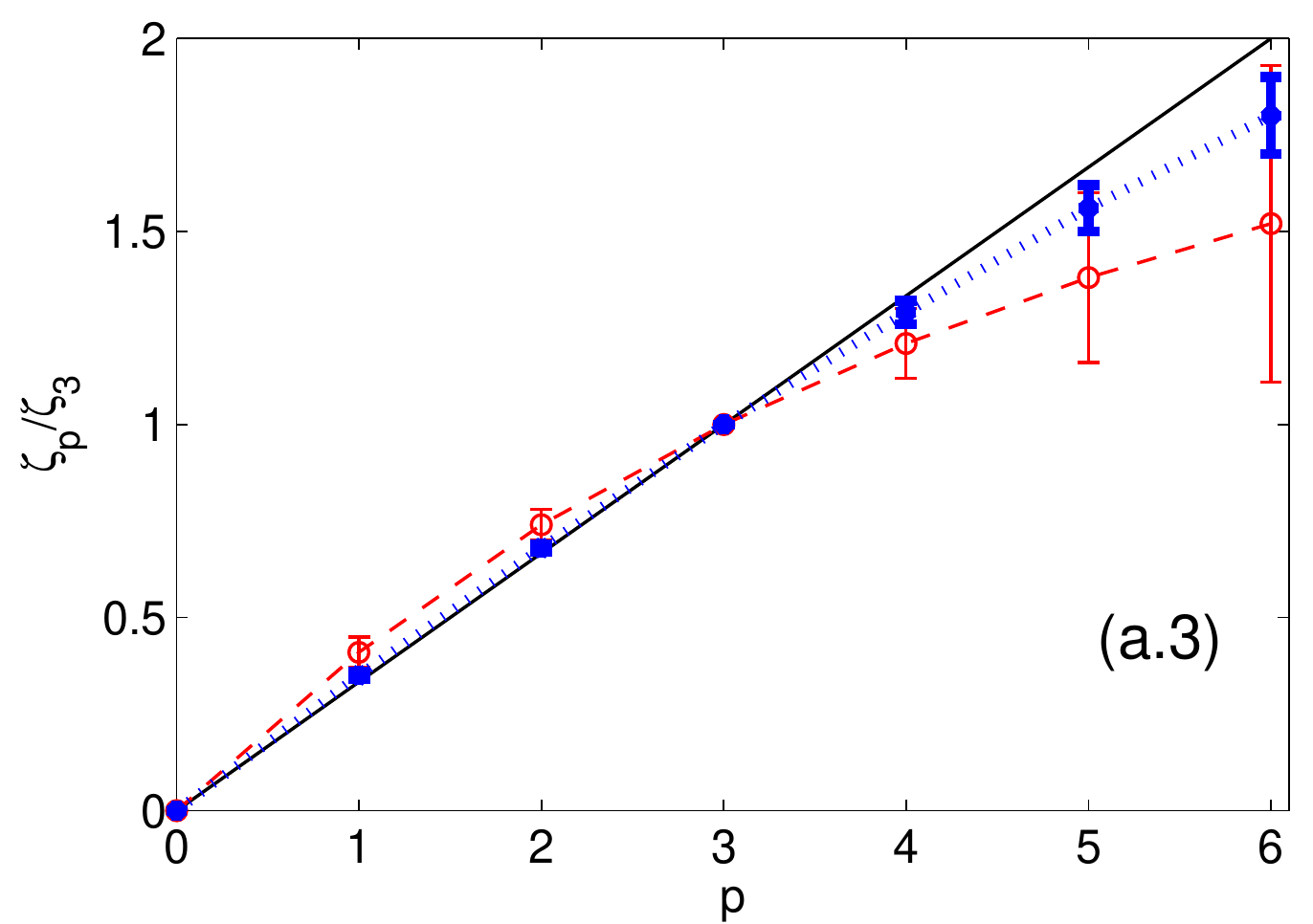}
\includegraphics[width=0.23\textwidth]{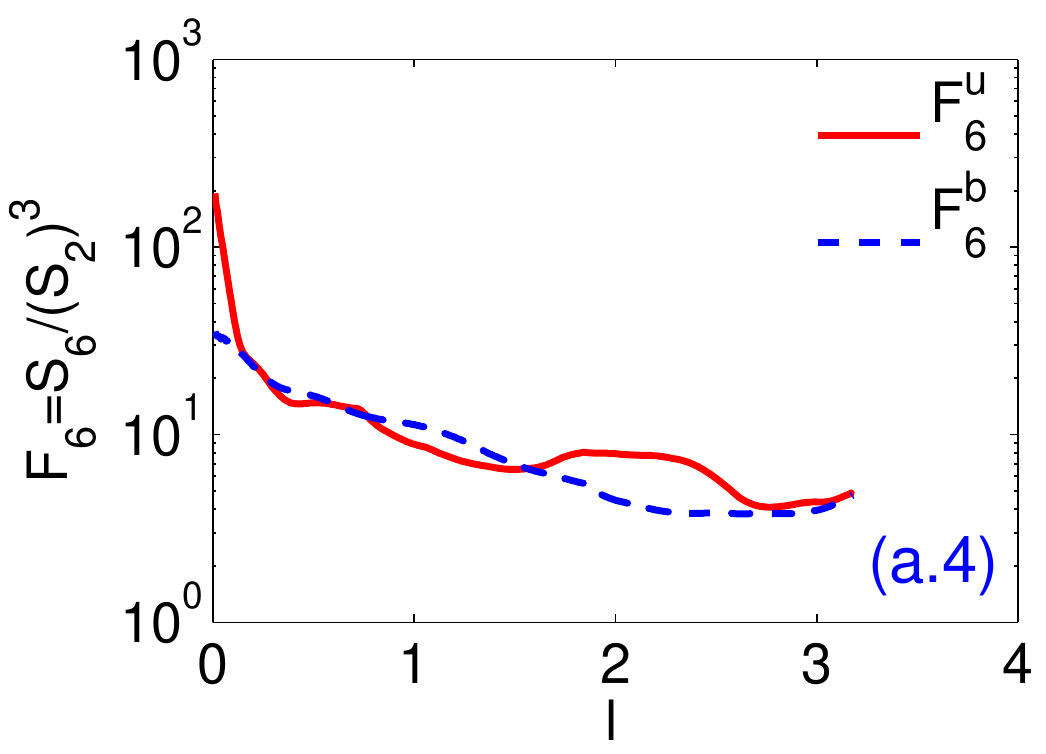}\\
\includegraphics[width=0.23\textwidth]{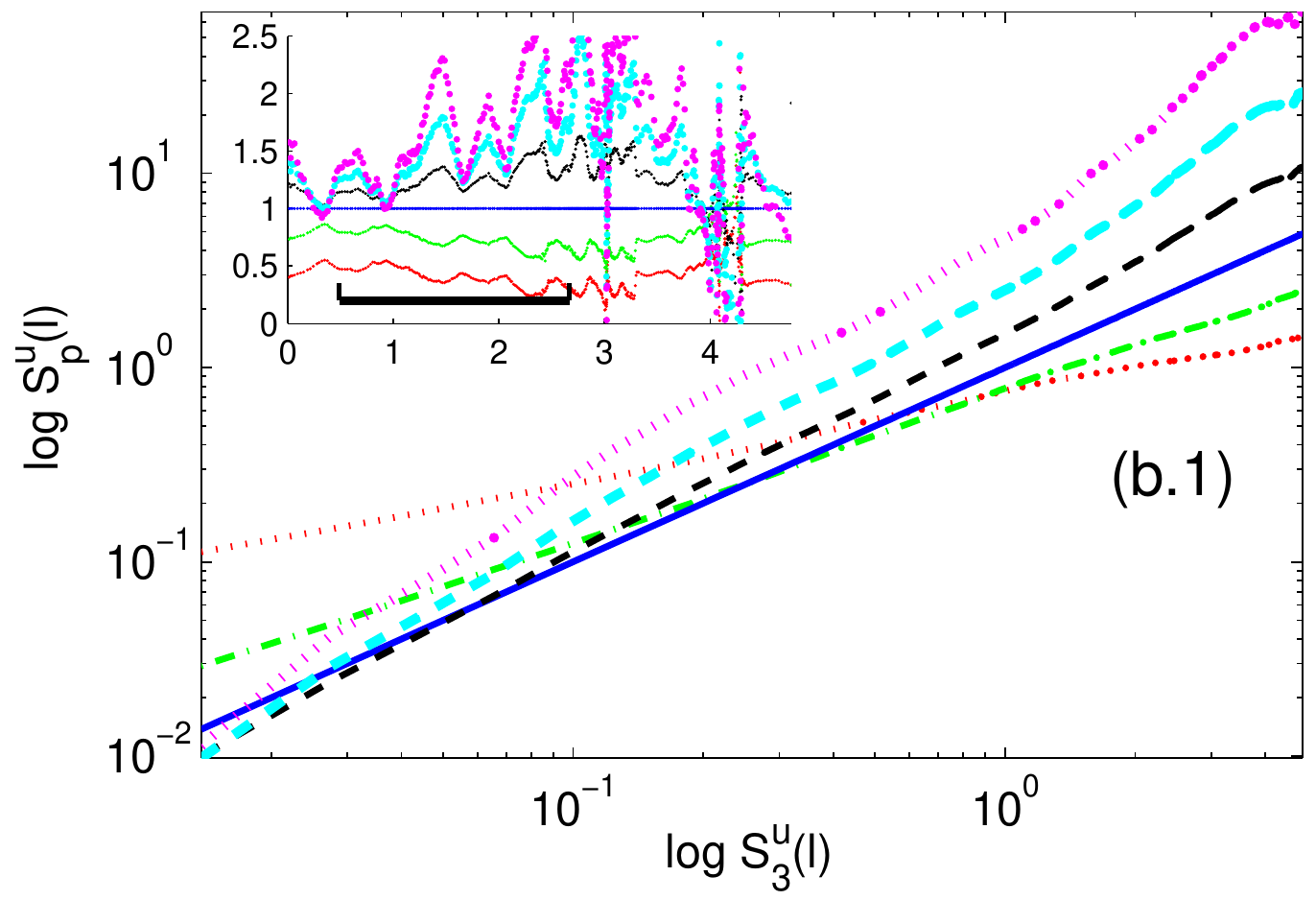}
\includegraphics[width=0.23\textwidth]{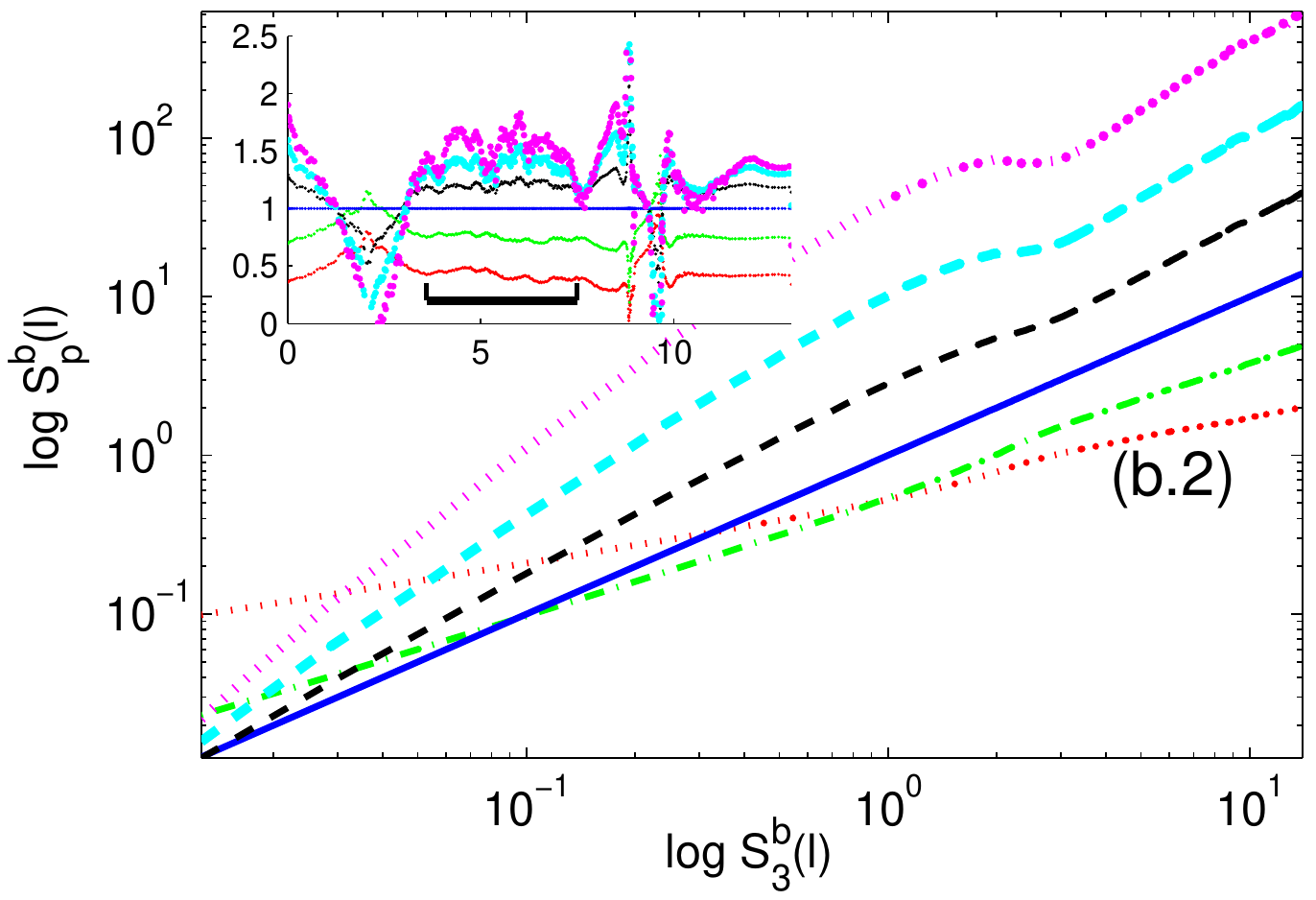}
\includegraphics[width=0.23\textwidth]{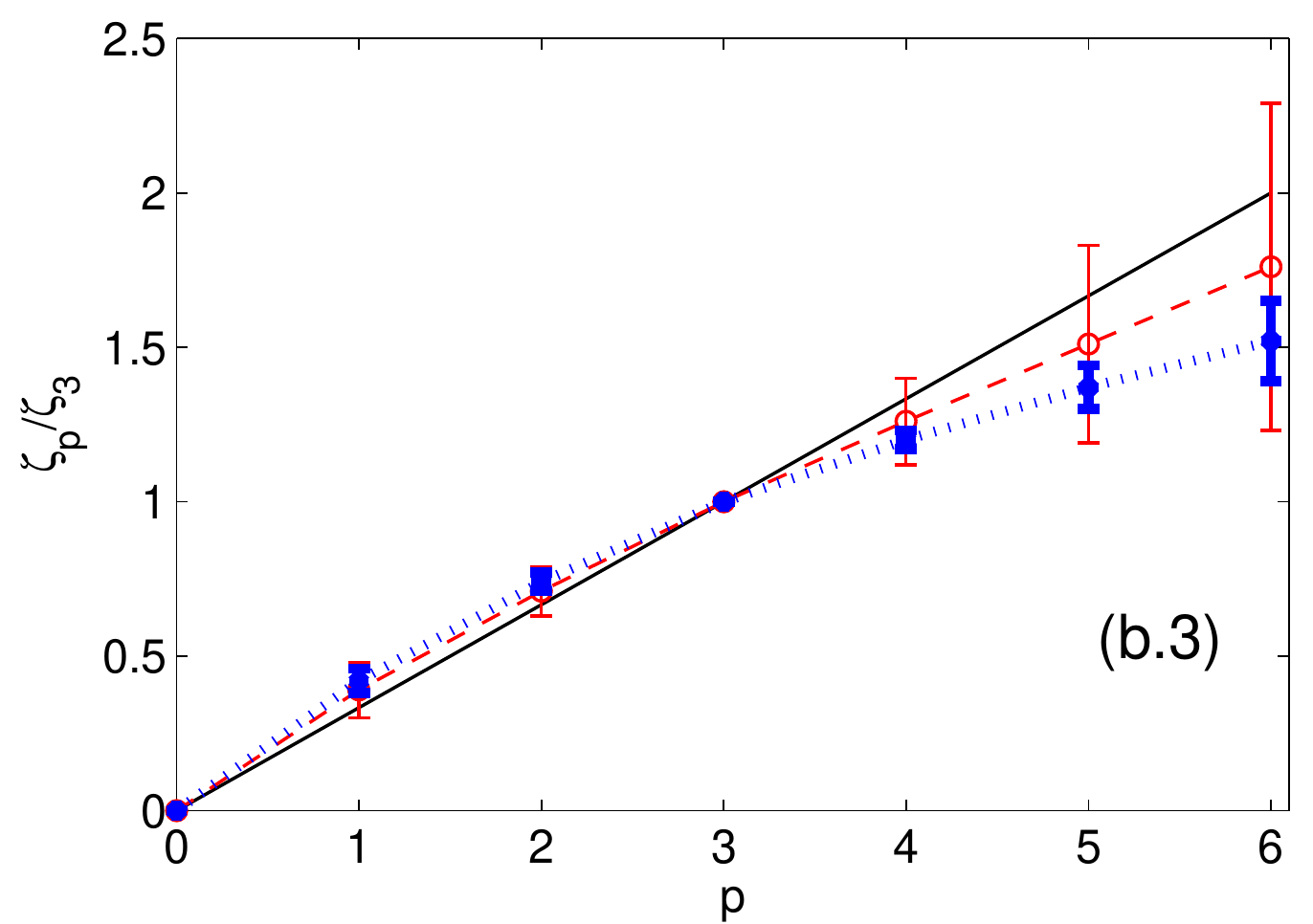}
\includegraphics[width=0.23\textwidth]{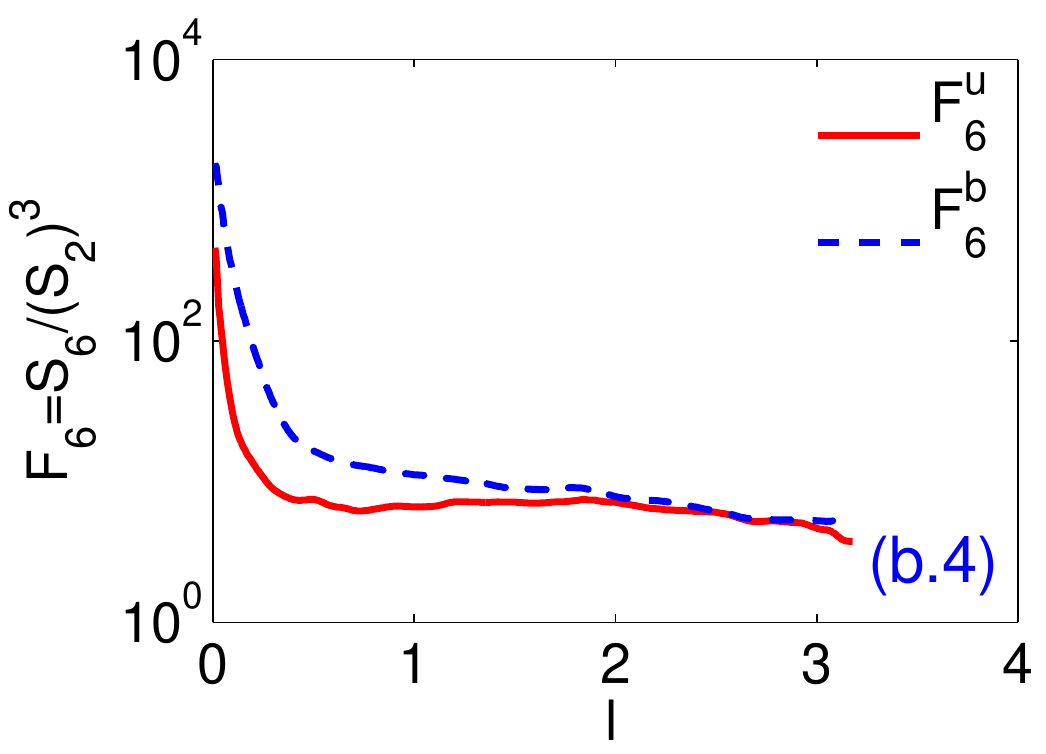}\\
\includegraphics[width=0.23\textwidth]{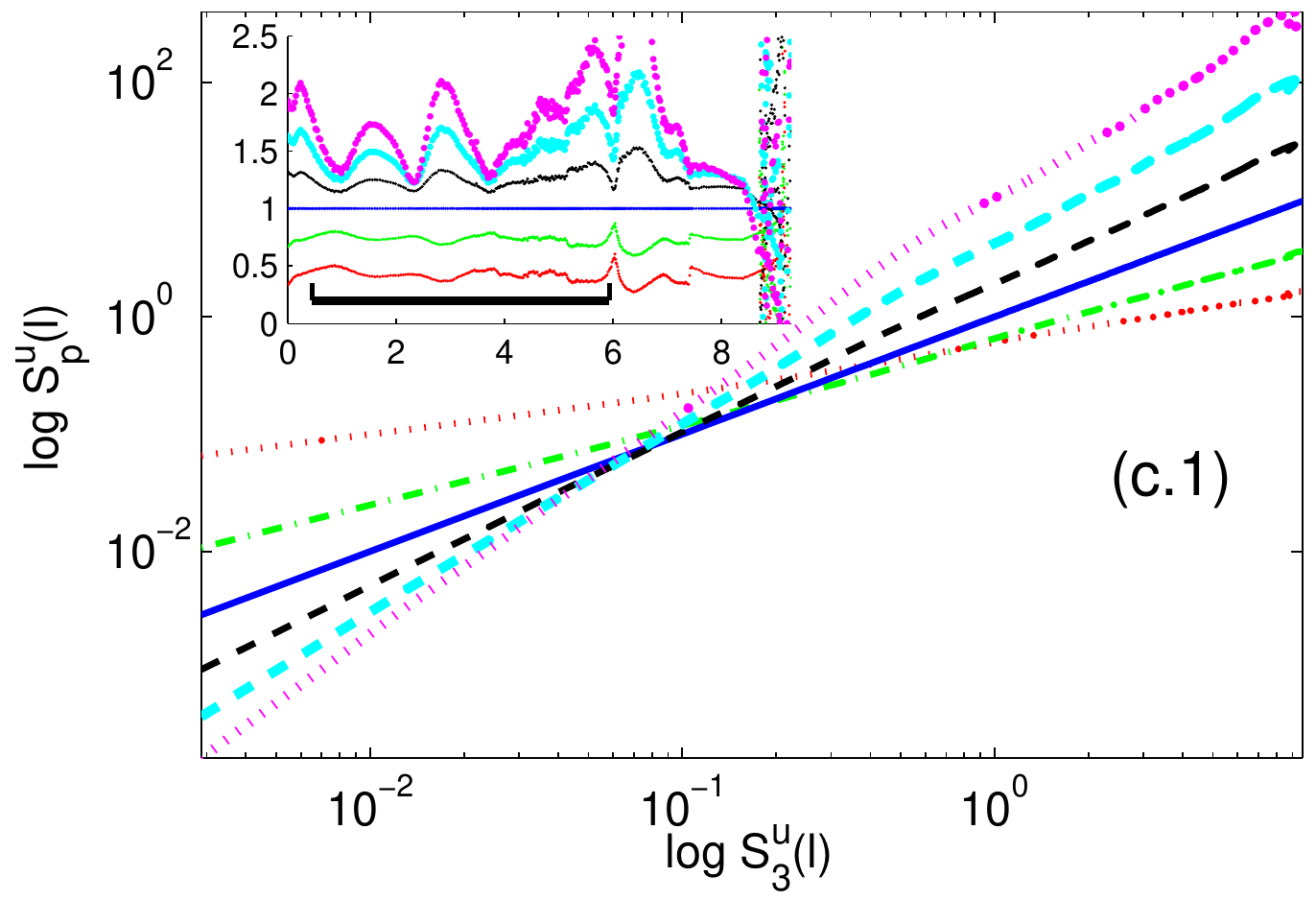}
\includegraphics[width=0.23\textwidth]{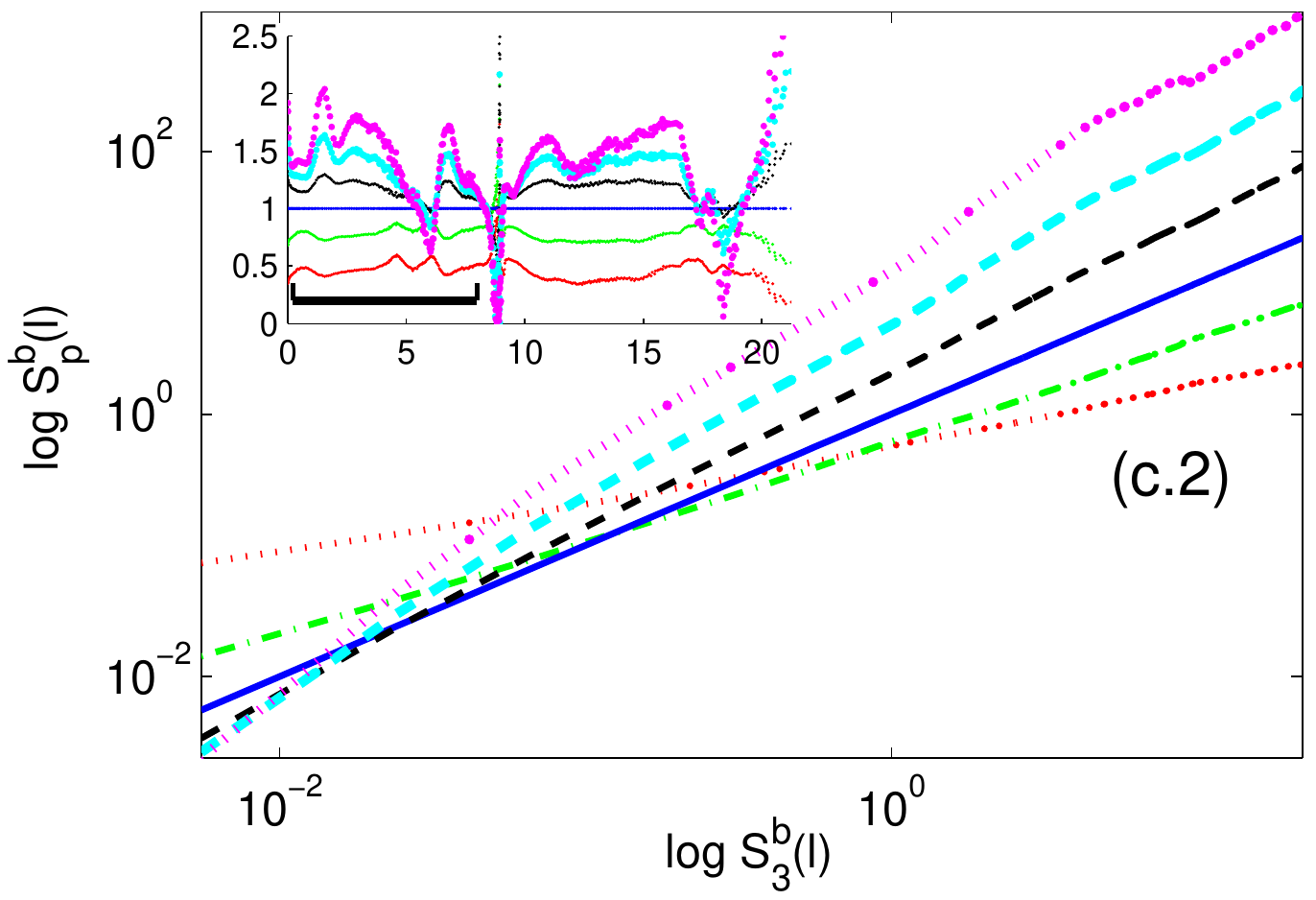}
\includegraphics[width=0.23\textwidth]{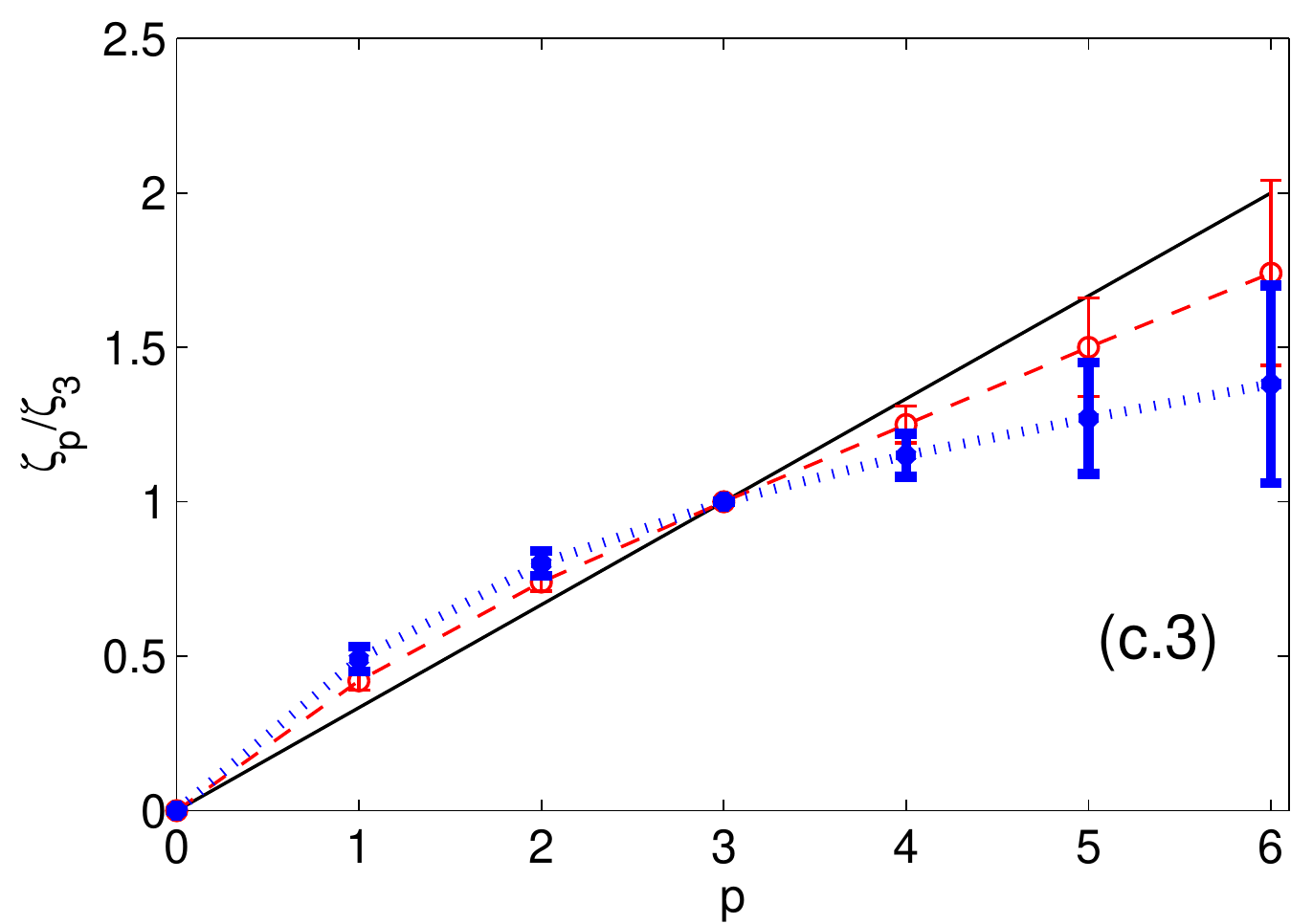}
\includegraphics[width=0.23\textwidth]{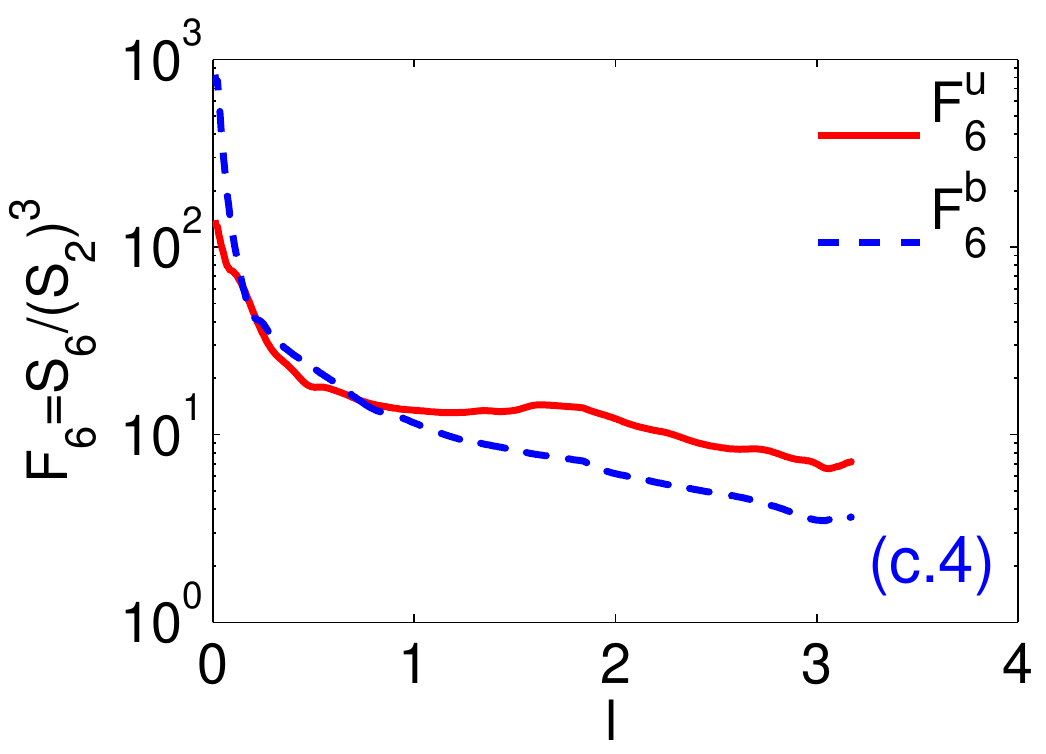}\\
\includegraphics[width=0.23\textwidth]{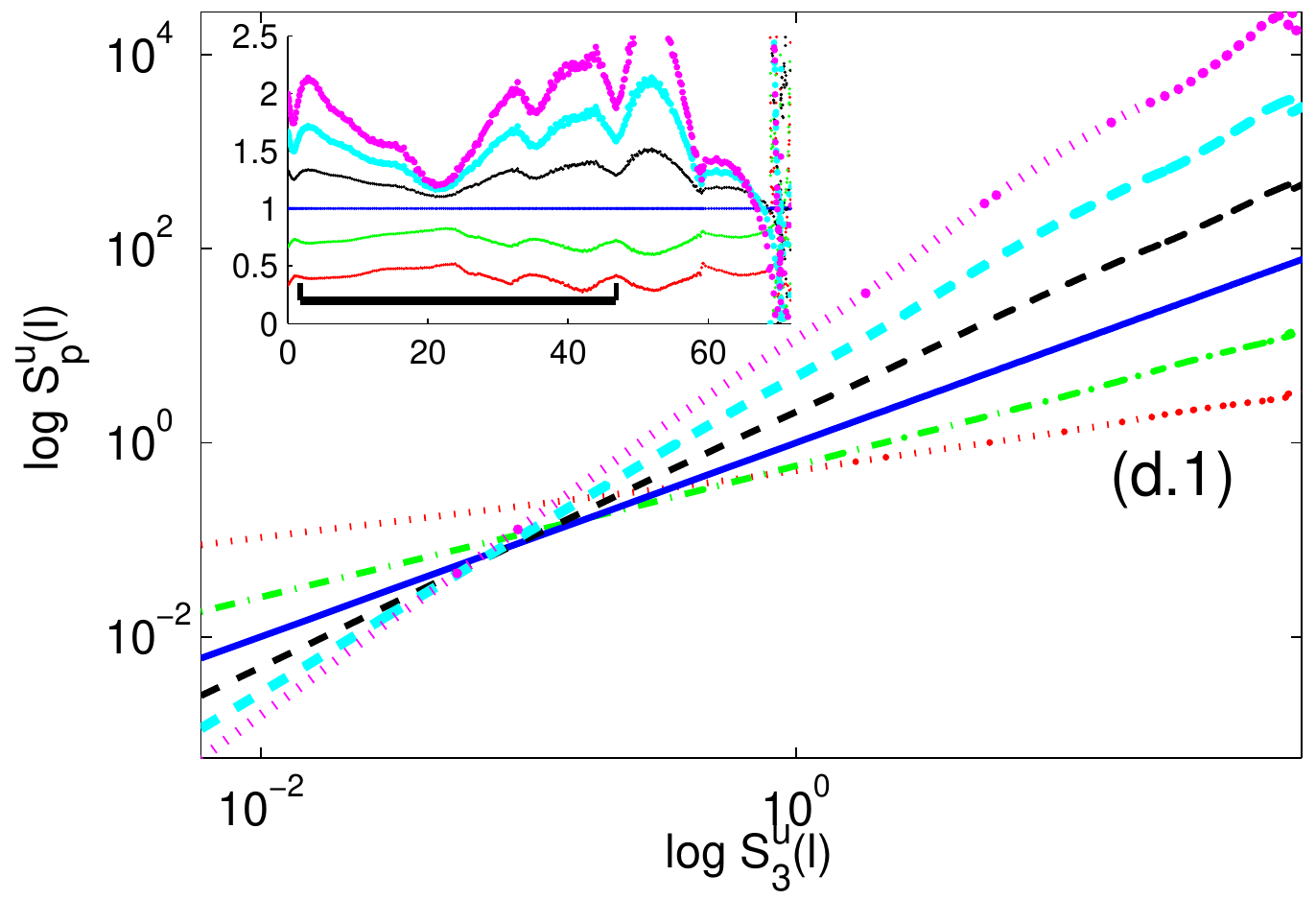}
\includegraphics[width=0.23\textwidth]{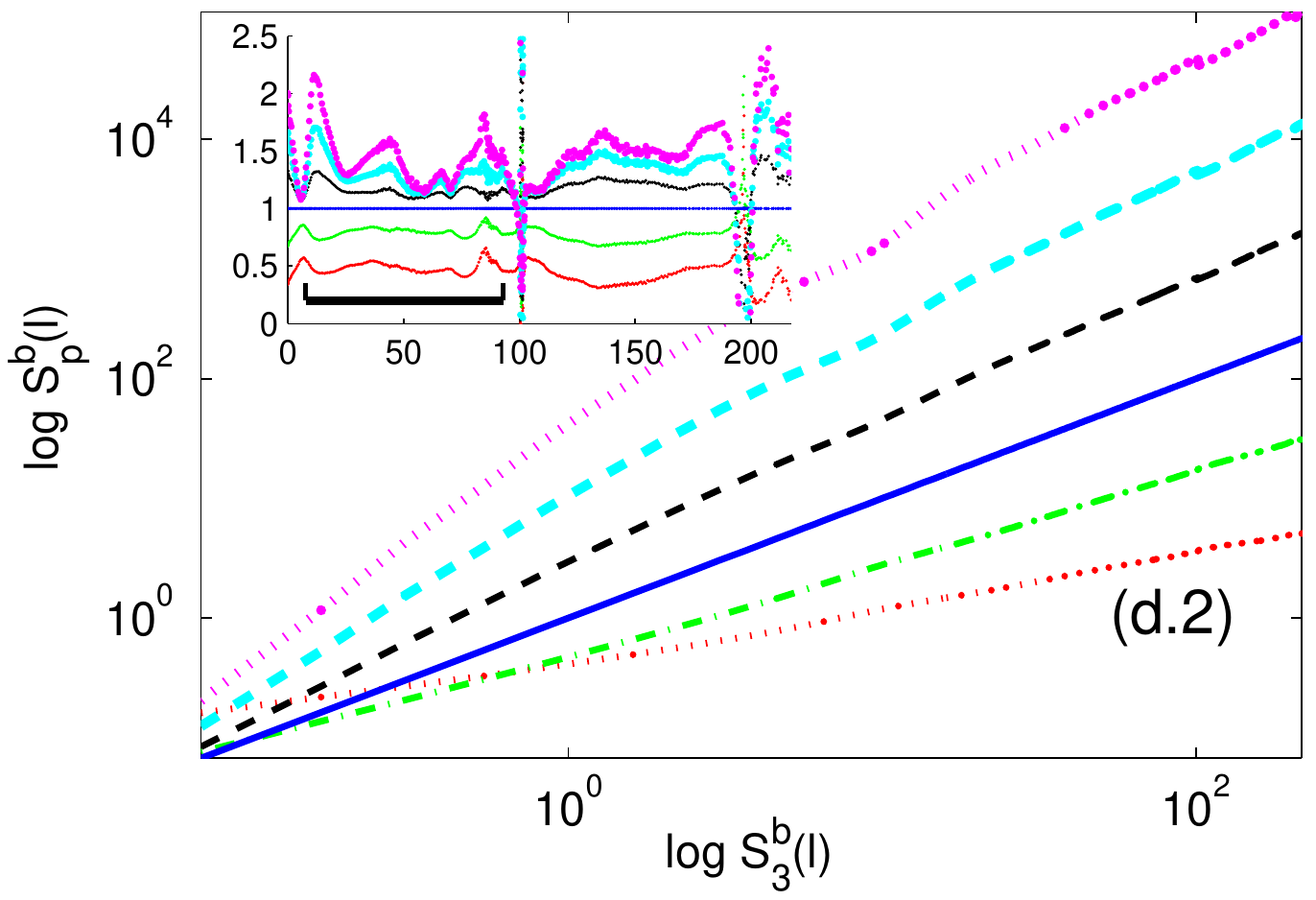}
\includegraphics[width=0.23\textwidth]{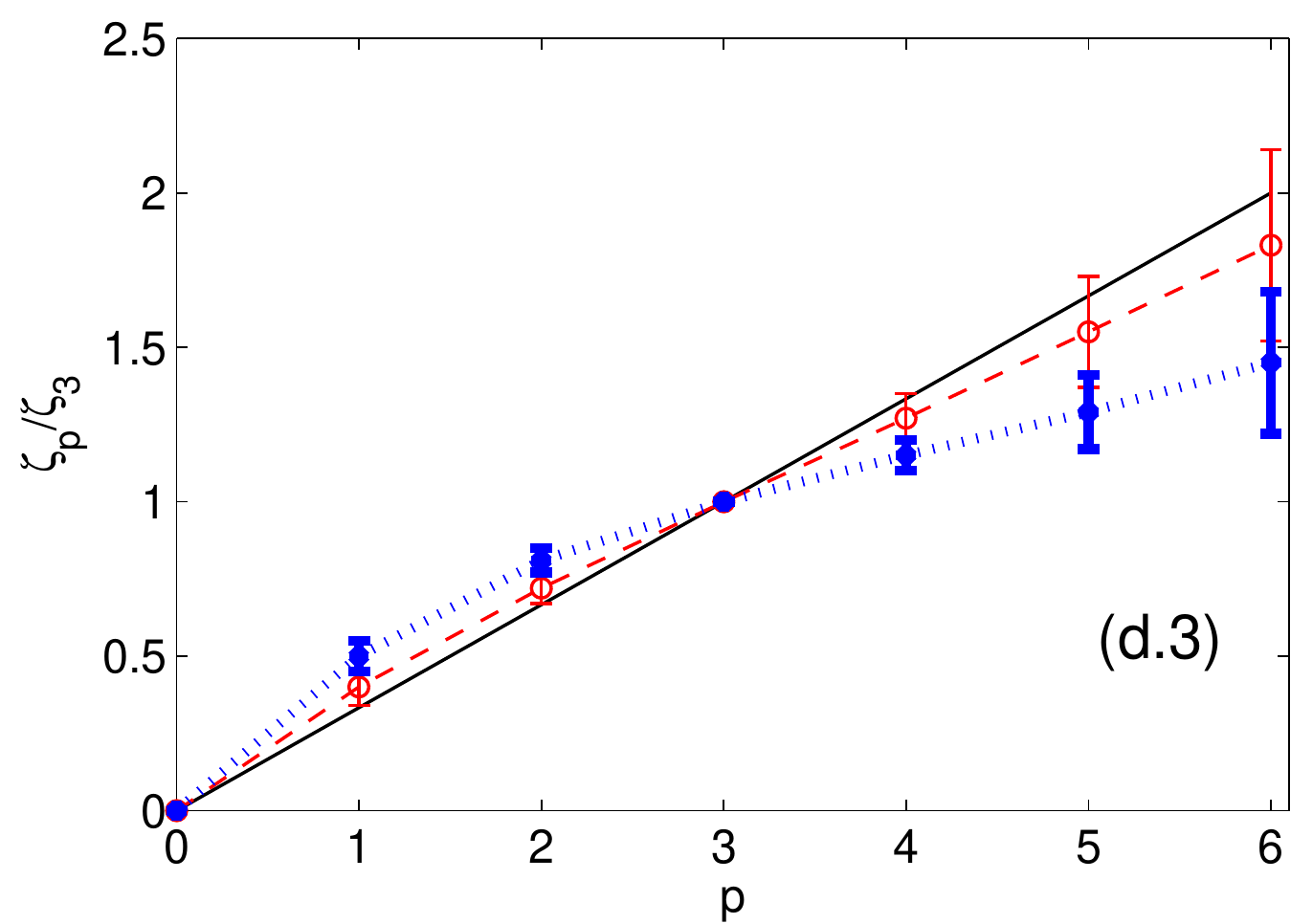}
\includegraphics[width=0.23\textwidth]{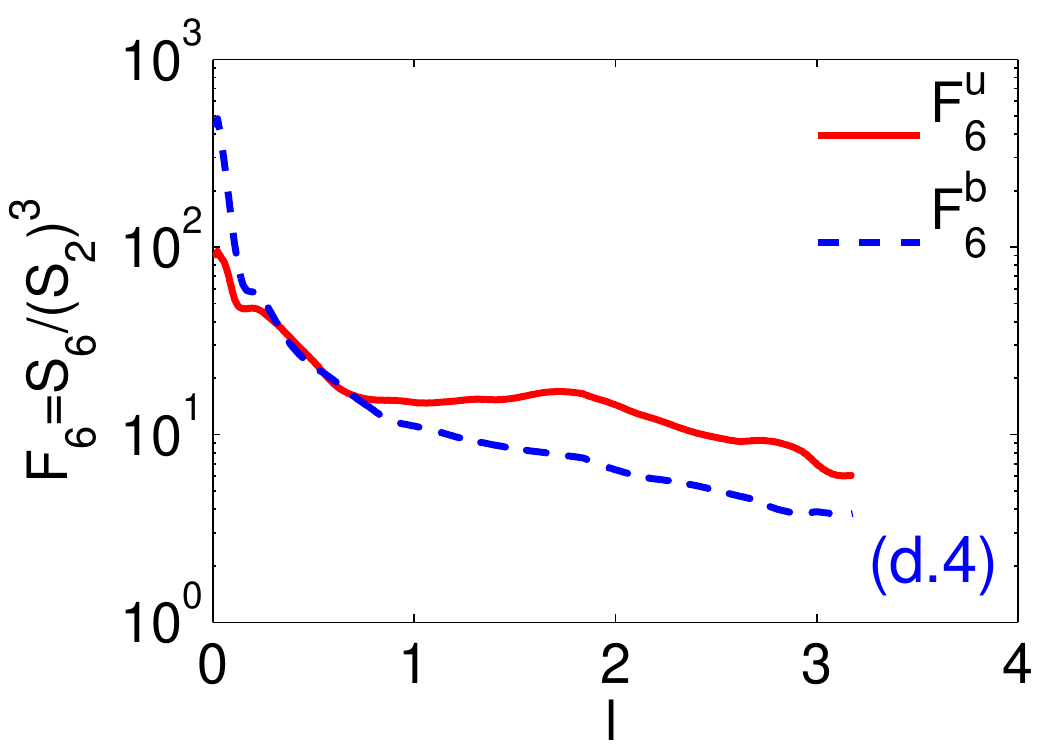}
\end{center}
\caption[]{Log–log (base 10) ESS plots of order-$p$
structure functions of the velocity $S^u_p(l)$ [(a.1)-(d.1)] and
magnetic-field $S^b_p(l)$ [(a.2)-(d.2)] versus $S^u_3(l)$ and $S^b_3(l)$,
respectively; plots of the local slopes of these curves are shown in the inset. 
The black horizontal lines, with vertical ticks at their ends,
show the inertial range over which we have averaged the exponent ratios
$\zeta^u_p/\zeta_p^u$ and $\zeta^u_p/\zeta_p^u$; plots are shown for
$p=1$ (red small-dotted line), $p=2$ (green dot-dashed line),
$p=3$ (blue line), $p=4$ (black thin-dashed line), 
$p=5$ (cyan thick-dashed line), and $p=6$ (magenta large-dotted
line). Subplots (a.3)-(d.3) show the exponent ratios
$\zeta_p/\zeta_3$ versus $p$ for the velocity (red dashed line with thin
errorbars) and magnetic fields (blue dotted line with
thick errorbars); the black solid line shows the K41 result
$\zeta_p^{K41}=p/3$. The semilog (base 10) plots (a.4)-(d.4) 
show the hyperflatnesses $F^u_6(l)$ (red line) and $F^b_6(l)$ 
(blue dashed line) versus $l$. Subplots in panels (a), (b), (c), and (d) 
are from our decaying-turbulence runs R1C, R2C, R3C, and R4C, respectively, 
with ${\rm Pr_M} = 0.01,\, 0.1,\, 1$, and $10$.}
\label{fig:stfn-1024}
\end{figure}
\begin{table*}
\caption{Multiscaling exponent ratios $\zeta_p^u/\zeta_3^u$ 
and $\zeta_p^b/\zeta_3^b$ from our decaying-MHD-turbulence runs
R1C-R4C.}
\label{table:zetap-decaying}
\begin{center}
\begin{tabular}{l|l|l}
\hline\noalign{\smallskip}
$p$ & $\zeta_p^u/\zeta_3^u$; $\zeta_p^b/\zeta_3^b ({\rm Pr_M}=0.01)$ &
$\zeta_p^u/\zeta_3^u$; $\zeta_p^b/\zeta_3^b ({\rm Pr_M}=0.1)$ \\  
\noalign{\smallskip}\hline\noalign{\smallskip}
1 & 0.41 $\pm$ 0.04; 0.35 $\pm$ 0.01 & 0.39 $\pm$ 0.09; 0.42 $\pm$ 0.04 \\ 
2 & 0.74 $\pm$ 0.04; 0.68 $\pm$ 0.01 & 0.71 $\pm$ 0.08; 0.74 $\pm$ 0.03 \\ 
3 & 1.00 $\pm$ 0.00; 1.00 $\pm$ 0.00 & 1.00 $\pm$ 0.00; 1.00 $\pm$ 0.00 \\    
4 & 1.21 $\pm$ 0.09; 1.29 $\pm$ 0.03 & 1.26 $\pm$ 0.14; 1.20 $\pm$ 0.03 \\ 
5 & 1.38 $\pm$ 0.22; 1.56 $\pm$ 0.06 & 1.51 $\pm$ 0.32; 1.37 $\pm$ 0.07 \\ 
6 & 1.52 $\pm$ 0.41; 1.80 $\pm$ 0.10 & 1.76 $\pm$ 0.53; 1.52 $\pm$ 0.13 \\ 
\noalign{\smallskip}\hline\noalign{\smallskip}
$p$ & $\zeta_p^u/\zeta_3^u$; $\zeta_p^b/\zeta_3^b ({\rm Pr_M}=1)$  &
$\zeta_p^u/\zeta_3^u$; $\zeta_p^b/\zeta_3^b ({\rm Pr_M}=10)$  \\
\noalign{\smallskip}\hline\noalign{\smallskip}
1 & 0.42 $\pm$ 0.03; 0.49 $\pm$ 0.04 & 0.40 $\pm$ 0.06; 0.50 $\pm$ 0.05 \\
2 & 0.74 $\pm$ 0.03; 0.80 $\pm$ 0.04 & 0.72 $\pm$ 0.05; 0.81 $\pm$ 0.04 \\
3 & 1.00 $\pm$ 0.00; 1.00 $\pm$ 0.00 & 1.00 $\pm$ 0.00; 1.00 $\pm$ 0.00 \\
4 & 1.25 $\pm$ 0.06; 1.15 $\pm$ 0.07 & 1.27 $\pm$ 0.08; 1.15 $\pm$ 0.05 \\
5 & 1.50 $\pm$ 0.16; 1.27 $\pm$ 0.18 & 1.55 $\pm$ 0.18; 1.29 $\pm$ 0.12 \\
6 & 1.74 $\pm$ 0.30; 1.38 $\pm$ 0.32 & 1.83 $\pm$ 0.31; 1.45 $\pm$ 0.23 \\
\noalign{\smallskip}\hline
\end{tabular}
\end{center}
\end{table*}

We begin with data from our decaying-MHD-turbulence runs R1C-R4C, which use
$1024^3$ collocation points and span the ${\rm Pr_M}$ range $0.01 - 10$.
Figures~\ref{fig:stfn-1024}(a.1)-\ref{fig:stfn-1024}(d.1) show ESS plots for
$S^u_p(r)$ for runs R1C-R4C, respectively, for $p=1$ (red small-dotted line),
$p=2$ (green dot-dashed line), $p=3$ (blue line), $p=4$ (black thin-dashed
line), $p=5$ (cyan thick-dashed line), and $p=6$ (magenta large-dotted line);
their analogues for $S^b_p(r)$ are given in
Figs.~\ref{fig:stfn-1024}(a.2)-\ref{fig:stfn-1024}(d.2);  the local slopes of
these ESS curves are shown in the insets of these figures. Flat portions in
these plots of local slopes help us to identify the inertial ranges. The
regions that we have chosen for our fits are indicated by black horizontal
lines with vertical ticks at their ends.  In such a region, the mean value
and the standard deviation of the local slope of the ESS plot for $S^u_p(r)$
(or $S^b_p(r)$) yield, respectively, our estimates for the exponent ratio
$\zeta_p^u/\zeta_3^u$ (or $\zeta_p^b/\zeta_3^b$) and its errorbars.
Figures~\ref{fig:stfn-1024}(a.3)-\ref{fig:stfn-1024}(d.3) show plots of these
exponent ratios versus $p$ for the velocity field (blue dotted line with
thick errorbars) and the magnetic field (red dashed line with thin error
bars); the black solid line shows the K41 result for comparison. Though
earlier studies~\cite{biskamp00,mininni07} have obtained such exponents from
DNS studies, they have done so, to the best of our knowledge, only for ${\rm
Pr_M}=1$; furthermore, they have not reported errorbars. Although our
(conservative) errorbars are large, our plots of exponent ratios suggest the
following: (a) deviations from the K41 result are significant, especially for
$p > 3$, as in fluid turbulence; (b) at large values of ${\rm Pr_M}$ the
magnetic field is more intermittent than the velocity field, in so far as the
deviations of $\zeta_p^b/\zeta_3^b$ from the K41 result $p/3$ are larger than
those of $\zeta_p^u/\zeta_3^u$; (c) as we reduce ${\rm Pr_M}$ this difference
in intermittency reduces until, at ${\rm Pr_M} = 0.01$, the velocity field
shows signs of becoming more intermittent than the magnetic field.  This
trend in intermittency is corroborated by plots versus $l$ of the
the hyperflatnesses $F^u_6(l)=\frac{S^u_6(l)}{S^u_2(l)^3}$ (red line) and
$F^b_6(l)=\frac{S^b_6(l)}{S^b_2(l)^3}$ (blue dashed line) in
Figs.~\ref{fig:stfn-1024}(a.4)-\ref{fig:stfn-1024}(d.4) for runs R1C-R4C,
respectively: As $l$ decreases, $F^b_6(l)$ rises more rapidly than $F^u_6(l)$
except at ${\rm Pr_M} = 0.01$.

\begin{figure}[htb]
\begin{center}
\includegraphics[width=0.23\textwidth]{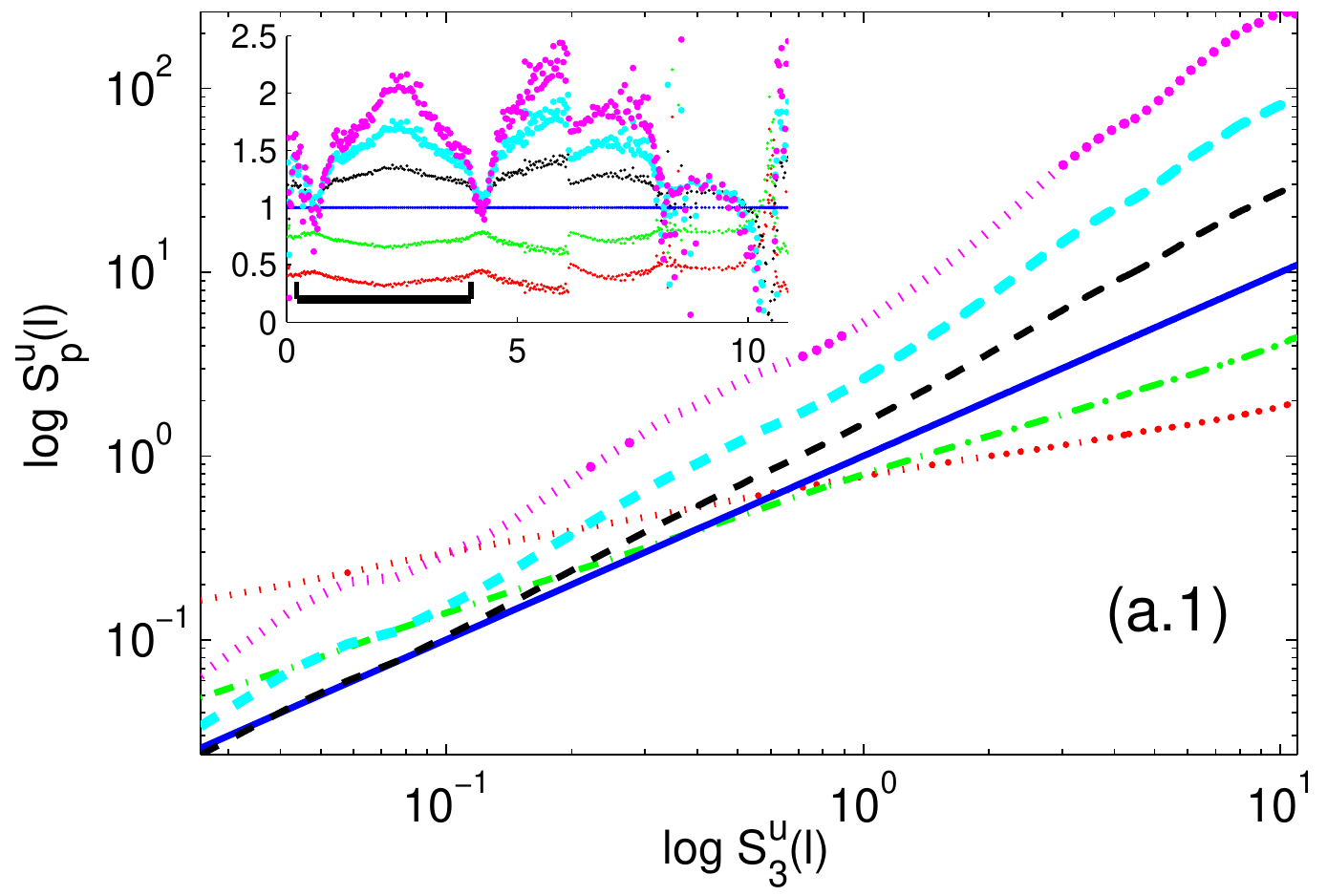}
\includegraphics[width=0.23\textwidth]{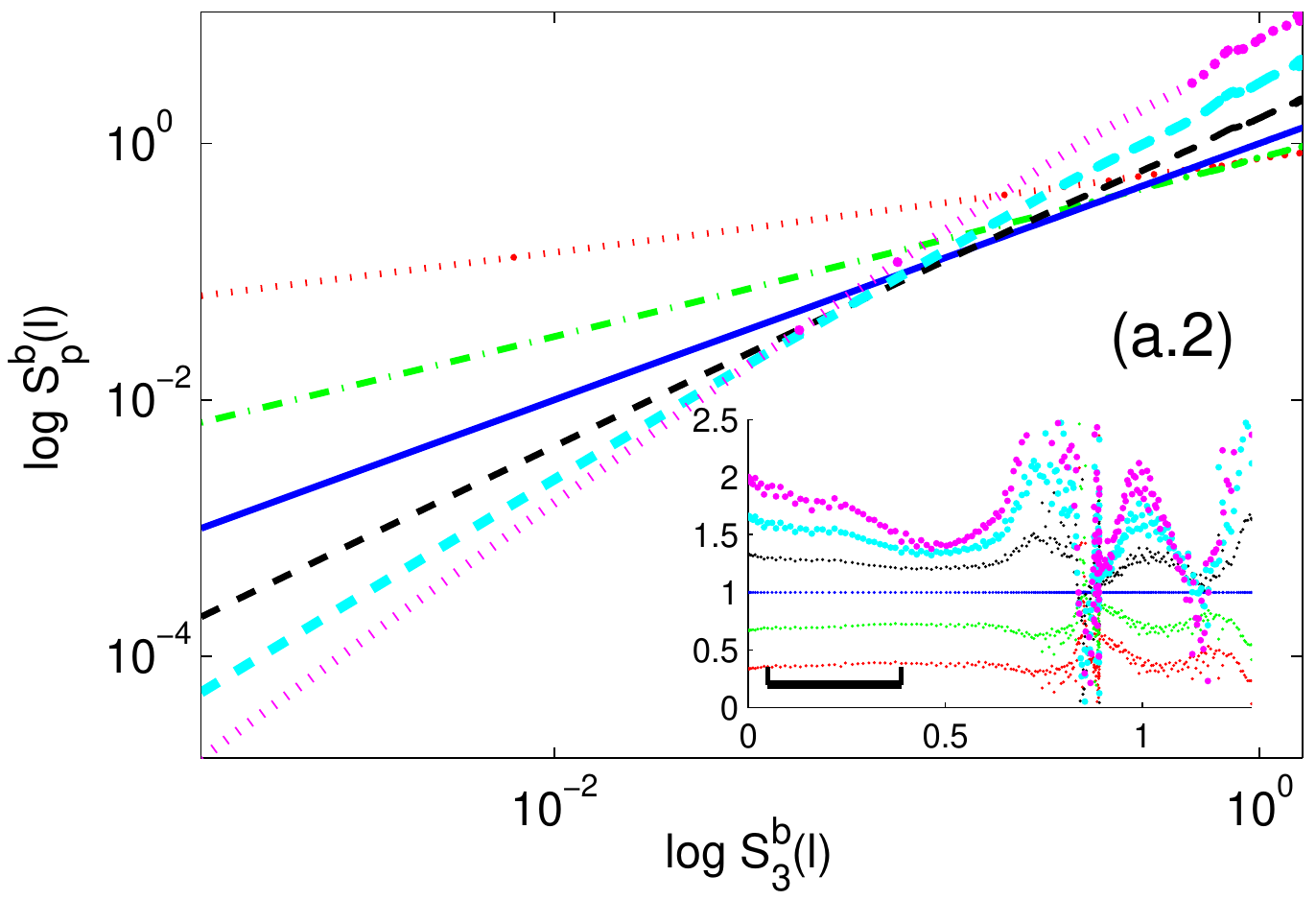}
\includegraphics[width=0.23\textwidth]{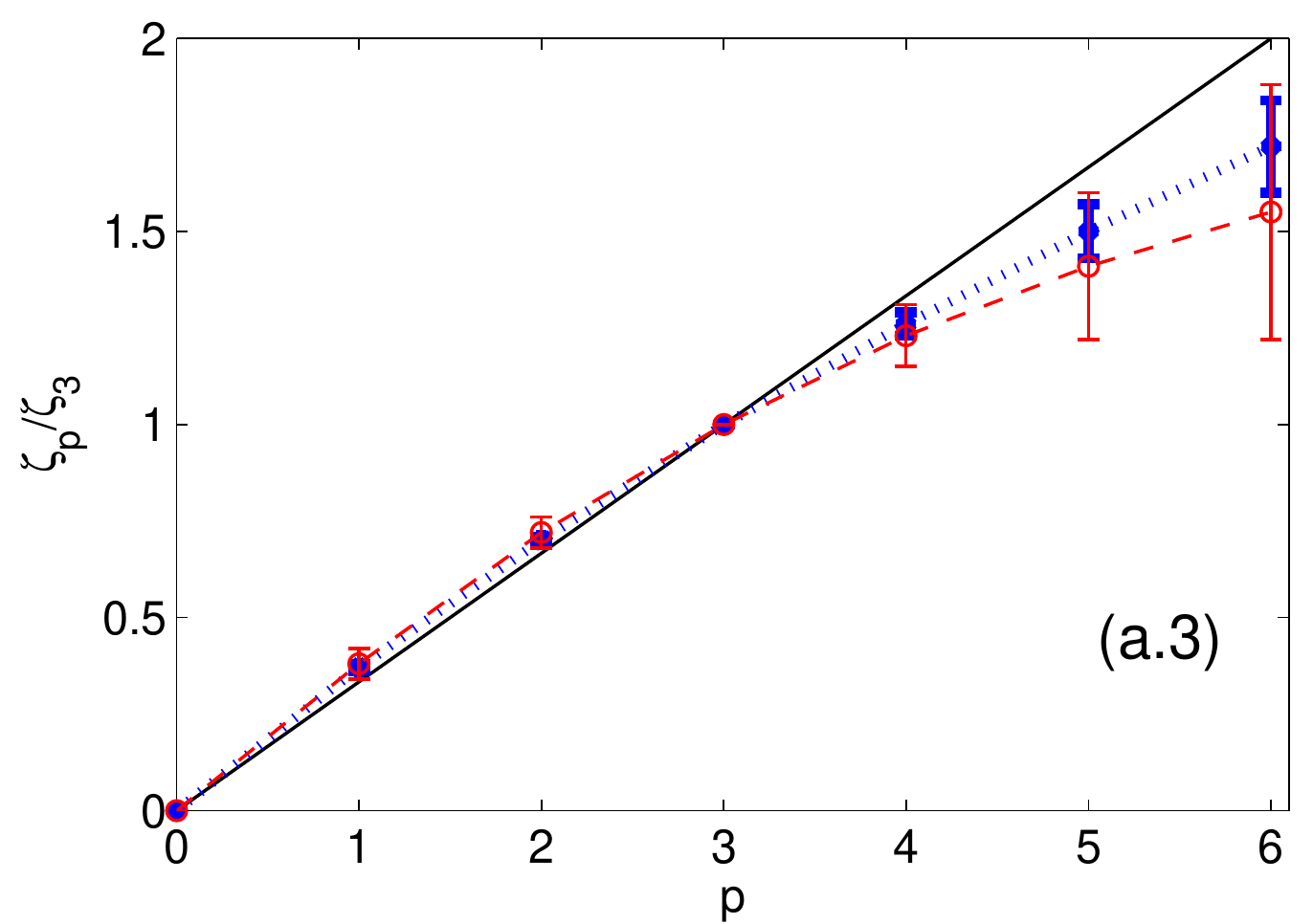}
\includegraphics[width=0.23\textwidth]{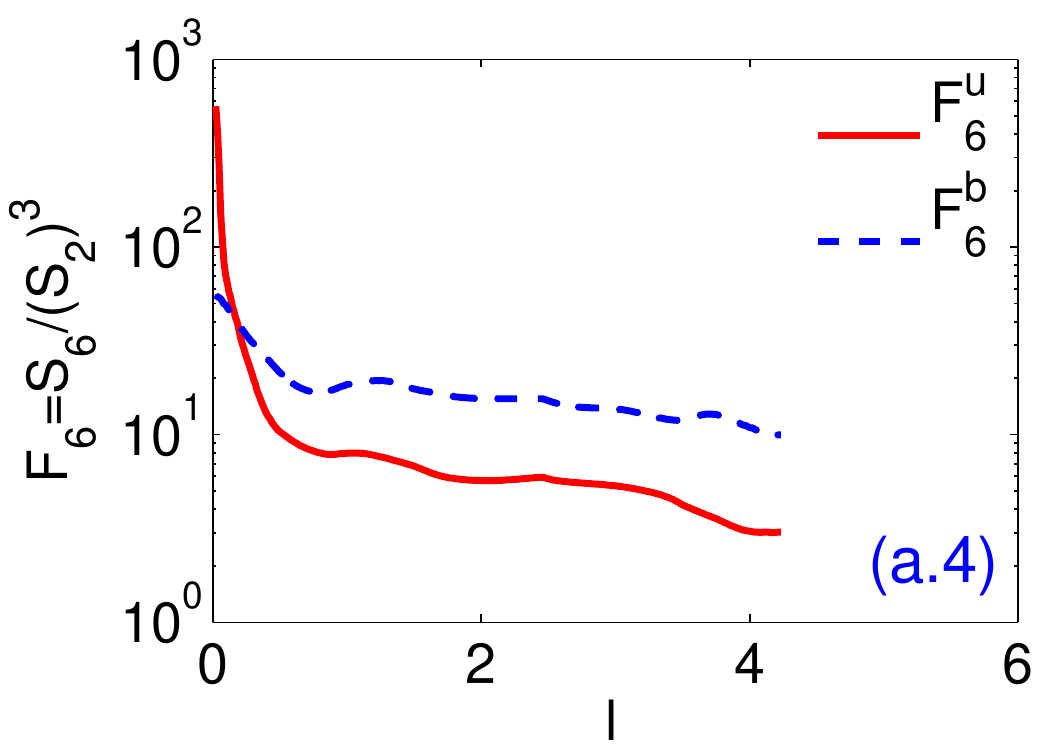}\\
\includegraphics[width=0.23\textwidth]{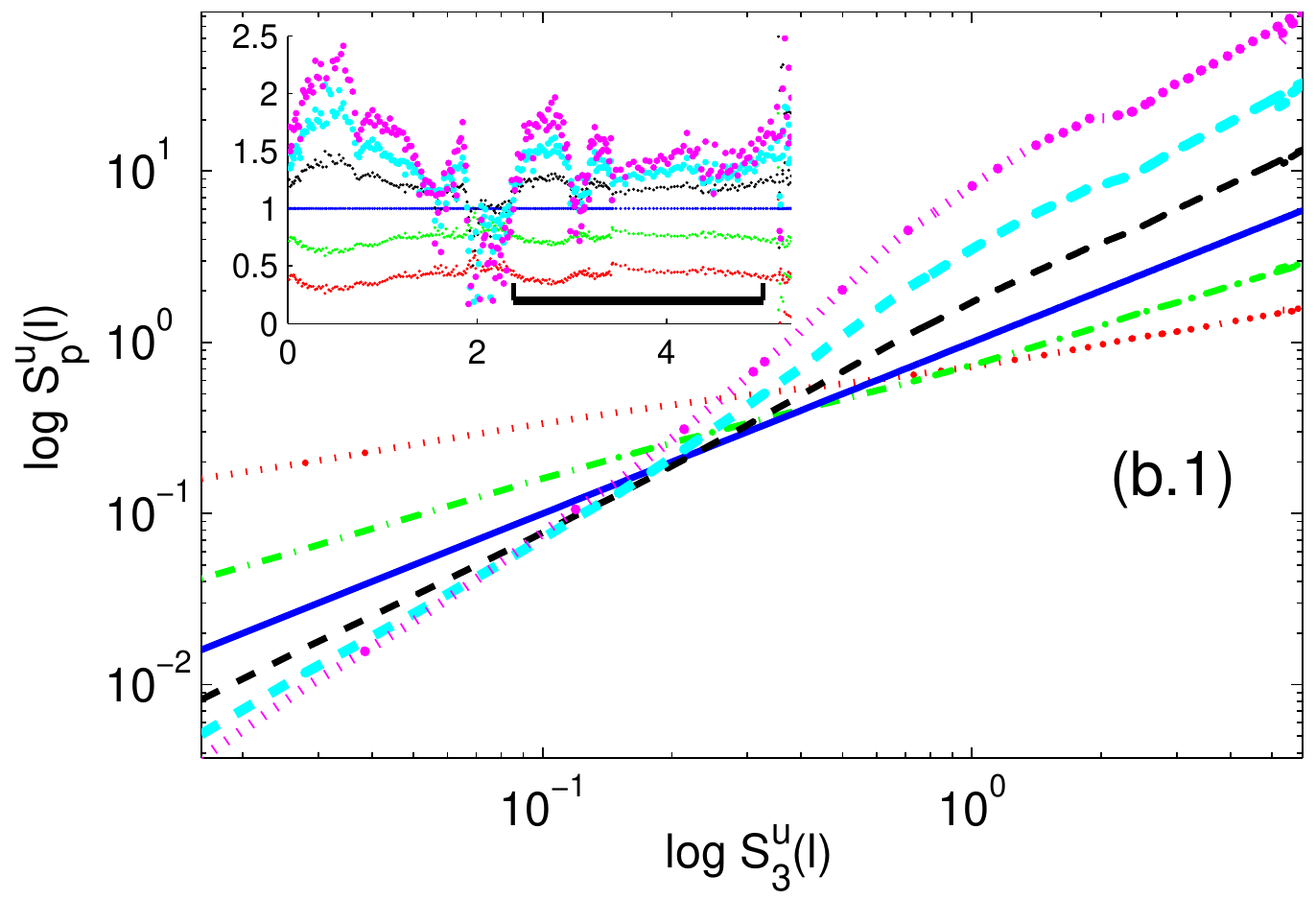}
\includegraphics[width=0.23\textwidth]{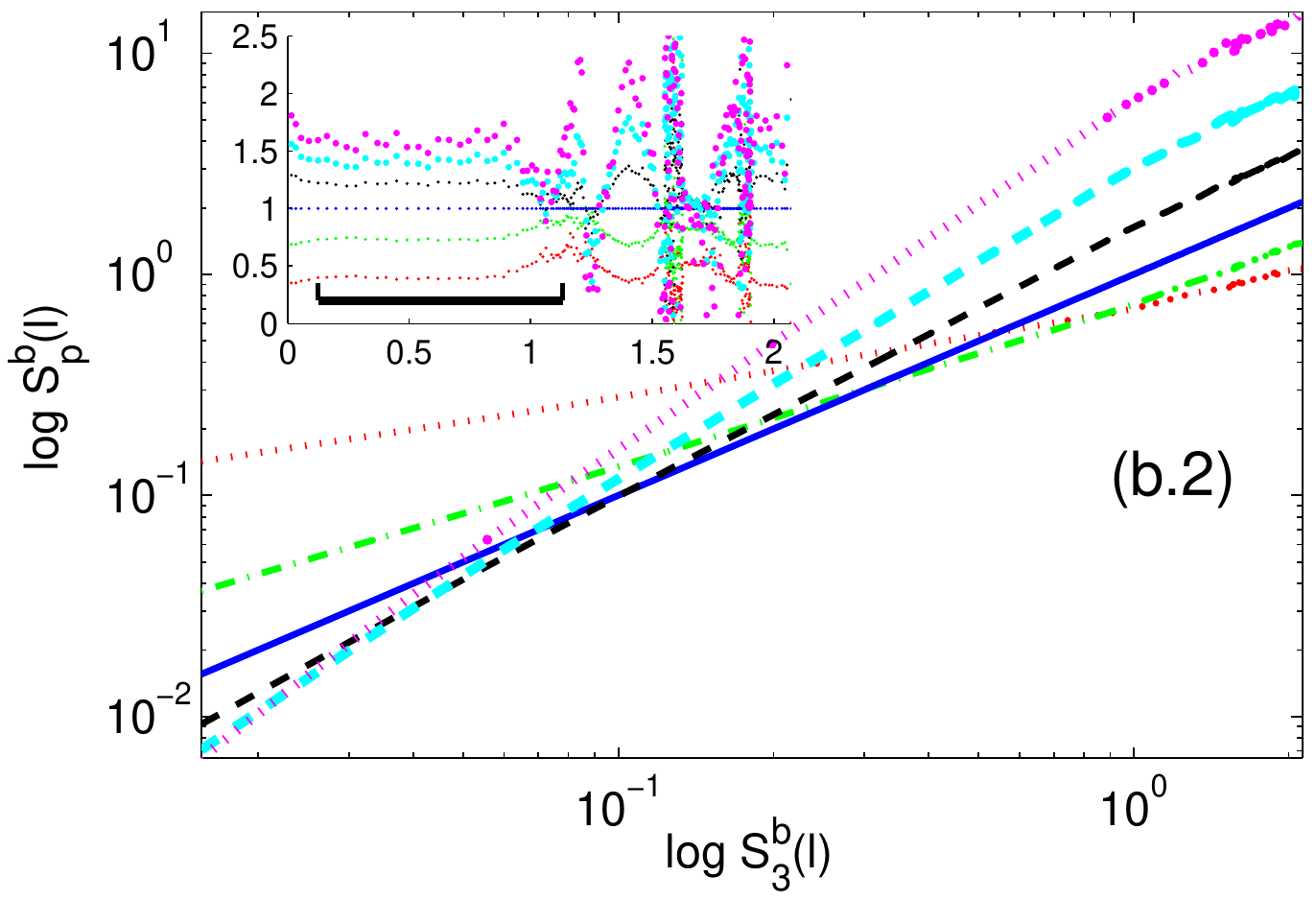}
\includegraphics[width=0.23\textwidth]{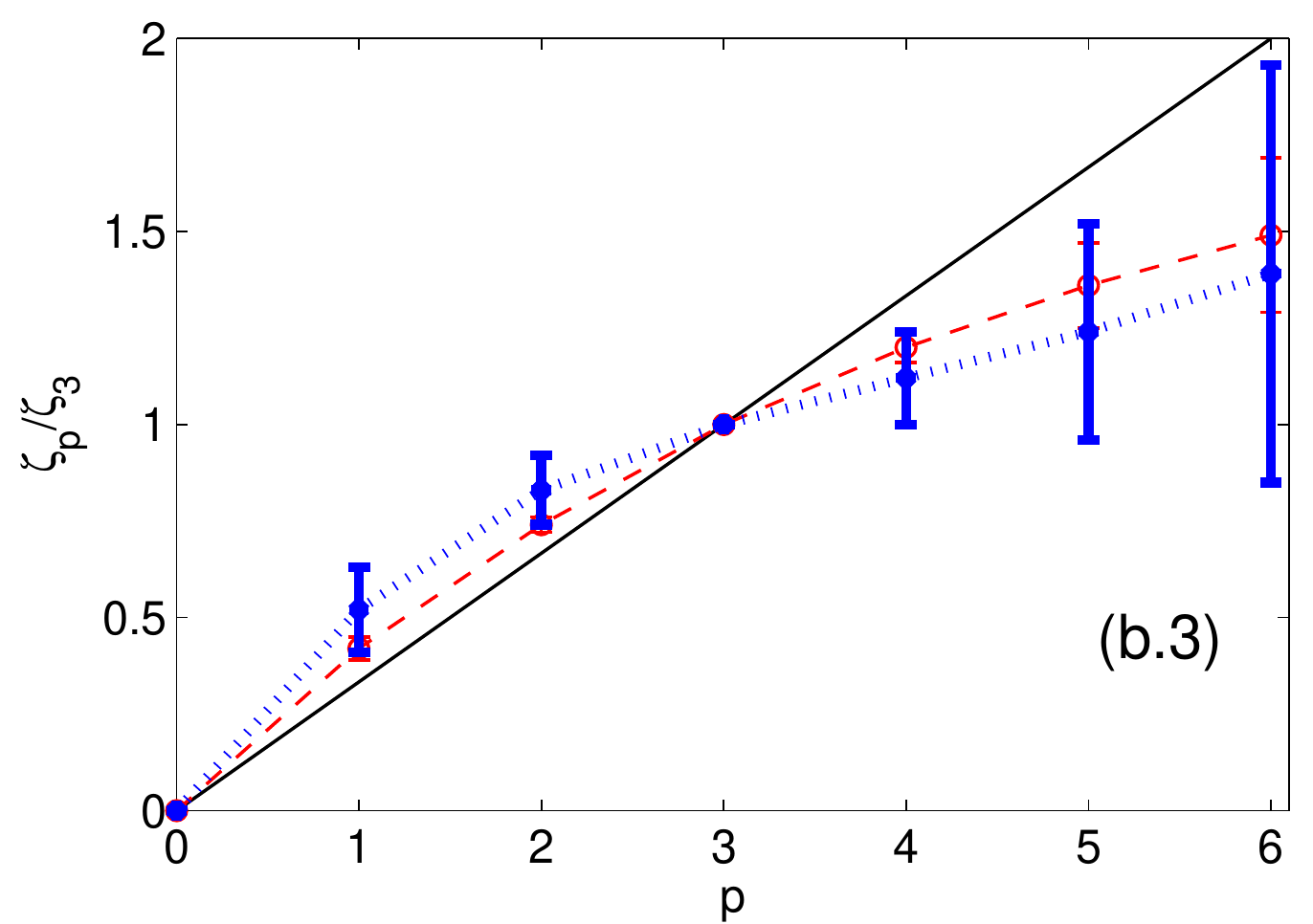}
\includegraphics[width=0.23\textwidth]{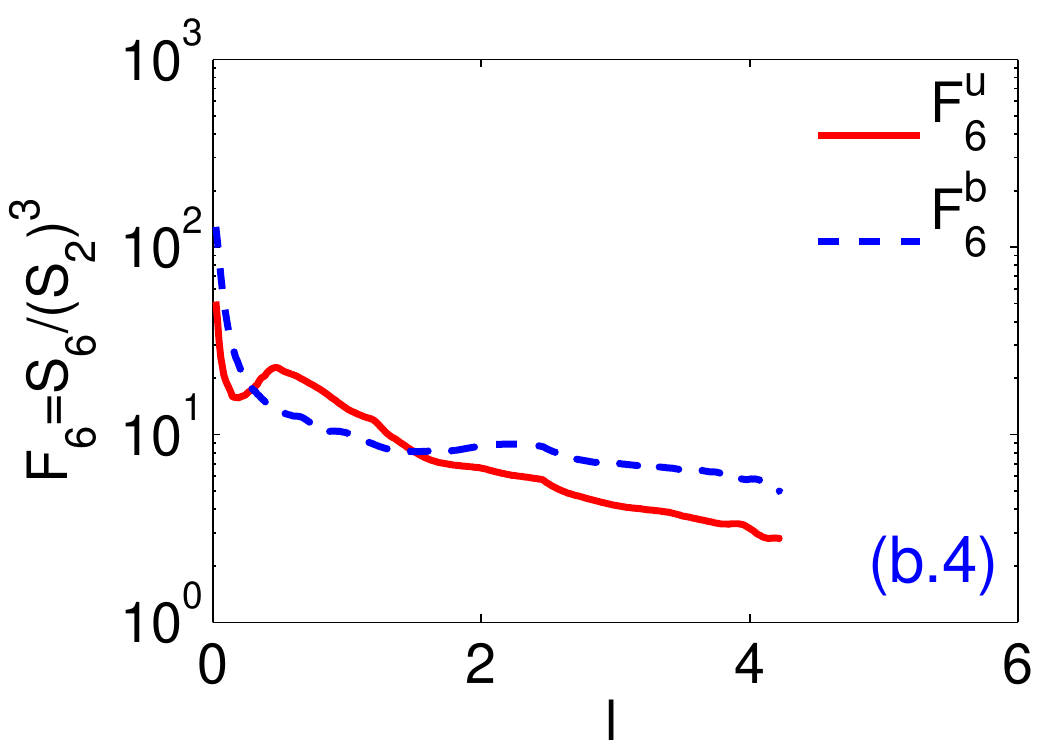}\\
\includegraphics[width=0.23\textwidth]{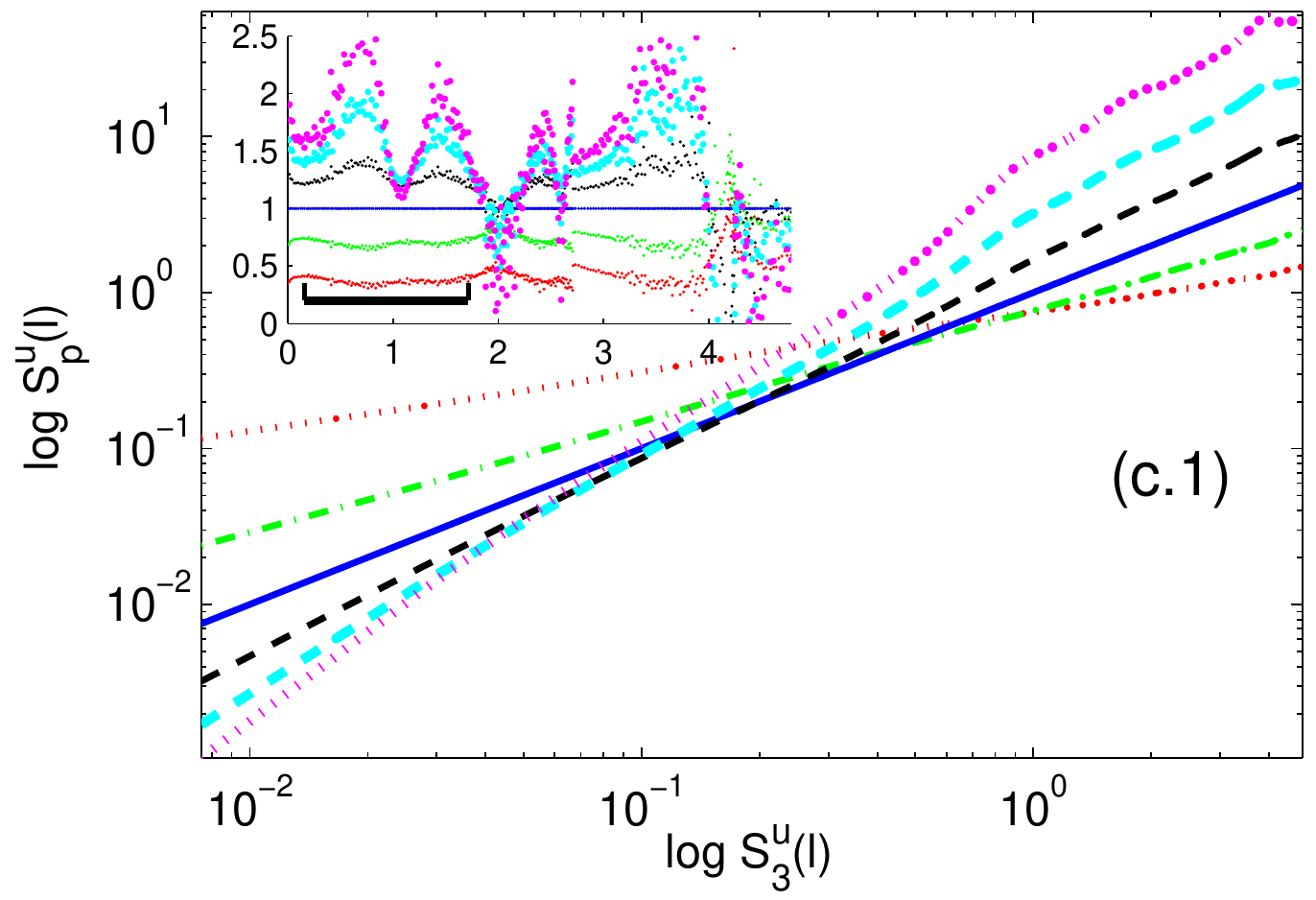}
\includegraphics[width=0.23\textwidth]{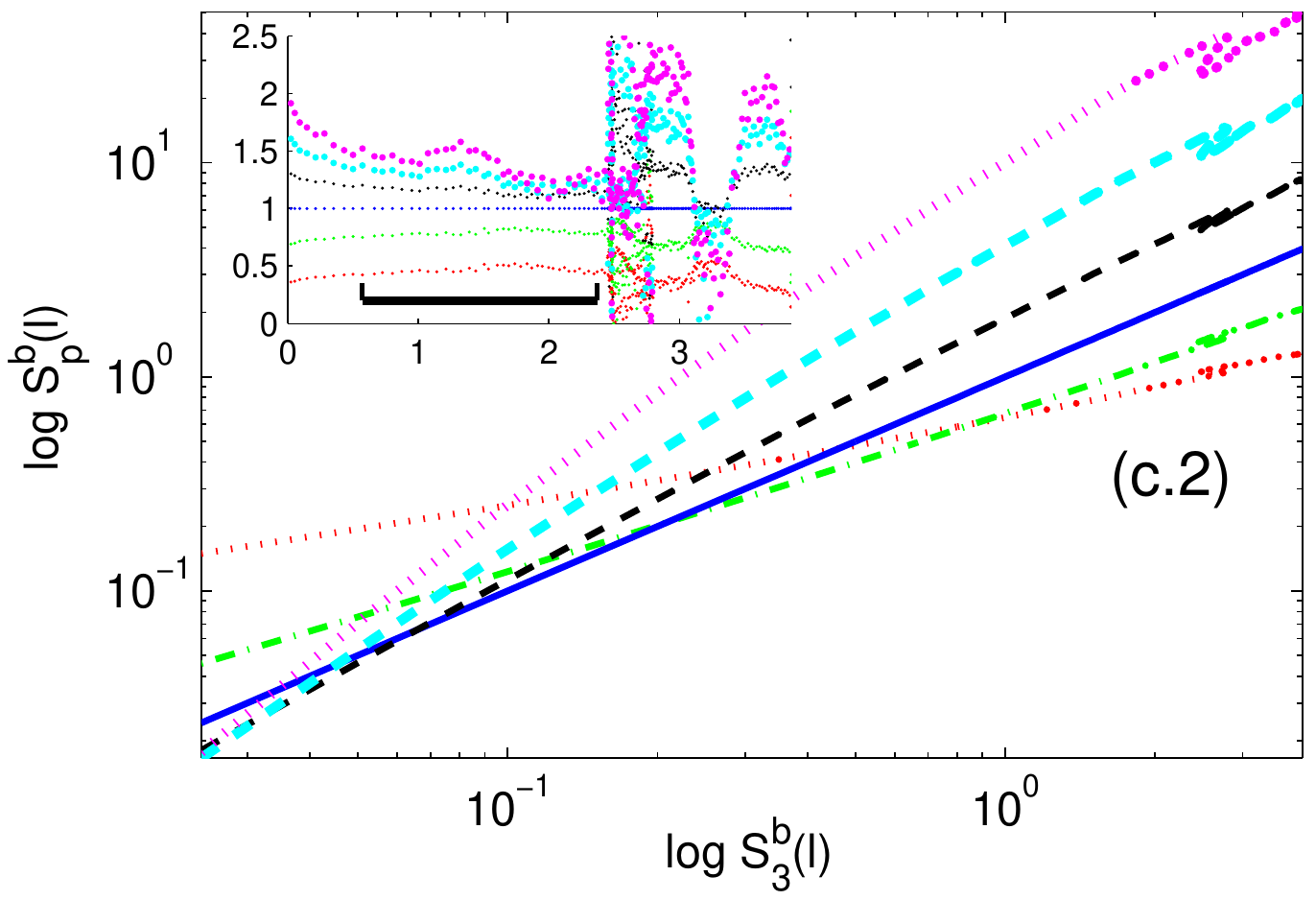}
\includegraphics[width=0.23\textwidth]{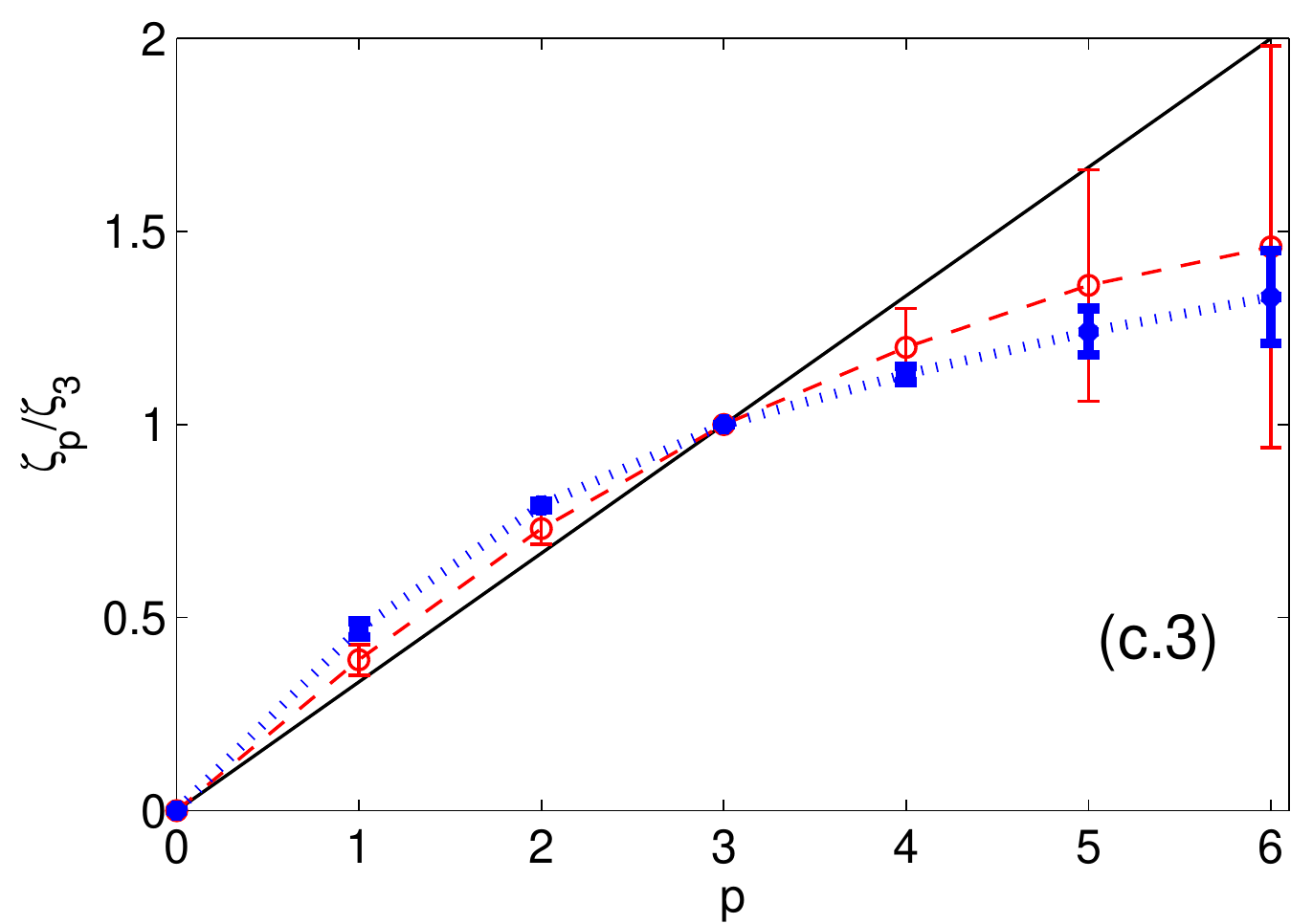}
\includegraphics[width=0.23\textwidth]{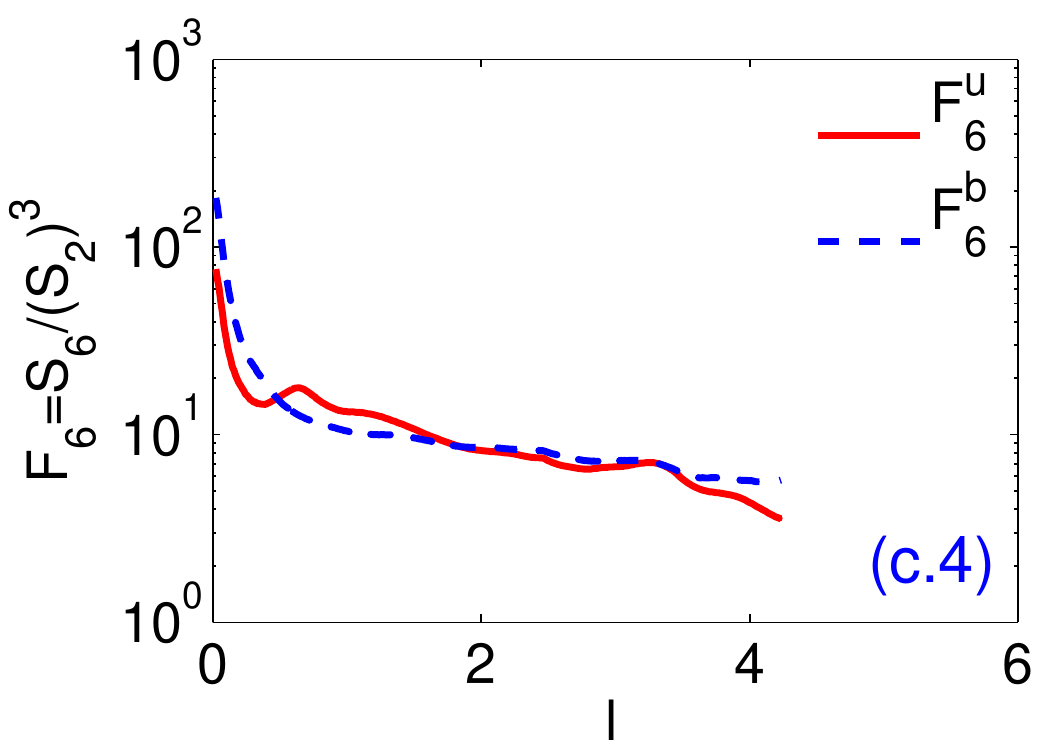}\\
\includegraphics[width=0.23\textwidth]{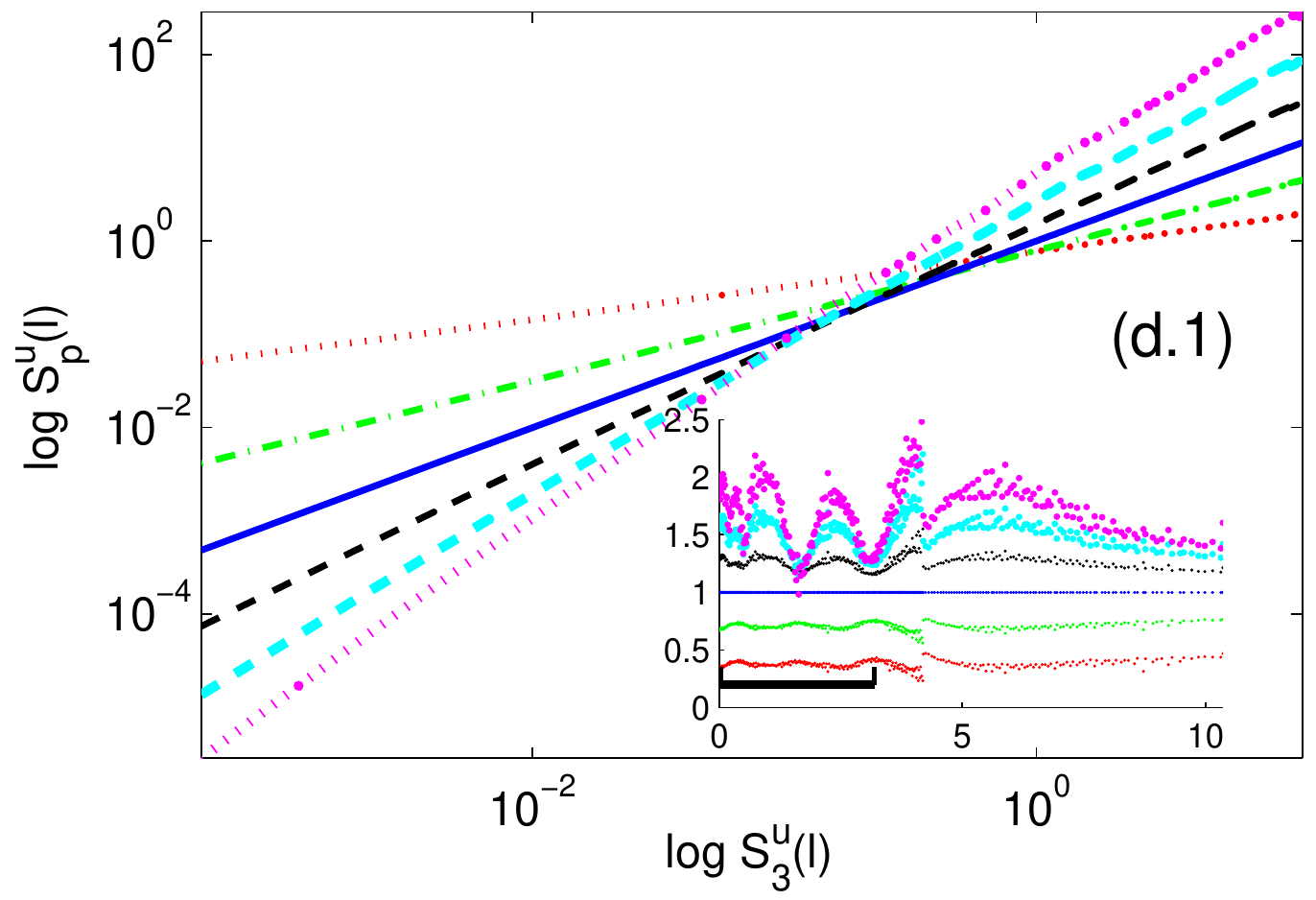}
\includegraphics[width=0.23\textwidth]{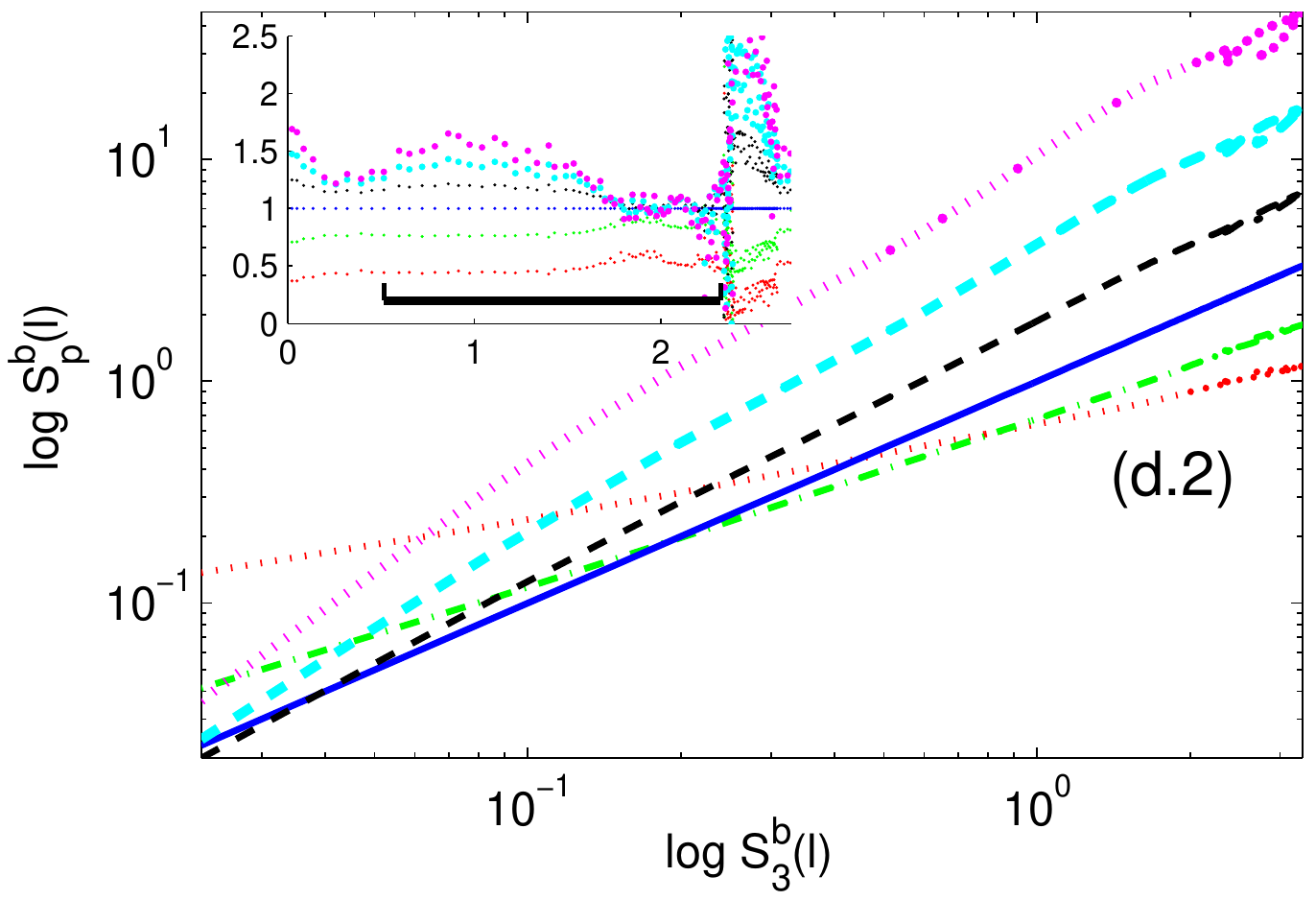}
\includegraphics[width=0.23\textwidth]{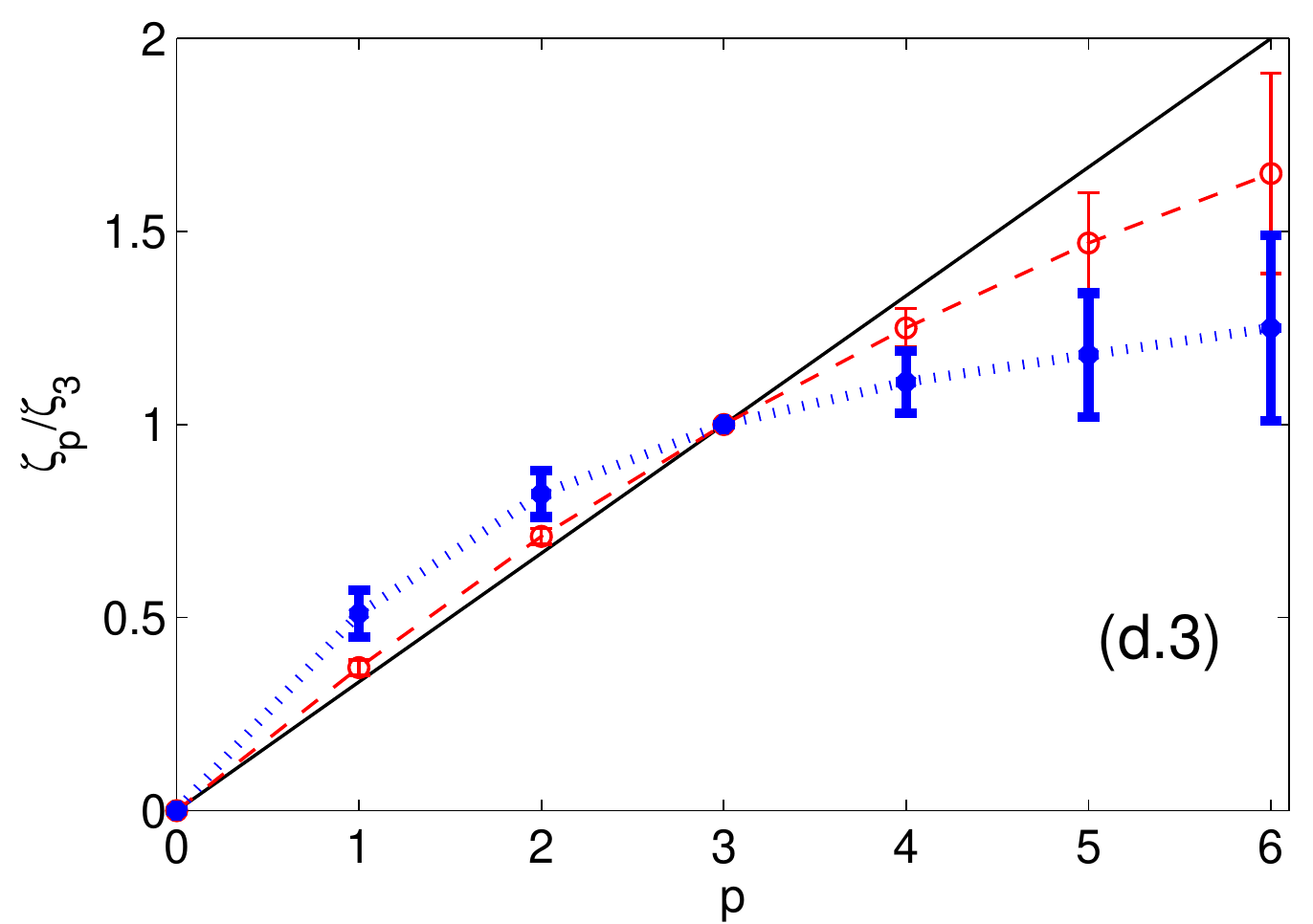}
\includegraphics[width=0.23\textwidth]{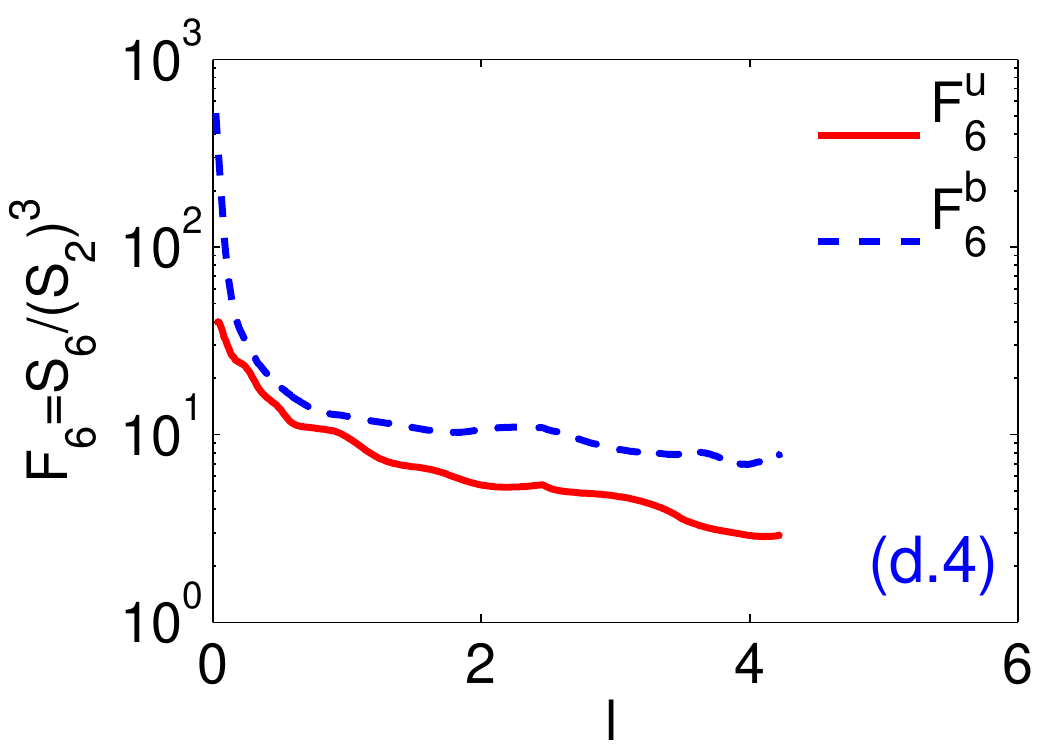}
\end{center}
\caption[]{Log–log (base 10) ESS plots of order-$p$
structure functions of the velocity $S^u_p(l)$ [(a.1)-(d.1)] and
magnetic-field $S^b_p(l)$ [(a.2)-(d.2)] versus $S^u_3(l)$ and $S^b_3(l)$,
respectively; plots of the local slopes of these curves are shown in the inset. 
The black horizontal lines, with vertical ticks at their ends,
show the inertial range over which we have averaged the exponent ratios
$\zeta^u_p/\zeta_p^u$ and $\zeta^u_p/\zeta_p^u$; plots are shown for
$p=1$ (red small-dotted line), $p=2$ (green dot-dashed line),
$p=3$ (blue line), $p=4$ (black thin-dashed line), 
$p=5$ (cyan thick-dashed line), and $p=6$ (magenta large-dotted
line). Subplots (a.3)-(d.3) show the exponent ratios
$\zeta_p/\zeta_3$ versus $p$ for the velocity (red dashed line with thin
errorbars) and magnetic fields (blue dotted line with
thick errorbars); the black solid line shows the K41 result
$\zeta_p^{K41}=p/3$. The semilog (base 10) plots (a.4)-(d.4) 
show the hyperflatnesses $F^u_6(l)$ (red line) and $F^b_6(l)$ 
(blue dashed line) versus $l$. Subplots in panels (a), (b), (c), and (d), are from our
statistically steady MHD-turbulence runs R1D, R2D, R3D, and R4D, 
respectively, with ${\rm Pr_M} = 0.01,\, 0.1,\, 1$, and $10$.}
\label{fig:stfn-forced}
\end{figure}
\begin{table}
\caption{Multiscaling exponent ratios $\zeta_p^u/\zeta_3^u$ 
and $\zeta_p^b/\zeta_3^b$ from our statistically steady 
MHD-turbulence runs R1D-R4D.}
\label{table:zetap-forced}
\begin{center}
\begin{tabular}{l|l|l}
\hline\noalign{\smallskip}
$p$ & $\zeta_p^u/\zeta_3^u$; $\zeta_p^b/\zeta_3^b ({\rm Pr_M}=0.01)$ &
$\zeta_p^u/\zeta_3^u$; $\zeta_p^b/\zeta_3^b ({\rm Pr_M}=0.1)$ \\
\noalign{\smallskip}\hline\noalign{\smallskip}
1 & 0.38 $\pm$ 0.04; 0.37 $\pm$ 0.01 & 0.42 $\pm$ 0.03; 0.52 $\pm$ 0.11 \\
2 & 0.72 $\pm$ 0.04; 0.70 $\pm$ 0.01 & 0.74 $\pm$ 0.02; 0.83 $\pm$ 0.09 \\
3 & 1.00 $\pm$ 0.00; 1.00 $\pm$ 0.00 & 1.00 $\pm$ 0.00; 1.00 $\pm$ 0.00 \\
4 & 1.23 $\pm$ 0.08; 1.26 $\pm$ 0.03 & 1.20 $\pm$ 0.04; 1.12 $\pm$ 0.12 \\
5 & 1.41 $\pm$ 0.19; 1.50 $\pm$ 0.07 & 1.36 $\pm$ 0.11; 1.24 $\pm$ 0.28 \\
6 & 1.55 $\pm$ 0.33; 1.72 $\pm$ 0.12 & 1.49 $\pm$ 0.20; 1.39 $\pm$ 0.54 \\
\noalign{\smallskip}\hline\noalign{\smallskip}
$p$ & $\zeta_p^u/\zeta_3^u$; $\zeta_p^b/\zeta_3^b ({\rm Pr_M}=1)$  &
$\zeta_p^u/\zeta_3^u$; $\zeta_p^b/\zeta_3^b ({\rm Pr_M}=10)$  \\
\noalign{\smallskip}\hline\noalign{\smallskip}
1 & 0.39 $\pm$ 0.04; 0.47 $\pm$ 0.02 & 0.37 $\pm$ 0.02; 0.51 $\pm$ 0.06 \\
2 & 0.73 $\pm$ 0.04; 0.79 $\pm$ 0.01 & 0.71 $\pm$ 0.02; 0.82 $\pm$ 0.06 \\
3 & 1.00 $\pm$ 0.00; 1.00 $\pm$ 0.00 & 1.00 $\pm$ 0.00; 1.00 $\pm$ 0.00 \\
4 & 1.20 $\pm$ 0.10; 1.13 $\pm$ 0.02 & 1.25 $\pm$ 0.05; 1.11 $\pm$ 0.08 \\
5 & 1.36 $\pm$ 0.30; 1.24 $\pm$ 0.06 & 1.47 $\pm$ 0.13; 1.18 $\pm$ 0.16 \\
6 & 1.46 $\pm$ 0.52; 1.33 $\pm$ 0.12 & 1.65 $\pm$ 0.26; 1.25 $\pm$ 0.24 \\
\noalign{\smallskip}\hline
\end{tabular}
\end{center}
\end{table}

Similar results follow from our studies of statistically steady MHD
turbulence in runs R1D-R4D, which use $512^3$ collocation points and span the
${\rm Pr_M}$ range $0.01 - 10$.
Figures~\ref{fig:stfn-forced}(a.1)-\ref{fig:stfn-forced}(d.1) show ESS plots
for $S^u_p(r)$ for runs R1D-R4D, respectively, for $p=1$ (red small-dotted
line), $p=2$ (green dot-dashed line), $p=3$ (blue line), $p=4$ (black
thin-dashed line), $p=5$ (cyan thick-dashed line), and $p=6$ (magenta
large-dotted line); their analogues for $S^b_p(r)$ are given in
Figs.~\ref{fig:stfn-forced}(a.2)-\ref{fig:stfn-forced}(d.2); the local slopes
of these ESS curves are shown in the insets of these figures.  We obtain
estimates for the exponent ratio $\zeta_p^u/\zeta_3^u$ and
$\zeta_p^b/\zeta_3^b$ and their errorbars as in Fig.~\ref{fig:stfn-1024}.
Figures~\ref{fig:stfn-forced}(a.3)-\ref{fig:stfn-forced}(d.3) show plots of
these exponent ratios versus $p$ for the velocity field (blue dotted line
with thick errorbars) and the magnetic field (red dashed line with thin
errorbars); the black solid line shows the K41 result for comparison.  Plots
versus $l$ of the hyperflatnesses $F^u_6(l)$ (red line) and $F^b_6(l)$
(blue dashed line) are given in
Figs.~\ref{fig:stfn-forced}(a.4)-\ref{fig:stfn-forced}(d.4) for runs R1D-R4D,
respectively. All the trends here as exactly as in the
decaying-MHD-turbulence plots in Fig.~\ref{fig:stfn-1024}.

Tables~\ref{table:zetap-decaying} and \ref{table:zetap-forced} summarise,
respectively, our results for multiscaling exponent ratios for our
decaying-MHD-turbulence runs R1C-R4C and our statistically steady
MHD-turbulence runs R1D-R4D.  The trends of these ratios with ${\rm Pr_M}$
have been discussed above.  By comparing corresponding entries in the columns
and rows of these tables, we see that exponent ratios from decaying and
statistically steady MHD turbulence agree, given our (conservative) error
bars. Thus, at least at this level of resolution and accuracy, we have strong
universality of these exponent ratios, for a given value of ${\rm Pr_M}$, in
as much as the ratios from decaying-MHD turbulence agree with those from the
statistically steady case.  The dependence on ${\rm Pr_M}$ will be examined
in Sec.~\ref{sec:conclusions}. 

\begin{table}
\caption{A comparison of multiscaling exponent ratios $\zeta_p^u/\zeta_3^u$ 
and $\zeta_p^b/\zeta_3^b$ from our statistically steady 
and decaying MHD simulations and from decaying MHD simulations
by Mininni and Pouquet (Ref.~\cite{mininni09}), for ${\rm Pr_M}=1$.}
\label{table:compare}
\begin{center}
\begin{tabular}{l|l|l|l|l}
\hline\noalign{\smallskip}
$p$ & $\zeta^u_p$ Ref.~\cite{mininni09} & $\zeta^u_p/\zeta^u_3$ Ref.~\cite{mininni09} &
$\zeta^u_p/\zeta^u_3$ (R3D) & $\zeta^u_p/\zeta^u_3$ (R3C) \\ 
\noalign{\smallskip}\hline\noalign{\smallskip}
1 & 0.30 & 0.40 & 0.39 $\pm$ 0.04 & 0.42 $\pm$ 0.03 \\
2 & 0.55 & 0.74 & 0.73 $\pm$ 0.04 & 0.74 $\pm$ 0.03 \\
3 & 0.74 & 1.00 & 1.00 $\pm$ 0.00 & 1.00 $\pm$ 0.00 \\
4 & 0.91 & 1.22 & 1.20 $\pm$ 0.10 & 1.25 $\pm$ 0.06 \\
5 & 1.04 & 1.39 & 1.36 $\pm$ 0.30 & 1.50 $\pm$ 0.16 \\
6 & 1.17 & 1.56 & 1.46 $\pm$ 0.52 & 1.74 $\pm$ 0.30 \\
\noalign{\smallskip}\hline\noalign{\smallskip}
$p$ & $\zeta^b_p$ Ref.~\cite{mininni09} & $\zeta^b_p/\zeta^b_3$ Ref.~\cite{mininni09} &
$\zeta^b_p/\zeta^b_3$ (R3D) & $\zeta^b_p/\zeta^b_3$ (R3C) \\ 
\noalign{\smallskip}\hline\noalign{\smallskip}
1 & 0.36 & 0.43 & 0.47 $\pm$ 0.02 & 0.49 $\pm$ 0.04 \\
2 & 0.63 & 0.76 & 0.79 $\pm$ 0.01 & 0.80 $\pm$ 0.04 \\
3 & 0.83 & 1.00 & 1.00 $\pm$ 0.00 & 1.00 $\pm$ 0.00 \\
4 & 0.97 & 1.16 & 1.13 $\pm$ 0.02 & 1.15 $\pm$ 0.07 \\
5 & 1.07 & 1.28 & 1.24 $\pm$ 0.06 & 1.27 $\pm$ 0.18 \\
6 & 1.14 & 1.36 & 1.33 $\pm$ 0.12 & 1.38 $\pm$ 0.32 \\
\noalign{\smallskip}\hline
\end{tabular}
\end{center}
\end{table}

\subsection{Isosurfaces \label{sec:isosurf}} 

As we have mentioned in our discussion of fluid turbulence, isosurface plots
of quantities such as $\omega$, the modulus of the vorticity, give us a
visual appreciation of small-scale structures in a turbulent flow; in fluid
turbulence, iso-$\omega$ surfaces are slender tubes if $\omega$ is chosen to
be well above its mean value~\cite{pramana09,okamoto07}.  For the case of MHD
turbulence it is natural to consider isosurface plots~\cite{yoshimatsu09} of
$\omega$, the modulus $j$ of the current density, energy dissipation rates,
and the effective pressure.

\begin{figure}[htb]
\begin{center}
\includegraphics[width=0.23\textwidth]{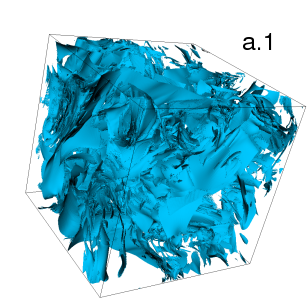}
\includegraphics[width=0.23\textwidth]{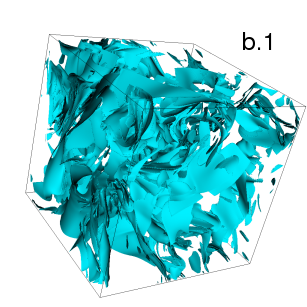}
\includegraphics[width=0.23\textwidth]{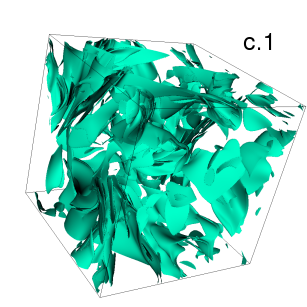}
\includegraphics[width=0.23\textwidth]{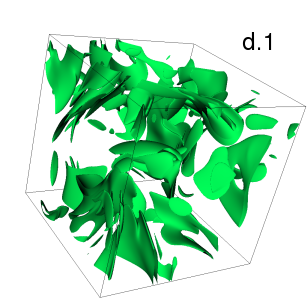}
\includegraphics[width=0.23\textwidth]{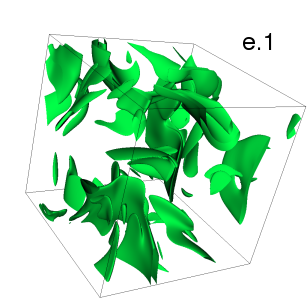}
\includegraphics[width=0.23\textwidth]{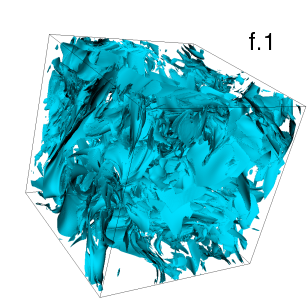}
\includegraphics[width=0.23\textwidth]{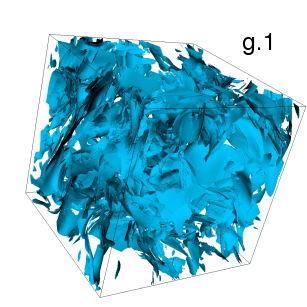}
\includegraphics[width=0.23\textwidth]{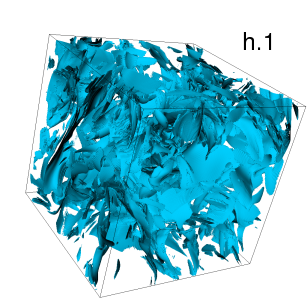}
\includegraphics[width=0.23\textwidth]{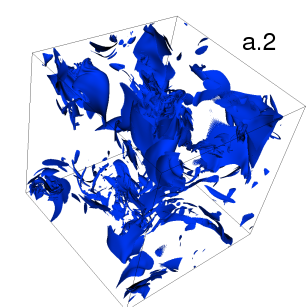}
\includegraphics[width=0.23\textwidth]{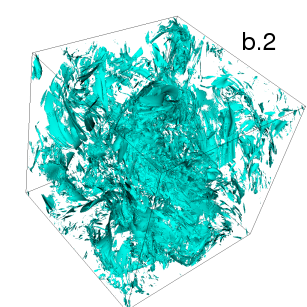}
\includegraphics[width=0.23\textwidth]{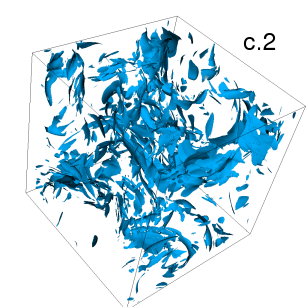}
\includegraphics[width=0.23\textwidth]{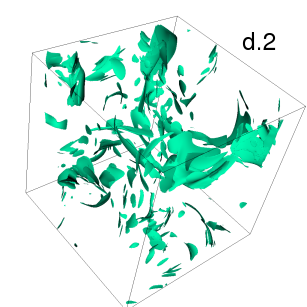}
\includegraphics[width=0.23\textwidth]{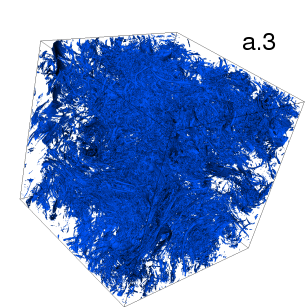}
\includegraphics[width=0.23\textwidth]{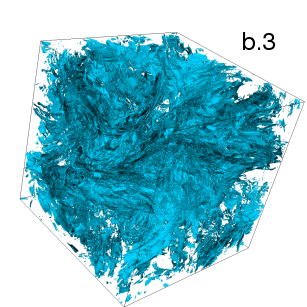}
\includegraphics[width=0.23\textwidth]{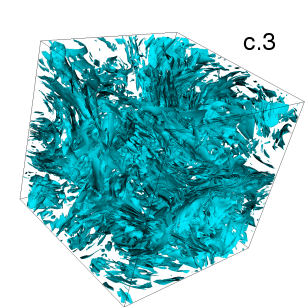}
\includegraphics[width=0.23\textwidth]{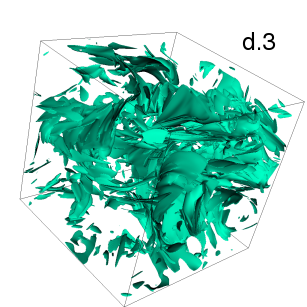}
\end{center}
\caption[]{Isosurfaces of the modulus $\omega$ of the vorticity:
(a.1) ${\rm Pr_M}=0.1$ (R1), (b.1) ${\rm Pr_M}=0.5$ (R2), (c.1) ${\rm
Pr_M}=1.0$ (R3), (d.1) ${\rm Pr_M}=5.0$ (R4), (e.1) ${\rm Pr_M}=10.0$ (R5),
(f.1) ${\rm Pr_M}=1.0$ (R3B), (g.1) ${\rm Pr_M}=5.0$ (R4B), (h.1) ${\rm
Pr_M}=10.0$ (R5B), (a.2) ${\rm Pr_M}=0.01$ (R1C), (b.2) ${\rm Pr_M}=0.1$
(R2C), (c.2) ${\rm Pr_M}=1.0$ (R3C), and (d.2) ${\rm Pr_M}=10.0$ (R4C) for
decaying MHD turbulence; and for statistically steady MHD turbulence (a.3)
${\rm Pr_M}=0.01$ (R1D), (b.3) ${\rm Pr_M}=0.1$ (R2D), (c.3) ${\rm Pr_M}=1.0$
(R3D), and (d.3) ${\rm Pr_M}=10.0$ (R4D); these isosurfaces go through
points at which the value of $\omega$ is two standard deviations above its
mean value (for any given plot).}
\label{fig:iso-w}
\end{figure}
\begin{figure}[htb]
\begin{center}
\includegraphics[width=0.23\textwidth]{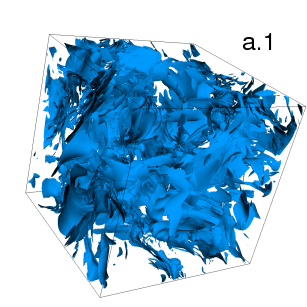}
\includegraphics[width=0.23\textwidth]{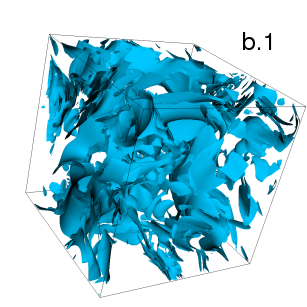}
\includegraphics[width=0.23\textwidth]{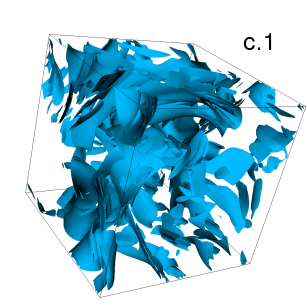}
\includegraphics[width=0.23\textwidth]{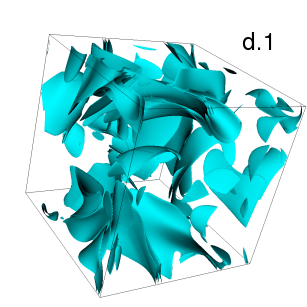}
\includegraphics[width=0.23\textwidth]{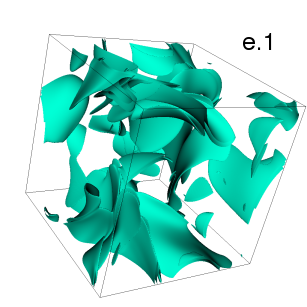}
\includegraphics[width=0.23\textwidth]{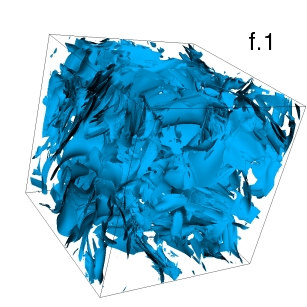}
\includegraphics[width=0.23\textwidth]{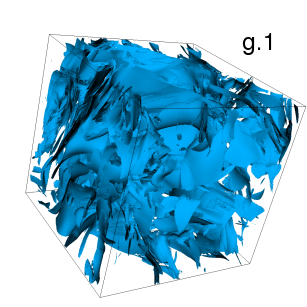}
\includegraphics[width=0.23\textwidth]{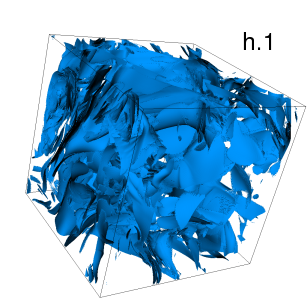}
\includegraphics[width=0.23\textwidth]{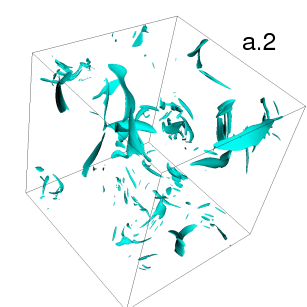}
\includegraphics[width=0.23\textwidth]{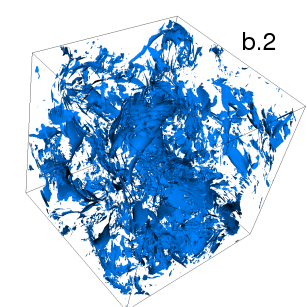}
\includegraphics[width=0.23\textwidth]{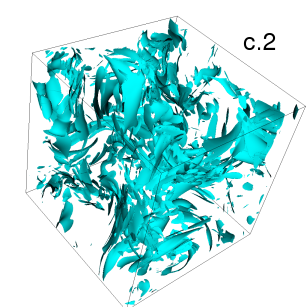}
\includegraphics[width=0.23\textwidth]{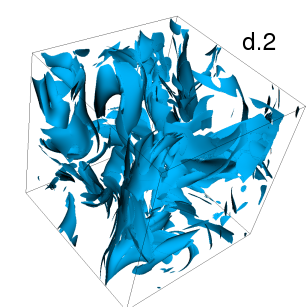}
\includegraphics[width=0.23\textwidth]{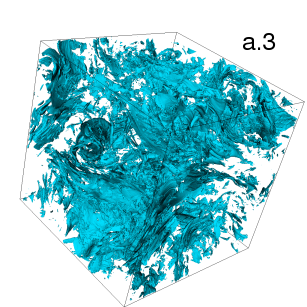}
\includegraphics[width=0.23\textwidth]{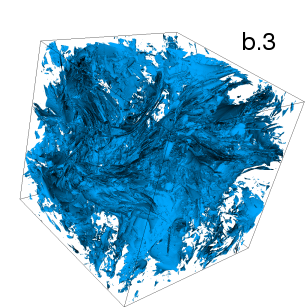}
\includegraphics[width=0.23\textwidth]{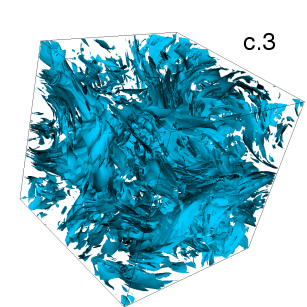}
\includegraphics[width=0.23\textwidth]{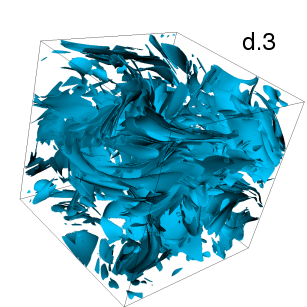}
\end{center}
\caption[]{Isosurfaces of the modulus $j$ of the current density:
(a.1) ${\rm Pr_M}=0.1$ (R1), (b.1) ${\rm Pr_M}=0.5$ (R2), (c.1) ${\rm
Pr_M}=1.0$ (R3), (d.1) ${\rm Pr_M}=5.0$ (R4), (e.1) ${\rm Pr_M}=10.0$ (R5),
(f.1) ${\rm Pr_M}=1.0$ (R3B), (g.1) ${\rm Pr_M}=5.0$ (R4B), (h.1) ${\rm
Pr_M}=10.0$ (R5B), (a.2) ${\rm Pr_M}=0.01$ (R1C), (b.2) ${\rm Pr_M}=0.1$
(R2C), (c.2) ${\rm Pr_M}=1.0$ (R3C), and (d.2) ${\rm Pr_M}=10.0$ (R4C) for
decaying MHD turbulence; and for statistically steady MHD turbulence (a.3)
${\rm Pr_M}=0.01$ (R1D), (b.3) ${\rm Pr_M}=0.1$ (R2D), (c.3) ${\rm Pr_M}=1.0$
(R3D), and (d.3) ${\rm Pr_M}=10.0$ (R4D); these isosurfaces go through
points at which the value of $j$ is two standard deviations above its
mean value (for any given plot).}
\label{fig:iso-j}
\end{figure}

Isosurfaces of $\omega$ are shown at $t_c$ for runs R1-R5 in
Figs.~\ref{fig:iso-w}(a.1)-\ref{fig:iso-w}(e.1), runs R3B-R5B in
Figs.~\ref{fig:iso-w}(f.1)-\ref{fig:iso-w}(h.1), and runs R1C-R4C in
Figs.~\ref{fig:iso-w}(a.2)-\ref{fig:iso-w}(d.2) for decaying MHD turbulence;
and for statistically steady MHD turbulence they are shown in
Figs.~\ref{fig:iso-w}(a.3)-\ref{fig:iso-w}(d.3) for runs R1D-R4D; these
isosurfaces go through points at which the value of $\omega$ is two standard
deviations above its mean value (for any given plot). For ${\rm Pr_M}=1$ it
has been noted in several DNS studies that such isosurfaces are
sheets~\cite{book-biskamp,biskamp00,yoshimatsu09,grauer00} and that there is
a general tendency for such sheet formation in MHD turbulence; our results
show that this tendency persists even when ${\rm Pr_M}\neq1$. The number of
high-intensity isosurfaces of $\omega$ shrink as we increase ${\rm Pr_M}$
[Figs.~\ref{fig:iso-w}(a.1)-\ref{fig:iso-w}(e.1) for runs R1-R5,
respectively], by increasing $\nu$ while holding the initial energy fixed.
However, if we compensate for the increase in $\nu$ by increasing the energy
in the initial condition such that $k_{\rm max}\eta_d^u$ and $k_{\rm
max}\eta_d^b$ are both $\simeq 1$, we see that high-$\omega$ sheets reappear
[Figs.~\ref{fig:iso-w}(f.1)-\ref{fig:iso-w}(h.1) for runs R3B-R5B,
respectively]. These trends are also visible in our high-resolution,
decaying-MHD-turbulence runs R1C-R4C
[Figs.~\ref{fig:iso-w}(a.2)-\ref{fig:iso-w}(d.2)] and the statistically steady
ones, namely, R1D-R4D [Figs.~\ref{fig:iso-w}(a.3)-\ref{fig:iso-w}(d.3)].  One
interesting point that has not been noticed before is that some tube-type
structures appear along with the sheets at small values of ${\rm Pr_M}$ as
can be seen by enlarging Fig.~\ref{fig:iso-w}(a.3) for run R1D. 

Similar features and trends appear in isosurfaces of $j$ that are shown at
$t_c$ for runs R1-R5 in Figs.~\ref{fig:iso-j}(a.1)-\ref{fig:iso-j}(e.1), runs
R3B-R5B in Figs.~\ref{fig:iso-j}(f.1)-\ref{fig:iso-j}(h.1), and runs R1C-R4C
in Figs.~\ref{fig:iso-j}(a.2)-\ref{fig:iso-j}(d.2) for decaying MHD
turbulence; and for statistically steady MHD turbulence they are shown in
Figs.~\ref{fig:iso-j}(a.3)-\ref{fig:iso-j}(d.3) for runs R1D-R4D; these
isosurfaces go through points at which the value of $j$ is two standard
deviations above its mean value (for any given plot). Again the dominant
features in these isosurface plots are sheets; their number goes down as
${\rm Pr_M}$ increases with $\nu$ while the initial energy is held constant;
but if this energy is increased, the number of high-intensity sheets
increase.

\begin{figure}[htb]
\begin{center}
\includegraphics[width=0.23\textwidth]{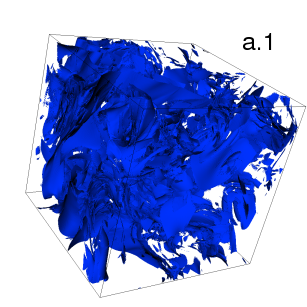}
\includegraphics[width=0.23\textwidth]{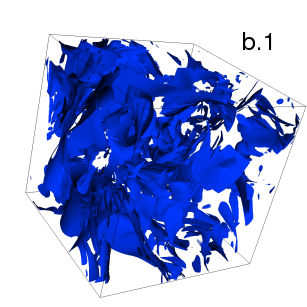}
\includegraphics[width=0.23\textwidth]{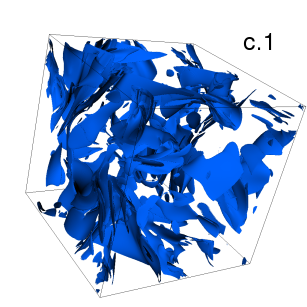}
\includegraphics[width=0.23\textwidth]{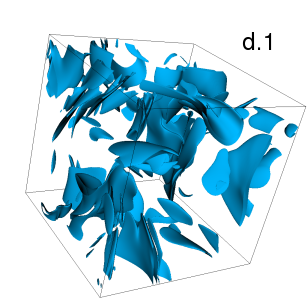}
\includegraphics[width=0.23\textwidth]{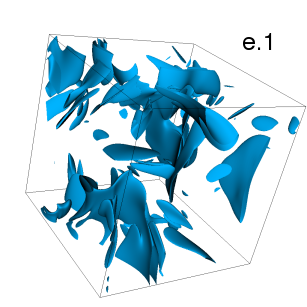}
\includegraphics[width=0.23\textwidth]{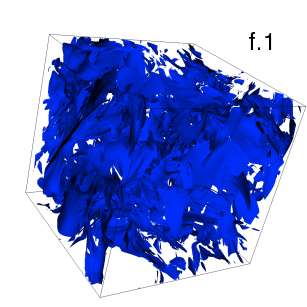}
\includegraphics[width=0.23\textwidth]{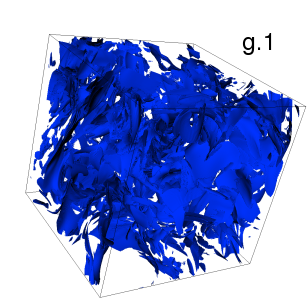}
\includegraphics[width=0.23\textwidth]{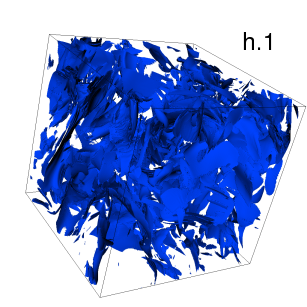}
\includegraphics[width=0.23\textwidth]{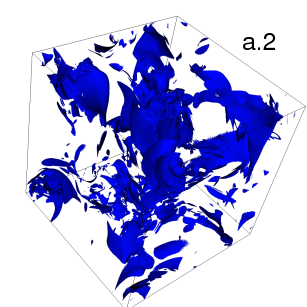}
\includegraphics[width=0.23\textwidth]{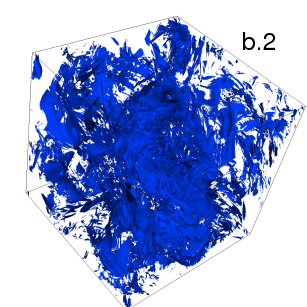}
\includegraphics[width=0.23\textwidth]{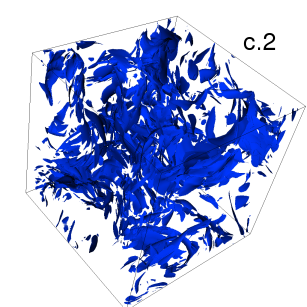}
\includegraphics[width=0.23\textwidth]{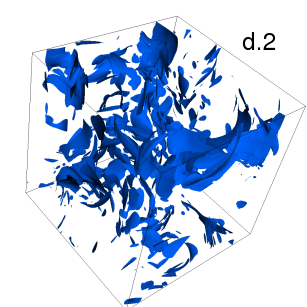}
\includegraphics[width=0.23\textwidth]{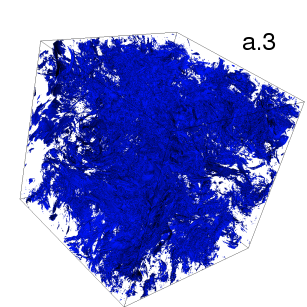}
\includegraphics[width=0.23\textwidth]{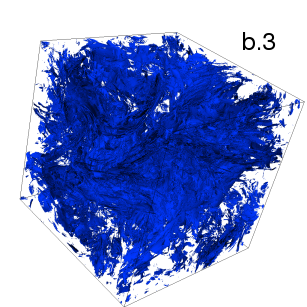}
\includegraphics[width=0.23\textwidth]{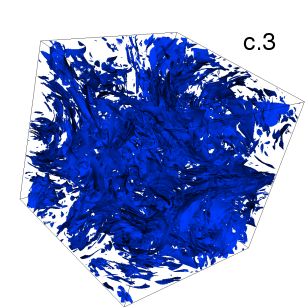}
\includegraphics[width=0.23\textwidth]{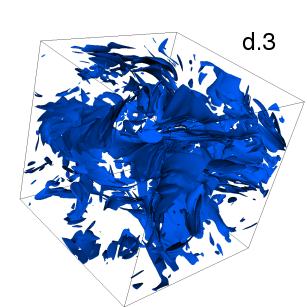}
\end{center}
\caption[]{Isosurfaces of the local fluid energy dissipation rate
$\epsilon_u$: (a.1) ${\rm Pr_M}=0.1$ (R1), (b.1) ${\rm Pr_M}=0.5$ (R2), (c.1)
${\rm Pr_M}=1.0$ (R3), (d.1) ${\rm Pr_M}=5.0$ (R4), (e.1) ${\rm Pr_M}=10.0$
(R5), (f.1) ${\rm Pr_M}=1.0$ (R3B), (g.1) ${\rm Pr_M}=5.0$ (R4B), (h.1) ${\rm
Pr_M}=10.0$ (R5B), (a.2) ${\rm Pr_M}=0.01$ (R1C), (b.2) ${\rm Pr_M}=0.1$
(R2C), (c.2) ${\rm Pr_M}=1.0$ (R3C), and (d.2) ${\rm Pr_M}=10.0$ (R4C) for
decaying MHD turbulence; and for statistically steady MHD turbulence (a.3)
${\rm Pr_M}=0.01$ (R1D), (b.3) ${\rm Pr_M}=0.1$ (R2D), (c.3) ${\rm Pr_M}=1.0$
(R3D), and (d.3) ${\rm Pr_M}=10.0$ (R4D); these isosurfaces go through
points at which the value of $\epsilon_u$ is two standard deviations above 
its mean value (for any given plot).}
\label{fig:iso-epsv}
\end{figure}

\begin{figure}[htb]
\begin{center}
\includegraphics[width=0.23\textwidth]{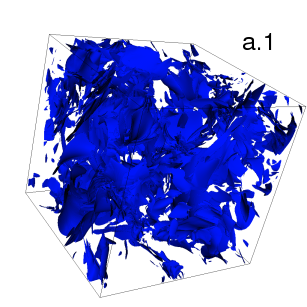}
\includegraphics[width=0.23\textwidth]{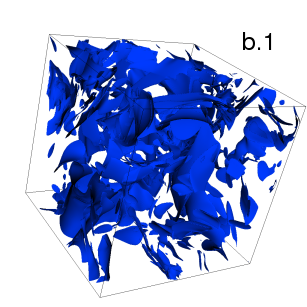}
\includegraphics[width=0.23\textwidth]{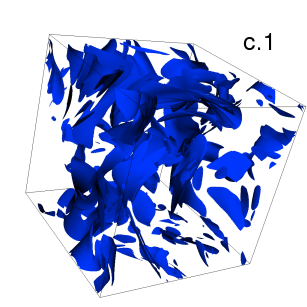}
\includegraphics[width=0.23\textwidth]{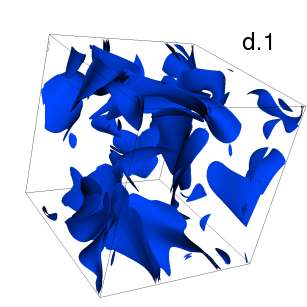}
\includegraphics[width=0.23\textwidth]{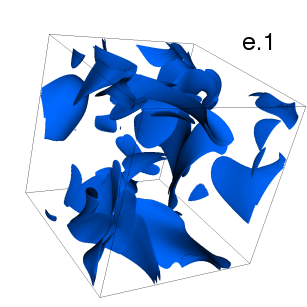}
\includegraphics[width=0.23\textwidth]{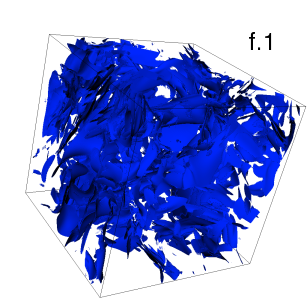}
\includegraphics[width=0.23\textwidth]{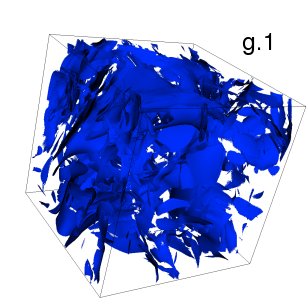}
\includegraphics[width=0.23\textwidth]{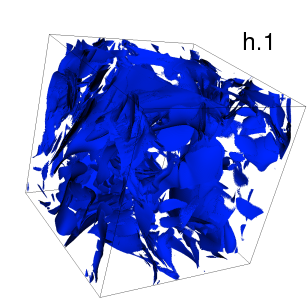}
\includegraphics[width=0.23\textwidth]{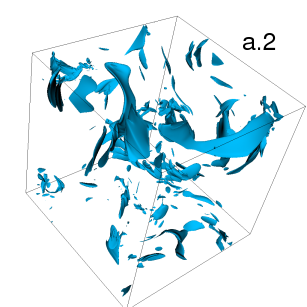}
\includegraphics[width=0.23\textwidth]{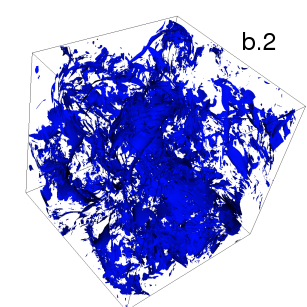}
\includegraphics[width=0.23\textwidth]{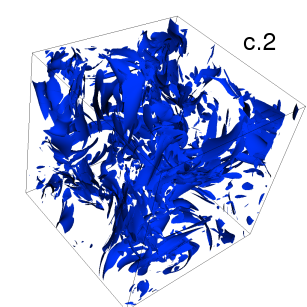}
\includegraphics[width=0.23\textwidth]{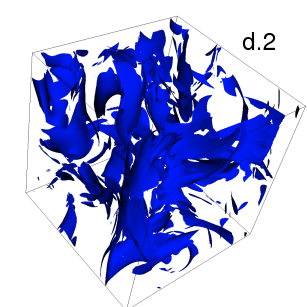}
\includegraphics[width=0.23\textwidth]{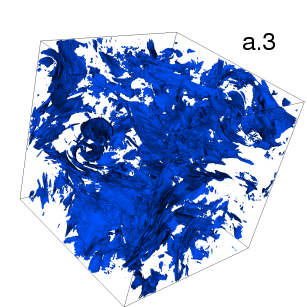}
\includegraphics[width=0.23\textwidth]{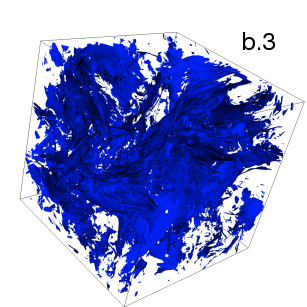}
\includegraphics[width=0.23\textwidth]{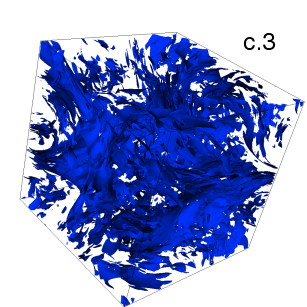}
\includegraphics[width=0.23\textwidth]{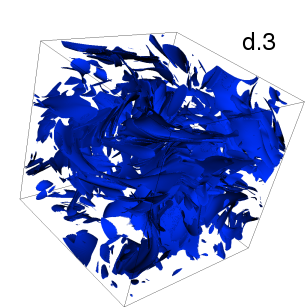}
\end{center}
\caption[]{Isosurfaces of the local magnetic-energy dissipation rate
$\epsilon_b$: (a.1) ${\rm Pr_M}=0.1$ (R1), (b.1) ${\rm Pr_M}=0.5$ (R2), (c.1)
${\rm Pr_M}=1.0$ (R3), (d.1) ${\rm Pr_M}=5.0$ (R4), (e.1) ${\rm Pr_M}=10.0$
(R5), (f.1) ${\rm Pr_M}=1.0$ (R3B), (g.1) ${\rm Pr_M}=5.0$ (R4B), and (h.1)
${\rm Pr_M}=10.0$ (R5B).  (a.2) ${\rm Pr_M}=0.01$ (R1C), (b.2) ${\rm
Pr_M}=0.1$ (R2C), (c.2) ${\rm Pr_M}=1.0$ (R3C), and (d.2) ${\rm Pr_M}=10.0$
(R4C) for decaying MHD turbulence; and for statistically steady MHD
turbulence (a.3) ${\rm Pr_M}=0.01$ (R1D), (b.3) ${\rm Pr_M}=0.1$ (R2D), (c.3)
${\rm Pr_M}=1.0$ (R3D), and (d.3) ${\rm Pr_M}=10.0$ (R4D); these isosurfaces 
go through points at which the value of $\epsilon_b$ is two standard 
deviations above its mean value (for any given plot).}
\label{fig:iso-epsb}
\end{figure}

Isosurfaces of $\epsilon_u$ are shown at $t_c$ for runs R1-R5 in
Figs.~\ref{fig:iso-epsv}(a.1)-\ref{fig:iso-epsv}(e.1), runs R3B-R5B in
Figs.~\ref{fig:iso-epsv}(f.1)-\ref{fig:iso-epsv}(h.1), and runs R1C-R4C in
Figs.~\ref{fig:iso-epsv}(a.2)-\ref{fig:iso-epsv}(d.2) for decaying MHD
turbulence; and for statistically steady MHD turbulence they are shown in
Figs.~\ref{fig:iso-epsv}(a.3)-\ref{fig:iso-epsv}(d.3) for runs R1D-R4D; the
isosurfaces go through points at which the value of $\epsilon_u$ is two
standard deviations above its mean value (for any given plot). Similar
isosurfaces of $\epsilon_b$ are shown at $t_c$ for runs R1-R5 in
Figs.~\ref{fig:iso-epsb}(a.1)-\ref{fig:iso-epsb}(e.1), runs R3B-R5B in
Figs.~\ref{fig:iso-epsb}(f.1)-\ref{fig:iso-epsb}(h.1), and runs R1C-R4C in
Figs.~\ref{fig:iso-epsb}(a.2)-\ref{fig:iso-epsb}(d.2) for decaying MHD
turbulence; and for statistically steady MHD turbulence they are shown in
Figs.~\ref{fig:iso-epsb}(a.3)-\ref{fig:iso-epsb}(d.3) for runs R1D-R4D; the
isosurfaces go through points at which the value of $\epsilon_b$ is two
standard deviations above its mean value (for any given plot). Here too the
isosurfaces are sheets; they lie close to, but are not coincident with,
isosurfaces of $\omega$ and $j$; changes in ${\rm Pr_M}$ affect these
isosurfaces much as they affect isosurfaces of $\omega$ and $j$.

\begin{figure}[htb]
\begin{center}
\includegraphics[width=0.23\textwidth]{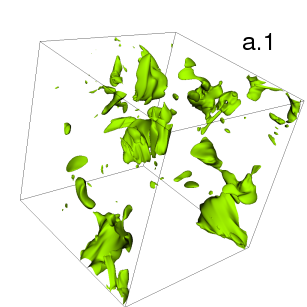}
\includegraphics[width=0.23\textwidth]{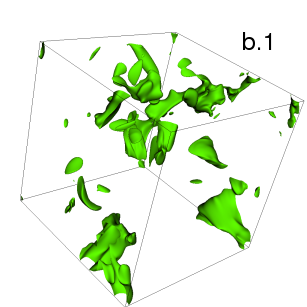}
\includegraphics[width=0.23\textwidth]{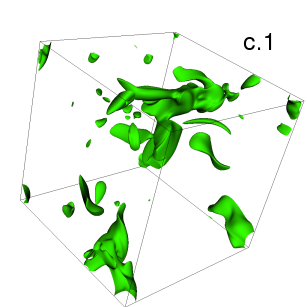}
\includegraphics[width=0.23\textwidth]{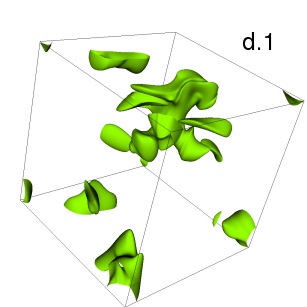}
\includegraphics[width=0.23\textwidth]{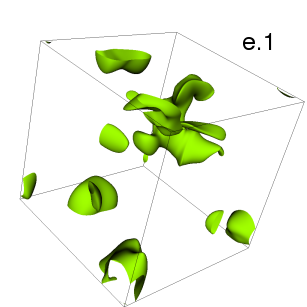}
\includegraphics[width=0.23\textwidth]{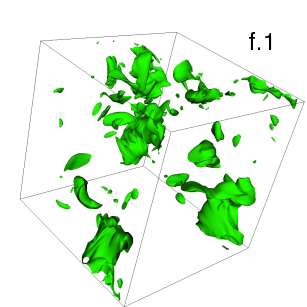}
\includegraphics[width=0.23\textwidth]{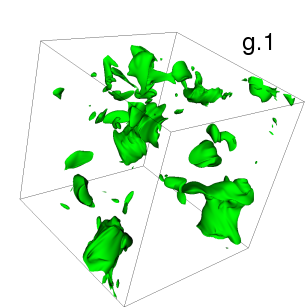}
\includegraphics[width=0.23\textwidth]{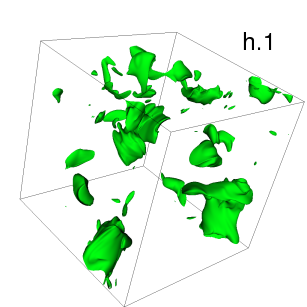}
\includegraphics[width=0.23\textwidth]{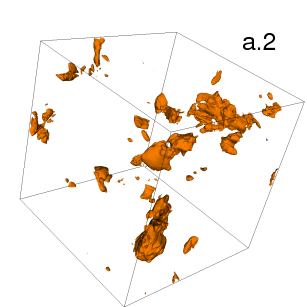}
\includegraphics[width=0.23\textwidth]{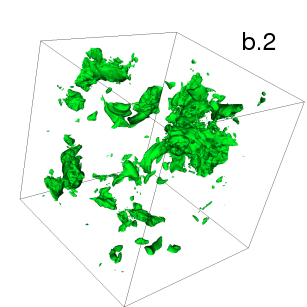}
\includegraphics[width=0.23\textwidth]{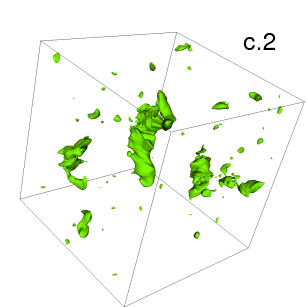}
\includegraphics[width=0.23\textwidth]{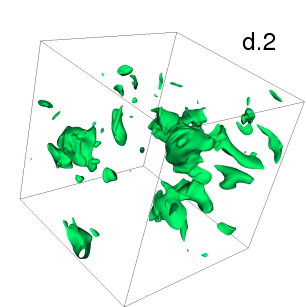}
\end{center}
\caption[]{Isosurfaces of the local effective pressure $\bar{p}$:
(a.1) ${\rm Pr_M}=0.1$ (R1), (b.1) ${\rm Pr_M}=0.5$ (R2), (c.1) ${\rm
Pr_M}=1.0$ (R3), (d.1) ${\rm Pr_M}=5.0$ (R4), (e.1) ${\rm Pr_M}=10.0$ (R5),
(f.1) ${\rm Pr_M}=1.0$ (R3B), (g.1) ${\rm Pr_M}=5.0$ (R4B), and (h.1) ${\rm
Pr_M}=10.0$ (R5B) for decaying MHD turbulence; and for statistically steady
MHD turbulencs (a.2) ${\rm Pr_M}=0.01$ (R1D), (b.2) ${\rm Pr_M}=0.1$ (R2D),
(c.2) ${\rm Pr_M}=1.0$ (R3D), and (d.2) ${\rm Pr_M}=10.0$ (R4D); 
these isosurfaces go through points at which the value of $\bar{p}$ is two 
standard deviations above its mean value (for any given plot).}
\label{fig:iso-press}
\end{figure}

Isosurfaces of $\bar{p}$ are shown at $t_c$ for runs R1-R5 in
Figs.~\ref{fig:iso-press}(a.1)-\ref{fig:iso-press}(e.1) and runs R3B-R5B in
Figs.~\ref{fig:iso-press}(f.1)-\ref{fig:iso-press}(h.1) for decaying MHD
turbulence; and for statistically steady MHD turbulence they are shown in
Figs.~\ref{fig:iso-press}(a.3)-\ref{fig:iso-press}(d.3) for runs R1D-R4D; the
isosurfaces go through points at which the value of $\bar{p}$ is two standard
deviations above its mean value (for any given plot). The general form of
these isosurfaces is cloud-type, to borrow the term that has been used for
isosurfaces of the pressure in fluid turbulence~\cite{schumann78}. Here also
changes in ${\rm Pr_M}$ affect these isosurfaces much as they affect
isosurfaces of $\omega$ and $j$, in as much as high-intensity isosurfaces are
suppressed as ${\rm Pr_M}$ increases via an increase in $\nu$, unless this is
compensated for by an increase in the initial energy (in the case of decaying
MHD turbulence) or ${\rm Re}_{\lambda}$.

\subsection{Joint probability distribution functions \label{sec:jpdf}}

In this Subsection we present three sets of joint PDFs that have, to the best
of our knowledge, not been used to characterise MHD turbulence. The first of
these is a $QR$ plot that is often used in studies of fluid turbulence as
we have discussed in Subsections~\ref{sec:statmeasures} and \ref{sec:NS}; the
next is a joint PDF of $\omega$ and $j$; and the last is a joint PDF of
$\epsilon_u$ and $\epsilon_b$.

\begin{figure}[htb]
\begin{center}
\includegraphics[width=0.23\textwidth]{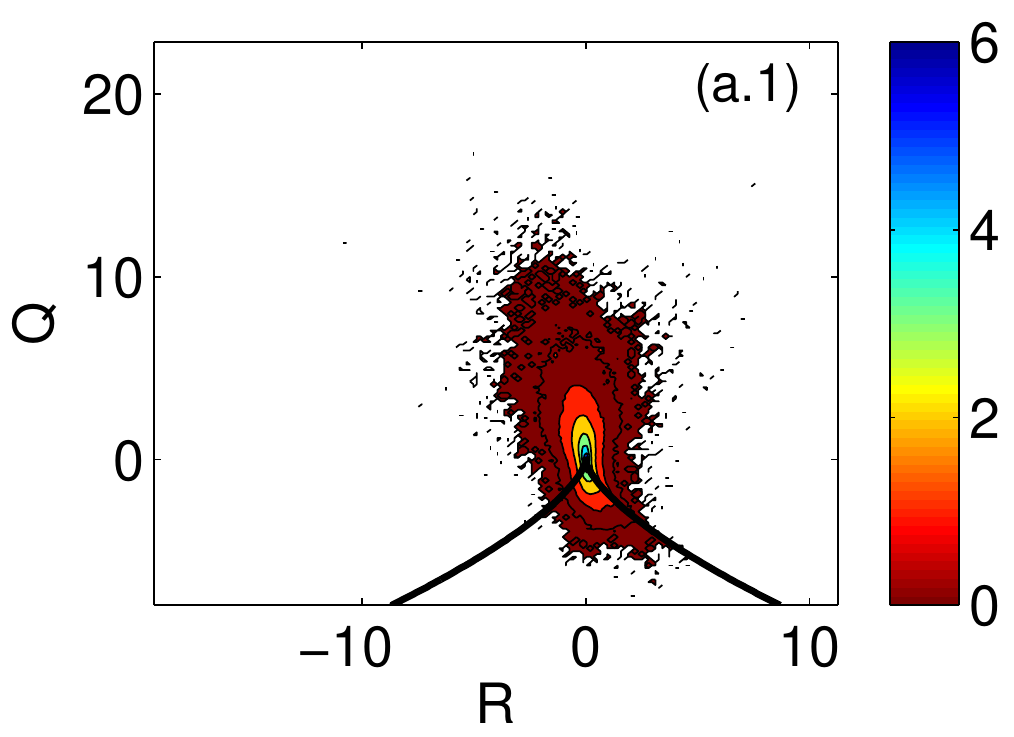}
\includegraphics[width=0.23\textwidth]{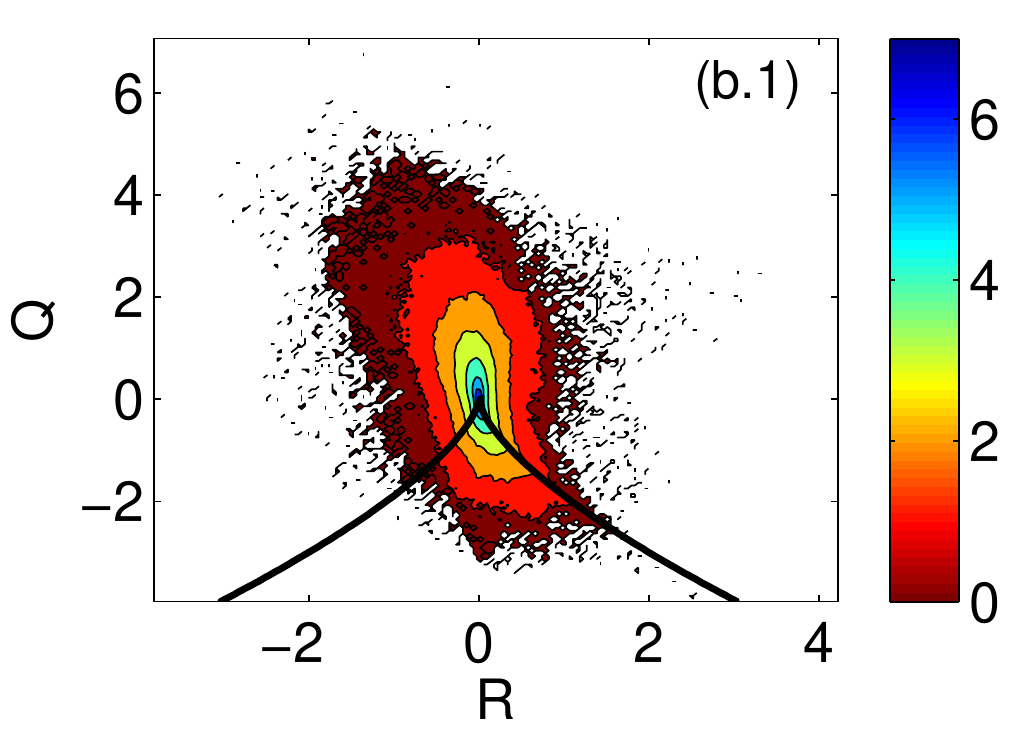}
\includegraphics[width=0.23\textwidth]{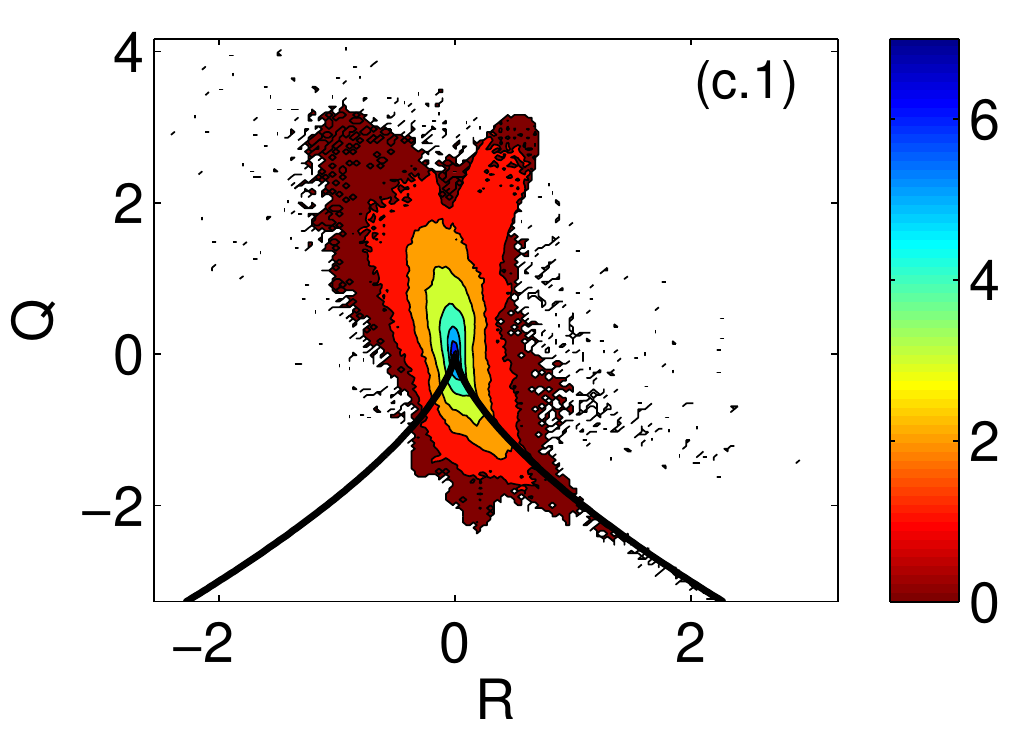}
\includegraphics[width=0.23\textwidth]{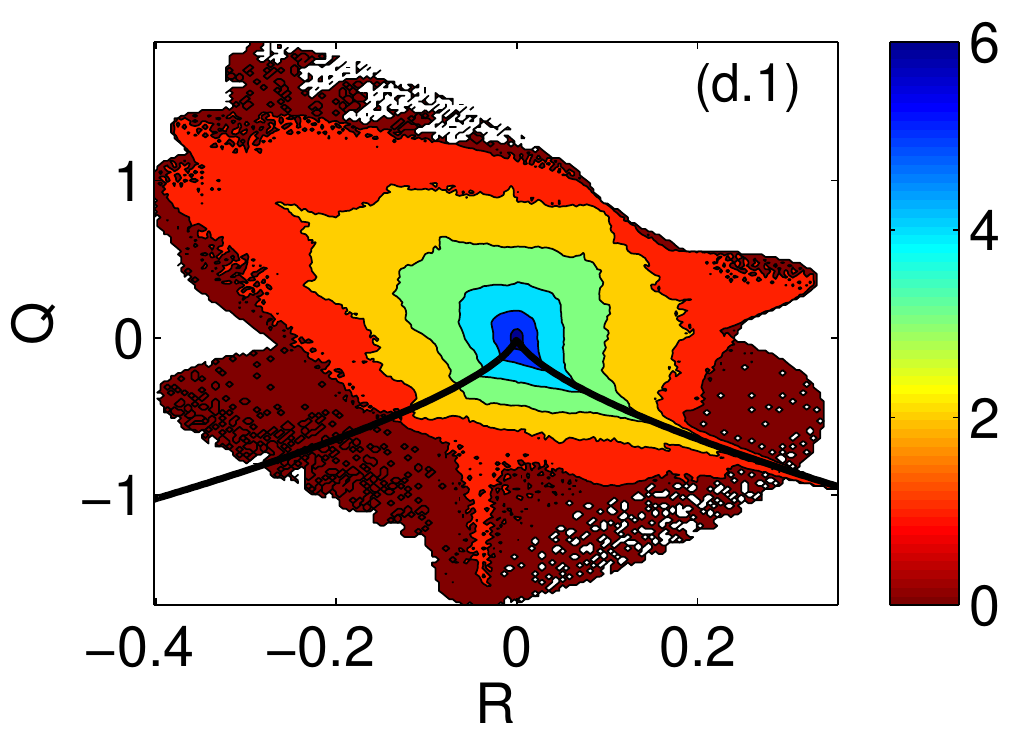}\\
\includegraphics[width=0.23\textwidth]{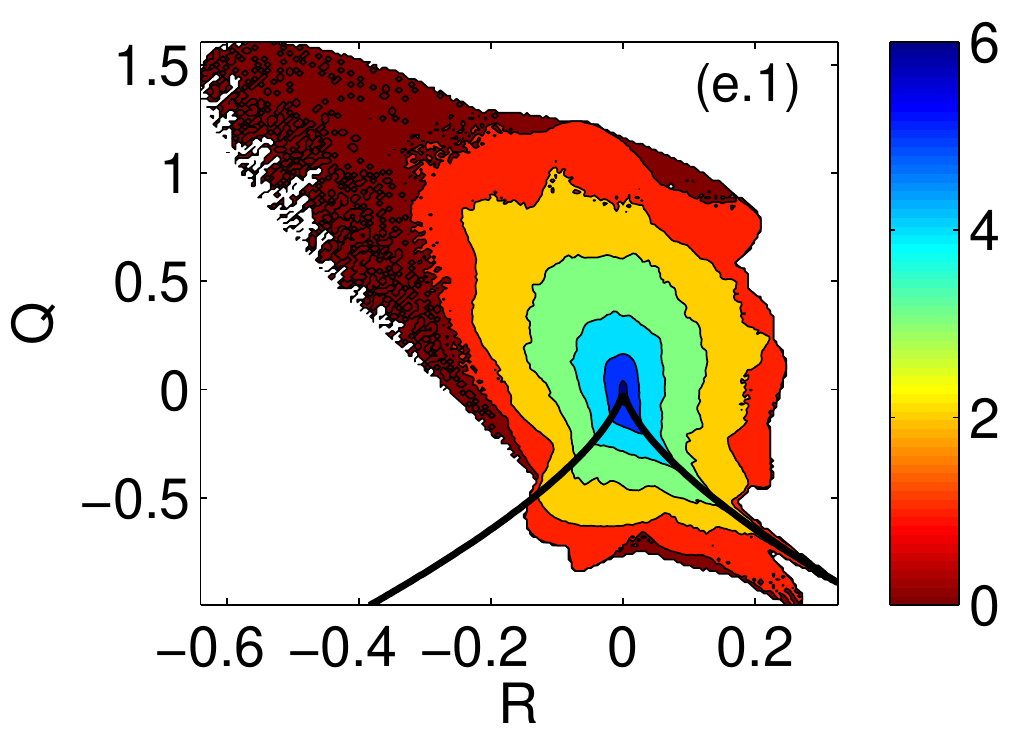}
\includegraphics[width=0.23\textwidth]{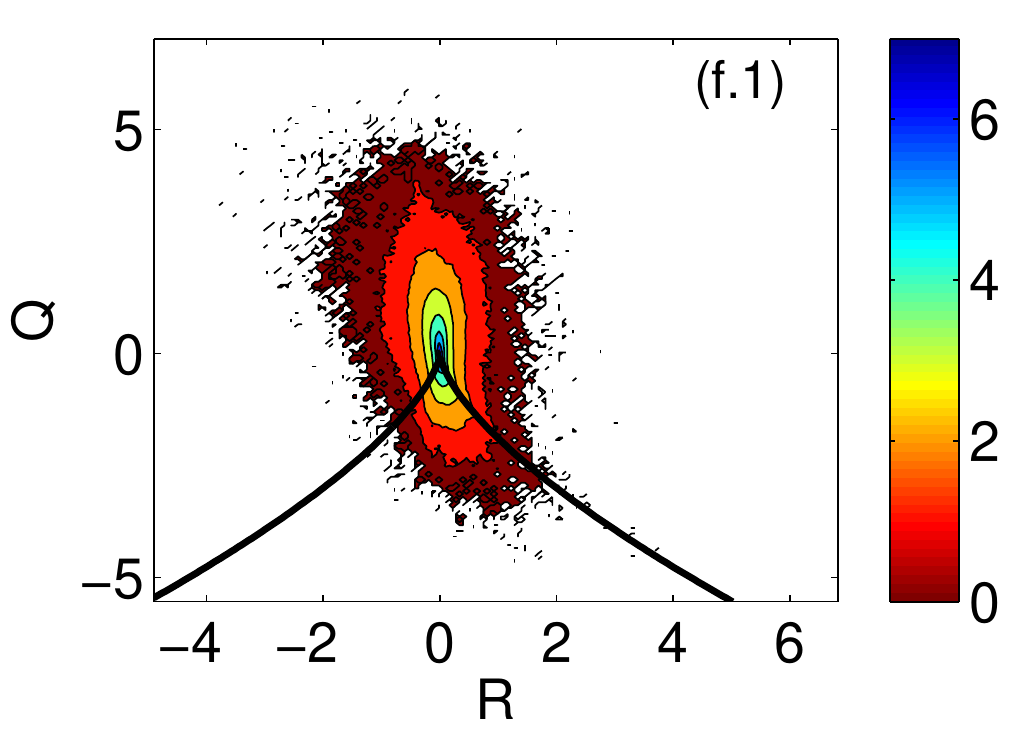}
\includegraphics[width=0.23\textwidth]{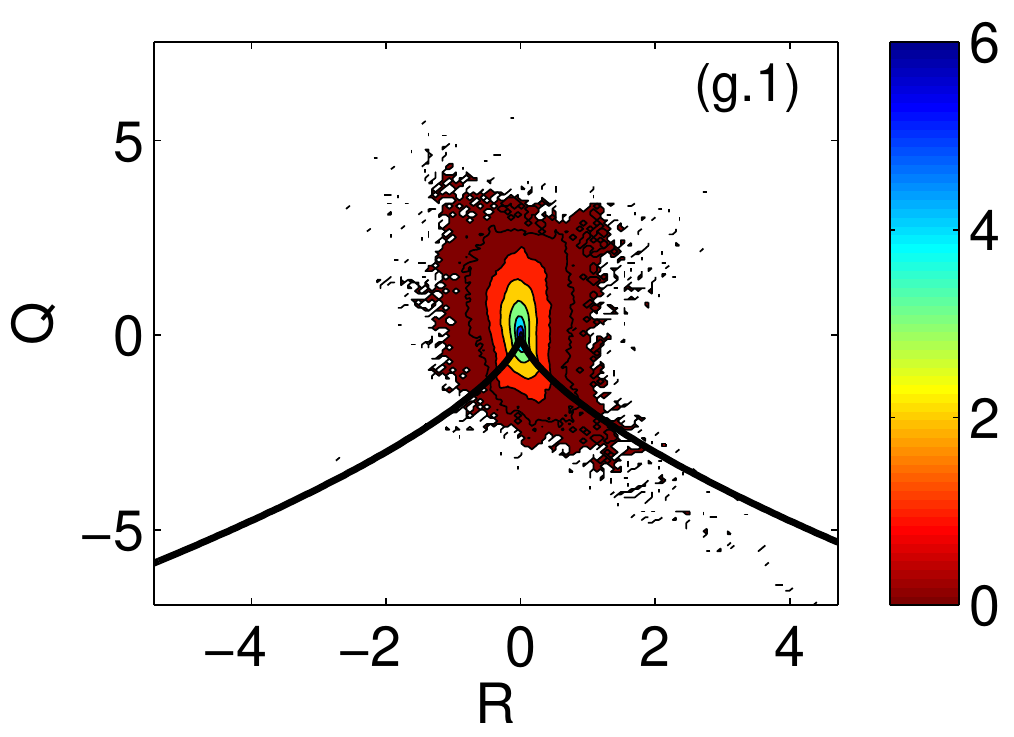}
\includegraphics[width=0.23\textwidth]{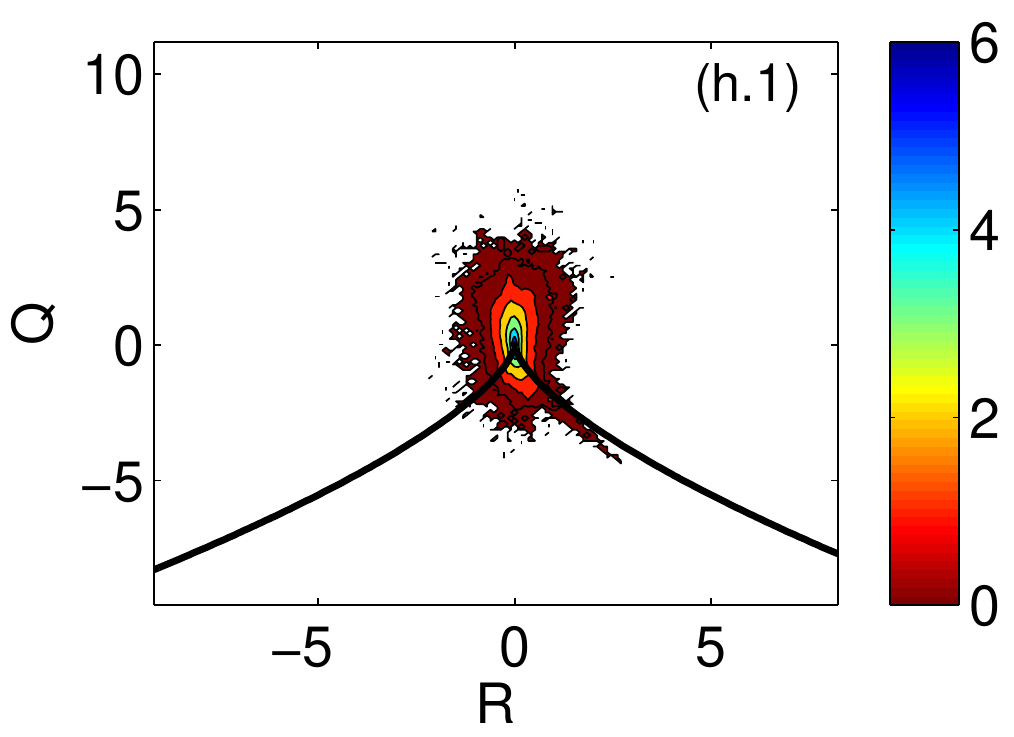}\\
\includegraphics[width=0.23\textwidth]{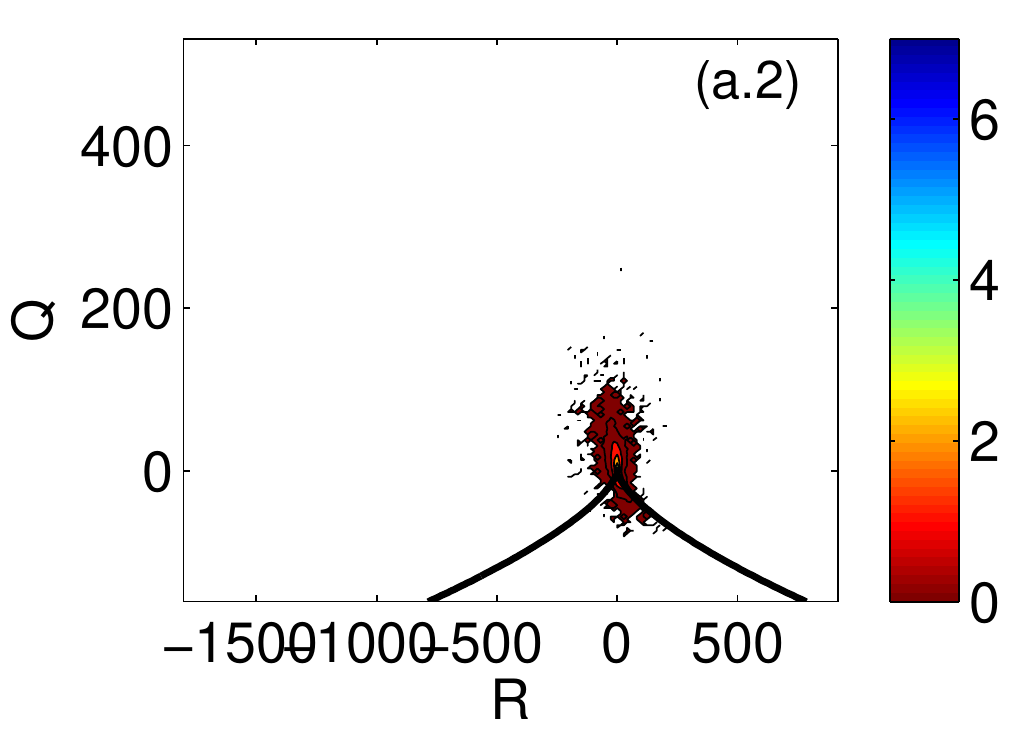}
\includegraphics[width=0.23\textwidth]{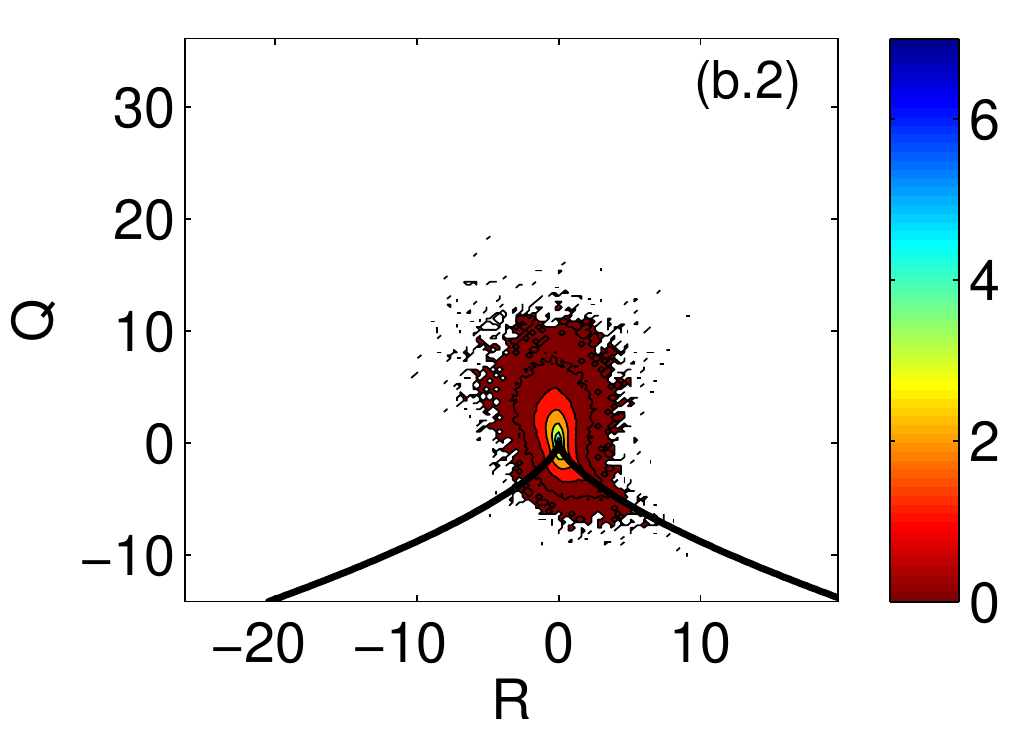}
\includegraphics[width=0.23\textwidth]{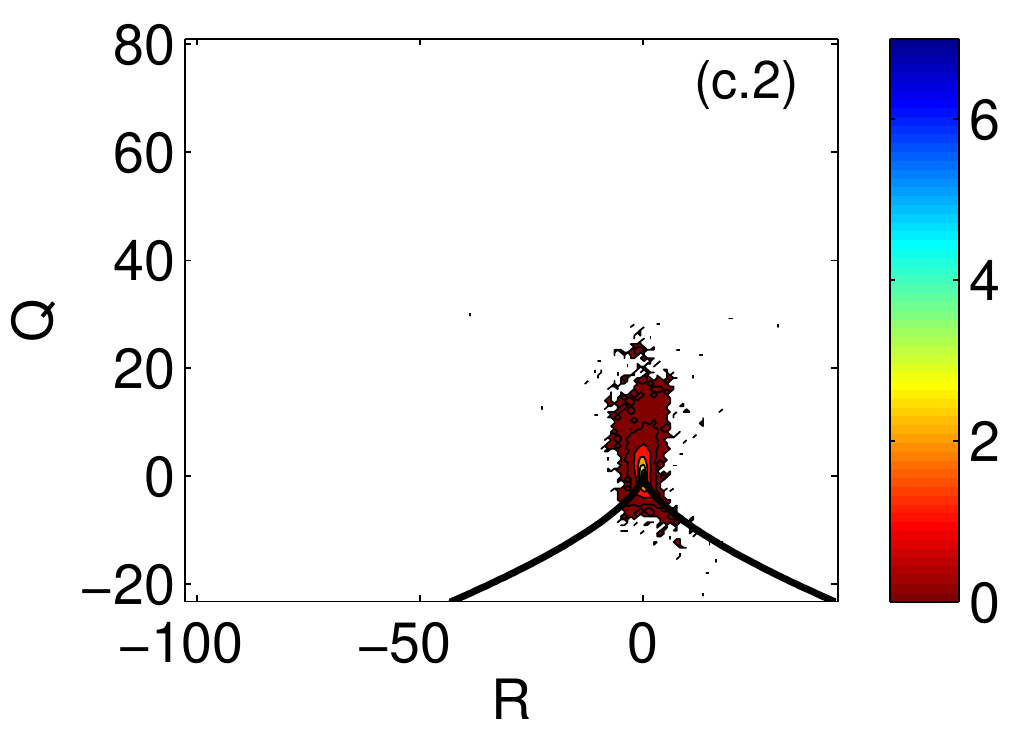}
\includegraphics[width=0.23\textwidth]{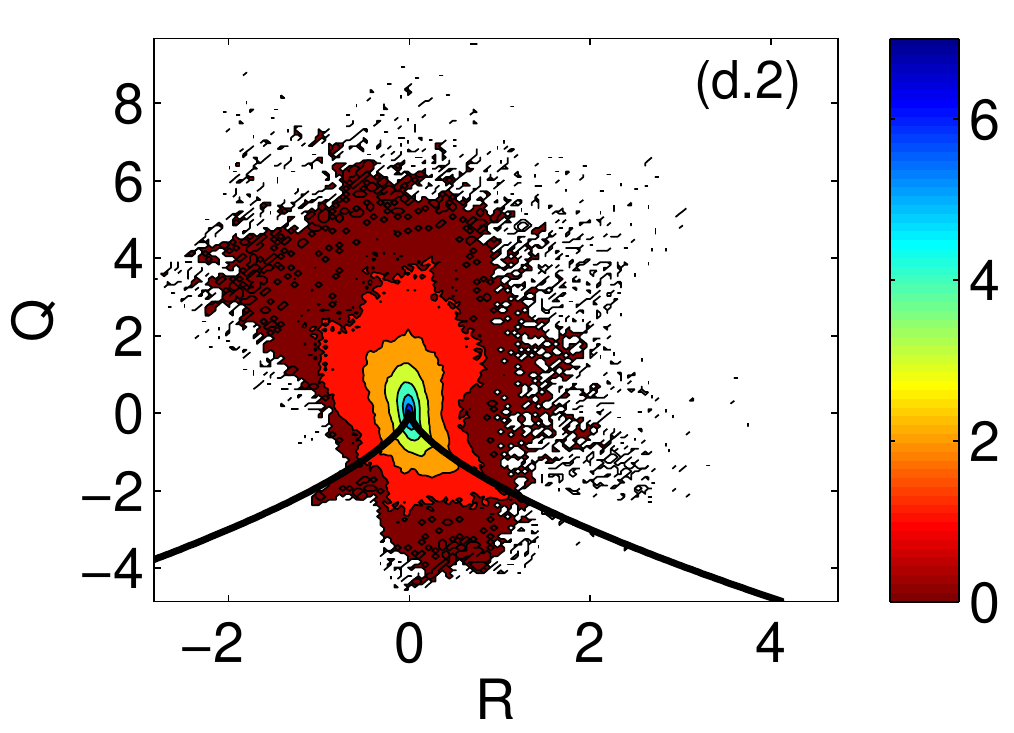}
\includegraphics[width=0.23\textwidth]{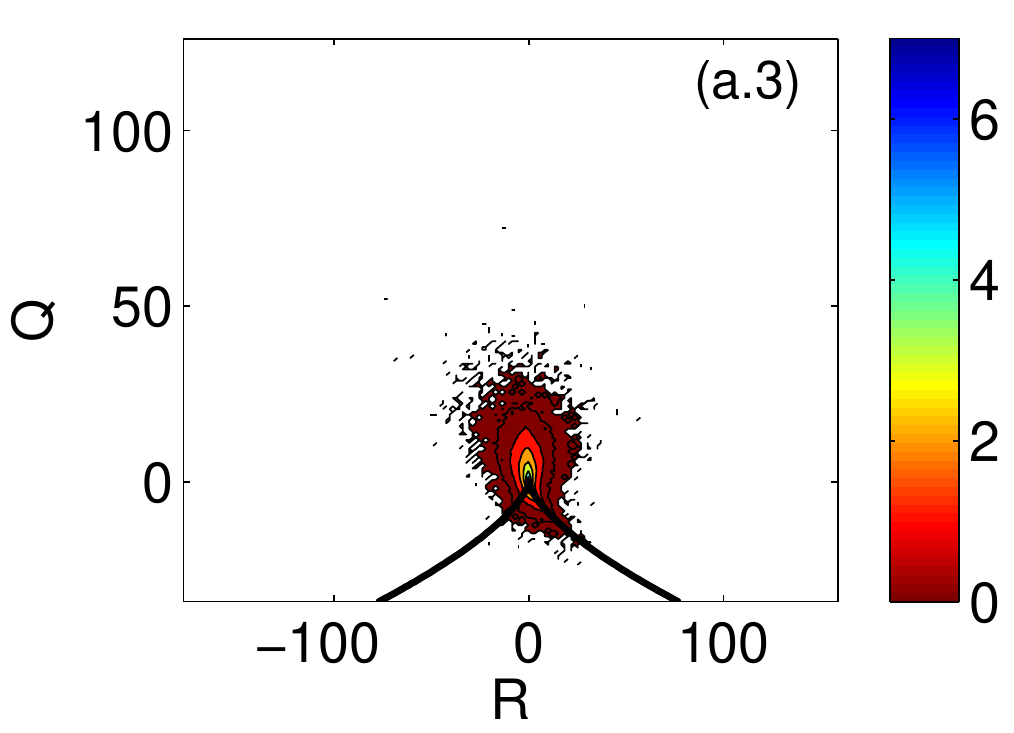}
\includegraphics[width=0.23\textwidth]{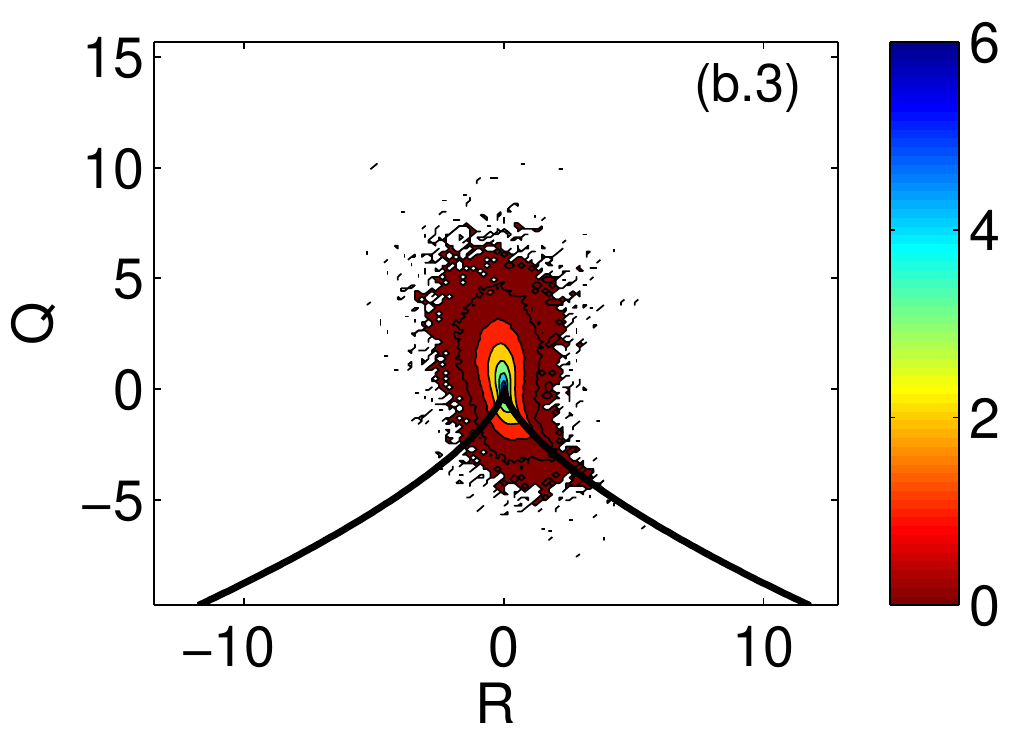}
\includegraphics[width=0.23\textwidth]{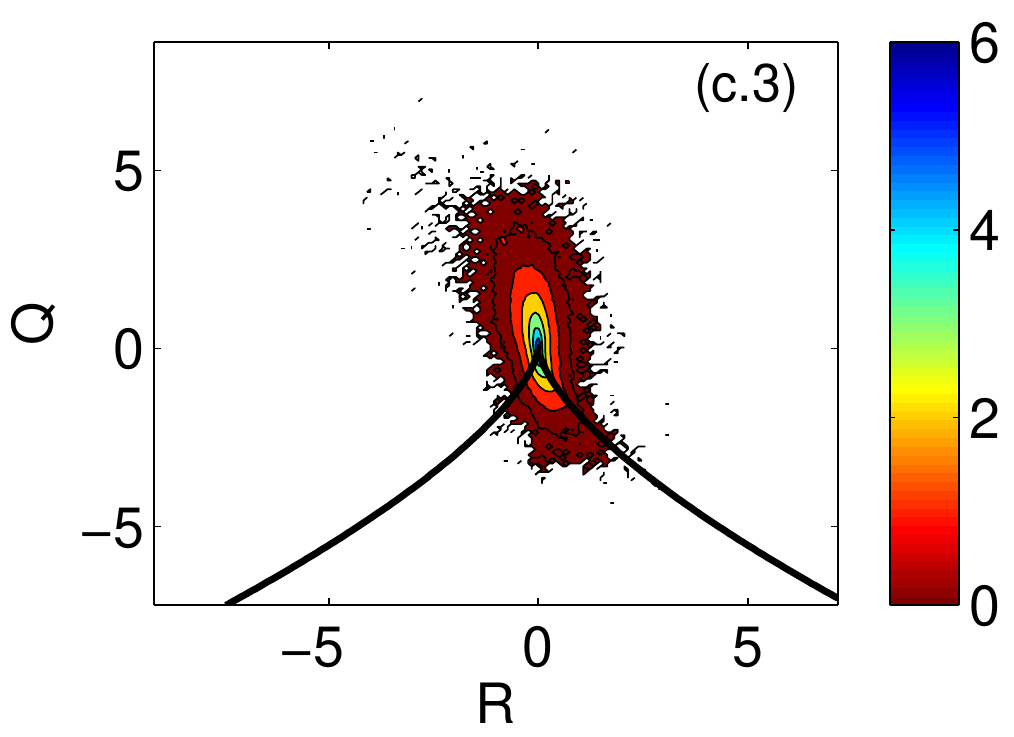}
\includegraphics[width=0.23\textwidth]{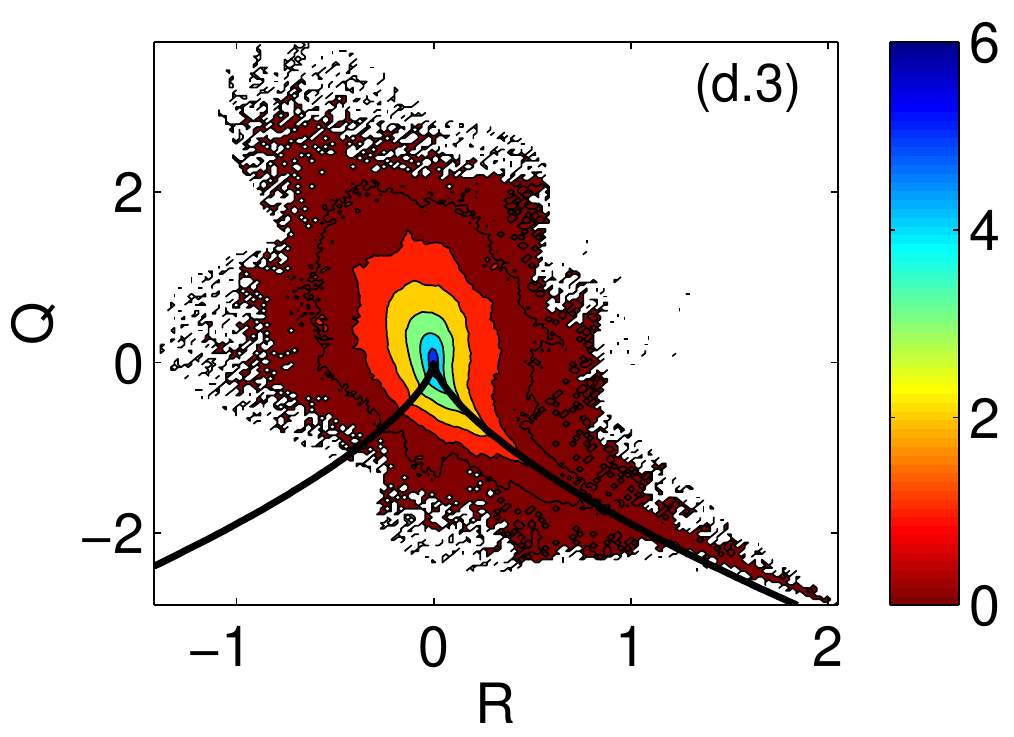}
\end{center}
\caption[]{$QR$ plots, i.e., joint PDFs of $Q$ and $R$ shown as filled
contour plots on a logarithmic scale for (a.1) ${\rm Pr_M}=0.1$ (R1), (b.1)
${\rm Pr_M}=0.5$ (R2), (c.1) ${\rm Pr_M}=1.0$ (R3), (d.1) ${\rm Pr_M}=5.0$
(R4), (e.1) ${\rm Pr_M}=10.0$ (R5), (f.1) ${\rm Pr_M}=1.0$ (R3B), (g.1) ${\rm
Pr_M}=5.0$ (R4B), (h.1) ${\rm Pr_M}=10.0$ (R5B), (a.2) ${\rm Pr_M}=0.01$
(R1C), (b.2) ${\rm Pr_M}=0.1$ (R2C), (c.2) ${\rm Pr_M}=1.0$ (R3C), and (d.2)
${\rm Pr_M}=10.0$ (R4C) for decaying MHD turbulence; and for statistically
steady MHD turbulence (a.3) ${\rm Pr_M}=0.01$ (R1D), (b.3) ${\rm Pr_M}=0.1$
(R2D), (c.3) ${\rm Pr_M}=1.0$ (R3D), and (d.3) ${\rm Pr_M}=10.0$ (R4D). The
arguments $Q$ and $R$ of the $QR$ plots are normalised by
$\langle\omega^2\rangle$ and $\langle\omega^2\rangle^{3/2}$, respectively. The
black curve is the zero-discriminant line $D\equiv \frac{27}{4}R^2+Q^3=0$.}
\label{fig:qrplots}
\end{figure}

We show $QR$ plots, i.e., joint PDFs of $Q$ and $R$, via filled contour
plots; these are obtained at $t_c$ for runs R1-R5 in
Figs.~\ref{fig:qrplots}(a.1)-\ref{fig:qrplots}(e.1), runs R3B-R5B in
Figs.~\ref{fig:qrplots}(f.1)-\ref{fig:qrplots}(h.1), and runs R1C-R4C in
Figs.~\ref{fig:qrplots}(a.2)-\ref{fig:qrplots}(d.2) for decaying MHD
turbulence; and for statistically steady MHD turbulence they are shown in
Figs.~\ref{fig:qrplots}(a.3)-\ref{fig:qrplots}(d.3) for runs R1D-R4D; the
black curve in these plots is the zero-discriminant line $D\equiv
\frac{27}{4}R^2+Q^3=0$.  These $QR$ plots retain overall, aside from some
distortions, the characteristic tear-drop structure familiar from fluid
turbulence (see Subsection~\ref{sec:NS} and Fig.~\ref{fig:ns-qrplot}). If we
recall our discussion of $QR$ plots in Subsection~\ref{sec:statmeasures} and
we notice that, as we increase ${\rm Pr_M}$
[Figs.~\ref{fig:qrplots}(a.1)-\ref{fig:qrplots}(e.1) for runs R1-R5,
respectively] while holding $\eta$ and the initial energy fixed, there is a
general decrease in the probability of having large values of $Q$ and $R$,
i.e., regions of large strain or vorticity are suppressed; this corroborates
what we have found from the PDFs and isosurfaces discussed above. However, if
we compensate for the increase in $\nu$ by increasing the initial energy, or
${\rm Re}_{\lambda}$, so that $k_{\rm max}\eta_d^u$ and $k_{\rm max}\eta_d^b$ are
both $\simeq 1$, we see that $Q$ and $R$ can increase again. Note that when
${\rm Pr_M}$ is very small as in run R1D [Fig.~\ref{fig:qrplots}(a.3)], the
tear-drop structure is very much like its fluid-turbulence counterpart
Fig.~\ref{fig:ns-qrplot}, which might well correlate with the appearance of
some tube-type structures in the $\omega$ isosurface in enlarged versions of
Fig.~\ref{fig:iso-w}(a.3).

\begin{figure}[htb]
\begin{center}
\includegraphics[width=0.23\textwidth]{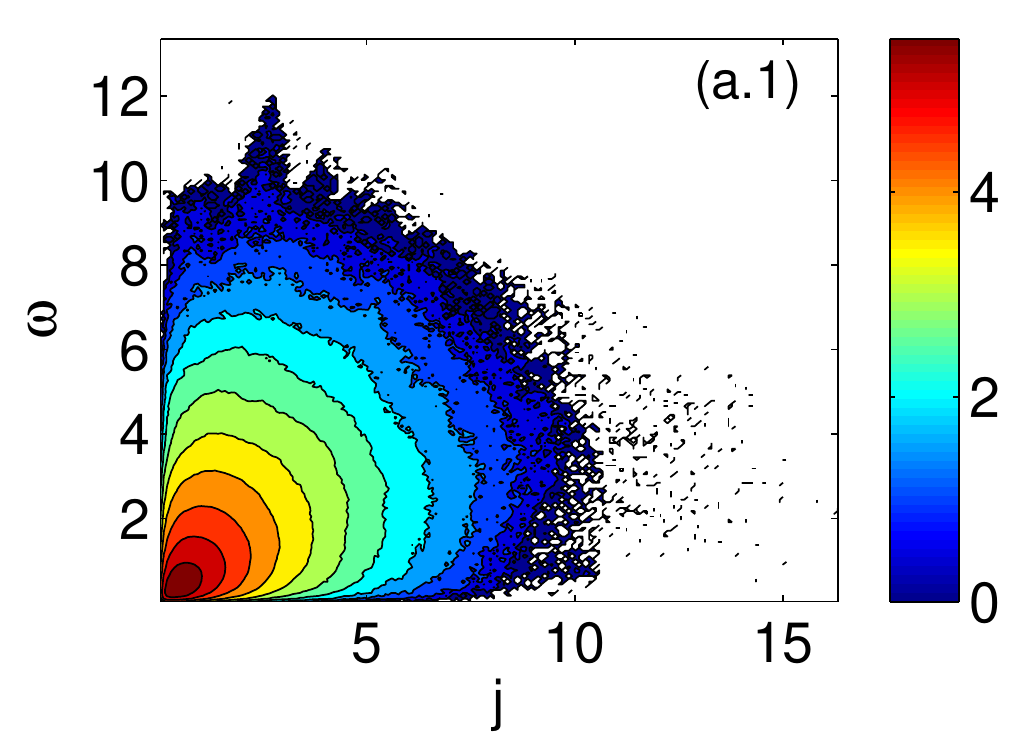}
\includegraphics[width=0.23\textwidth]{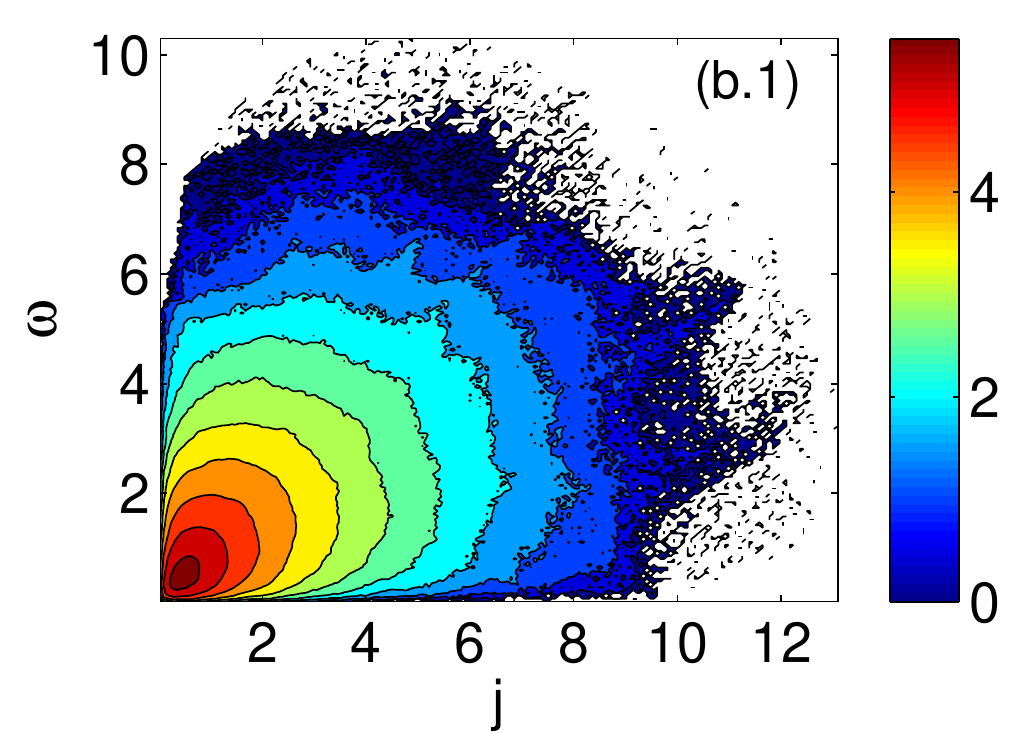}
\includegraphics[width=0.23\textwidth]{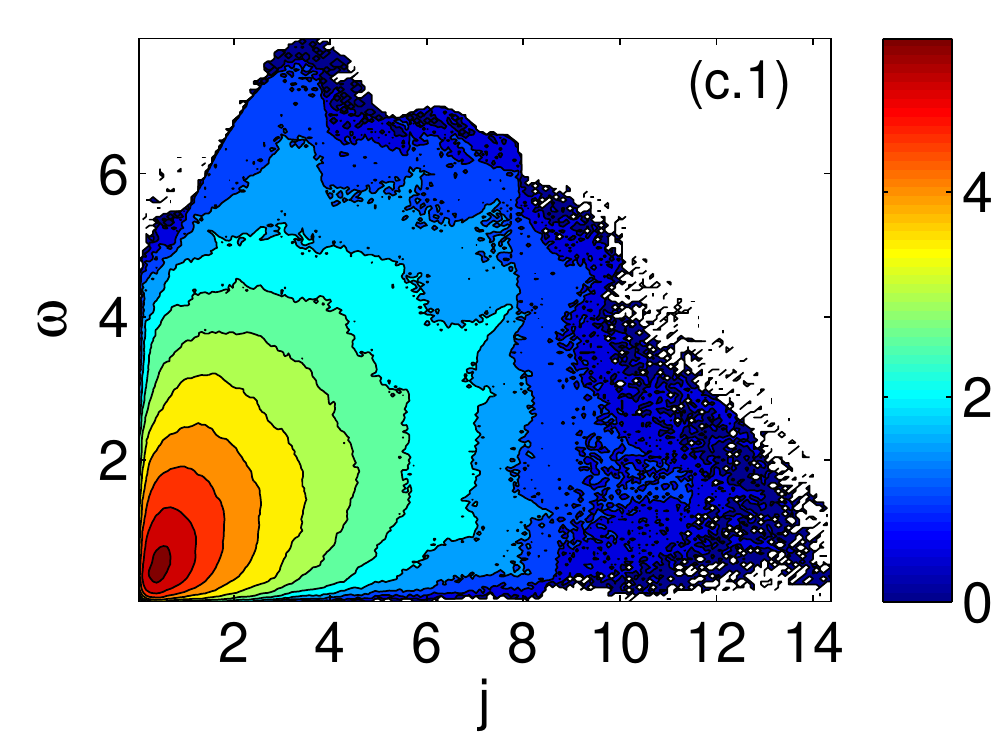}
\includegraphics[width=0.23\textwidth]{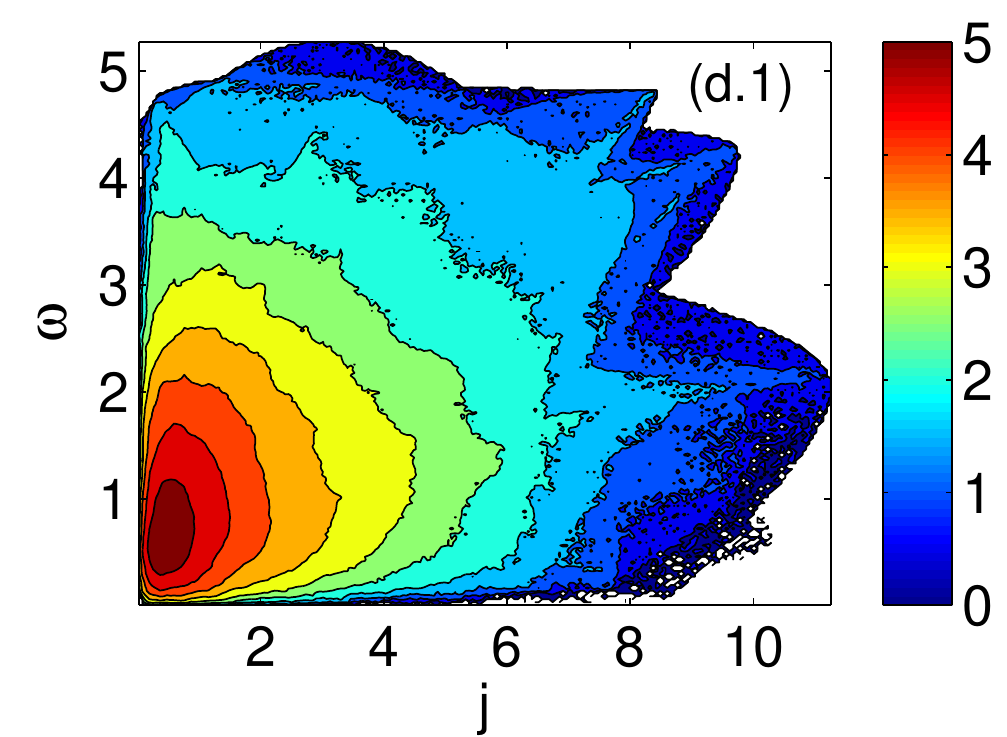}
\includegraphics[width=0.23\textwidth]{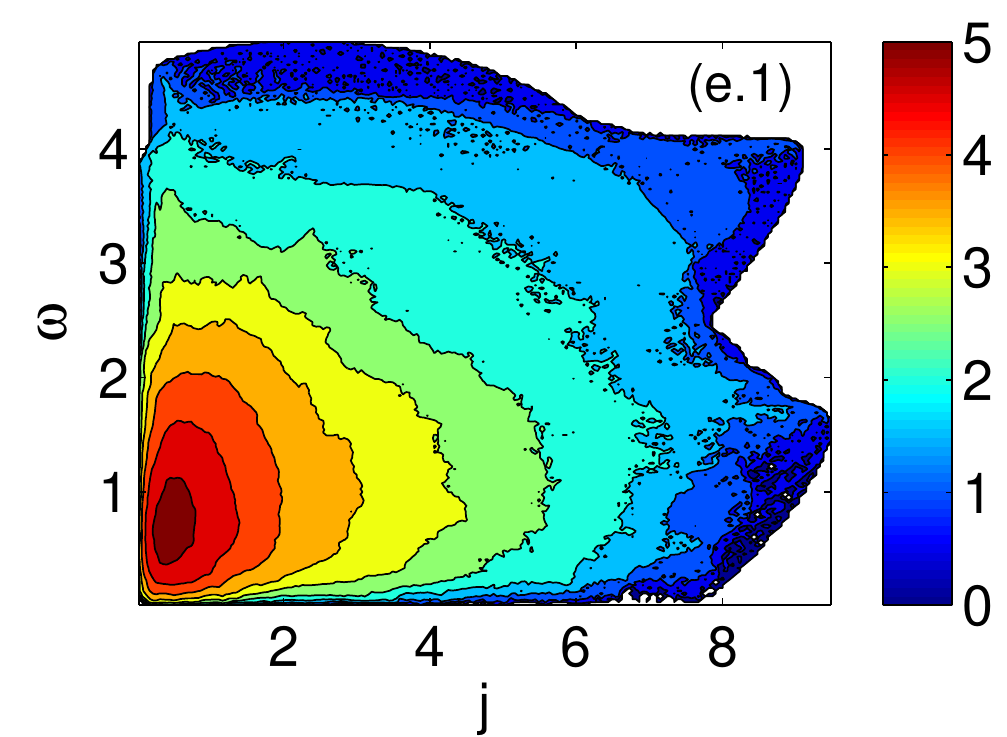}
\includegraphics[width=0.23\textwidth]{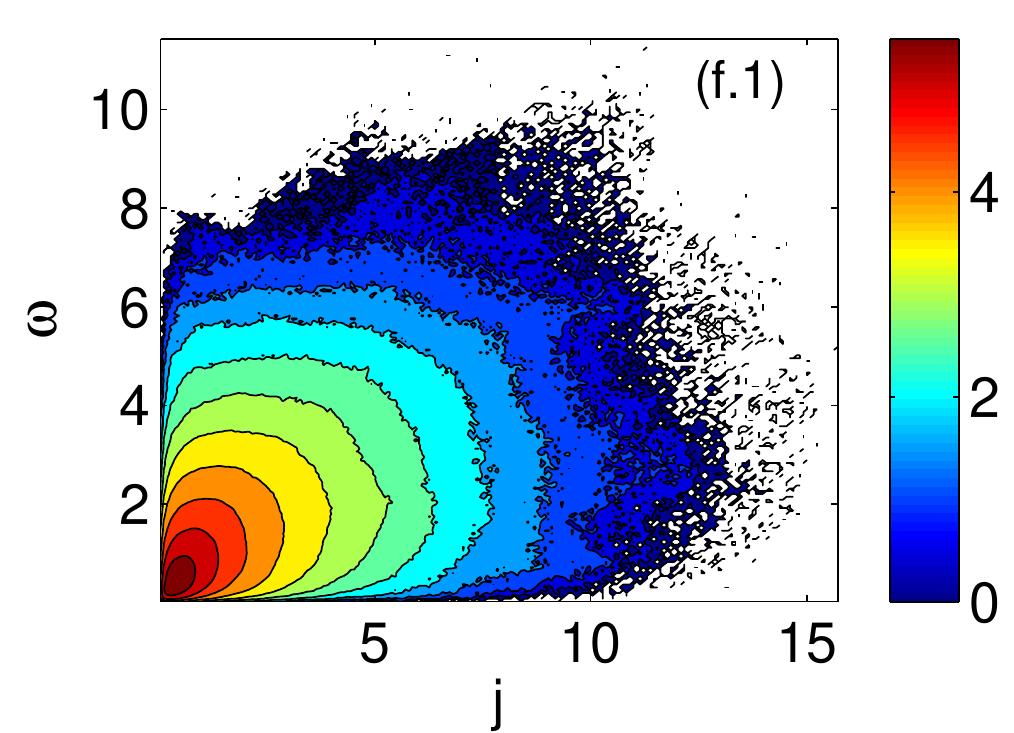}
\includegraphics[width=0.23\textwidth]{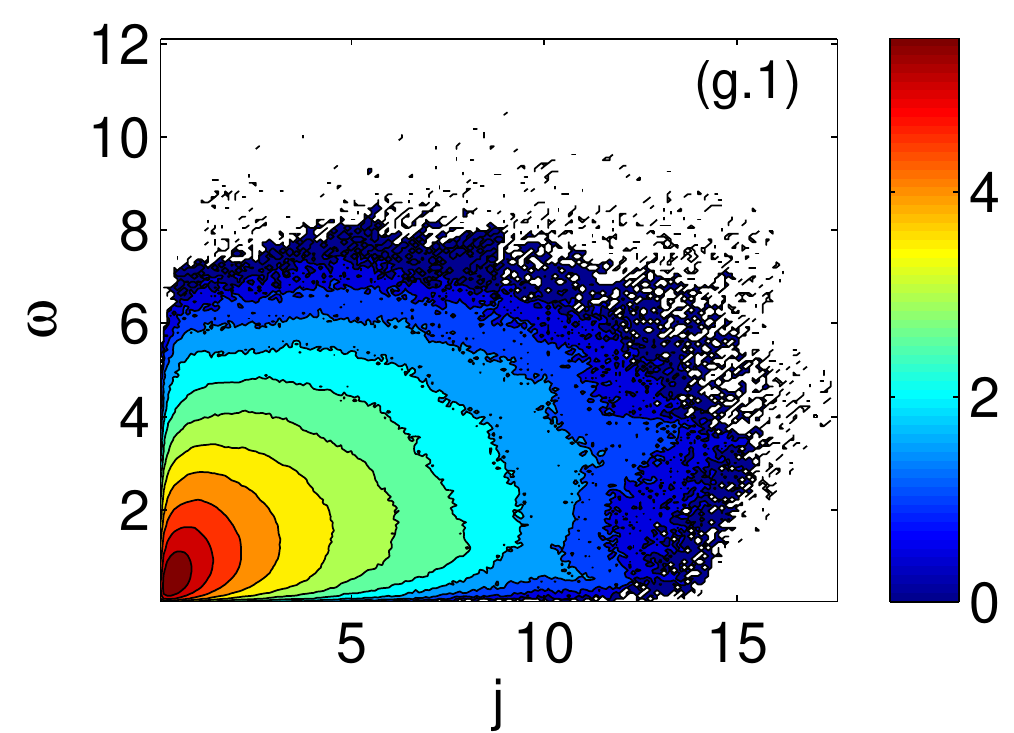}
\includegraphics[width=0.23\textwidth]{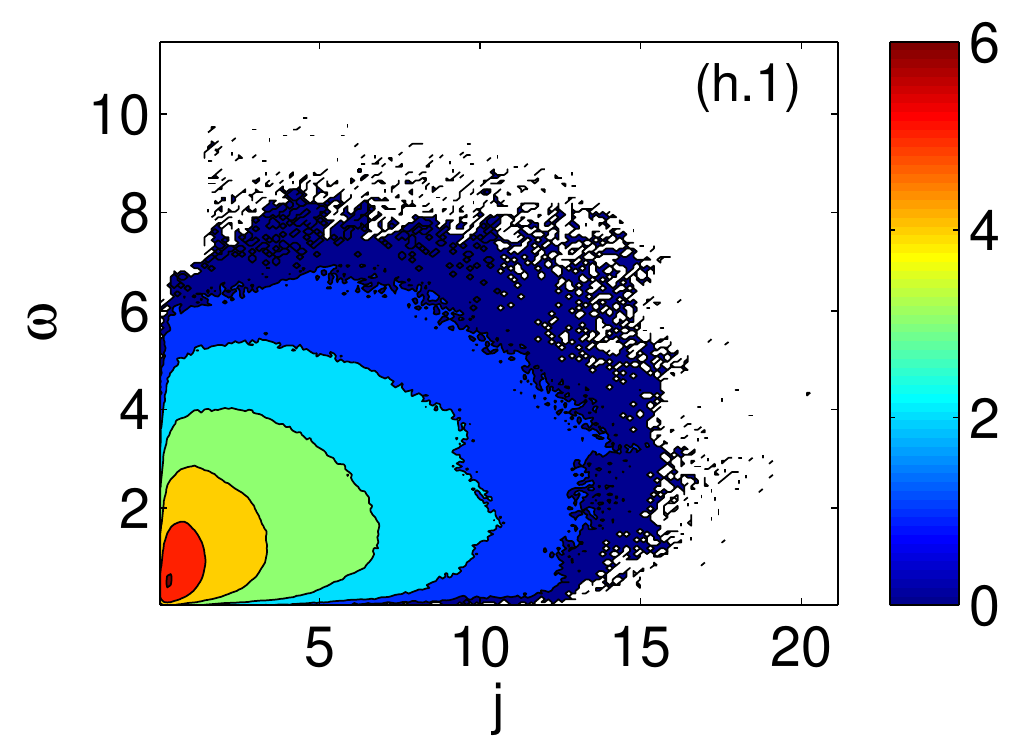}\\
\includegraphics[width=0.23\textwidth]{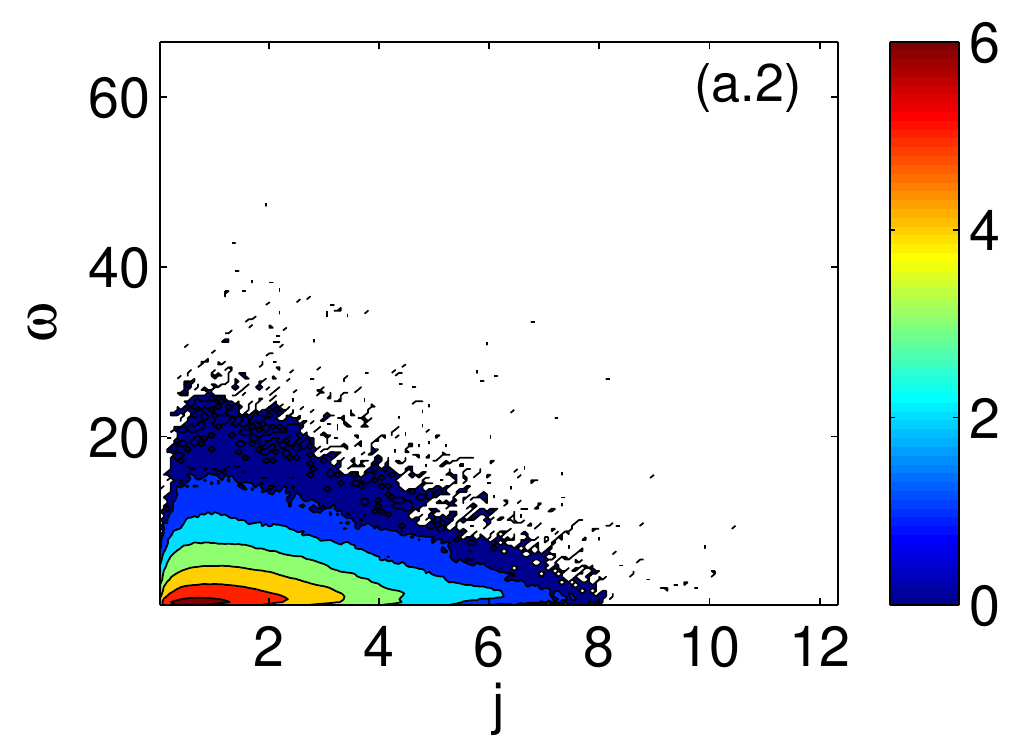}
\includegraphics[width=0.23\textwidth]{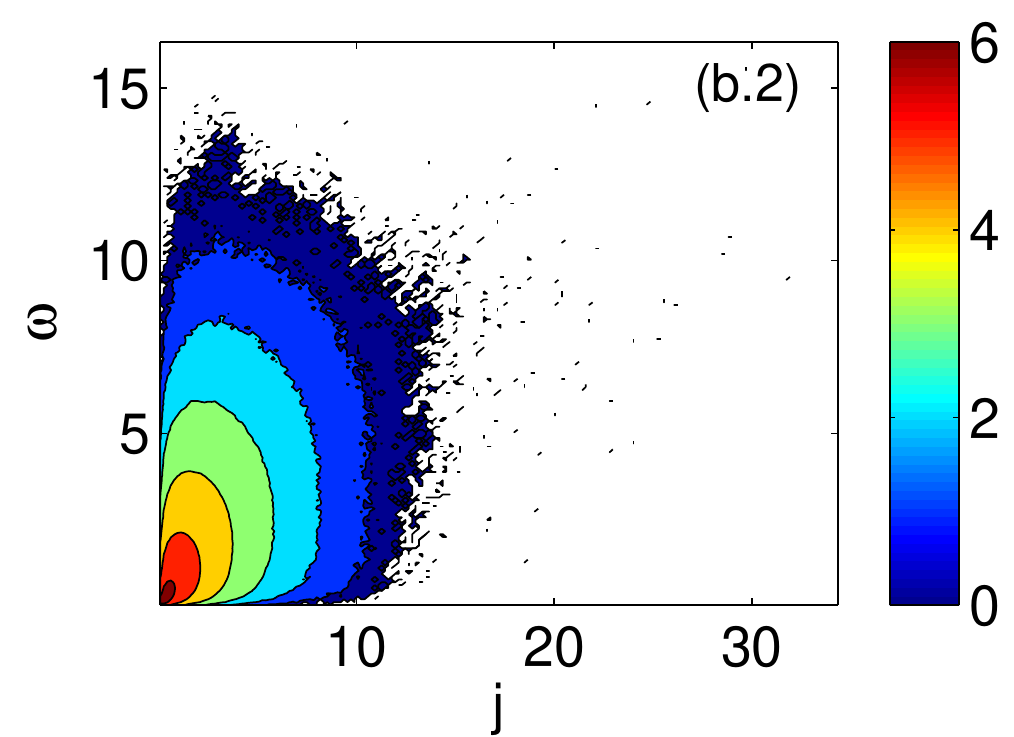}
\includegraphics[width=0.23\textwidth]{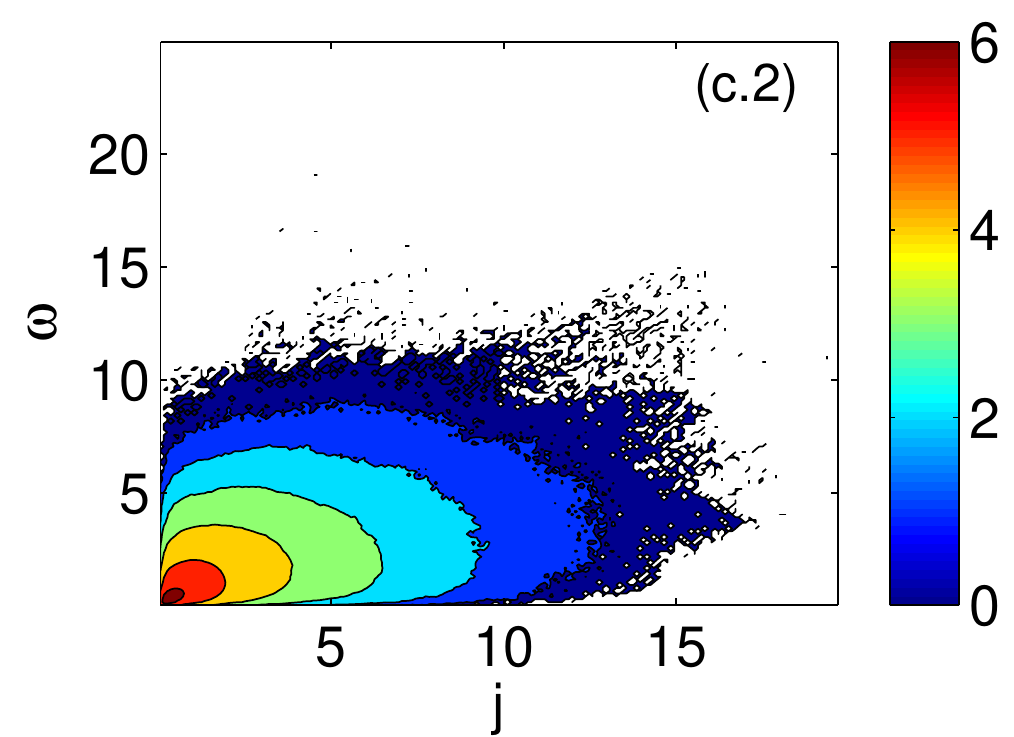}
\includegraphics[width=0.23\textwidth]{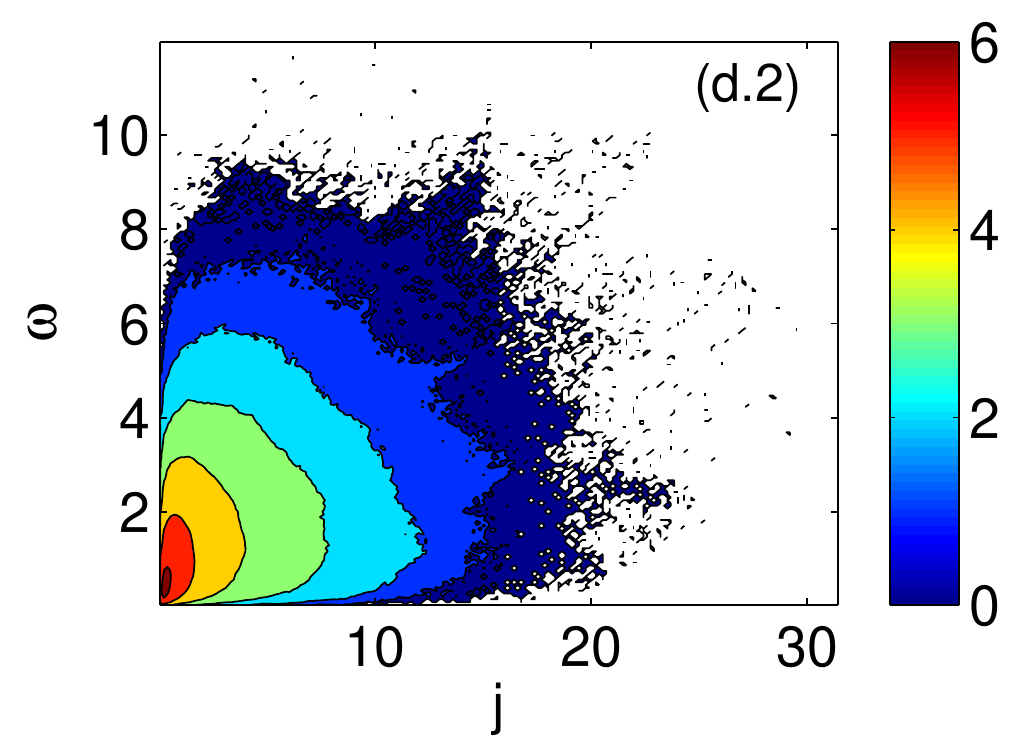}
\includegraphics[width=0.23\textwidth]{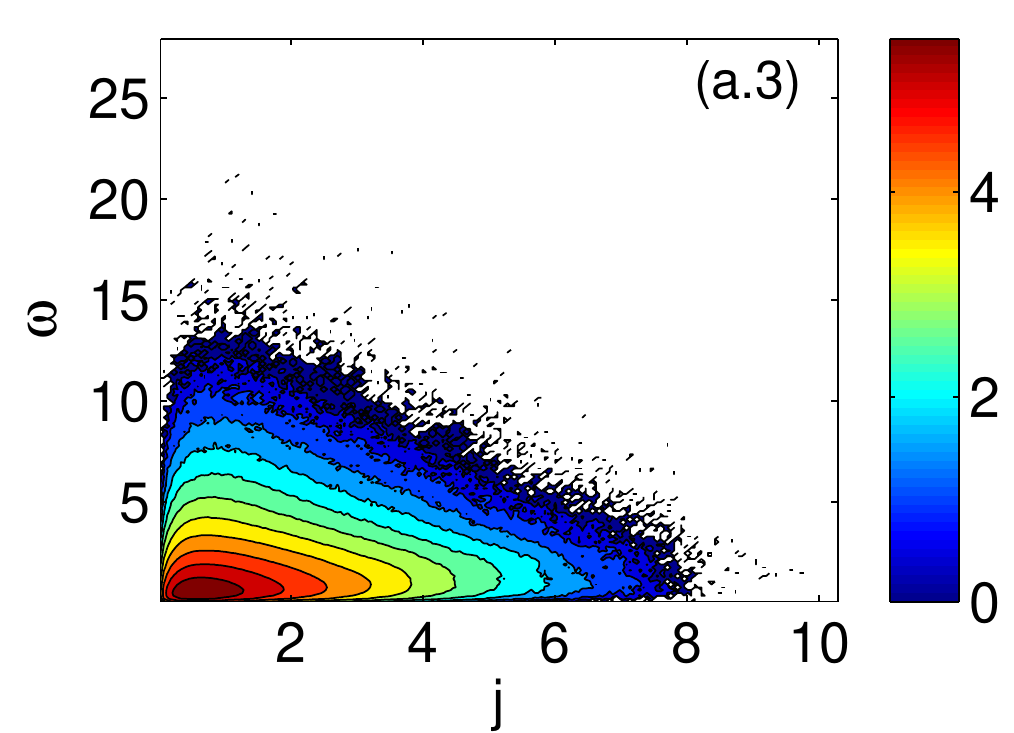}
\includegraphics[width=0.23\textwidth]{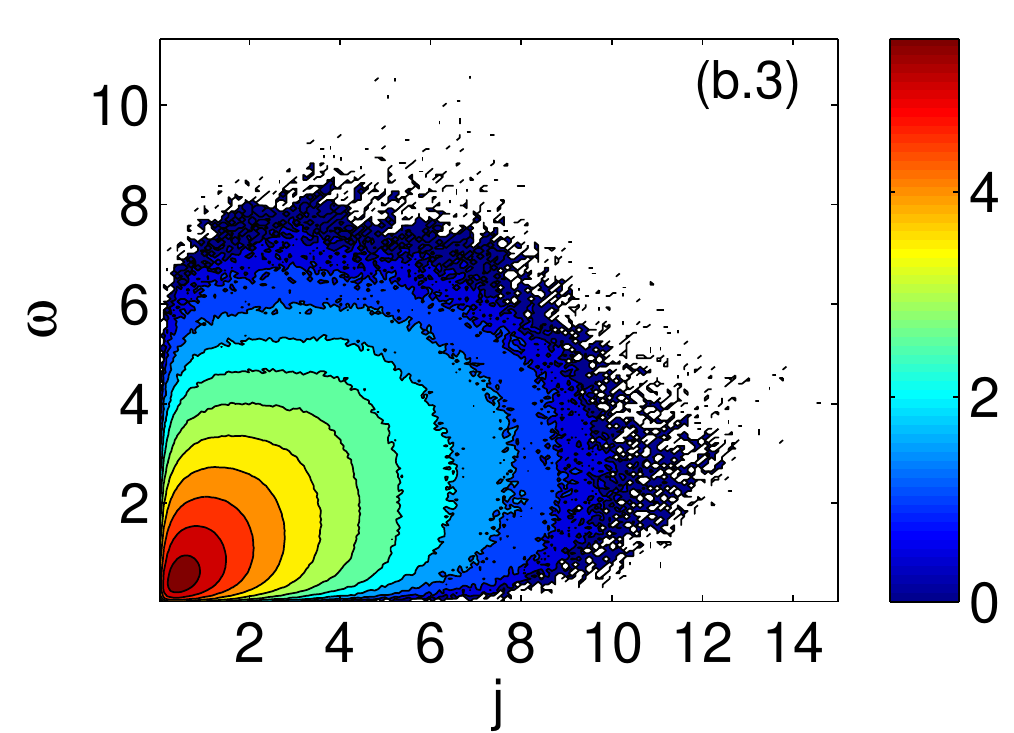}
\includegraphics[width=0.23\textwidth]{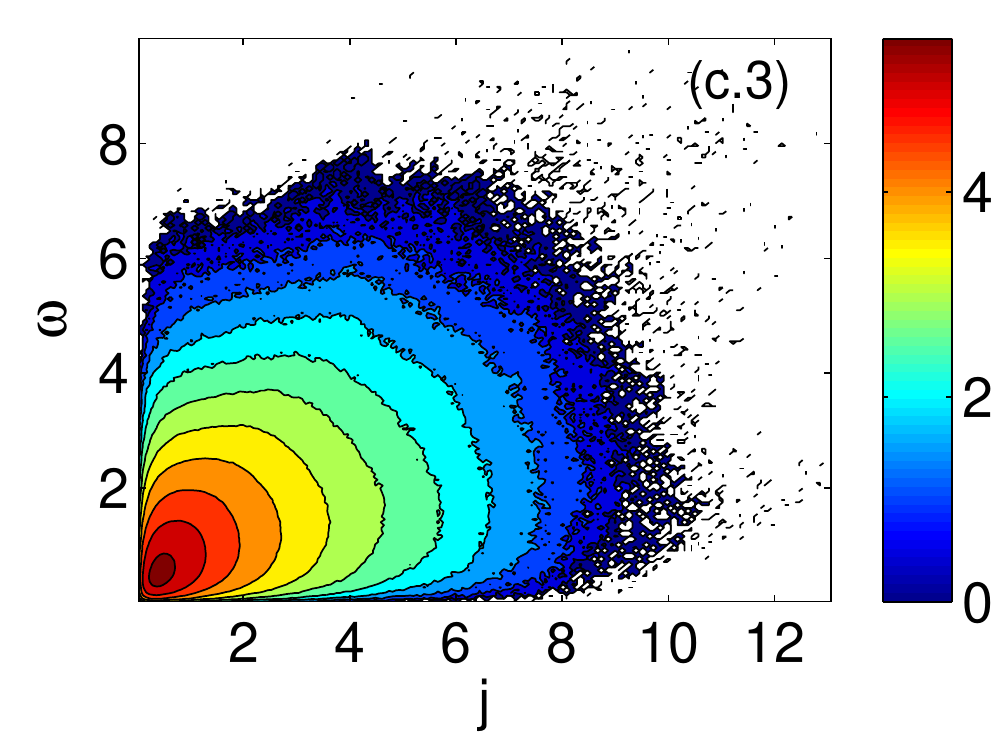}
\includegraphics[width=0.23\textwidth]{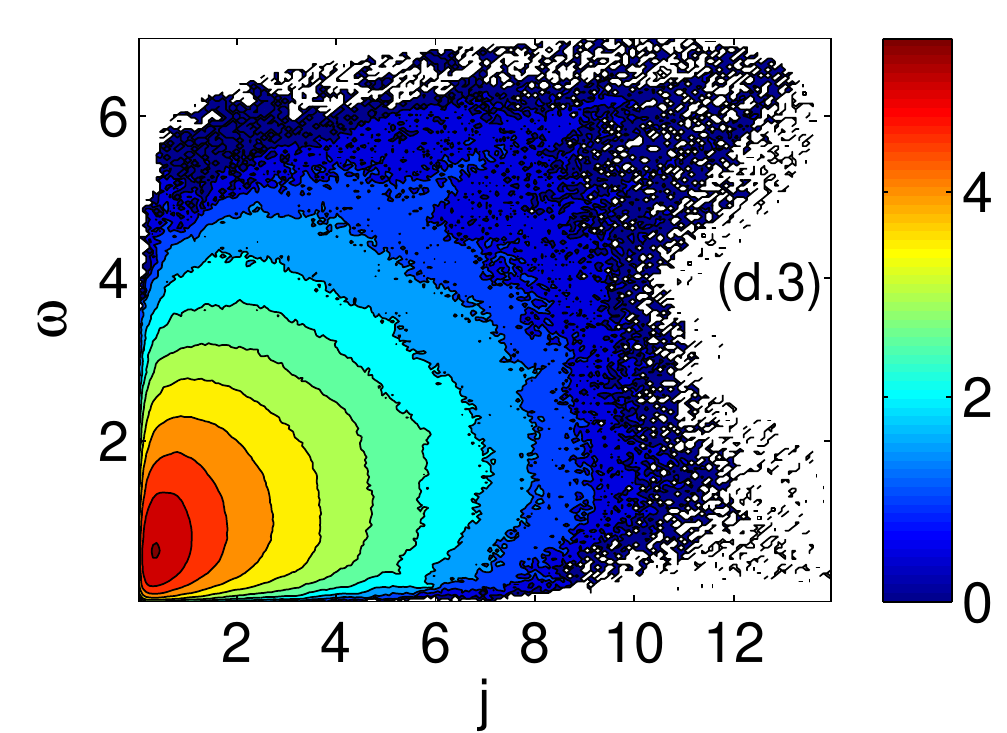}
\end{center}
\caption[]{Joint PDFs of $\omega$ and $j$ shown as filled
contour plots on a logarithmic scale for (a.1) ${\rm Pr_M}=0.1$ (R1), (b.1)
${\rm Pr_M}=0.5$ (R2), (c.1) ${\rm Pr_M}=1.0$ (R3), (d.1) ${\rm Pr_M}=5.0$
(R4), (e.1) ${\rm Pr_M}=10.0$ (R5), (f.1) ${\rm Pr_M}=1.0$ (R3B), (g.1) ${\rm
Pr_M}=5.0$ (R4B), (h.1) ${\rm Pr_M}=10.0$ (R5B), (a.2) ${\rm Pr_M}=0.01$
(R1C), (b.2) ${\rm Pr_M}=0.1$ (R2C), (c.2) ${\rm Pr_M}=1.0$ (R3C), and (d.2)
${\rm Pr_M}=10.0$ (R4C) for decaying MHD turbulence; and for statistically
steady MHD turbulence (a.3) ${\rm Pr_M}=0.01$ (R1D), (b.3) ${\rm Pr_M}=0.1$
(R2D), (c.3) ${\rm Pr_M}=1.0$ (R3D), and (d.3) ${\rm Pr_M}=10.0$ (R4D). The
arguments of the joint PDFs are normalised by their standard deviations.}
\label{fig:jpdf-wj}
\end{figure}

We now consider joint PDFs of $\omega$ and $j$ that are obtained at $t_c$ for
runs R1-R5 in Figs.~\ref{fig:jpdf-wj}(a.1)-\ref{fig:jpdf-wj}(e.1), runs
R3B-R5B in Figs.~\ref{fig:jpdf-wj}(f.1)-\ref{fig:jpdf-wj}(h.1), and runs
R1C-R4C  in Figs.~\ref{fig:jpdf-wj}(a.2)-\ref{fig:jpdf-wj}(d.2) for decaying
MHD turbulence; and for statistically steady MHD turbulence they are shown in
Figs.~\ref{fig:jpdf-wj}(a.3)-\ref{fig:jpdf-wj}(d.3) for runs R1D-R4D. All
these joint PDFs have long tails; as we move away from ${\rm Pr_M} = 1$ they
become more and more asymmetrical. Furthermore, as we expect, the tails of
these PDFs are drawn in towards small values of $\omega$ and $j$ as we
increase ${\rm Pr_M}$ [Figs.~\ref{fig:jpdf-wj}(a.1)-\ref{fig:jpdf-wj}(e.1)
for runs R1-R5, respectively] while holding $\eta$ and the initial energy
fixed. However, if we compensate for the increase in $\nu$ by increasing the
initial energy or ${\rm Re}_{\lambda}$, so that $k_{\rm max}\eta_d^u$ and $k_{\rm
max}\eta_d^b$ are both $\simeq 1$, we see that the tails of the PDFs get
elongated again.

\begin{figure}[htb]
\begin{center}
\includegraphics[width=0.23\textwidth]{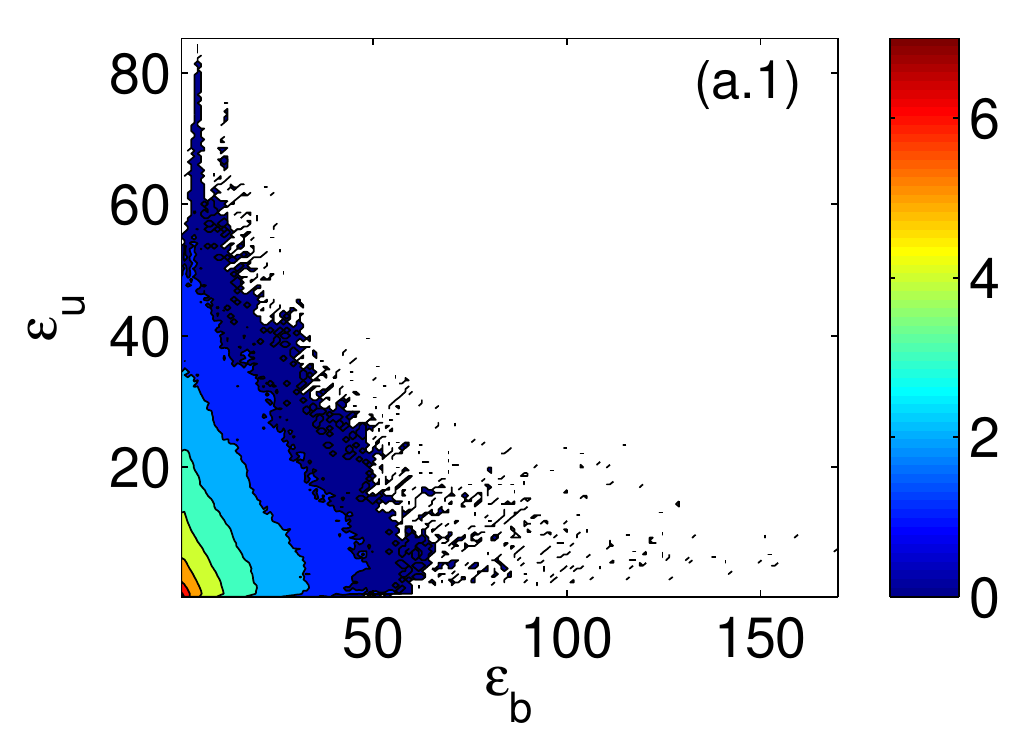}
\includegraphics[width=0.23\textwidth]{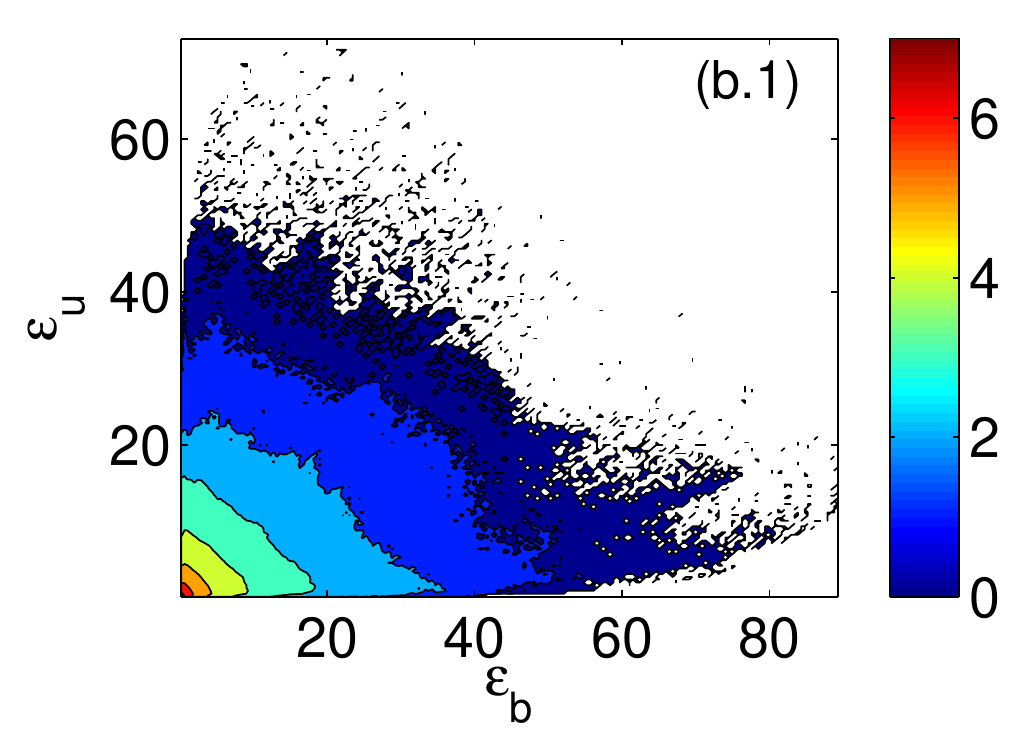}
\includegraphics[width=0.23\textwidth]{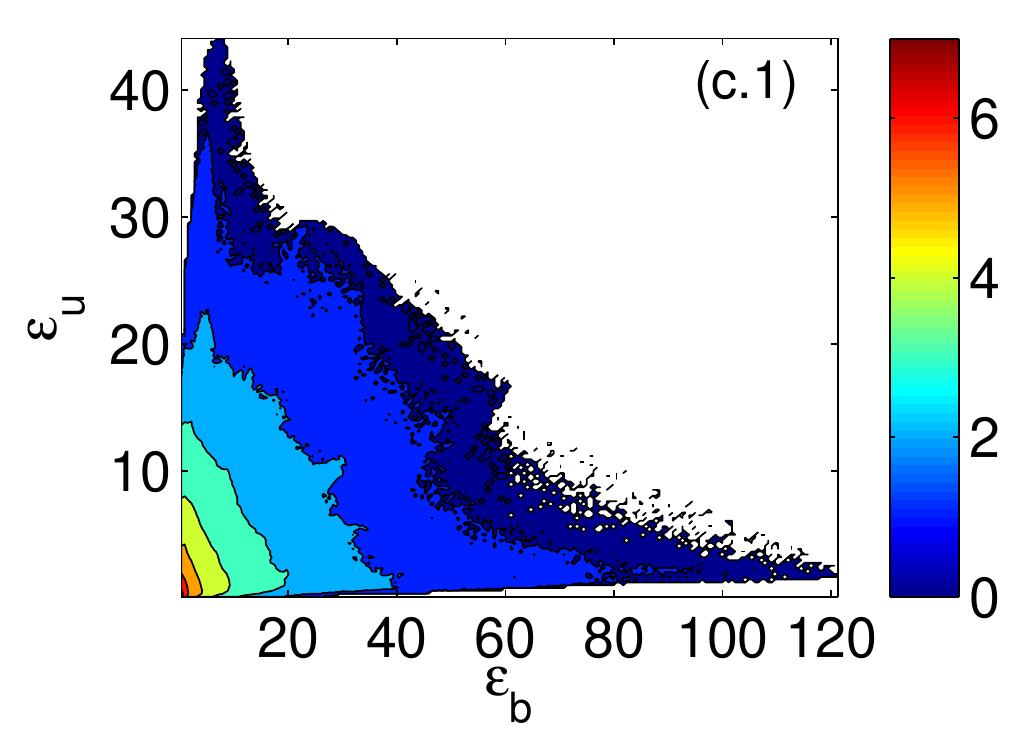}
\includegraphics[width=0.23\textwidth]{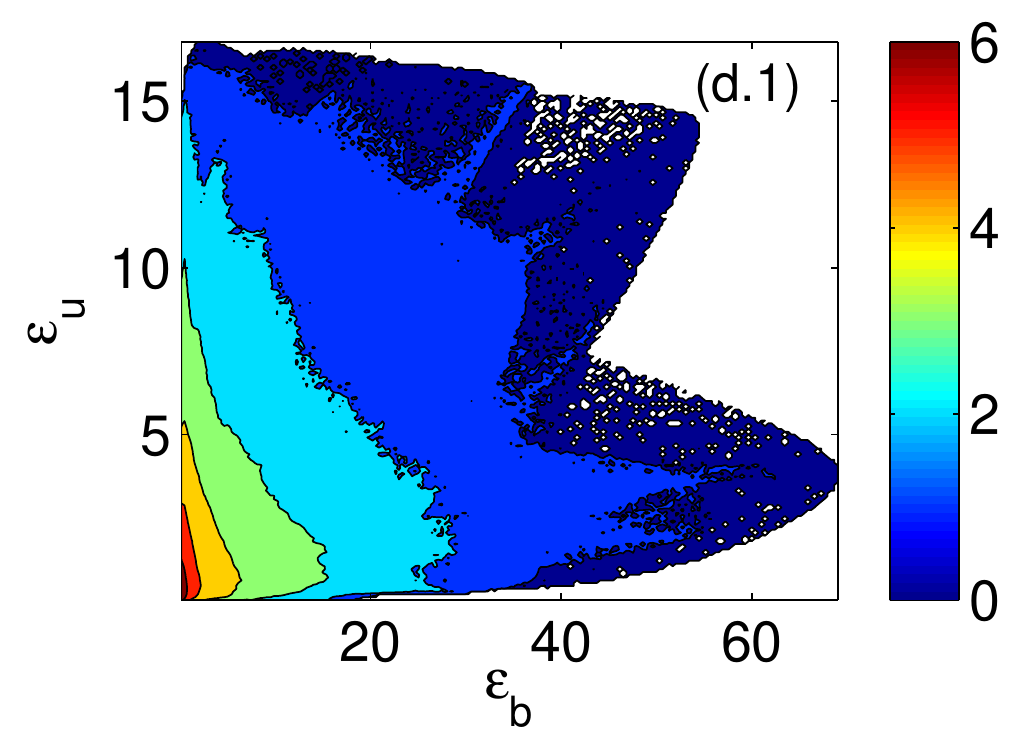}
\includegraphics[width=0.23\textwidth]{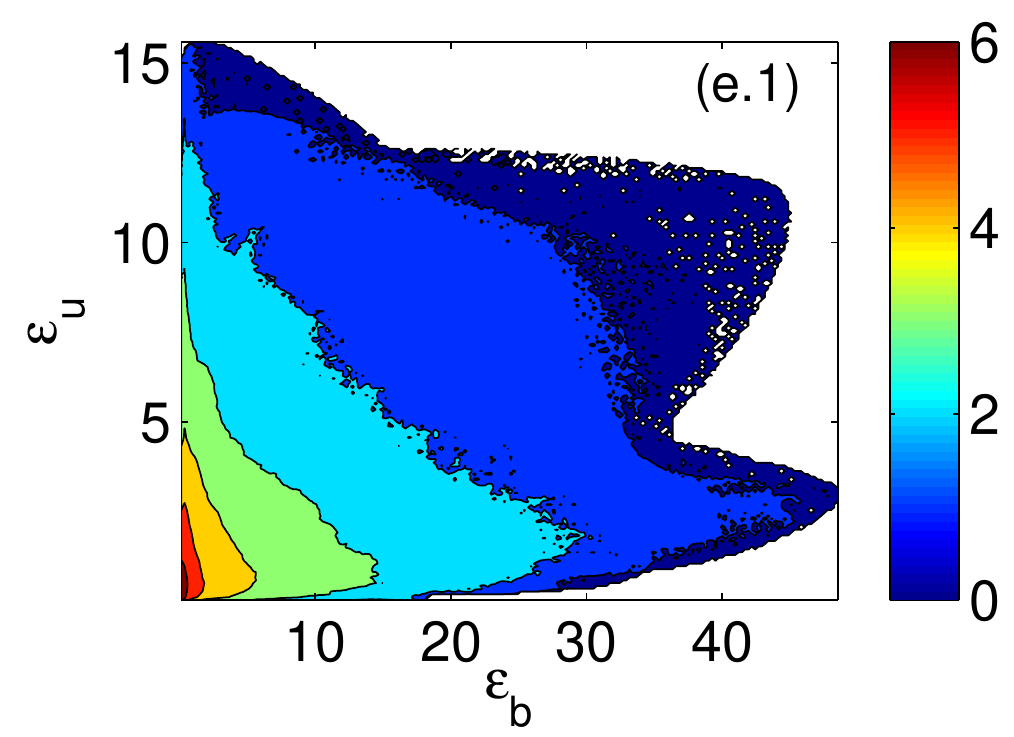}
\includegraphics[width=0.23\textwidth]{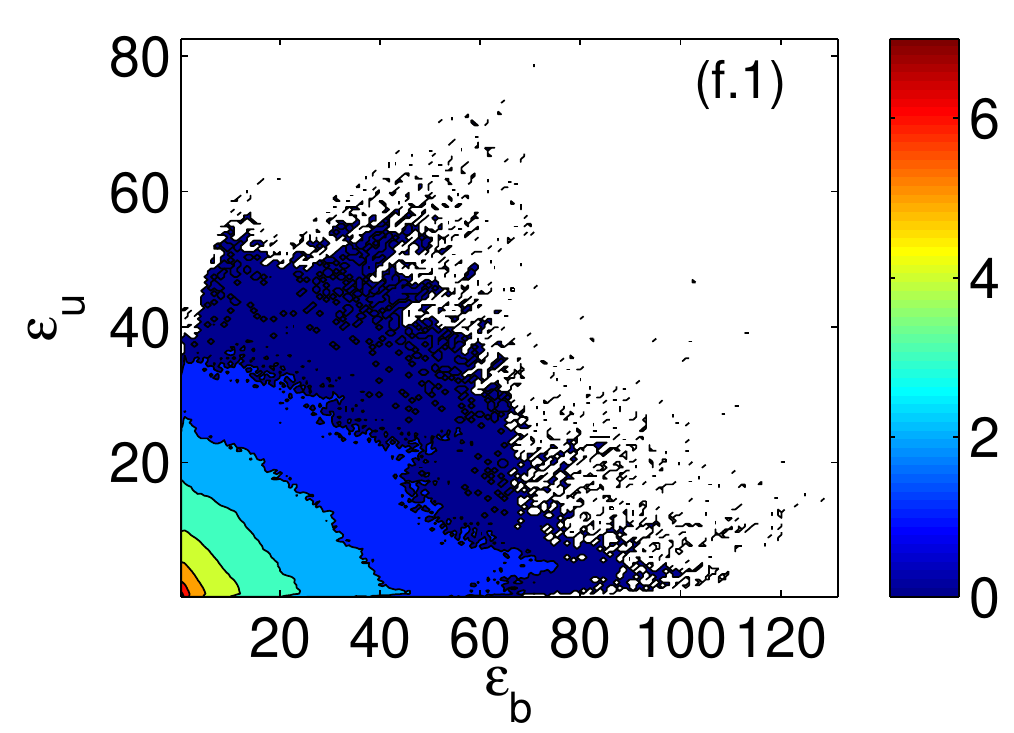}
\includegraphics[width=0.23\textwidth]{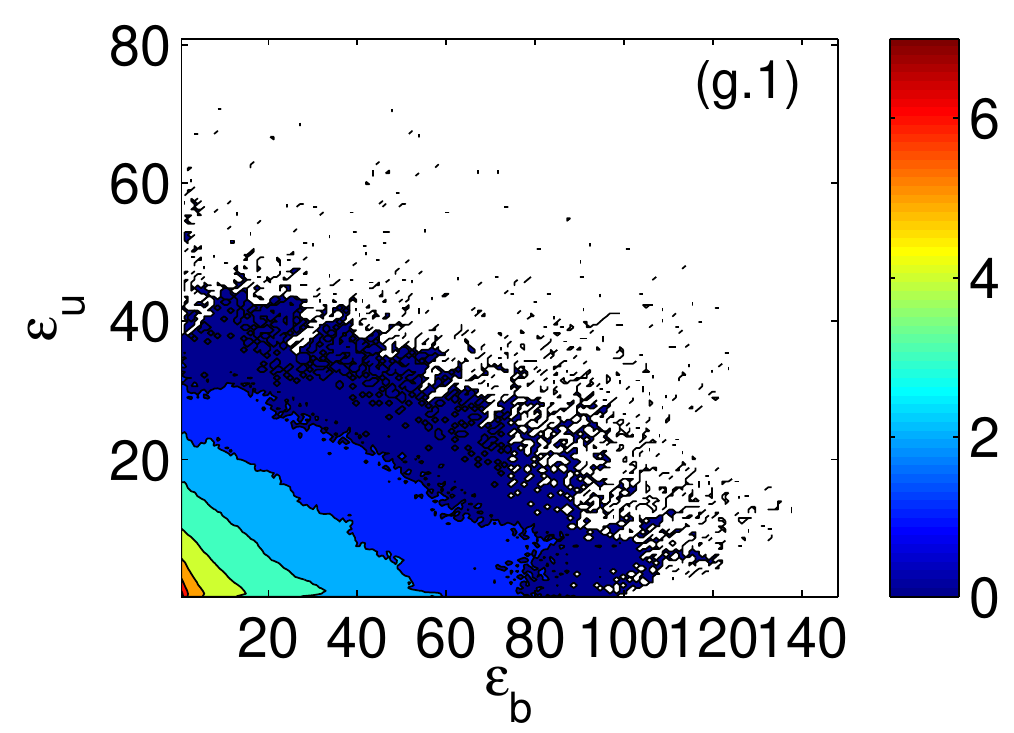}
\includegraphics[width=0.23\textwidth]{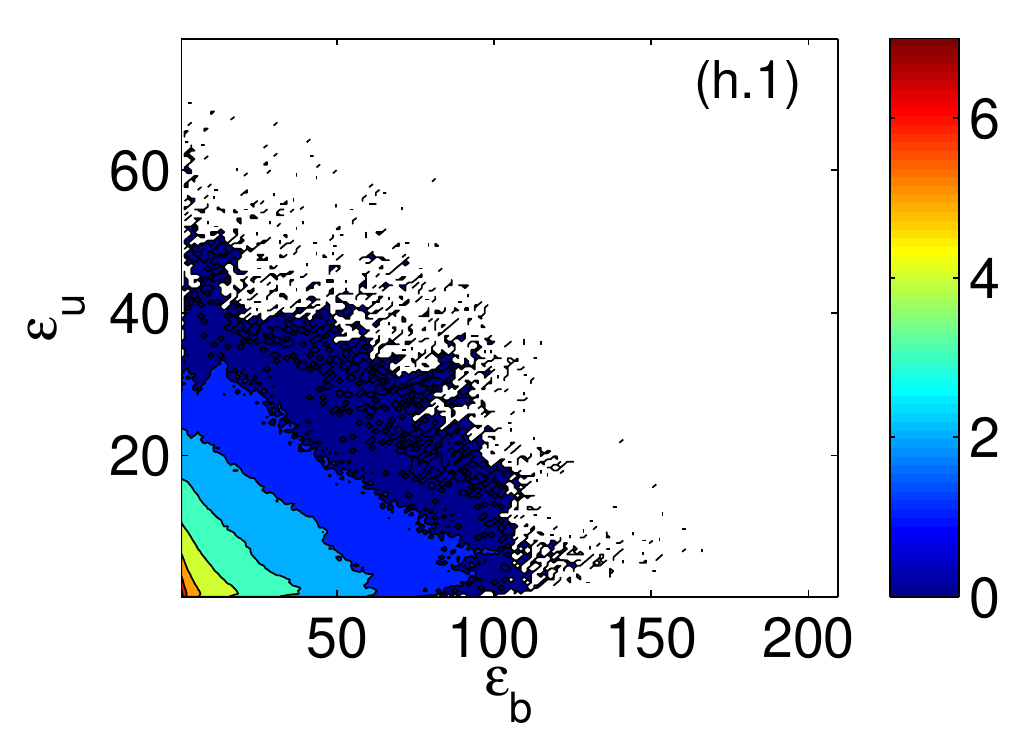}
\includegraphics[width=0.23\textwidth]{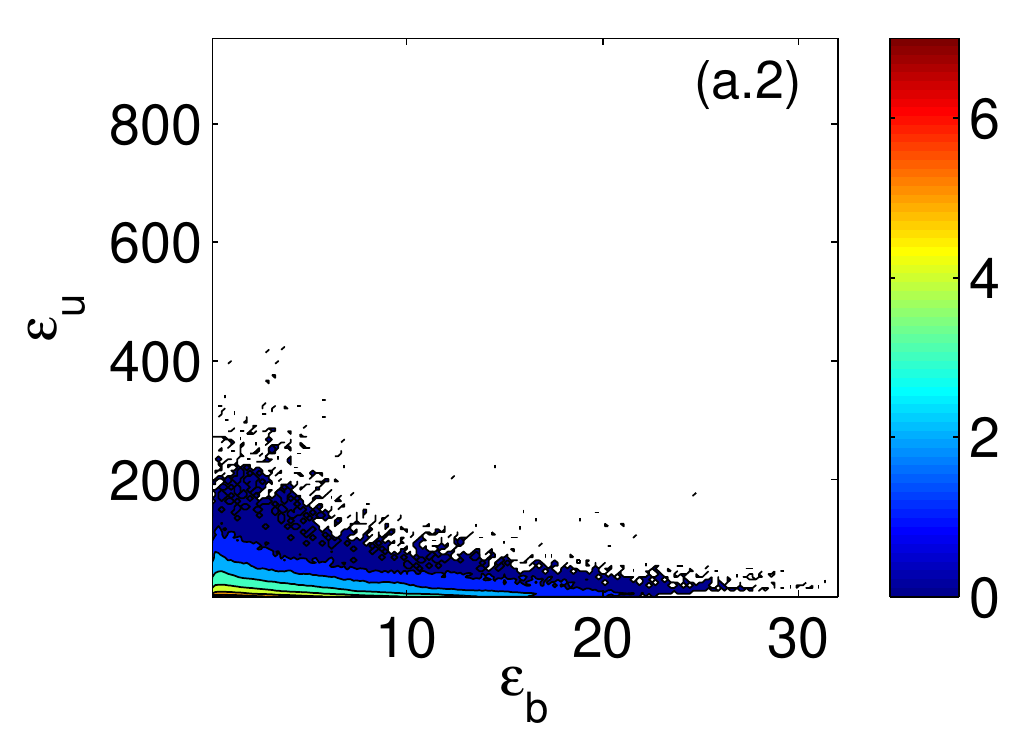}
\includegraphics[width=0.23\textwidth]{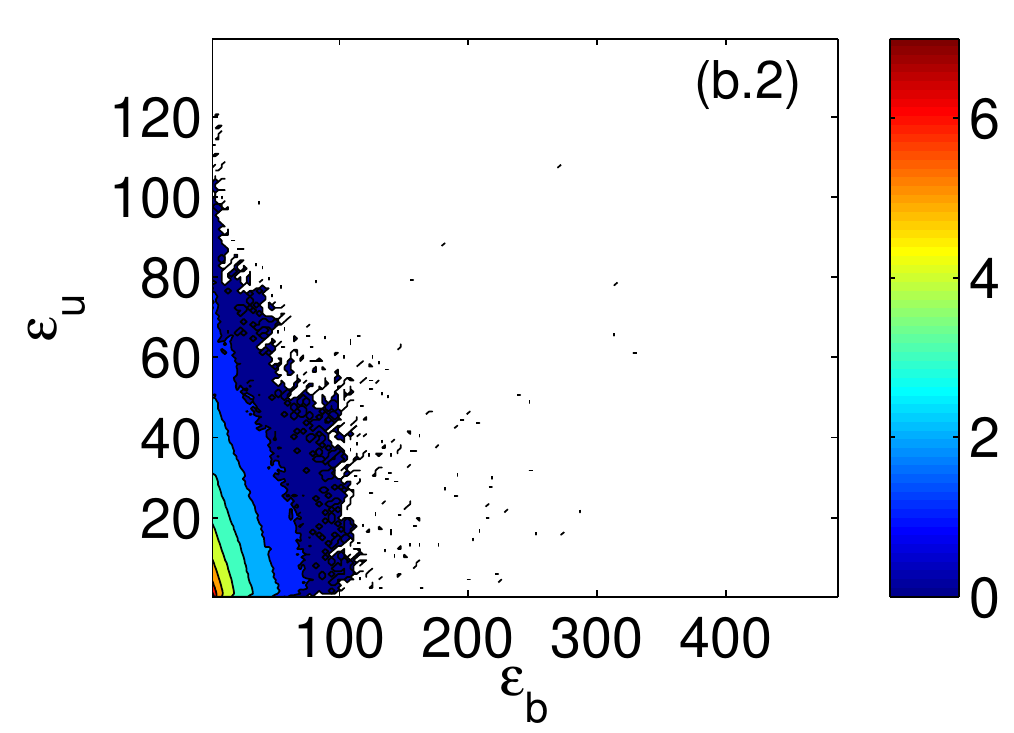}
\includegraphics[width=0.23\textwidth]{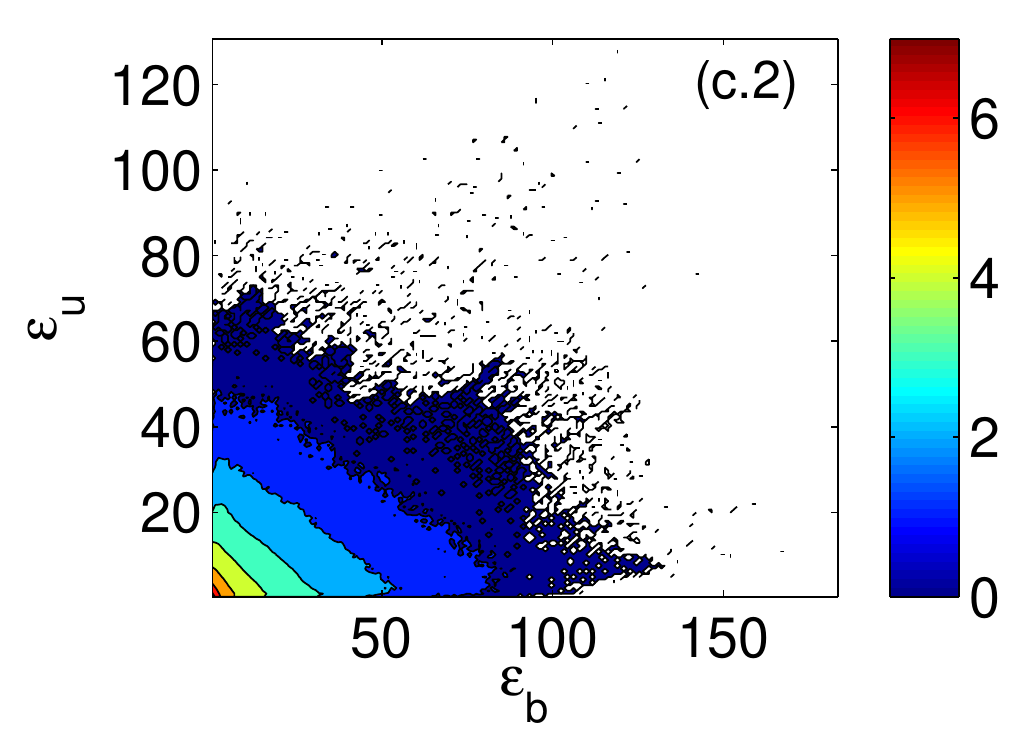}
\includegraphics[width=0.23\textwidth]{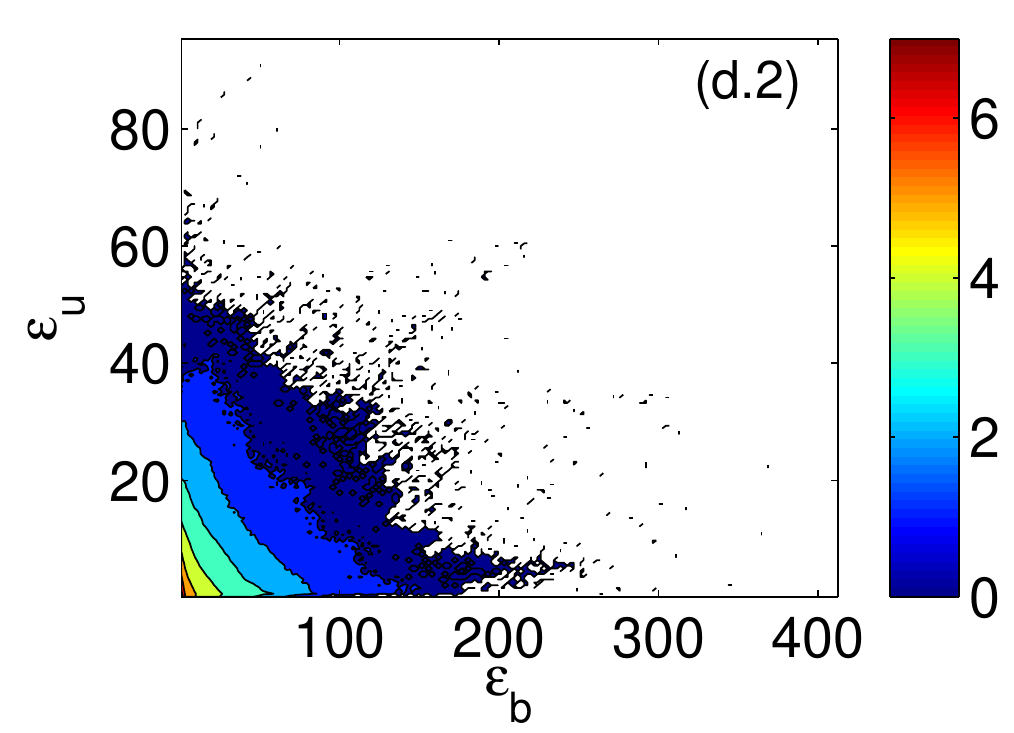}
\includegraphics[width=0.23\textwidth]{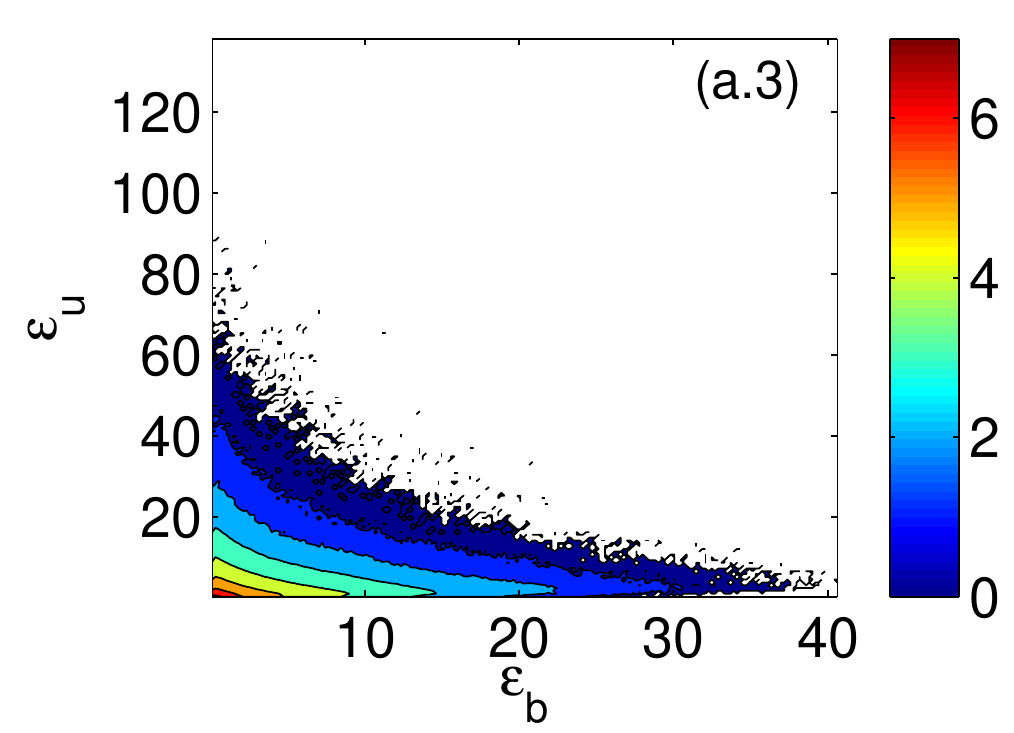}
\includegraphics[width=0.23\textwidth]{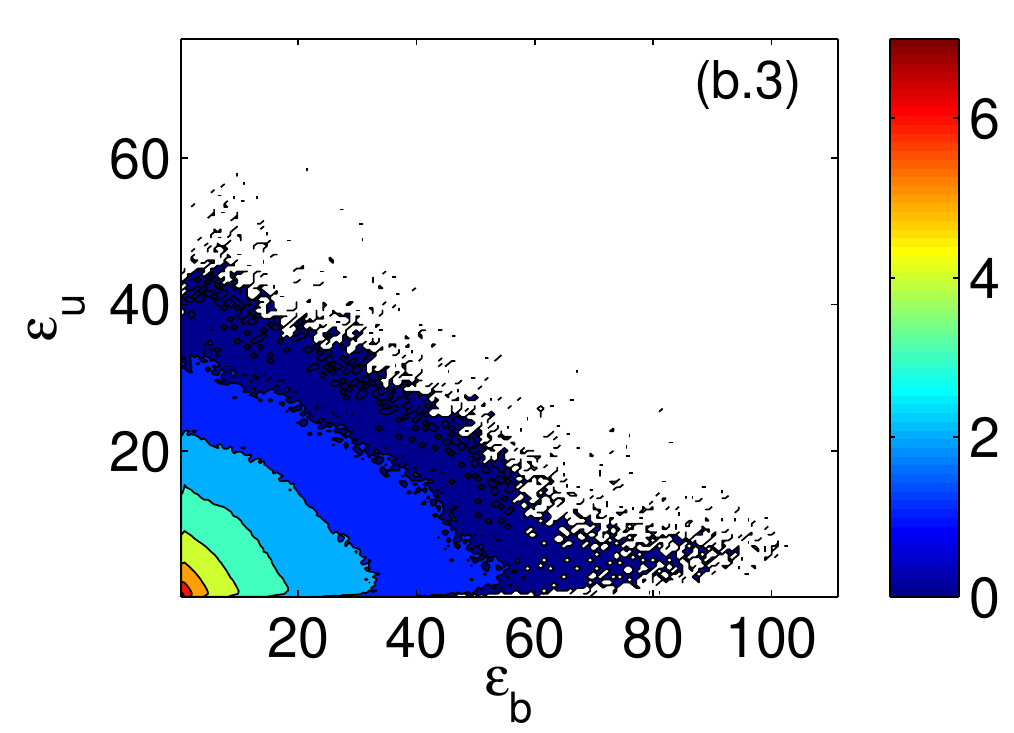}
\includegraphics[width=0.23\textwidth]{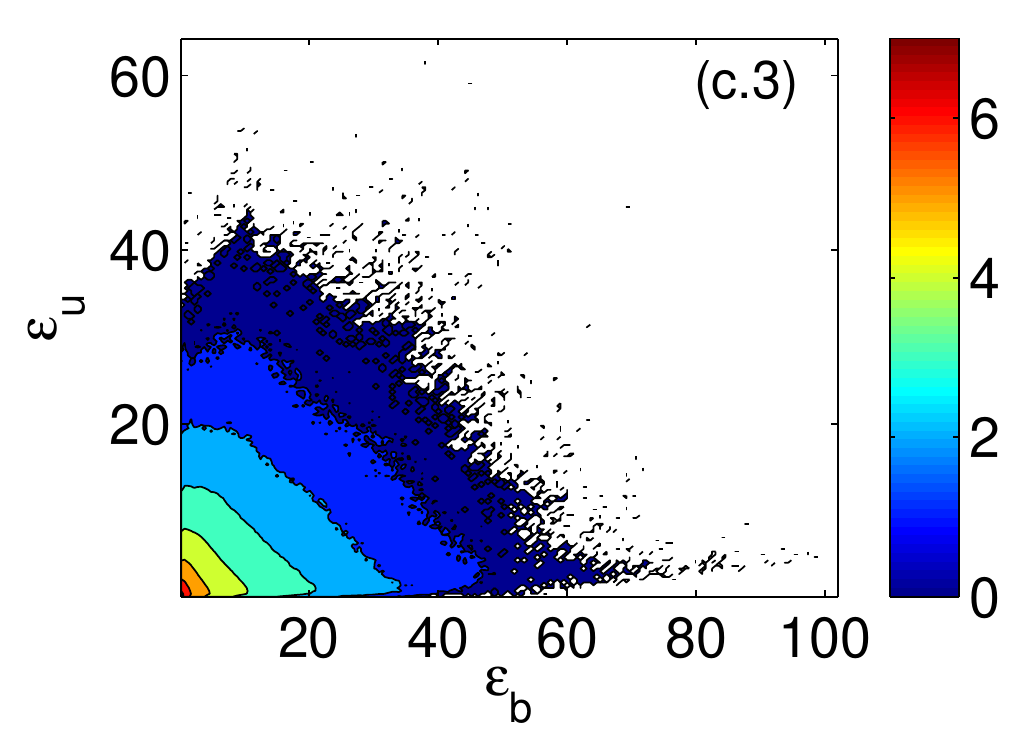}
\includegraphics[width=0.23\textwidth]{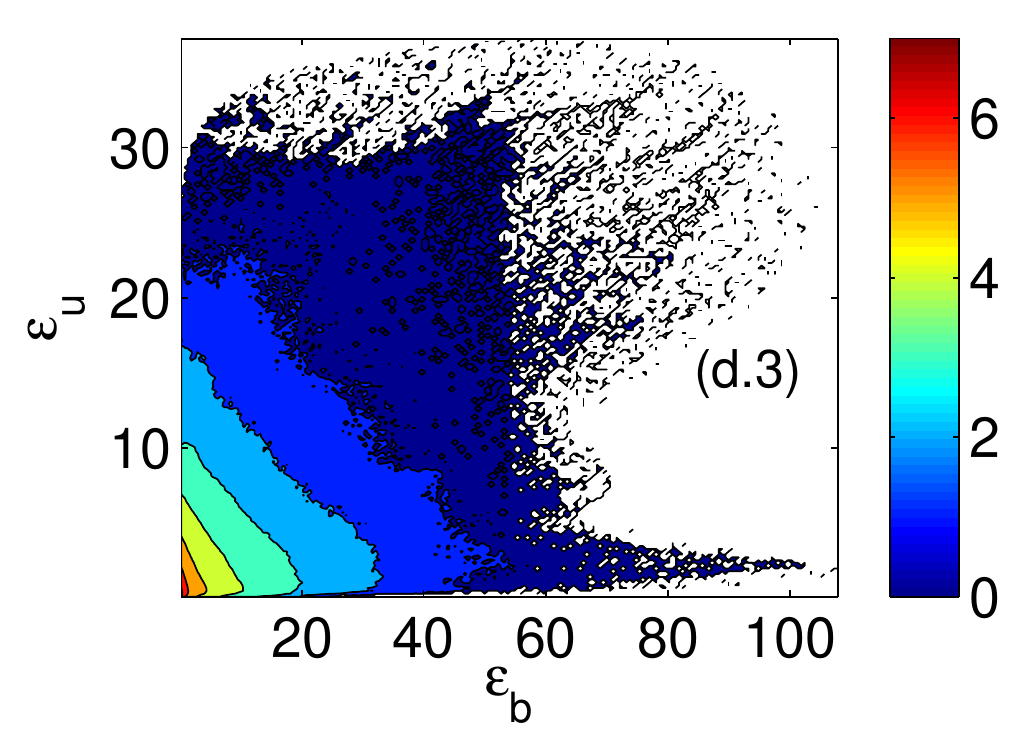}
\end{center}
\caption[]{Joint PDFs of $\epsilon_u$ and $\epsilon_b$ shown as filled
contour plots on a logarithmic scale for (a.1) ${\rm Pr_M}=0.1$ (R1), (b.1)
${\rm Pr_M}=0.5$ (R2), (c.1) ${\rm Pr_M}=1.0$ (R3), (d.1) ${\rm Pr_M}=5.0$
(R4), (e.1) ${\rm Pr_M}=10.0$ (R5), (f.1) ${\rm Pr_M}=1.0$ (R3B), (g.1) ${\rm
Pr_M}=5.0$ (R4B), (h.1) ${\rm Pr_M}=10.0$ (R5B), (a.2) ${\rm Pr_M}=0.01$
(R1C), (b.2) ${\rm Pr_M}=0.1$ (R2C), (c.2) ${\rm Pr_M}=1.0$ (R3C), and (d.2)
${\rm Pr_M}=10.0$ (R4C) for decaying MHD turbulence; and for statistically
steady MHD turbulence (a.3) ${\rm Pr_M}=0.01$ (R1D), (b.3) ${\rm Pr_M}=0.1$
(R2D), (c.3) ${\rm Pr_M}=1.0$ (R3D), and (d.3) ${\rm Pr_M}=10.0$ (R4D). The
arguments of the joint PDFs are normalised by their standard deviations.}
\label{fig:jpdf-eps}
\end{figure}

In the end we consider joint PDFs of $\epsilon_u$ and $\epsilon_b$ that are
obtained at $t_c$ for runs R1-R5 in
Figs.~\ref{fig:jpdf-eps}(a.1)-\ref{fig:jpdf-eps}(e.1), runs R3B-R5B in
Figs.~\ref{fig:jpdf-eps}(f.1)-\ref{fig:jpdf-eps}(h.1), and runs R1C-R4C in
Figs.~\ref{fig:jpdf-eps}(a.2)-\ref{fig:jpdf-eps}(d.2) for decaying MHD
turbulence; and for statistically steady MHD turbulence they are shown in
Figs.~\ref{fig:jpdf-eps}(a.3)-\ref{fig:jpdf-eps}(d.3) for runs R1D-R4D.  The
trends here are similar to the ones discussed in the previous paragraph.  In
particular, these joint PDFs have long tails; as we move away from ${\rm
Pr_M} = 1$ they become more and more asymmetrical; and the tails of these
PDFs are drawn in towards small values of $\epsilon_u$ and $\epsilon_b$ as we
increase ${\rm Pr_M}$ [Figs.~\ref{fig:jpdf-eps}(a.1)-\ref{fig:jpdf-eps}(e.1)
for runs R1-R5, respectively] while holding $\eta$ and the initial energy
fixed.  But, if we make up for the increase in $\nu$ by increasing the
initial energy or ${\rm Re}_{\lambda}$ so that $k_{\rm max}\eta_d^u$ and $k_{\rm
max}\eta_d^b$ are both $\simeq 1$, we see that the tails of the PDFs get
elongated again.

\section{Discussions and Conclusion\label{sec:conclusions}}

We have carried out an extensive study of the statistical properties of both
decaying and statistically steady homogeneous, isotropic MHD turbulence.  Our
study, which has been designed specifically to study the systematics of the
dependence of these properties on the magnetic Prandtl number ${\rm Pr_M}$,
uses a large number of statistical measures to characterise the statistical
properties of both decaying and statistically steady MHD turbulence.  Our
study is restricted to incompressible MHD turbulence;  we do not include a
mean magnetic field as, e.g., in Refs.~\cite{goldreich95}; furthermore we do
not study Lagrangian properties considered, e.g., in Ref.~\cite{homann07}. In
our studies we obtain (a) various PDFs, such as those of the moduli of the
vorticity and current density, the energy dissipation rates, of cosines of
angles between various vectors, and scale-dependent velocity and
magnetic-field increments, (b) spectra, e.g., those of the energy and the
effective pressure, (c) velocity and magnetic-field structure functions that
can be used to characterise intermittency, (d) isosurfaces
of quantities such as the moduli of the vorticity and current, and (e) joint
PDFs such as $QR$ plots.  The evolution of these properties with ${\rm
Pr_M}$ has been described in detail in the previous Section. 

To the best of our knowledge, such a comprehensive study of the ${\rm
Pr_M}-$dependence of incompressible, homogeneous, isotropic MHD turbulence,
both decaying and statistically steady, has not been attempted before.
Studies that draw their inspiration from astrophysics often consider
anisotropic flows~\cite{montgomery81,veltri82,kinney97,galtier05,chandran08,
shebalin83,brandenburg07, bigot08}, flows that are
compressible~\cite{brandenburg95,christensson01}, or flows that include a mean magnetic
field~\cite{goldreich95,shebalin83,alemany79,oughton94}. Yet other studies
concentrate on the alignment between various vectors such as $\bfu$ and
$\bfb$ as, e.g., in Refs.~\cite{mason08,brandenburg95,matthaeus08}; some of these include
a few, but not all, of the PDFs we have studied; and, typically, these studies
are restricted to the case ${\rm Pr_M}=1$.  Some of the spectra we study have
been obtained in earlier DNS studies but, typically, only for the case ${\rm
Pr_M}=1$; a notable exception is Ref.~\cite{chou01}, which examines the ${\rm
Pr_M}-$dependence of energy spectra but with a relatively low resolution.
References~\cite{schekochihin02,brandenburg09,iskakov07} have also considered
some ${\rm Pr_M}-$dependence but not for low ${\rm Pr_M}$.  Isosurfaces of
the moduli of the vorticity and current density have been obtained
earlier~\cite{biskamp00,mininni07,yoshimatsu09} for the case ${\rm Pr_M}=1$.
The ${\rm Pr_M}$ dependence of these and other isosurfaces is presented here
for the first time. The joint PDFs we have shown above have also not been
investigated before.

Here we wish to highlight, and examine in detail, the implications of our
study for intermittency. Some earlier DNS studies, such as
Refs.~\cite{mininni09}, had noted that, for the case ${\rm Pr_M}=1$,
the magnetic field is more intermittent than the velocity field. This is why
we have concentrated on velocity and magnetic-field structure functions. Our
study confirms this finding, for the case ${\rm Pr_M}=1$. This can be seen
clearly from the comparison of our exponent ratios, for ${\rm Pr_M}=1$, with
those of the recent DNS of decaying-MHD-turbulence in Ref.~\cite{mininni09}
in Table~\ref{table:compare}; the errorbars that we quote for our exponent
ratios have been calculated as described in the previous Section; we have
obtained exponent ratios for Ref.~\cite{mininni09} by
digitising~\cite{g3data} the data in their plot [Fig.~3 of
Ref.~\cite{mininni09}] of multiscaling exponents versus the order $p$ (error
bars are not given in their plot).  Thus, at least given our errorbars,
there is agreement between our exponent ratios, both for decaying and
statistically steady MHD turbulence, and those of Ref.~\cite{mininni09} for
${\rm Pr_M}=1$. We note in passing that the latter DNS is one of decaying MHD
turbulence but with a very special initial condition, which allows an
effective resolution greater than that we have obtained; however, the initial
condition we use in our decaying-MHD-turbulence DNS is more generic than that
of Ref.~\cite{mininni09}. It is our expectation that nonuniversal effects,
associated with different initial conditions~\cite{kalelkar04,lee10}, might
not affect multiscaling exponent ratios, except if we use nongeneric,
power-law initial conditions~\cite{kalelkar04} in which $E(k)$ grows with $k$
(at least until some large-$k$ cutoff). 

Direct numerical simulations of decaying-MHD-turbulence, e.g., those of
Refs.~\cite{biskamp00,mininni09}, often average data obtained from field
configurations at different times that are close to the time at which the peak
appears in plots of the energy-dissipation rate. This is a reasonable
procedure, for ${\rm Pr_M}=1$, because the temporal evolution of the system
is slow in the vicinity of this peak. We have not adopted this procedure here
because, as we move away from ${\rm Pr_M}=1$, the cascade-completion peaks
occur at different times in plots of $\epsilon_u$ and $\epsilon_b$ as we have
discussed in detail in earlier sections of this paper. 

%
%

Let us now turn to the ${\rm Pr_M}-$dependence of the multiscaling exponent
ratios shown in Tables~\ref{table:zetap-decaying} and
\ref{table:zetap-forced} and in
Figs.\ref{fig:stfn-1024}(a.3)-\ref{fig:stfn-1024}(d.3) and
\ref{fig:stfn-forced}(a.3)-\ref{fig:stfn-forced}(d.3). Even though our error
bars are large, given the conservative, local-slope error analysis we have
described in the previous Section, a trend emerges: at large values
of ${\rm Pr_M}$ the magnetic field is clearly more intermittent than the
velocity field, in as much as the deviations of $\zeta_p^b$ from
the simple-scaling prediction are stronger than their counterparts
for $\zeta_p^u$.
However, the velocity field becomes more intermittent than the magnetic field
as we lower ${\rm Pr_M}$.  Could this result, namely, the dependence of our
multiscaling exponent ratios on ${\rm Pr_M}$, be an artifact? We
believe not.  As we have discussed above, dissipation ranges in our spectra
are adequately resolved; furthermore, we have determined exponent ratios from
a a rather stringent local-slope analysis, which is rarely presented in
earlier DNS studies of MHD turbulence. Ultimately, of course, this ${\rm
Pr_M}$ dependence of multiscaling exponents in MHD turbulence must be tested
in detail in very-high-resolution DNS studies of MHD turbulence; such studies
should become possible with the next generation of supercomputers.

It is useful to note at this stage that a recent experimental study of MHD
turbulence in the solar wind~\cite{salem09} provides evidence for velocity
fields that are more strongly intermittent than the magnetic field; this
study does not give the value of ${\rm Pr_M}$.  However, their data for
multiscaling exponents are qualitatively similar to those we obtain at low
values of ${\rm Pr_M}$. Furthermore, PDFs of $H_C$ have also been obtained from solar-wind
data~\cite{podesta07}; these are similar to the PDFs we obtain for $H_C$.
Of course, we must exercise caution in comparing
results from DNS studies of homogeneous, isotropic, incompressible MHD
turbulence with measurements on the solar wind where anisotropy and
compressibility can be significant; and, for the solar wind, we might
also have to consider kinetic effects that are not captured by the MHD
equations. 

The last point we wish to address is the issue of strong universality of
exponent ratios. In the fluid-turbulence context such strong
universality~\cite{lvov03,ray08} implies the equality of exponents
(and, therefore, their ratios) determined from decaying-turbulence studies
(say at the cascade-completion time) or from studies of statistically steady
turbulence. Does such strong universality have an analogue in MHD turbulence?
Our data, for any fixed value of ${\rm Pr_M}$ in
Tables~\ref{table:zetap-decaying} and \ref{table:zetap-forced}, are
consistent with such strong universality of multiscaling exponent ratios in
MHD turbulence; but, of course, our large errorbars imply that a definitive
confirmation of such strong universality in MHD turbulence must await DNS
studies that might become possible in the next generation of
high-performance-computing facilities.

\section*{Acknowledgements}

We thank C. Kalelkar, V. Krishan, S. Ramaswamy, D. Mitra, and
S.S. Ray for discussions, SERC(IISc) for computational resources
and DST, UGC and CSIR India for support. Two of us (PP and RP)
are members of the International Collaboration for Turbulence
Research (ICTR). RP thanks the Observatoire de la C\^ote d'Azur
for their hospitality while the last parts of this paper were
written. GS thanks the JNCASR for support.

\section*{References}

\end{document}